\documentclass[traditabstract]{aa}  

\usepackage{amsmath}
\usepackage{fixltx2e}
\usepackage[english]{babel}
\usepackage{graphicx}
\usepackage{epstopdf}
\usepackage{epsf,color}
\usepackage[mathscr]{eucal}
\usepackage{amsmath} 
\usepackage{amssymb,amsfonts}
\usepackage{natbib}
\usepackage{graphicx}
\usepackage{txfonts}
\usepackage{dsfont}
\definecolor{Mygreen}{rgb}{0.0, 0.80, 0.0}
\definecolor{Mypink}{rgb}{1.0, 0.0, 0.5}
\definecolor{Myred}{rgb}{0.7, 0.0, 0.0}
\usepackage[breaklinks, citecolor=blue, linkcolor=Myred, urlcolor=Myred, colorlinks=true, debug, baseurl=' ']{hyperref}
\usepackage{float} 
\usepackage{color} 
\usepackage{ulem}
\usepackage{subcaption}

\bibpunct{(}{)}{;}{a}{}{,}
\begin{document}

\title{Gamma-ray detection toward the Coma cluster with \textit{Fermi}-LAT: \\
Implications for the cosmic ray content in the hadronic scenario}

\author{R.~Adam \inst{\ref{LLR}}\thanks{Corresponding author: R\'emi Adam, \url{remi.adam@llr.in2p3.fr}}
\and H.~Goksu \inst{\ref{LLR}}
\and S.~Brown \inst{\ref{Iowa}}
\and L.~Rudnick \inst{\ref{Minesota}}
\and C.~Ferrari \inst{\ref{OCA}}}

\institute{
Laboratoire Leprince-Ringuet (LLR), CNRS, \'Ecole polytechnique, Institut Polytechnique de Paris, 91120 Palaiseau, France
\label{LLR}
\and
Department of Physics \& Astronomy, The University of Iowa, Iowa City, IA, 52245
\label{Iowa}
\and
Minnesota Institute for Astrophysics, University of Minnesota,  Minneapolis, MN 55455
\label{Minesota}
\and
Universit\'e C\^ote d'Azur, Observatoire de la C\^ote d'Azur, CNRS, Laboratoire Lagrange, France
\label{OCA}
}

\date{Received \today \ / Accepted --}
\abstract {
The presence of relativistic electrons within the diffuse gas phase of galaxy clusters is now well established, thanks to deep radio observations obtained over the last decade, but their detailed origin remains unclear. Cosmic ray protons are also expected to accumulate during the formation of clusters. They may explain part of the radio signal and would lead to $\gamma$-ray emission through hadronic interactions within the thermal gas. Recently, the detection of $\gamma$-ray emission has been reported toward the Coma cluster with \textit{Fermi}-LAT.
Assuming that this $\gamma$-ray emission arises essentially from pion decay produced in proton-proton collisions within the intracluster medium (ICM), we aim at exploring the implication of this signal on the cosmic ray content of the Coma cluster and comparing it to observations at other wavelengths.
We use the {\tt MINOT} software to build a physical model of the Coma cluster, which includes the thermal target gas, the magnetic field strength, and the cosmic rays, to compute the corresponding expected $\gamma$-ray signal. We apply this model to the \textit{Fermi}-LAT data using a binned likelihood approach, together with constraints from X-ray and Sunyaev-Zel'dovich observations. We also consider contamination from compact sources and the impact of various systematic effects on the results.
We confirm that a significant $\gamma$-ray signal is observed within the characteristic radius $\theta_{500}$ of the Coma cluster, with a test statistic ${\rm TS} \simeq 27$ for our baseline model. The presence of a possible point source (4FGL~J1256.9+2736) may account for most of the observed signal. However, this source could also correspond to the peak of the diffuse emission of the cluster itself as it is strongly degenerate with the expected ICM emission, and extended models match the data better. Given the \textit{Fermi}-LAT angular resolution and the faintness of the signal, it is not possible to strongly constrain the shape of the cosmic ray proton spatial distribution when assuming an ICM origin of the signal, but preference is found in a relatively flat distribution elongated toward the southwest, which, based on data at other wavelengths, matches the spatial distribution of the other cluster components well. Assuming that the whole $\gamma$-ray signal is associated with hadronic interactions in the ICM, we constrain the cosmic ray to thermal energy ratio within $R_{500}$ to $X_{\rm CRp} = 1.79^{+1.11}_{-0.30}$\% and the slope of the energy spectrum of cosmic rays to $\alpha = 2.80^{+0.67}_{-0.13}$ ($X_{\rm CRp} = 1.06^{+0.96}_{-0.22}$\% and $\alpha = 2.58^{+1.12}_{-0.09}$ when including both the cluster and 4FGL~J1256.9+2736 in our model). Finally, we compute the synchrotron emission associated with the secondary electrons produced in hadronic interactions assuming steady state. This emission is about four times lower than the overall observed radio signal (six times lower when including 4FGL J1256.9+2736), so that primary cosmic ray electrons or reacceleration of secondary electrons is necessary to explain the total emission. We constrain the amplitude of the primary to secondary electrons, or the required boost from reacceleration with respect to the steady state hadronic case, depending on the scenario, as a function of radius.
Our results confirm that $\gamma$-ray emission is detected in the direction of the Coma cluster. Assuming that the emission is due to hadronic interactions in the intracluster gas, they provide the first quantitative measurement of the cosmic ray proton content in a galaxy cluster and its implication for the cosmic ray electron populations.}
\titlerunning{$\gamma$-ray emission toward Coma}
\authorrunning{R. Adam et al.}
\maketitle

\section{Introduction}\label{sec:Introduction}
Galaxy clusters form hierarchically via the smooth accretion of surrounding material and the merging of subclusters and groups \citep{Kravtsov2012}. During these processes, most of the binding gravitational energy is dissipated into the hot, thermal, ionized gas phase. However, shock waves propagating in the intracluster medium (ICM) and turbulence are also expected to accelerate both electrons and protons at relativistic energies, leading to a nonthermal population of cosmic rays (CR) that are confined within the cluster magnetic fields \citep[see][for a review]{Brunetti2014}. In addition, CR may also arise from direct injection in the ICM, trough active galactic nucleus (AGN) outbursts \citep[e.g.,][]{Bonafede2014}, or galactic winds associated with star formation activity in cluster member galaxies \citep[e.g.,][]{Rephaeli2016}. While high energy cosmic ray electrons (CRe) should quickly lose their energy \citep[$\sim 10^8$ years at 10 GeV; see e.g.,][]{Sarazin1999}, cosmic ray protons (CRp) should accumulate over the cluster formation history \citep[see, e.g.,][for the simulation of CR in clusters]{Pfrommer2007}. Nevertheless, the CR physics and content of galaxy clusters remain largely unknown to date.

In fact, the presence of CRe and magnetic fields in galaxy clusters is now well established, thanks to the deep observations of diffuse radio synchrotron emission obtained over the last decade \citep[see][for a review]{vanWeeren2019}. We can distinguish between radio relics, believed to trace merger shock acceleration at the periphery of clusters \citep[e.g.,][]{vanWeeren2010}, and radio halos, which are spatially coincident with the thermal gas in massive clusters. Radio halos are generally further divided into mini-halos, associated with relaxed cool-core clusters \citep{Giacintucci2017}, and giant radio halos, which are megaparsec-size sources found in merging clusters \citep{Cassano2010}, although the distinction is not necessarily clear \citep{Ferrari2011,Bonafede2014b,Savini2019}. Two main mechanisms have been proposed in the literature to explain the origin of CRe that generate radio halos: the hadronic model \citep{Dennison1980,Blasi1999,Dolag2000}, in which secondary CRe are the products of pion decay generated in collisions between the CRp and the thermal gas, and the turbulent reacceleration of a seed population of electrons \citep[possibly secondary CRe;][]{Brunetti2007,Brunetti2011}. In both scenarios, CR might show up as a $\gamma$-ray signal due to the inverse Compton interaction of CRe with background light. In the hadronic model, the decays of pions should also lead to additional $\gamma$-ray (and neutrino) emission, which is expected to be the dominant component of the $\gamma$-ray flux at energies above $\sim 100$ MeV \citep{Pinzke2010}.

Attempts to detect the $\gamma$-ray emission from galaxy clusters have been carried out over the last two decades using individual targets, stacking analysis, and cross correlations to external datasets \citep[see, e.g.,][]{Reimer2003,Ackermann2010,Aleksic2010,Dutson2013,Huber2013,Griffin2014,Branchini2017,Colavincenzo2020}. Despite the lack of detection, these searches proved extremely useful in constraining the CRp content of clusters to below a few percent relative to the thermal pressure \citep{Ackermann2014}.

Among the relevant individual targets, the Coma cluster has been one of the most promising sources to search for $\gamma$-ray emission. Indeed, it is a massive ($M_{500} \simeq 7 \times 10^{14}$ M$_{\odot}$) and nearby system (redshift $z=0.023$, about $100$ Mpc); as such its $\gamma$-ray flux is expected to be large \citep{Pinzke2010}. Additionally, it is located near the galactic north pole and thus benefits from a low galactic background. The signal is expected to be extended even for the \textit{Fermi}-LAT (characteristic radius $\theta_{500} \sim 0.75$ deg), but not so extended that the analysis suffer strong systematic effects in the background modeling. The region around Coma is not affected by the presence of bright $\gamma$-ray compact sources. Finally, the Coma cluster is a well-known ongoing merger, which has been extensively observed at various wavelengths using various probes \citep[e.g.,][]{Kent1982,Briel1992,Gavazzi2009,Bonafede2010,PlanckX2013,Mirakhor2020}. In particular, a well-measured giant radio halo and a radio relic have been observed, proving the presence of CRe \citep[e.g.,][]{Willson1970,Giovannini1993,Thierbach2003,Brown2011,Bonafede2020}.

Several analyses have focussed on the search for $\gamma$-rays toward the Coma cluster, using both ground-based and space-based instruments \citep[e.g.,][]{Perkins2008,Aharonian2009,Arlen2012,Prokhorov2014,Zandanel2014}. It is noteworthy that the Coma cluster was analyzed by the \textit{Fermi}-LAT collaboration \citep{Ackermann2016}, with six years of data, who found some residual emission in the direction of the cluster, though not enough to claim a detection. Despite these unsuccessful searches, the combination of the obtained $\gamma$-ray upper limits with radio synchrotron data was used to show that pure hadronic models cannot explain all the observed radio emission in the case of Coma \citep{Brunetti2012}, and this allowed constraints to be set on turbulent reacceleration models \citep{Brunetti2017}. Recently, \cite{Xi2018} claimed the first significant detection of $\gamma$-ray signal toward the Coma cluster, using \textit{Fermi}-LAT data. This detection was also discussed in the context of a dark matter interpretation of the signal in \cite{Liang2018}, but only minor investigations of the consequences for the CR physics were carried out. In 2020, the \textit{Fermi}-LAT collaboration released an update of the 4FGL catalog \citep[4FGL-DR2,][]{Abdollahi2020,Ballet2020}, also indicating that a source was detected in the direction of the Coma cluster, named 4FGL~J1256.9+2736. While this source could be a contaminant, it could also be associated with the cluster's diffuse emission itself.

In this paper, we explore the consequences on the CR physics of the $\gamma$-ray emission observed in the direction of the Coma cluster under the assumption that this signal is associated with the diffuse ICM. To do so, we present an analysis of the \textit{Fermi}-LAT data \citep{Atwood2009}, to search for $\gamma$-ray emission within the Coma cluster region, and use this measurement to constrain the properties of the CR content of the cluster. We construct a model for the expected signal, in which we set the thermal gas properties using thermal Sunyaev-Zel'dovich (tSZ) effect and X-ray observations. The \textit{Fermi}-LAT data are analyzed, and we extract the spectral energy distribution (SED) of the emission assuming different spatial templates. The \textit{Fermi}-LAT maps are compared to data at other wavelengths for better interpretation of the results. In particular, we compare the $\gamma$-ray signal to the radio synchrotron emission, to the thermal pressure traced by the tSZ signal, to the galaxy distribution, and to the thermal gas distribution from X-ray data. We use our model to constrain the CRp population in the Coma cluster assuming that all the signal observed is due to the hadronic interactions between CRp and the ambient thermal gas, but also consider possible point source contamination. Finally, we compute the amount of secondary CRe associated with the $\gamma$-ray emission assuming steady state. We use this model to constrain the fraction of primary to secondary CRe in the absence of reacceleration, and the boost required from reacceleration with respect to the steady state hadronic model, to explain the radio emission assuming that reacceleration of secondary electrons is dominant, respectively.

This paper is organized as follows. Section~\ref{sec:data} describes the multiwavelength dataset that is used through the paper. In Section~\ref{sec:modeling}, we present the physical modeling of the cluster and the computation of the observables. The \textit{Fermi}-LAT analysis is presented in Section~\ref{sec:Fermi_analysis} and the outputs are compared to multiwavelength data in Section~\ref{sec:multiwavelength_comparison}. Section~\ref{sec:implication_for_CR_content} discusses the implications of the observed signal for the CR content of the Coma cluster. In Section~\ref{sec:Discussions}, we discuss the results presented in this paper and compare them to the literature. Finally, a summary and conclusions are given in Section~\ref{sec:Summary_and_conclusions}. Throughout this paper, we assume a flat $\Lambda$CDM cosmology according to \textit{Planck} results \citep{Planck2016XIII} with $H_0 = 67.8$ km s$^{-1}$ Mpc$^{-1}$, $\Omega_{\rm M} = 0.308$, and $\Omega_{\Lambda} = 0.692$. The reference coordinates of the Coma cluster are RA, Dec = 194.9118, 27.9537 degrees and the redshift $z = 0.0231$, based on the \textit{Planck} catalog \citep[PSZ2,][]{Planck2016XXVII}. Given the reference cosmological model, 1 degree corresponds to 1.73 Mpc at the redshift of Coma. We use the value of $R_{500} = 1310$ kpc based on \cite{PlanckX2013}, corresponding to $M_{500} = 6.13 \times 10^{14}$~M$_{\odot}$ given our cosmological model.

\section{Data}\label{sec:data}
While this paper focuses mainly on the $\gamma$-ray analysis of the Coma cluster, we also use complementary data both for multiwavelength comparison of the signal and in order to build templates used to model the $\gamma$-ray emission. In addition to the \textit{Fermi}-LAT $\gamma$-ray data, we thus use thermal Sunyaev-Zel'dovich data from \textit{Planck}, X-ray data from \textit{ROSAT}, radio data from the Westerbork Synthesis Radio Telescope and optical data from the Sloan Digital Sky Survey. This section presents a brief description of these data.

\subsection{\textit{Fermi} Large Area Telescope $\gamma$-ray data}
The $\gamma$-ray analysis is based on the publicly available\footnote{\url{https://fermi.gsfc.nasa.gov/ssc/data/analysis/}} \textit{Fermi} Large Area Telescope data \citep[\textit{Fermi}-LAT,][]{Atwood2009}, which has been collecting all sky $\gamma$-ray data at energies from about 20 MeV to more than 300 GeV since June 2009. In this paper, we use almost 12 years of Pass 8 data (P8R3), collected from August 4, 2008, to April 2, 2020. Events within a radius of 10 degrees from the cluster center were collected for the analysis (see also Section~\ref{sec:Fermi_analysis} for the data selection). As a description of the compact sources around the cluster, we use the second release of the 4FGL catalog (see \citealt{Abdollahi2020} for the 4FGL catalog and \citealt{Ballet2020} for the 4FGL-DR2 update). The 4FGL-DR2 catalog is an incremental version of the 4FGL catalog. It is based on 10 years of survey data in the 50 MeV - 1 TeV range \citep[see][for the analysis improvement relative to previous catalogs]{Abdollahi2020}. The diffuse isotropic background and the galactic interstellar emission are modeled using the latest available templates: {\tt iso\_P8R3\_SOURCE\_V2\_v1} and {\tt gll\_iem\_v07}, respectively.

\subsection{\textit{Planck} thermal Sunyav-Zel'dovich data}
The tSZ effect \citep{Sunyaev1972} is a relevant probe for comparison to the $\gamma$-ray signal as it provides a direct measurement of the integrated thermal gas pressure along the line-of-sight \citep[see][for reviews]{Birkinshaw1999,Mroczkowski2019}. We used the {\tt MILCA} \citep{Hurier2013} Compton parameter map obtained from \textit{Planck} \citep{PlanckXXII2016} to extract a high signal-to-noise image of the signal at an angular resolution of 10 arcmin (full width at half maximum, FWHM). In the region of Coma, the signal is not significantly affected by any artifacts, e.g., from compact sources, or remaining diffuse galactic emission \citep[see also][for a \textit{Planck} analysis of the Coma cluster]{PlanckX2013}. The data are publicly available at the \textit{Planck} Legacy Archive\footnote{\url{http://pla.esac.esa.int/pla/}}.

\subsection{\textit{ROSAT} All Sky Survey X-ray data}
The X-ray diffuse cluster emission traces the thermal gas density \citep[see][for a review]{Bohringer2010} and is expected to be correlated with the $\gamma$-ray signal. Given the large size of the region of interest (ROI) and the fact that the angular resolution is not critical for our purpose, we used \textit{ROSAT} \citep{Truemper1993} data as obtained from the \textit{ROSAT} All Sky Survey (RASS)\footnote{Collected from \url{http://cade.irap.omp.eu/dokuwiki/doku.php?id=rass}} to image the cluster at a resolution of 1.8 arcmin (FWHM). To do so, we subtracted the background and normalized the data using the exposure maps. In addition, we subtracted the compact sources present in the field using both the RASS bright and faint source catalogs \citep{Voges1999,Voges2000}.

\subsection{Westerbork Synthesis Radio Telescope data}
The radio emission traces relativistic electrons via synchrotron emission and could be associated with the $\gamma$-ray signal. We used the Westerbork Synthesis Radio Telescope (WSRT) radio data at 352 MHz \citep{Brown2011} in order to dispose of an image of the cluster halo and relic. Compact sources were subtracted using the method described in \cite{Rudnick2002}. The residual contamination was estimated by injecting fake point sources into the image, applying the filter to the image, and measuring the residual flux. For sources that were detected by more than $2 \sigma$, the residual flux was less than 3\% and as such the contamination is expected to be less than 3\%. Nevertheless, some of the emission from the bright central radio galaxy remains blended with the signal, and we masked it in the following analysis using a threshold of 100 mJy/beam. Additionally, the bright source Coma A north of the radio relic causes some artifacts that should be accounted for in the morphological comparison (Section~\ref{sec:multiwavelength_comparison}). The image angular resolution is $4 \times 3$ arcmin$^2$ (FWHM).

The WSRT map is also used to extract the average radial profile of the radio halo. To do so, we first subtract the zero level (i.e., offset), and define bins of 3 arcmin width. The zero level arises from the standard clean deconvolution that creates a negative offset around bright sources such as those found in the Coma cluster. After filtering out the point sources and convolving the image to a lower resolution, this resulted in a local background of -20 mJy/beam. Error bars on the profile are computed using inverse variance Gaussian weighting with a constant noise root-mean-square of 5 mJy/beam. The pixels from the masked regions are excluded from the profile. We also convert from Jy/beam to Jy/sr assuming that the beam is Gaussian. We stress that we are considering the total (i.e., average) profile, but the variations in profile from one direction to another \citep[][]{Brown2011} should not have a significant effect on our analysis given the purpose of this paper. The profile will be used in Section~\ref{sec:implication_for_CR_content} as a comparison to our model.

In addition, we also use the integrated fluxes of the Coma radio halo as compiled in \cite{Pizzo2010}. Given the diffuse nature of the signal, it is important to make sure that the flux we used was measured in a consistent aperture for the different observations. In this respect, we used a radius of $0.48 \times R_{500}$ \cite[i.e., 629 kpc or 0.36 deg, see][for details]{Brunetti2013}.

\subsection{Sloan Digital Sky Survey optical data}
The spatial distribution of galaxies is informative of the different groups in the cluster, and the $\gamma$-ray emission might be associated with CR escaping from cluster member galaxies. We thus use Sloan Digital Sky Survey \citep[SDSS,][]{York2000} optical data\footnote{\url{https://dr12.sdss.org/mosaics}} to construct a color image of the Coma region. In addition, we select galaxies with spectroscopic information from the SDSS database\footnote{\url{http://skyserver.sdss.org/CasJobs/}} and use this catalog to construct a galaxy density map of the Coma region. To to so, we bin the galaxy catalog on a grid, and we weight them with $w= {\rm exp}\left(-\frac{(z_{\rm gal} c - z_{\rm Coma} c)^2)}{2 \sigma_v^2}\right)$ to minimize the background. The quantity $c$ is the speed of light, $z_{\rm gal}$ and $z_{\rm Coma}$ the redshift of each galaxy and that of the cluster, respectively, and $\sigma_v = 2128$ km s$^{-1}$ (i.e., FWHM of 5000 km s$^{-1}$). The map is convolved with a Gaussian kernel with a FWHM of 10 arcmin to enhance the signal-to-noise. The background is then estimated as the mean of the map at radii above 90 arcmin from the cluster, and subtracted from the map.

\section{Modeling}\label{sec:modeling}
The physical modeling of the signal, as expected in the $\gamma$-ray band at \textit{Fermi}-LAT energies, is necessary for two reasons. First, we aim at constructing a template (spatial and spectral) of the diffuse emission in order to include the cluster when fitting the data (as done in Section~\ref{sec:Fermi_analysis}). Then, the physical model is needed to connect the observations to the CR physics and will be used to constrain the CR content of the Coma cluster (Section~\ref{sec:implication_for_CR_content}). Additionally, the synchrotron emission should be included in the modeling when comparing the implication of the \textit{Fermi}-LAT constraint on the radio signal. 

In this paper, we model the Coma cluster using the {\tt MINOT} software \citep{Adam2020}\footnote{\url{https://github.com/remi-adam/minot}}. This allows us to have a complete description of the cluster from the CR and thermal physics to the $\gamma$-ray signal, under the approximation of spherical symmetry. 

In addition, we also use spatial templates constructed using multiwavelength data. While this approach may provide a better match to the spatial shape of the signal due to deviations from spherical symmetry, it does not allow us to constrain the CR content of the cluster because of sky projection and degeneracies between the different physical components.

\begin{figure*}
	\centering
	\includegraphics[width=0.45\textwidth]{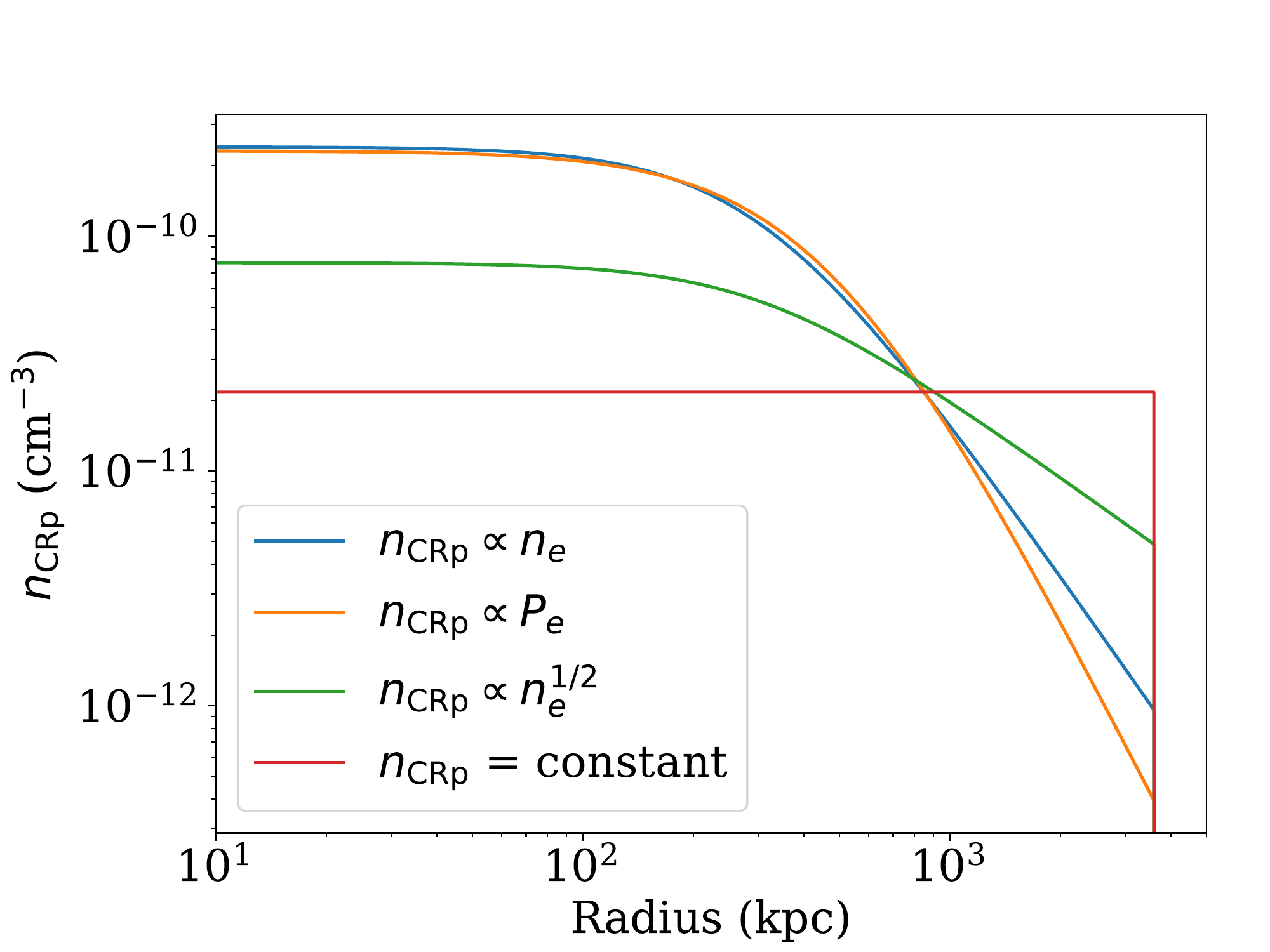}
	\includegraphics[width=0.45\textwidth]{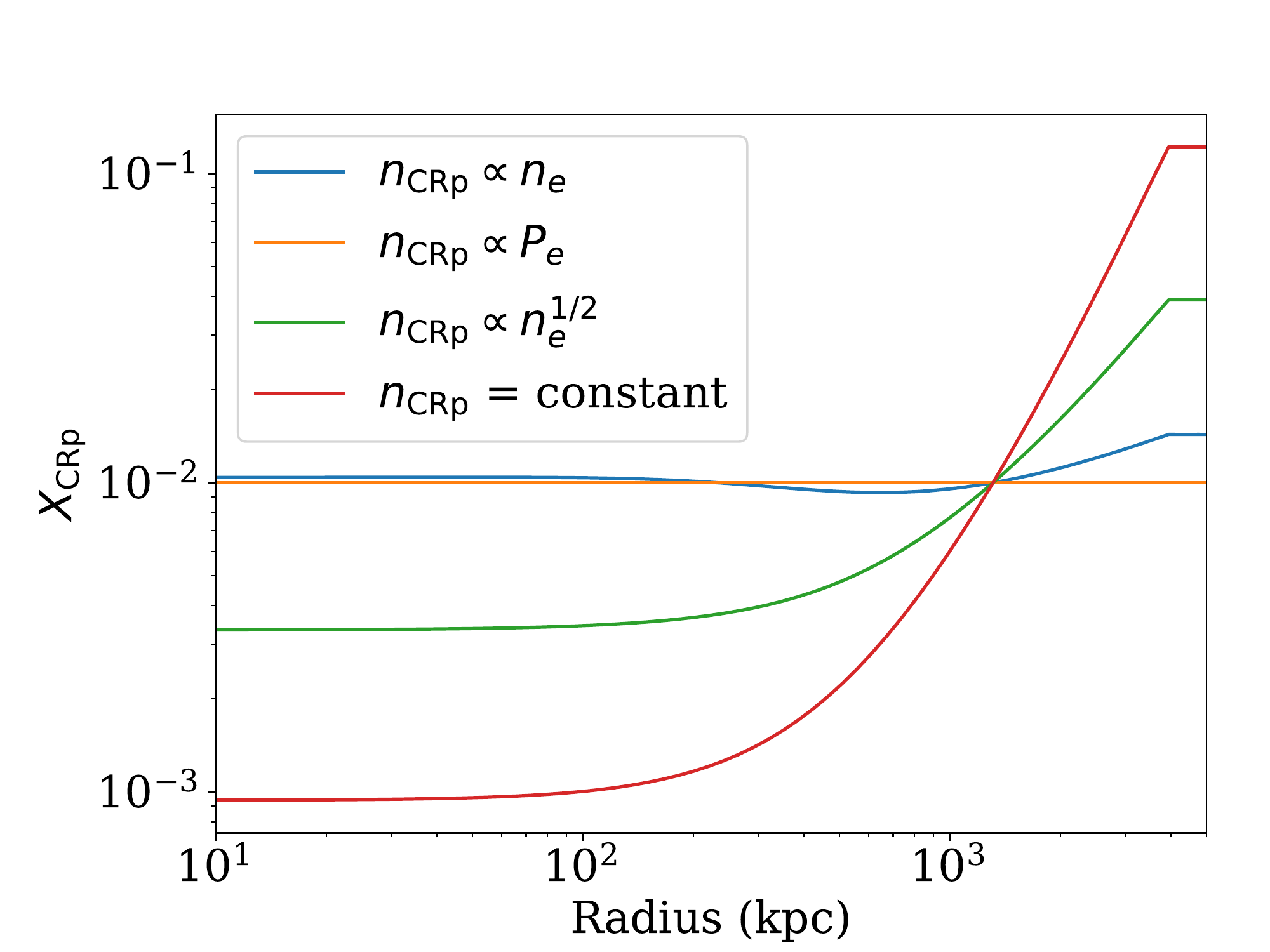}
	\includegraphics[width=0.45\textwidth]{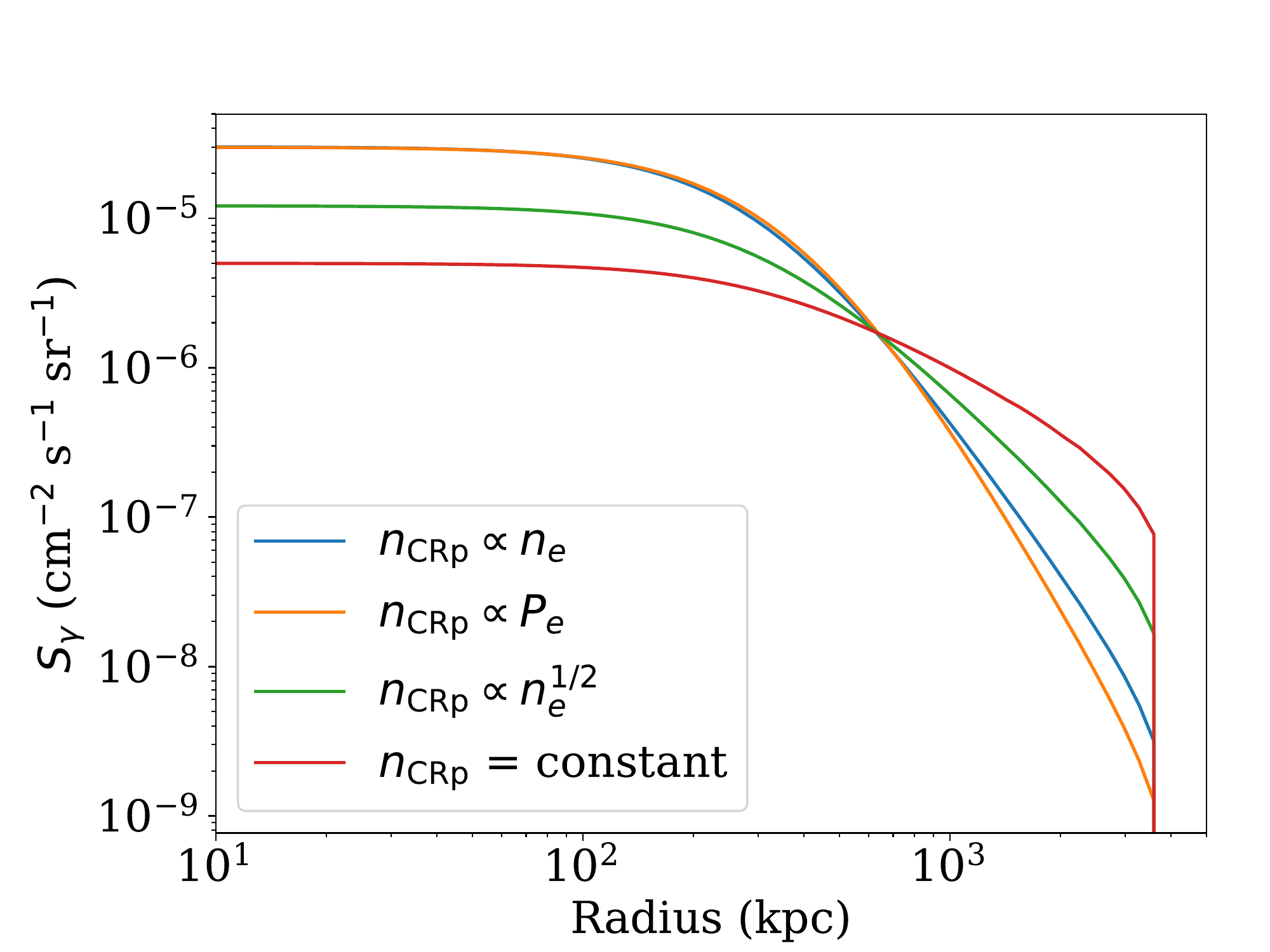}
	\includegraphics[width=0.45\textwidth]{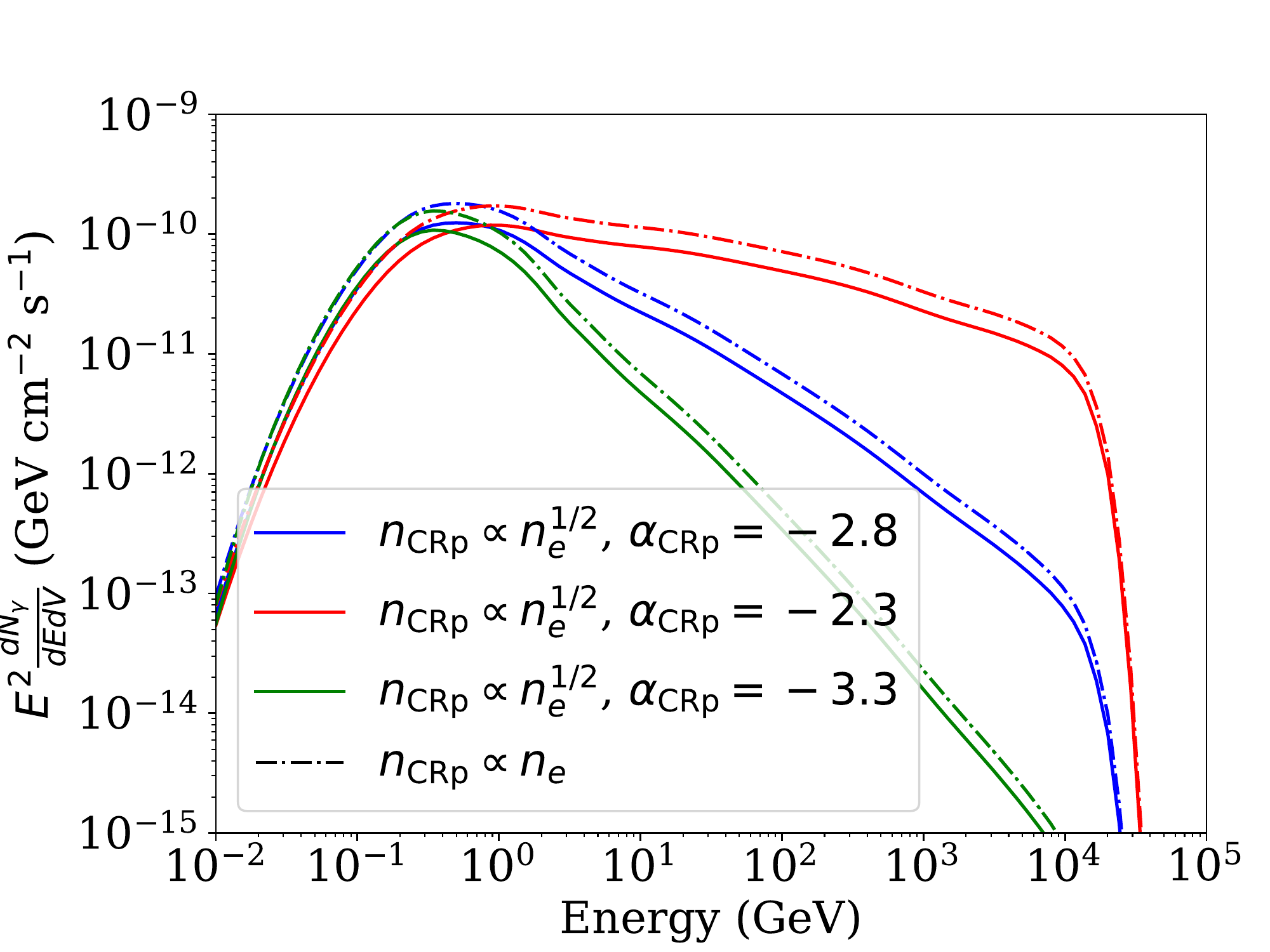}
	\caption{\small {\tt MINOT} templates used to describe the cluster, for different assumptions regarding the CRp distributions. All models are normalized so that $X_{\rm CRp}(R_{500}) = 10^{-2}$. {\bf Top left}: Radial profile of the CRp distribution. {\bf Top right}: Enclosed CRp to thermal energy profile. {\bf Bottom left}: $\gamma$-ray surface brightness profile. {\bf Bottom right}: $\gamma$-ray energy spectrum integrated within $R_{500}$.}
\label{fig:template_minot}
\end{figure*}

\begin{figure*}
	\centering
	\includegraphics[width=0.33\textwidth]{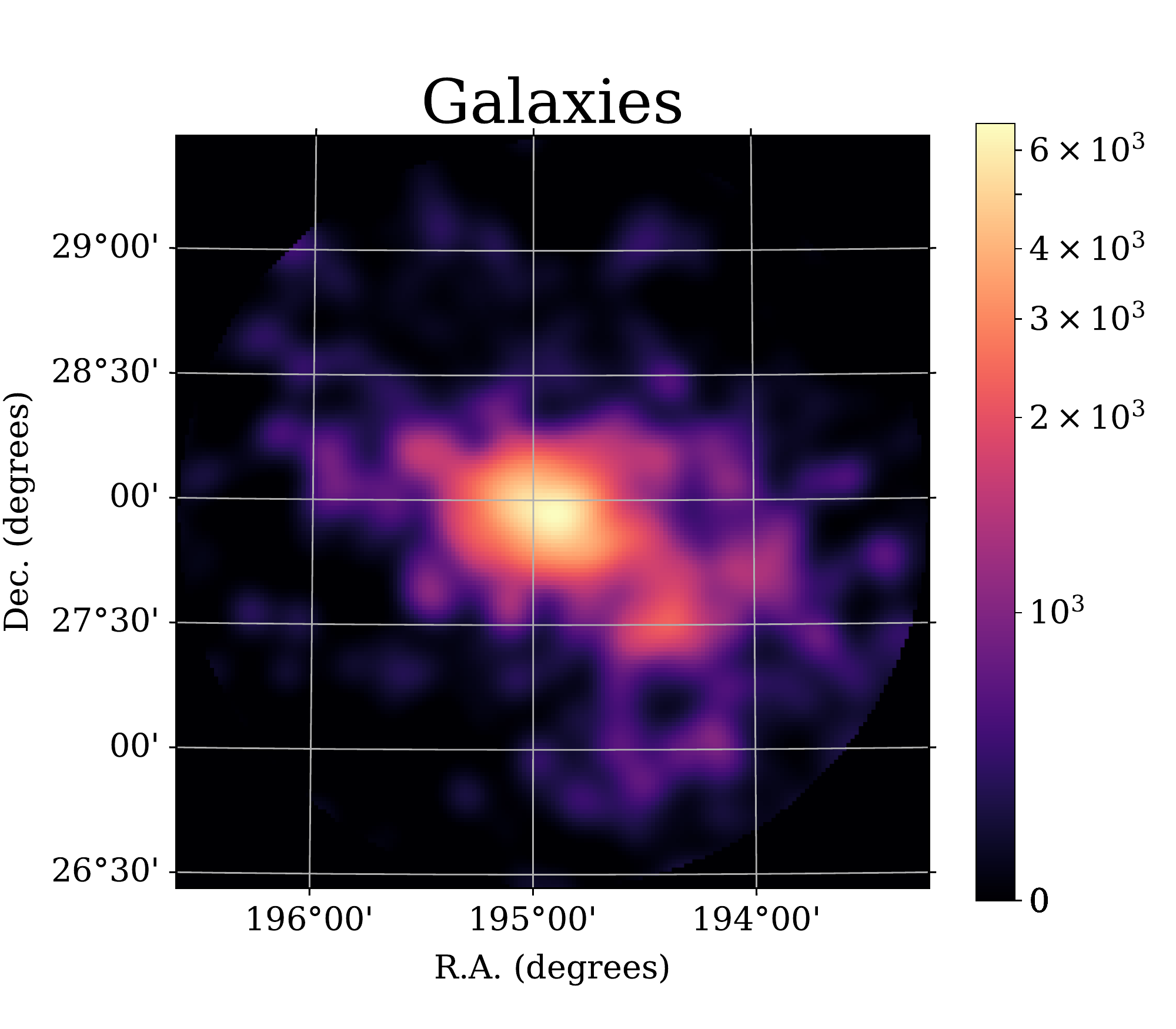}
	\includegraphics[width=0.33\textwidth]{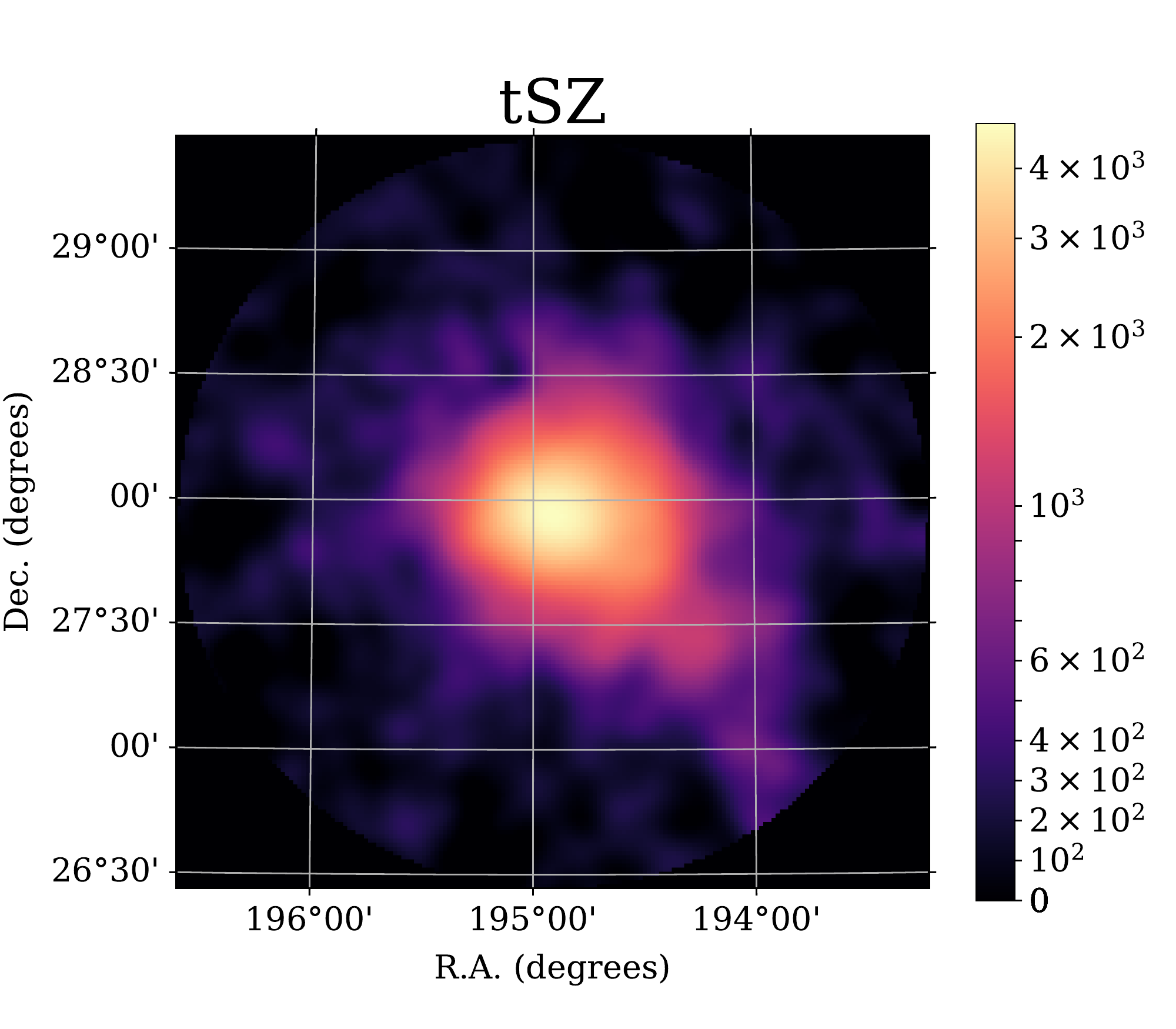}
	\includegraphics[width=0.33\textwidth]{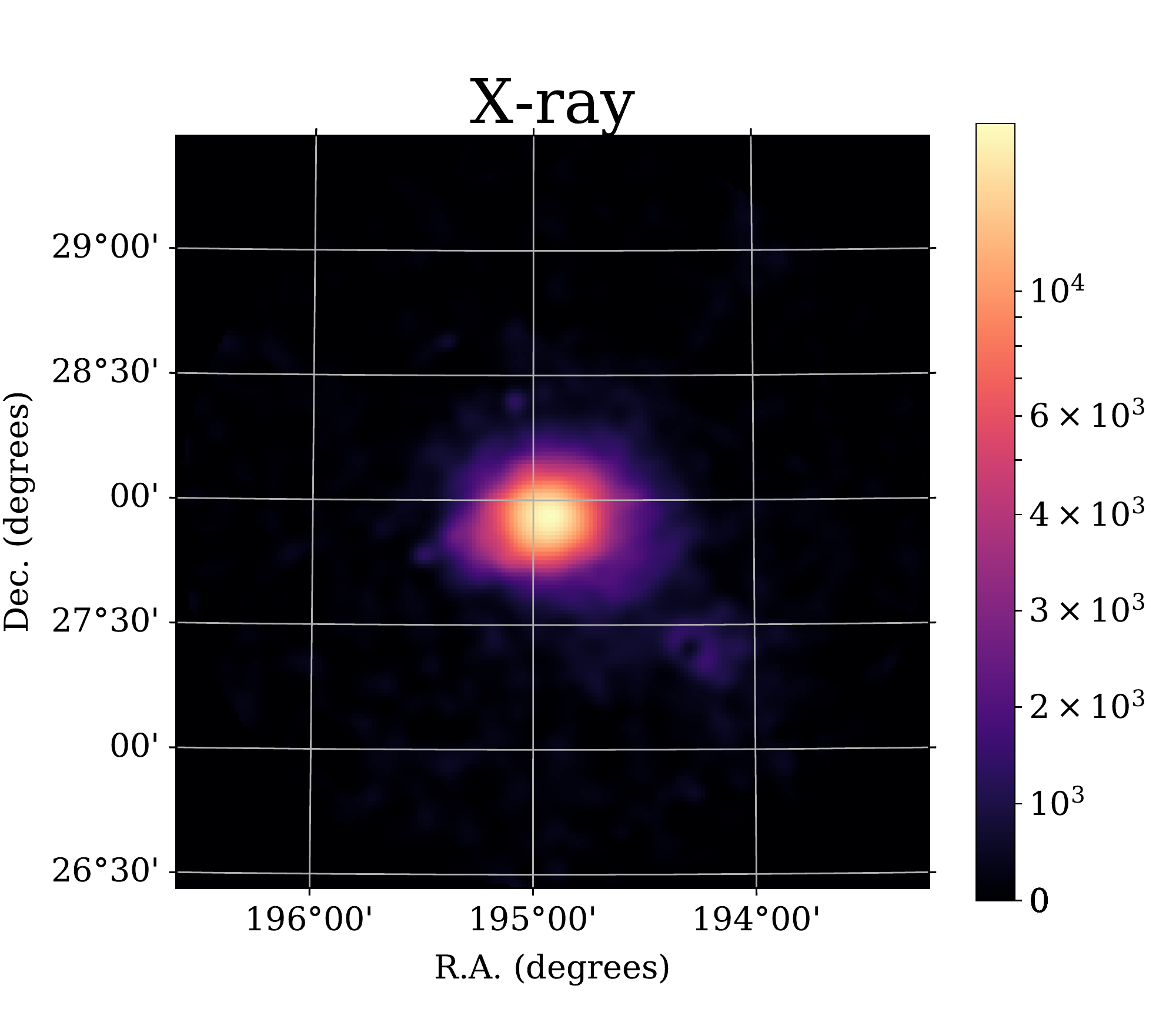}
	\includegraphics[width=0.33\textwidth]{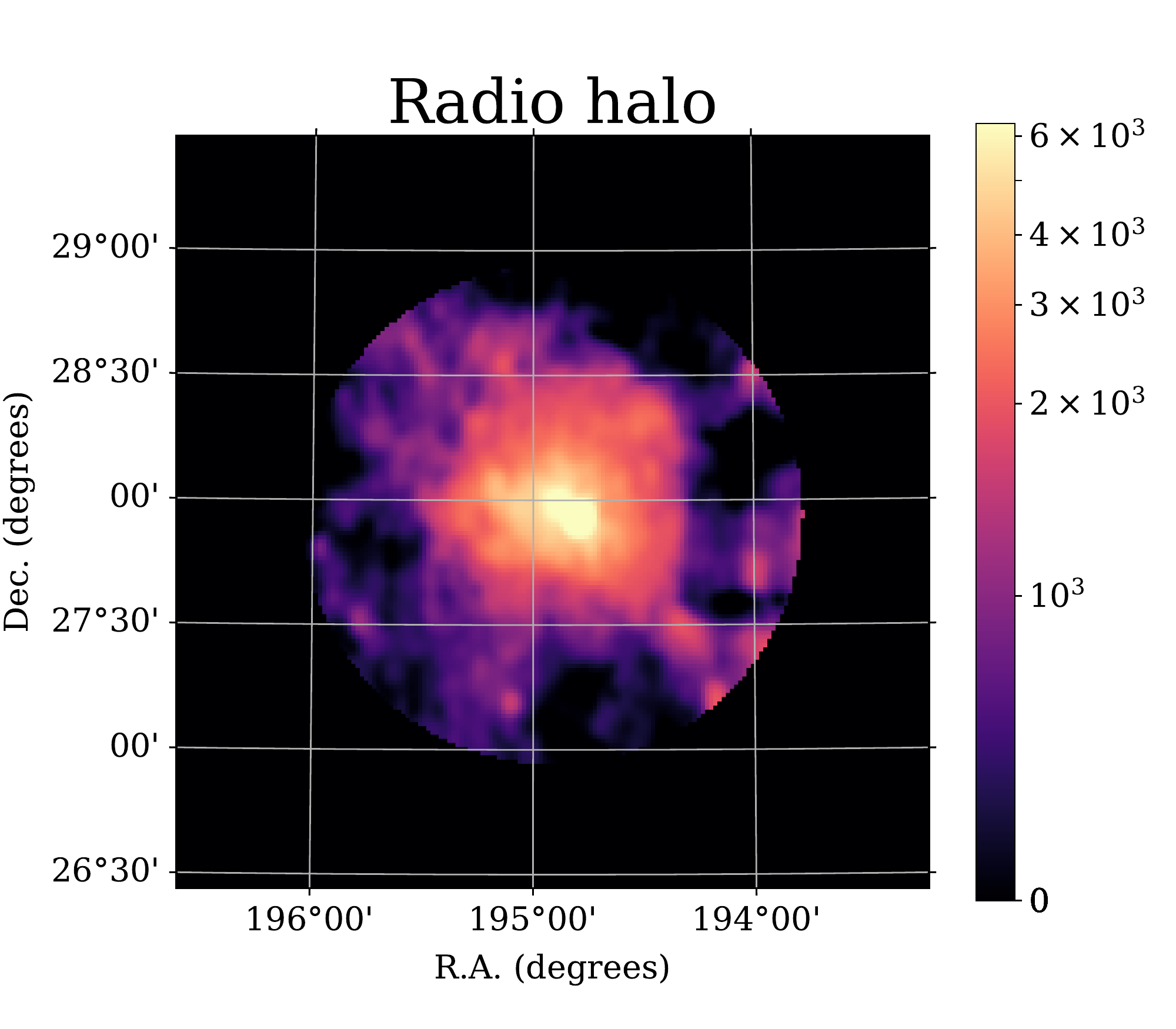}
	\includegraphics[width=0.33\textwidth]{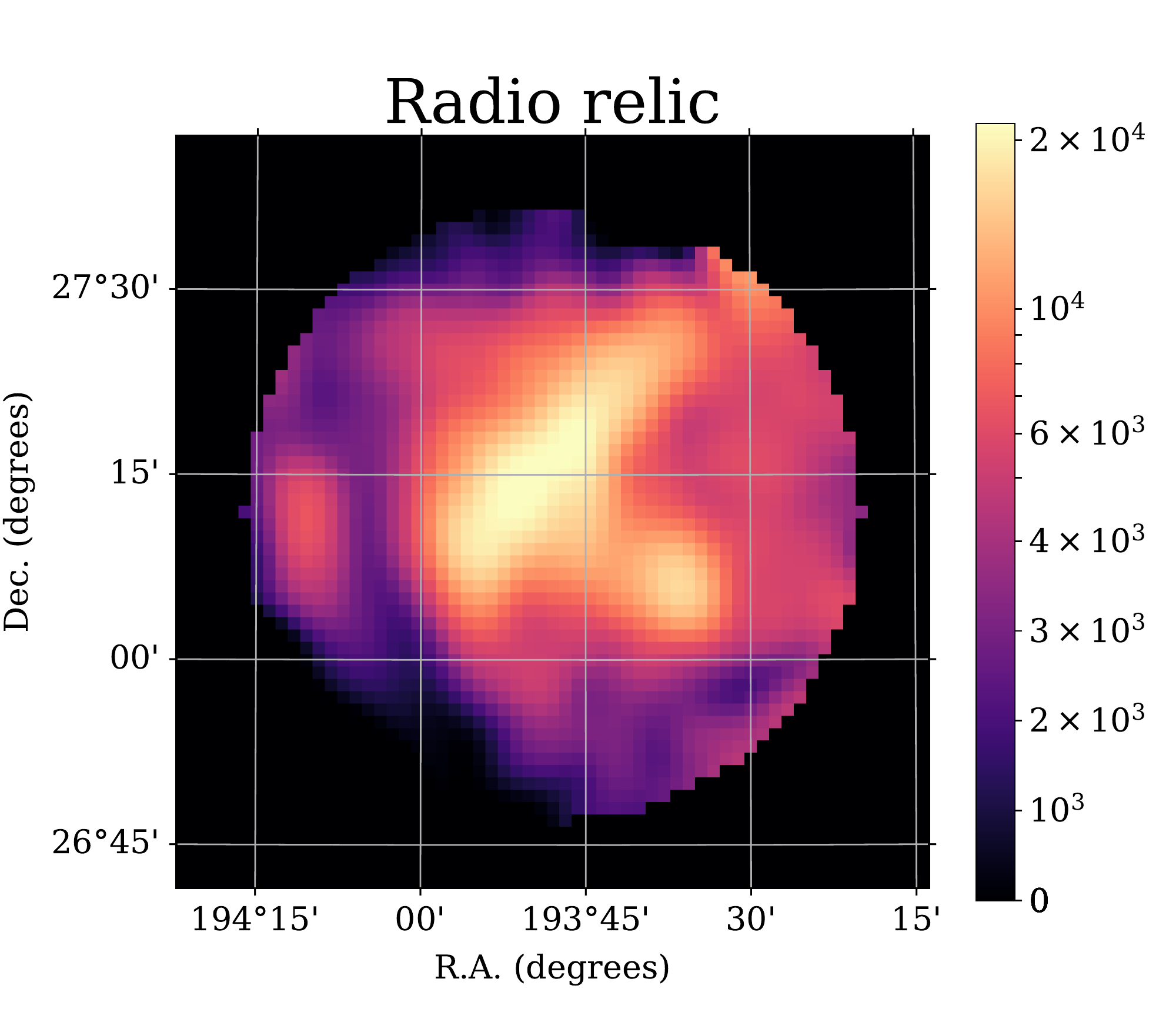}
	\caption{\small Template images used to describe the morphology of the $\gamma$-ray emission. From left to right and top to bottom, we present the templates based on the distribution of galaxies, the tSZ signal, the X-ray emission, the radio halo emission, and the radio relic emission. Images are $3\times3$ deg$^2$, except for the radio relic (bottom right), which is only $1\times1$ deg$^2$. Units are arbitrary, but the integral of all maps over the solid angle is set to unity. For display purpose, the scale is linear from zero to 20\% of the peak value, and logarithmic above.}
\label{fig:template_2d}
\end{figure*}

\subsection{Physically motivated modeling using {\tt MINOT}}
\subsubsection{Thermal component}
The {\tt MINOT} software models the thermal ICM component using the gas electron pressure and density profiles. The gas density profile is key to the $\gamma$-ray emission as it provides the target for proton-proton interactions. The thermal pressure is also particularly important, because it will be used to provide a normalization to the amount of CR.

The electron pressure radial profile, as a function of radius $r$, is described by a generalized Navarro Frenk White \citep[gNFW,][]{Nagai2007} model according to
\begin{equation}
P_e(r) = \frac{P_0}{\left(\frac{r}{r_p}\right)^c \left(1+\left(\frac{r}{r_p}\right)^a\right)^{\frac{b-c}{a}}},
\end{equation}
with parameters taken from \cite{PlanckX2013} and rescaled to our cosmological model ($P_0, r_p, a,b,c = 0.022 {\rm \ keV \ cm}^{-3}, 466.8 {\rm \ kpc}, 1.8, 3.1, 0.0$). 

The electron density is described by a $\beta$-model \citep{Cavaliere1978} according to
\begin{equation}
n_e(r) = n_0 \left(1+\left(\frac{r}{r_c}\right)^2\right)^{-3\beta/2},
\end{equation}
with parameters taken from \cite{Briel1992} and rescaled to our cosmological model ($n_0, r_c, \beta = 3.36 \times 10^{-3} {\rm \ cm}^{-3}, 310 {\rm \ kpc}, 0.75$). 

From the electron pressure and electron density, we also derive the total gas pressure and thermal proton (and heavier elements) density assuming a constant helium mass fraction of $Y_{\rm He} = 0.2735$ and metal abundances of $0.3$, relative to the solar value \citep[see][for more details]{Adam2020}. Finally, we note that the radial models are truncated at a radius $r > 3 \times R_{500}$. This also apply for nonthermal quantities discussed below.

\subsubsection{Magnetic field}
The synchrotron emission, from a given CRe population, depends on the magnetic field. The later is thus an important component of our model when comparing the implications of our constraints with radio data (Section~\ref{sec:implication_for_CR_electrons}). The magnetic field strength profile of the Coma cluster was inferred by \cite{Bonafede2010} using Faraday rotation measures and we used their result to model it as 
\begin{equation}
\left<B\right>(r) = \left<B_0\right> \left(\frac{n_e(r)}{n_{0}}\right)^{\eta_B},
\end{equation}
where $\left<B_0\right> = 4.7$ $\mu$G and $\eta_B = 0.5$. We note, however, that uncertainties on these parameters are relatively large and the parameters are degenerate \citep[see also the recent work by][]{Johnson2020}. They are constrained in the range $\left(\left<B_0\right> = 3.9 \ \mu {\rm G}; \eta_B = 0.4\right)$ to $\left(\left<B_0\right> = 5.4 \ \mu {\rm G}; \eta_B = 0.7\right)$ \citep[see][for more details]{Bonafede2010}. The impact of the magnetic field modeling will be further discussed in Section~\ref{sec:discussions_cr_physics}.

\subsubsection{Cosmic ray protons}
The CRp number density, per unit volume and energy, is modeled as
\begin{equation}
	J_{\rm CRp} (r, E) = A_{\rm CRp} f_1(E) f_2(r),
\label{eq:CRp_distribution}
\end{equation}
where $f_1(E)$ is the energy spectrum, $f_2(r)$ the radial dependence, and $A_{\rm CRp}$ is the normalization. The energy spectrum is modeled as a power law (expected from Fermi acceleration), as
\begin{equation}
	f_1(E) \propto E^{-\alpha_{\rm CRp}},
\end{equation}
with $\alpha_{\rm CRp}$ the slope of the CRp energy spectrum. The radial dependence is modeled assuming that the CRp number density scales with either the thermal density or pressure, as 
\begin{equation}
	f_2(r) \propto
	\left \{
	\begin{array}{r c l}
	n_e(r)^{\eta_{\rm CRp}} \\
	P_e(r)^{\eta_{\rm CRp}}.
	\end{array}
	\right.
\label{eq:CRp_thermal_scaling}
\end{equation}
Since the radial dependence of the CRp is not well-known, we test different values for the scaling, $\eta_{\rm CRp} = \left(0, \frac{1}{2}, 1\right)$, corresponding to flat, extended, and compact distributions. Finally, the normalization $A_{\rm CRp}$ is computed relative to the thermal energy enclosed within the radius $R_{500}$. In the following, we use the parameter
\begin{equation}
	X_{\rm CRp}(R) = \frac{U_{\rm CRp} (R)}{U_{\rm th} (R)},
\label{eq:CRp_thermal_normalization}
\end{equation}
where $U_{\rm CRp}(R)$ is the total CRp energy enclosed within $R$ computed by integrating Equation~\ref{eq:CRp_distribution} and $U_{\rm th} (R)$ the total thermal energy, obtained by integrating the gas pressure profile \citep[see][for more details]{Adam2020}.

In the end, three parameters are used to describe the CRp: the spatial scaling relative to the thermal gas $\eta_{\rm CRp}$, the CRp spectrum slope $\alpha_{\rm CRp}$, and the normalization encoded in $X_{\rm CRp}(R_{500})$. The value of these parameters will be further discussed in Section~\ref{sec:Fermi_analysis} and~\ref{sec:implication_for_CR_content}.

\subsubsection{Primary cosmic ray electrons}\label{sec:CRe1model}
At the \textit{Fermi}-LAT energies, the $\gamma$-ray emission is expected to be dominated by hadronic processes, and we neglect the inverse Compton emission due to secondary or primary CRe interacting with background light (see Appendix~\ref{app:neglecting_ic} for more discussions). However, CRe need to be included in our model when computing the radio synchrotron emission in Section~\ref{sec:implication_for_CR_content}. The primary cosmic ray electrons, CRe$_1$, are modeled similarly as the CRp. However, CRe$_1$ are affected by losses and we therefore account for this using three different models: the {\tt ExponentialCutoffPowerLaw}, the {\tt InitialInjection}, and the {\tt ContinuouslInjection} models as implemented in the {\tt MINOT} software \citep{Adam2020}. They are expressed as
\begin{equation}
	f_1(E) \propto \left(\frac{E}{E_0}\right)^{-\alpha_{\rm CRe_1}} \times \rm{exp} \left(-\frac{E}{E_{\rm cut, CRe_1}}\right),
\end{equation}
\begin{equation}
	f_1(E) \propto \left(\frac{E}{E_0}\right)^{-\alpha_{\rm CRe_1}} 
	\begin{cases} \left(1 - \frac{E}{E_{\rm cut, CRe_1}}\right)^{\alpha_{\rm CRe_1} - 2} &E < E_{\rm cut, CRe_1} \\
	0 &E \geqslant E_{\rm cut, CRe_1}
	\end{cases},
\end{equation}
and
\begin{equation}
	f_1(E) \propto \left(\frac{E}{E_0}\right)^{-\left(\alpha_{\rm CRe_1}+1\right)} 
	\begin{cases} \left(1 - \frac{E}{E_{\rm cut, CRe_1}}\right)^{\alpha_{\rm CRe_1} - 1} &E < E_{\rm cut, CRe_1} \\
	1 &E \geqslant E_{\rm cut, CRe_1}
	\end{cases},
\end{equation}
respectively. Using different parametrizations will allow us to estimate the systematic effects associated with the model.

There are therefore four parameters used to describe the primary CRe$_1$: the spatial scaling of the CRe$_1$ population relative to the thermal gas, $\eta_{\rm CRe_1}$ (as in Equation~\ref{eq:CRp_thermal_scaling}), the CRe$_1$ spectrum slope $\alpha_{\rm CRe_1}$ and break energy $E_{\rm cut, CRe_1}$, and the normalization encoded in $X_{\rm CRe_1}(R_{500})$ (as in Equation~\ref{eq:CRp_thermal_normalization}). The value of these parameters will be further discussed when comparing our model to radio data in Section~\ref{sec:implication_for_CR_content}. 

\subsubsection{$\gamma$-ray and synchrotron signal}
Given the thermal gas, magnetic field and CRp modeling, we compute the $\gamma$-ray emission due to hadronic collisions, the secondary cosmic ray electrons assuming a steady state scenario (CRe$_2$), and the radio synchrotron emission (from both CRe$_1$ and CRe$_2$) according to \cite{Adam2020}. We do not account for inverse Compton emission at \textit{Fermi}-LAT energies because it is expected to be negligible, as detailed in Appendix~\ref{app:neglecting_ic}.

In brief, the production rate of $\gamma$-rays and CRe$_2$ are computed as
\begin{equation}
	\frac{dN}{dE dV dt}(r,E) = \int_{E}^{+\infty} \frac{dN_{\rm col}}{dE_{\rm CRp}dVdt} F\left(E, E_{\rm CRp}\right) dE_{\rm CRp},
\label{eq:rate_crp}
\end{equation}
where $\frac{dN_{\rm col}}{dE_{\rm CRp}dVdt} \propto n_{\rm e}(r) \times J_{\rm CRp}(r,E)$ is the CRp--ICM collision rate and the function $F\left(E, E_{\rm CRp}\right)$ gives the number of secondary particles (electrons or $\gamma$-rays) produced in a collision as a function of the initial energy of the CRp. This computation is based on the parametrization by \cite{Kelner2006} and \cite{Kafexhiu2014} following the work by \cite{Zabalza2015}, as detailed in \cite{Adam2020}. The distribution of CRe$_2$ is then obtained accounting for losses under the steady state assumption. The $\gamma$-ray emission profile and spectrum are computed by integrating Equation~\ref{eq:rate_crp} along the line-of-sight, accounting for the distance to the Coma cluster, and integrating over the energy or the solid angle, respectively. The $\gamma$-ray attenuation due to interaction with the extragalactic background light (EBL) is also included and induces a cutoff in the spectrum above $E \gtrsim 10^4$ GeV. These spatial (profile) and spectral templates are used in Section~\ref{sec:Fermi_analysis} to account for the cluster in the modeling of the ROI.

The synchrotron emission rate is computed following the prescription given in \cite{Aharonian2010}. This assumes that the orientation of the magnetic field is randomized. Observable profile and spectra are then obtained as for the $\gamma$-rays.

In Figure~\ref{fig:template_minot}, we show how the different profiles and spectral slopes for the CRp translate into the $\gamma$-ray surface brightness profile and spectrum. The top left panel shows the four different radial distributions used for the CRp, in terms of $n_{\rm CRp}(r)~=~\int~J_{\rm CRp}(r, E)~dE$. As we can see, the difference between models based on the thermal density or thermal pressure are very close in the case of Coma as expected given the fact that the cluster is close to be isothermal. On the top right panel, we can see the correspondence in terms of the profile of the CRp energy to thermal energy ratio. We note that the normalization was fixed to $X_{\rm CRp}(R_{500}) = 10^{-2}$. The bottom left panel shows the $\gamma$-ray surface brightness for the different models. Because it arises from the product of the CRp and thermal gas density, the profiles are more peaked than the original CRp distributions. On the bottom right panel, we can see the energy spectrum integrated within $R_{500}$ for different slopes $\alpha_{\rm CRp}$. The CRp slope mainly drives the $\gamma$-ray slope between 1 GeV and $10^4$ GeV. At higher energies, the extragalactic background light induces a cutoff in the spectrum, but this is not in the range accessible by \textit{Fermi}-LAT. At lower energies, the energy threshold of the proton-proton collisions implies that the spectrum smoothly vanishes. Since the normalization is fixed, steeper is the spectrum and higher is the flux near the peak. As seen with the dash-dotted lines, the amplitude decreases when the CRp profiles become flatter because it leads to a decrease in the particle collision rate within $R_{500}$.

\subsection{Construction of spatial templates from multiwavelength data}
In addition to the spherically symmetric physical models, we build spatial templates based on the multiwavelength data discussed in Section~\ref{sec:data}. We consider templates based on the galaxy density, the tSZ Compton parameter, the X-ray surface brightness and the radio surface brightness (halo and relic). The maps are projected on $5 \times 5$ deg$^2$ maps with 1 arcmin pixel size (well below the \textit{Fermi}-LAT resolution). In order to reduce the noise, the maps are smoothed so that their angular resolution (FWHM) is 10, 11, 5, and 5 arcmin for the galaxy density, tSZ, X-ray, and radio maps, respectively. This smoothing remains well below the \textit{Fermi}-LAT angular resolution so that it will be negligible when convolving the maps to the instrument response function (Section~\ref{sec:Fermi_analysis}). In order to minimize bias from noise on large scales, we mask the pixels that are more than 90 arcmin away from the reference center for the galaxy density, tSZ and X-ray maps. For the radio map, two cases are considered: the halo for which pixels more than 60 arcmin from the center are masked (to avoid including the relic in the template) and the relic for which we mask pixels more than 25 arcmin from the coordinates (RA,Dec) = (193.8, 27.2) deg. The mask used for the relic allows us to exclude any bright radio galaxy focusing on the emission from the relic only.

The resulting templates are shown in Figure~\ref{fig:template_2d} on $3 \times 3$ deg$^2$ grids (except for the relic, for which it is $1\times1$ deg$^2$). In all cases, the cluster is elongated from the northeast to the southwest, with an excess associated with the NGC 4839 group on the southwest. The most compact template is the one based on the X-ray image as it is proportional to the gas density squared. The tSZ template, proportional to the gas pressure (temperature times density) is much more extended. The galaxy density template extension is in between the tSZ and X-ray ones, but it is more elongated. The radio halo template, on the other end, is very spherical and matches well the tSZ template in terms of the extension. The radio relic template is very elongated from the southeast to the northwest, much smaller than the other templates, and not spatially coincident (as will be further discussed in Section~\ref{sec:multiwavelength_comparison}).

\section{\textit{Fermi}-LAT analysis}\label{sec:Fermi_analysis}
In this Section, we describe the \textit{Fermi}-LAT $\gamma$-ray analysis, performed using Fermipy \citep{Wood2017}.  After the data selection, we model and fit the ROI in order to extract the signal in the Coma cluster region under different modeling assumptions (signal associated with the ICM, a point source, the combination of both) and compare these scenarios. The cluster model built in Section~\ref{sec:modeling} is used to extract the SED of the source. We also perform several tests to validate the global ROI model as a function of energy, radius and check the time stability. Finally, we discuss the systematic effects that might affect our findings.

\subsection{Data selection and binning}
Following the \textit{Fermi}-LAT collaboration recommendations for off galactic plane compact source analysis, we apply the {\tt P8R3\_SOURCE\_V2} selection (event class 128). The energy dispersion is also accounted for. We select {\tt FRONT+BACK} (event type 3) converting photons and apply a cut on zenith angles less than 90 degrees to effectively remove photons originating from the Earth limb. We used the recommended time selection, {\tt DATA\_QUAL>0 \&\& LAT\_CONFIG==1}, and also apply a cut on rocking angle, {\tt (ABS(ROCK\_ANGLE)<52)}.

The ROI was defined as a square of $12 \times 12$ deg$^2$ around the Coma center. As a baseline, the data were binned using a pixel resolution of $0.1$ deg and 8 energy bins per decade between 200 MeV and 300 GeV. The choice of the low energy threshold is a compromise between count statistics and robustness of the results since systematic effects increase at low energy (see also Section~\ref{sec:systematic_effects}).

\subsection{ROI modeling}\label{sec:ROI_modeling}
We start by modeling the ROI accounting for the diffuse background components, corresponding to the isotropic and galactic interstellar emission, as well as the 4FGL compact sources. Given the size of our ROI, the most distant pixels are 8.5 degrees away from the center. Nevertheless, we include all sources within a square with a width of 20 degrees from the reference center, because the point spread function (PSF) may extend up to about 10 degrees at low energies.

As discussed in Section~\ref{sec:Introduction}, the 4FGL-DR2 catalog includes a source named 4FGL~J1256.9+2736 that lies close to the Coma cluster peak. It is located at (RA,Dec) = 194.2417, 27.6076, corresponding to about 0.68 degrees from the cluster center and within $\theta_{500}$. This source is detected with a significance of 4.2, and is modeled by a point source with power law spectrum with index 2.73 in the 4FGL-DR2 catalog. Its origin is uncertain, but it is given as possibly associated with NGC~4839 (with a probability of 0.24). While this source could be a contaminant for the diffuse ICM signal, as, e.g., the AGN of NGC~4839, it could also corresponds to the peak of the diffuse emission associated with the ICM. This second scenario is motivated by the uncertain origin of the source due to the limited signal to noise ratio and angular resolution of the Fermi-LAT. We thus aim at testing and comparing the two hypothesis. To do so, we consider the three following scenarios for modeling the Coma cluster region. 1) We assume that 4FGL~J1256.9+2736 is a point source independent of the cluster diffuse emission. In this case, we include it in the ROI model as given by the 4FGL-DR2 catalog and do not include any extra diffuse emission associated with the ICM. 2) We assume that 4FGL~J1256.9+2736 is in fact associated with the diffuse cluster emission. In this case, the 4FGL-DR2 catalog model is not an appropriate description of the signal and we replace 4FGL~J1256.9+2736 by a diffuse emission model associated with the ICM. 3) We assume that 4FGL~J1256.9+2736 is a point source independent of the cluster diffuse emission, but we also consider a diffuse ICM component as $\gamma$-ray emission is expected from the cluster. In this case, we include 4FGL~J1256.9+2736 in the ROI model as given by the 4FGL-DR2 catalog and add an extra component to account for the diffuse emission associated with the ICM.

The diffuse Coma cluster ICM emission is modeled either using the {\tt MINOT} spatial and spectral physical templates, or using the {\tt MINOT} spectral template together with the spatial templates built from other wavelengths (see Section~\ref{sec:modeling}). We test different shapes for the profile of the CRp, and fix the CRp spectral index slope to $\alpha_{\rm CRp} = 2.8$. This value will be later constrained using the measured SED, in Section~\ref{sec:implication_for_CR_content}. The nonoptimal choice of this number will slightly reduce the significance associated with the Coma diffuse emission, but does not affect the SED fit performed afterward. As a baseline, and unless otherwise stated, we use the model with $\eta_{\rm CRp} = 1/2$ scaled with respect the gas thermal density (extended model).

\subsection{ROI fitting}\label{sec:ROI_fitting}
\begin{figure*}
\centering
\includegraphics[width=1\textwidth]{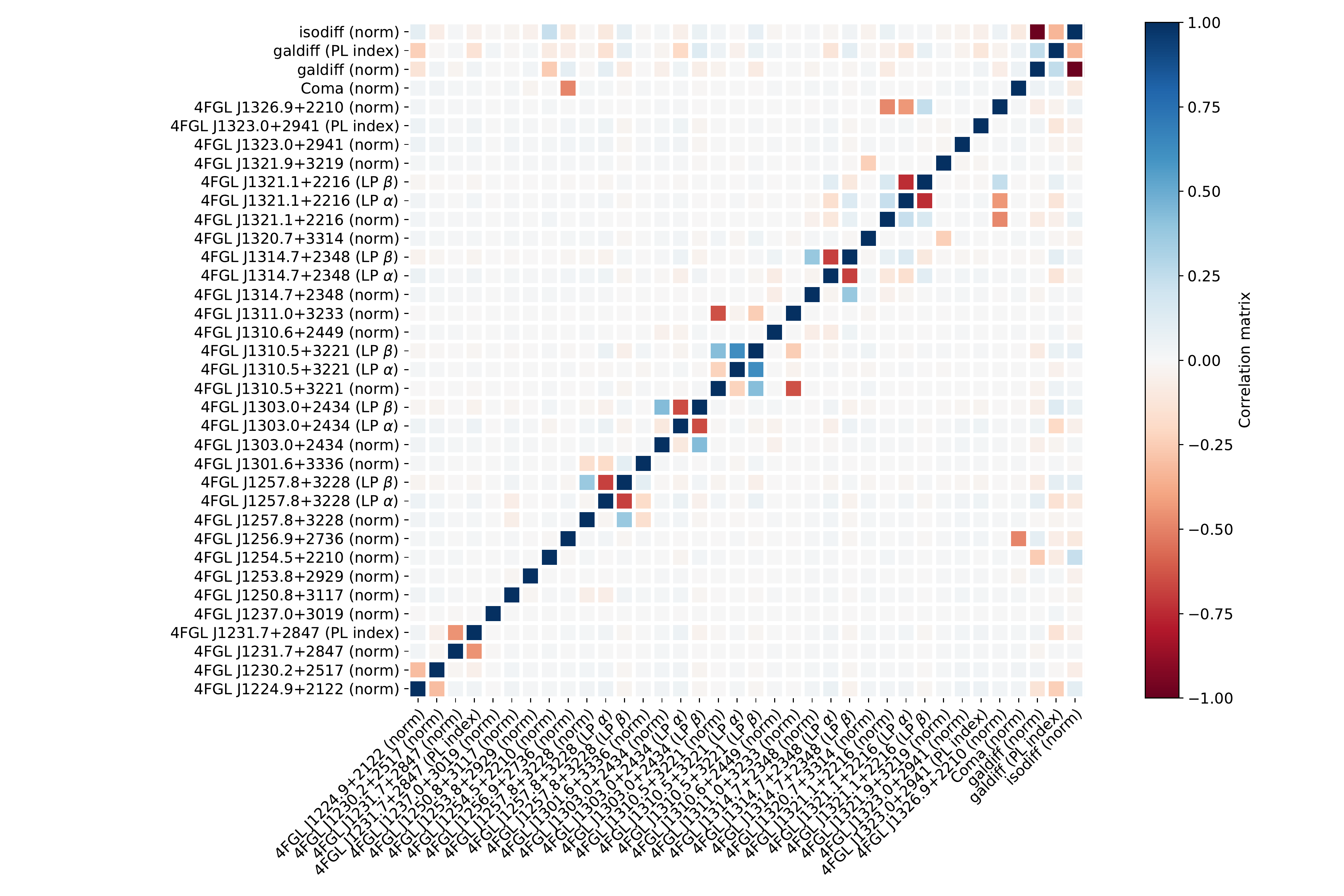}
\caption{\small Correlation matrix associated with the likelihood fit. Each entry corresponds to one free parameter of the model. The names isodiff and galdiff correspond to the isotropic and galactic diffuse backgrounds, Coma corresponds to the ICM cluster model, and all the other names to the 4FGL sources. The labels of the parameters are as follows: {Norm} stands for the spectrum normalization; {PL index} stands for the spectral index of sources described by a power law spectrum; 
{LP $\alpha$} and {LP $\beta$} stand for the spectral parameters of the sources described by a log-parabola spectrum. This correlation matrix corresponds to the case that includes 4FGL~J1256.9+2736, and for the {\tt MINOT} cluster model with $n_{\rm CRp} \propto n_e^{1/2}$.}
\label{fig:correlation_matrix}
\end{figure*}

Once the overall background model (large-scale diffuse plus point source emission) and cluster diffuse emission is defined, we use the following iterative procedure to fit the ROI. 1) We construct a first sky model using the diffuse background and the point sources spectral parametrization, with parameters from the 4FGL catalog, but exclude the cluster diffuse emission model at this stage. 2) We run the {\tt optimize} function of {\tt Fermipy}, including all sources in the model (except for the cluster diffuse emission). We thus obtain a first renormalization of the sources in the model and a first value of their test statistics \citep[TS,][]{Mattox1996}, defined as
\begin{equation}
{\rm TS} = -2 \left({\rm ln} \mathcal{L}_0 - {\rm ln} \mathcal{L} \right),
\label{eq:TS}
\end{equation}
with $\mathcal{L}_0$ the maximum likelihood value for the null hypothesis, and $\mathcal{L}$ the maximum likelihood when including the additional source. 3) To allow for variability, we free all the model parameters associated with sources with $\sqrt{\rm TS}>20$, and free only the normalization of sources with $5<\sqrt{\rm TS}<20$. 4) We use the {\tt fit} function of {\tt Fermipy} to perform the maximum likelihood model fit of all the free parameters in the sky model. 5) We add the cluster diffuse emission (except in the scenario 1, see subsection~\ref{sec:ROI_modeling}) in our model, with its normalization being its only free parameter, and refit the sky model as in step 4. In addition to the likelihood fit, we also use the {\tt tsmap} {\tt Fermipy} function to compute TS maps, in the case of radially symmetric cluster models.

In Figure~\ref{fig:correlation_matrix}, we present the correlation matrix recovered from the likelihood fit for all the fitted parameters included in our model (i.e., the free parameters). It corresponds to the case of the scenario 3 (see subsection~\ref{sec:ROI_modeling}), i.e., including both the Coma cluster diffuse emission and the contribution from 4FGL~J1256.9+2736. As we can observe the diffuse cluster emission is strongly degenerate (anti-correlated) with the normalization of 4FGL~J1256.9+2736, which is expected given its coordinate, as the signal from one component can be absorbed in the other one. This shows that a scenario in which 4FGL~J1256.9+2736 is in fact associated with the cluster diffuse emission should be considered, as done in the following. The Coma cluster diffuse emission is, on the other hand, not significantly correlated with any other compact source in the model, but is marginally anti-correlated with the isotropic background. We conclude that apart from the presence or not of 4FGL~J1256.9+2736, the Coma cluster model fit is expected to be robust with respect to mis-modeling of other components of the ROI (see also Section~\ref{sec:systematic_effects} for systematic effects associated with the diffuse background).

We present in Table~\ref{tab:table_fermi_analysis} the TS values obtained for the point source 4FGL~J1256.9+2736 and the diffuse cluster model component individually, for all the cases considered (we also include a point source model for the cluster as a spatial template for comparison). Because of the degeneracy between 4FGL~J1256.9+2736 and the diffuse cluster model, we also include the TS value for the total (i.e., both components included or excluded when computing the likelihood $\mathcal{L}$ and $\mathcal{L}_0$, see Equation~\ref{eq:TS}). Indeed, in the case of two degenerate sources, the maximum likelihood will not change much when removing one of the two source since the other source will absorb the missing component, in the null hypothesis. Nonetheless, removing the two sources simultaneously may lead to a large change in the maximum likelihood because at least one component is necessary to explain the data. Hence, the TS of each individual degenerate source is expected to be low, but that of the two sources taken together might be large. This is expected to be the case in scenario 3. All the considered {\tt MINOT} radial models give similar TS values, around ${\rm TS} = 24-27$, when replacing 4FGL~J1256.9+2736 (scenario 2). This is similar to the TS value for 4FGL~J1256.9+2736 alone, which is $25.61$ (scenario 1). On the other hand, the cluster emission modeled as a point source gives a significantly lower value, of $17.54$. Except for the radio relic, all the templates based on other data provide a better match to the signal, with the best match reached for the galaxy density and tSZ maps (${\rm TS} \sim 32-34$). This is likely because of the elongation of the Coma cluster, especially toward the southwest, where most of the $\gamma$-ray excess is observed. In the case both 4FGL~J1256.9+2736 and the diffuse cluster emission are included in the ROI model (scenario 3), the TS of each component drastically reduces, highlighting the fact that 4FGL~J1256.9+2736 and the diffuse emission are degenerate. For instance, the TS value reaches only $12.17$ and $12.44$ for the cluster and 4FGL~J1256.9+2736, respectively, in the case of the baseline model. However, the total TS value is higher compared to scenario 1 and 2, reaching about $32$-$36$, but the improvement is only marginal considering the fact that two components, instead of one, are included. Again, this highlight the degeneracy between 4FGL~J1256.9+2736 and the expected cluster signal. We conclude that while a diffuse cluster model (scenario 2) generally provides a better description of the data compared to 4FGL~J1256.9+2736 alone (scenario 1), the two scenarios cannot be significantly discriminate based on their likelihood fit. In Appendix~\ref{app:mc_realizations}, we also further discuss the agreement between scenario 2 and the data using a Monte Carlo realization.

Finally, we also consider the case of using a power law for the photon spectrum of Coma. In the case of our baseline spatial model ($n_{\rm CRp} \propto n_e^{1/2}$), in scenario 2, we obtain a best-fit photon index of $2.45 \pm 0.19$ for a TS value of 27.37. As expected, the overall spectrum is slightly flatter than in the case of our physical model. Indeed, the photon spectrum vanishes at low energy for a given CRp energy slope with our physical model, thus leading to a harder spectrum in this regime. The power law model thus averages the two regimes. Nevertheless, we note that the high energy photon spectrum does not strictly match the CRp energy spectrum \citep[see][for more details]{Adam2020}. The TS value is nearly the same as in the case of the physical spectrum and given the available signal-to-noise ratio, it is not possible to discriminate the two.

The modeling and fitting procedure described in Section~\ref{sec:ROI_modeling} and in this Section is also validated using a null test described in Appendix~\ref{app:null_test}. This shows that no cluster detection is obtained when including a cluster in the sky model close to other sources with similar background as around Coma.

\begin{table*}[]
	\caption{\small List of TS values and flux in the 200 MeV - 300 GeV band for all the models considered.}
	\begin{center}
	\resizebox{\textwidth}{!} {
	\begin{tabular}{c|cc|ccc|c}
	\hline
	\hline
	 Scenario$^{(\star)}$ & \multicolumn{2}{c|}{Sky model} &  \multicolumn{3}{c|}{TS} & Cluster flux \\
	& 4FGL~J1256.9+2736 & Cluster & Total$^{(\dagger)}$ & 4FGL~J1256.9+2736 & Cluster &  ($10^{-10}$ s$^{-1}$ cm$^{-2}$) \\
	\hline
	\hline	
	1 & included & None & 25.61 & 25.61 & -- & 0 \\
	\hline
	2 & replaced & Point-source & 17.54 & -- & 17.54 & $11.03 \pm 3.87$  \\
	2 & replaced & Compact model ($n_{\rm CRp} \propto n_{\rm gas}$) & 24.84 & -- & 24.84 & $13.38 \pm 4.05$   \\
	2 & replaced & Extended model ($n_{\rm CRp} \propto n_{\rm gas}^{1/2}$) & 27.00 & -- & 27.00 & $15.61 \pm 4.25$  \\
	2 & replaced & Flat model ($n_{\rm CRp} = {\rm constant}$) & 25.11 & -- & 25.11 & $18.94 \pm 4.78$  \\
	2 & replaced & Isobar ($n_{\rm CRp} = P_{\rm gas}$) & 24.56 & -- & 24.56 & $13.06 \pm 4.01$   \\
	\hline
	2 & replaced & tSZ & 32.18  & -- & 32.18 & $17.29 \pm 4.44$  \\
	2 & replaced & X-ray & 28.18 & -- & 28.18 & $14.66 \pm 4.20$  \\
	2 & replaced & Galaxies & 34.23 & -- & 34.23 & $17.37 \pm 4.45$  \\
	2 & replaced & Radio halo & 29.65 & -- & 29.65 & $15.80 \pm 4.25$   \\
	2 & replaced & Radio relic & 10.70 & -- & 10.70 & $8.02 \pm 3.99$ \\
	\hline
	3 & included & Point-source & 32.24 & 18.75 & 7.60 & $4.45 \pm 3.93$ \\
	3 & included & Compact model ($n_{\rm CRp} \propto n_{\rm gas}$) & 34.67 & 14.29 & 11.50 & $6.94 \pm 4.11$ \\
	3 & included & Extended model ($n_{\rm CRp} \propto n_{\rm gas}^{1/2}$) & 34.35 & 12.44 & 12.17 & $8.61 \pm 4.23$ \\
	3 & included & Flat model ($n_{\rm CRp} = {\rm constant}$) & 34.67 & 13.02 & 9.32 & $10.39 \pm 4.79$ \\
	3 & included & Isobar ($n_{\rm CRp} = P_{\rm gas}$) & 34.72 & 14.53 & 11.43 & $6.75 \pm 4.07$ \\
	\hline
	3 & included & tSZ & 34.35 & 8.06 & 15.79 & $10.77 \pm 4.50$ \\
	3 & included & X-ray & 35.23 & 11.79 & 13.11 & $7.86 \pm 4.26$ \\
	3 & included & Galaxies & 36.10 & 6.60 & 17.99 & $11.32 \pm 4.51$ \\
	3 & included & Radio halo & 35.28 & 10.53 & 13.95 & $9.20 \pm 4.31$ \\
	3 & included & Radio relic & 26.18 & 23.25 & 0.78 & $2.06 \pm 2.60$ \\
	\hline
	\end{tabular}
	}
	\end{center}
	{\small {\bf Notes.} 
	$^{(\star)}$ See subsection~\ref{sec:ROI_modeling} for the definition. 
	$^{(\dagger)}$ The total model TS corresponds to the TS of both the cluster and 4FGL~J1256.9+2736. It is equal to that of the cluster when 4FGL~J1256.9+2736 is excluded from the sky model, and that of 4FGL~J1256.9+2736 when no cluster is included in the sky model.}
	\label{tab:table_fermi_analysis}
\end{table*}

\subsection{Comparison between data and model}\label{sec:comparison_data_model}
\begin{figure*}
	\centering
	\rule{17cm}{0.01cm}
	\put(-450,5){\bf \footnotesize 200 MeV - 300 GeV}
	\put(-280,5){\bf \footnotesize  200 MeV - 1 GeV}
	\put(-120,5){\bf \footnotesize  1 GeV - 300 GeV}
	
	\rule{17cm}{0.01cm}
	\put(-350,2){\footnotesize Point source 4FGL~J1256.9+2736 replaced (scenario 2)}

	\includegraphics[width=0.3\textwidth]{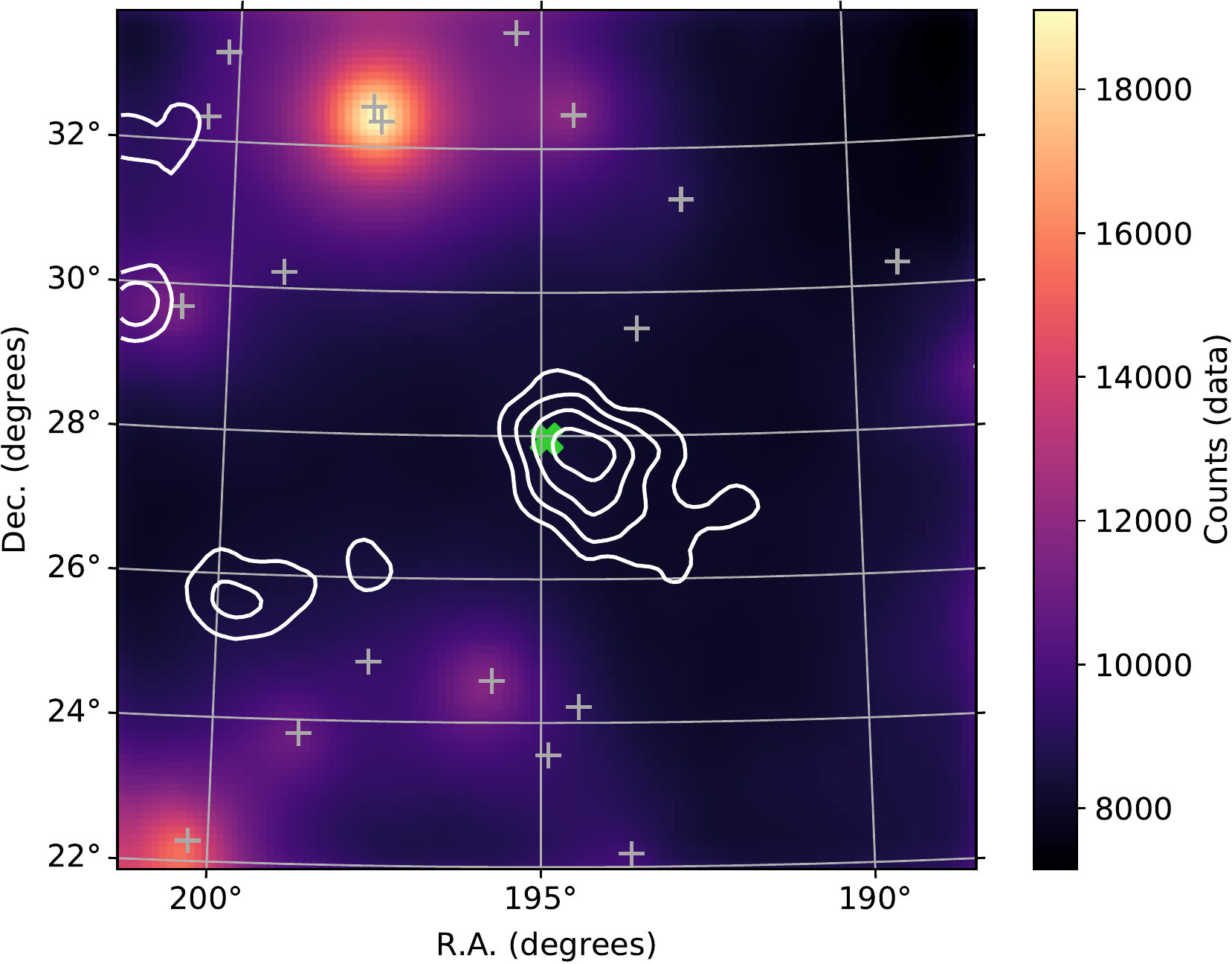}
	\put(-180,60){\rotatebox{90}{\bf \footnotesize Data}}
	\includegraphics[width=0.3\textwidth]{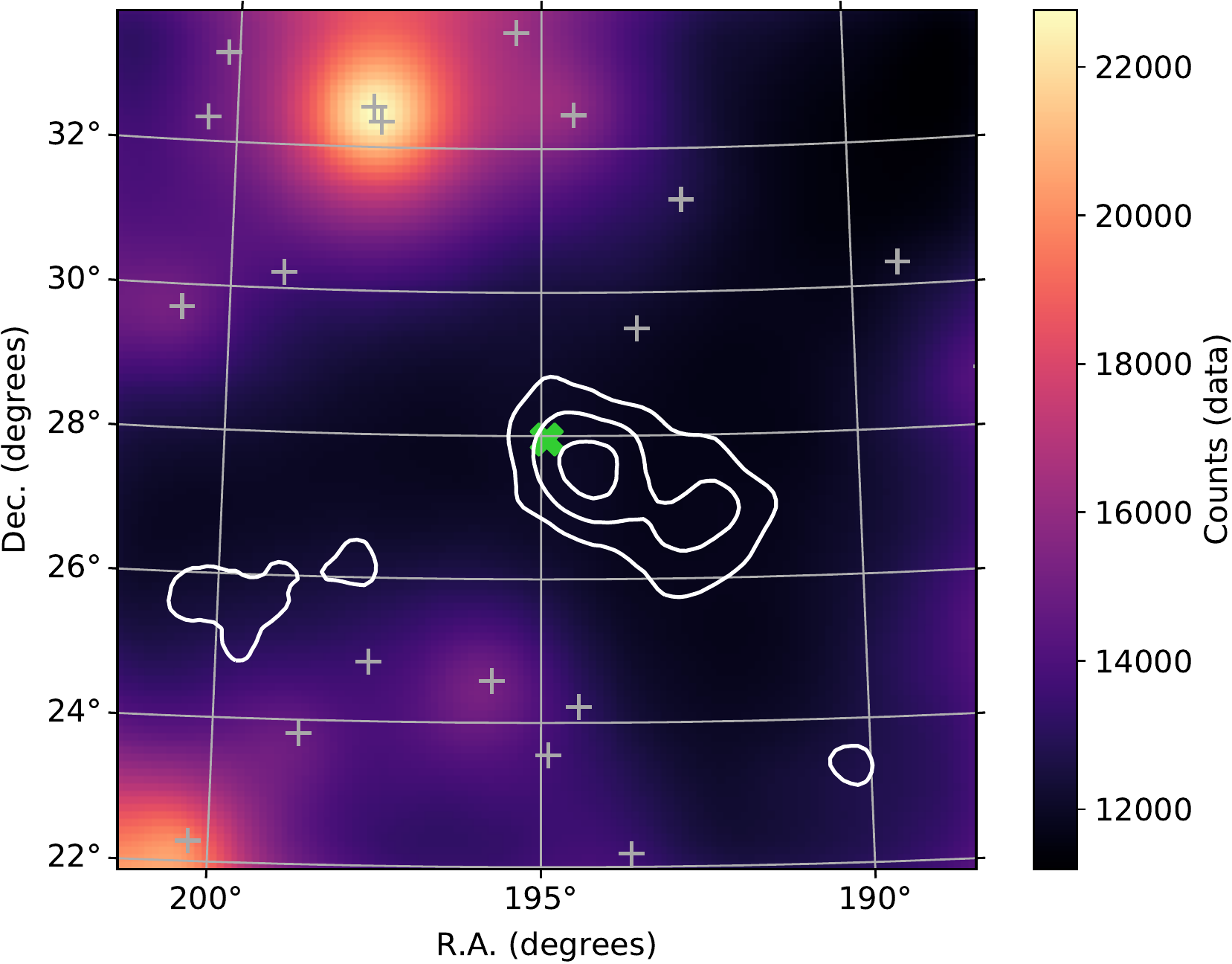}
	\includegraphics[width=0.3\textwidth]{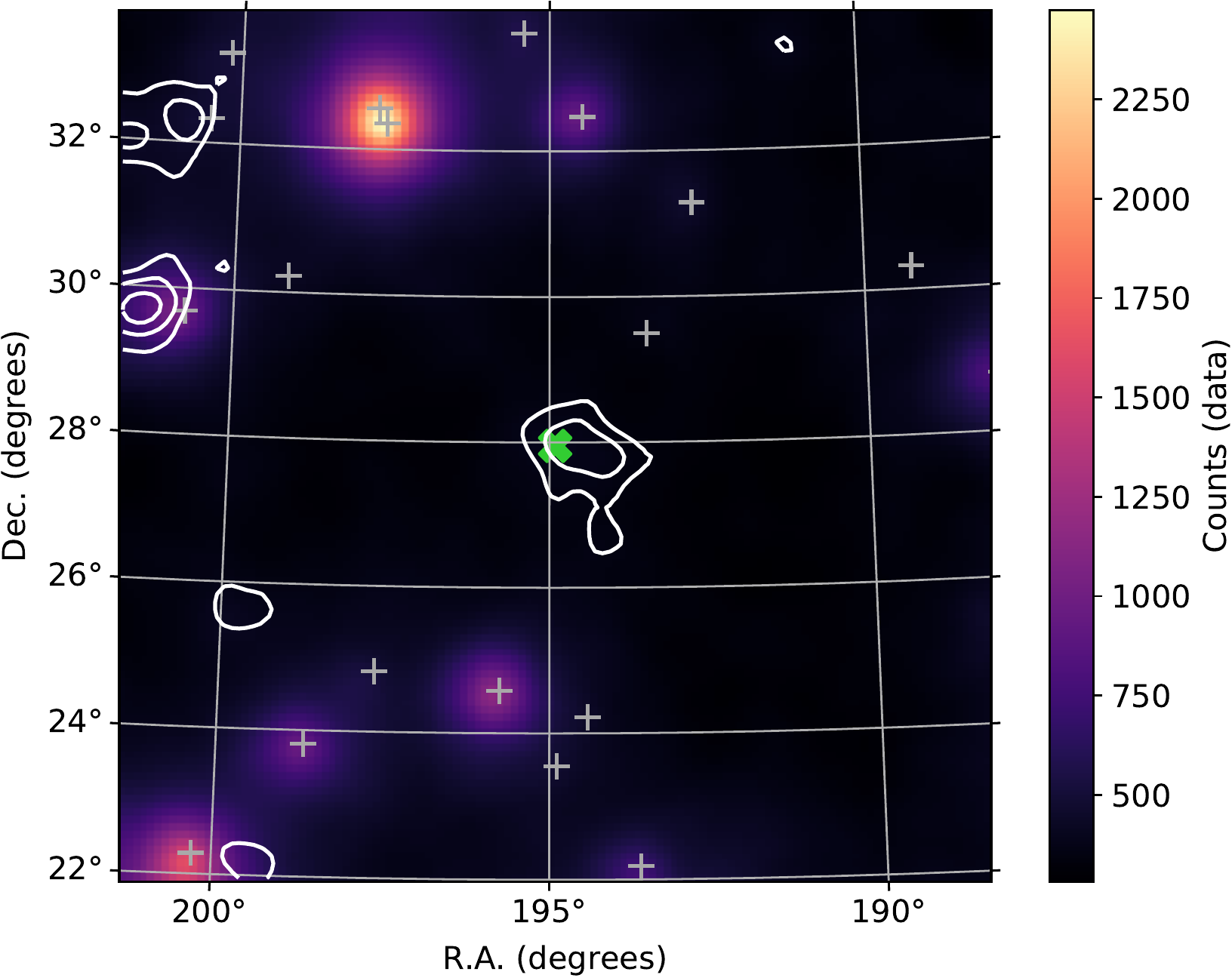}
	\includegraphics[width=0.3\textwidth]{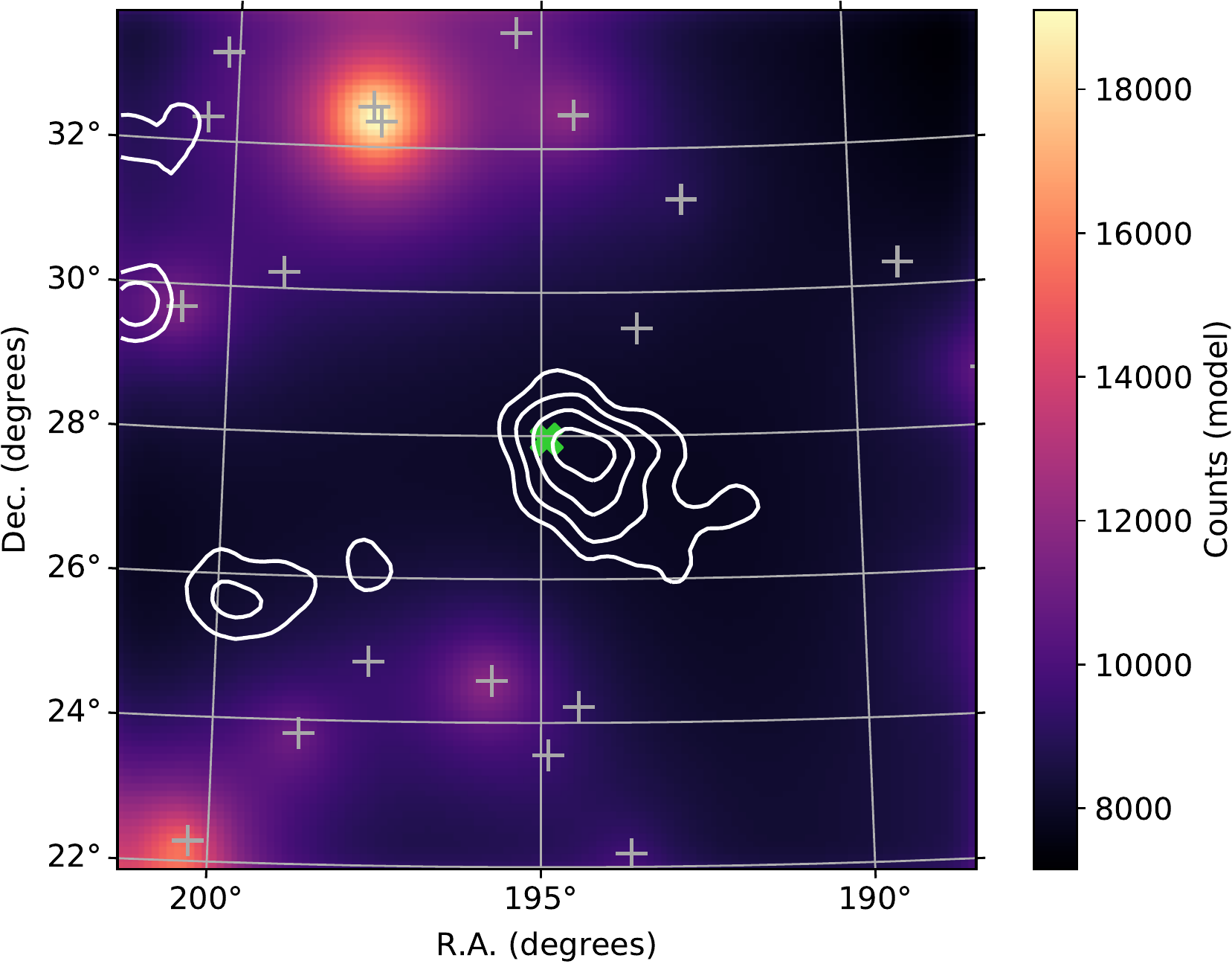}
	\put(-180,60){\rotatebox{90}{\bf \footnotesize Model}}
	\includegraphics[width=0.3\textwidth]{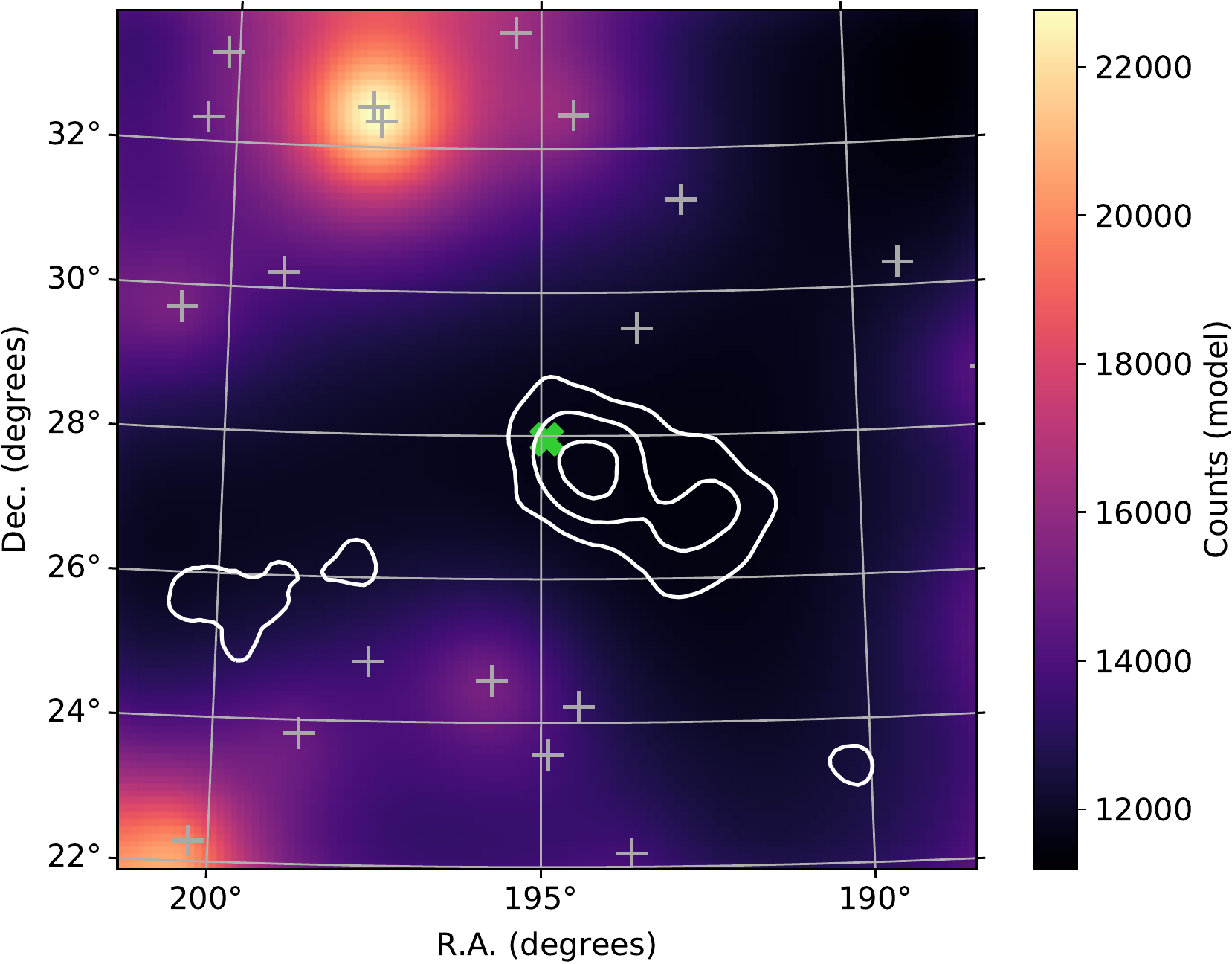}
	\includegraphics[width=0.3\textwidth]{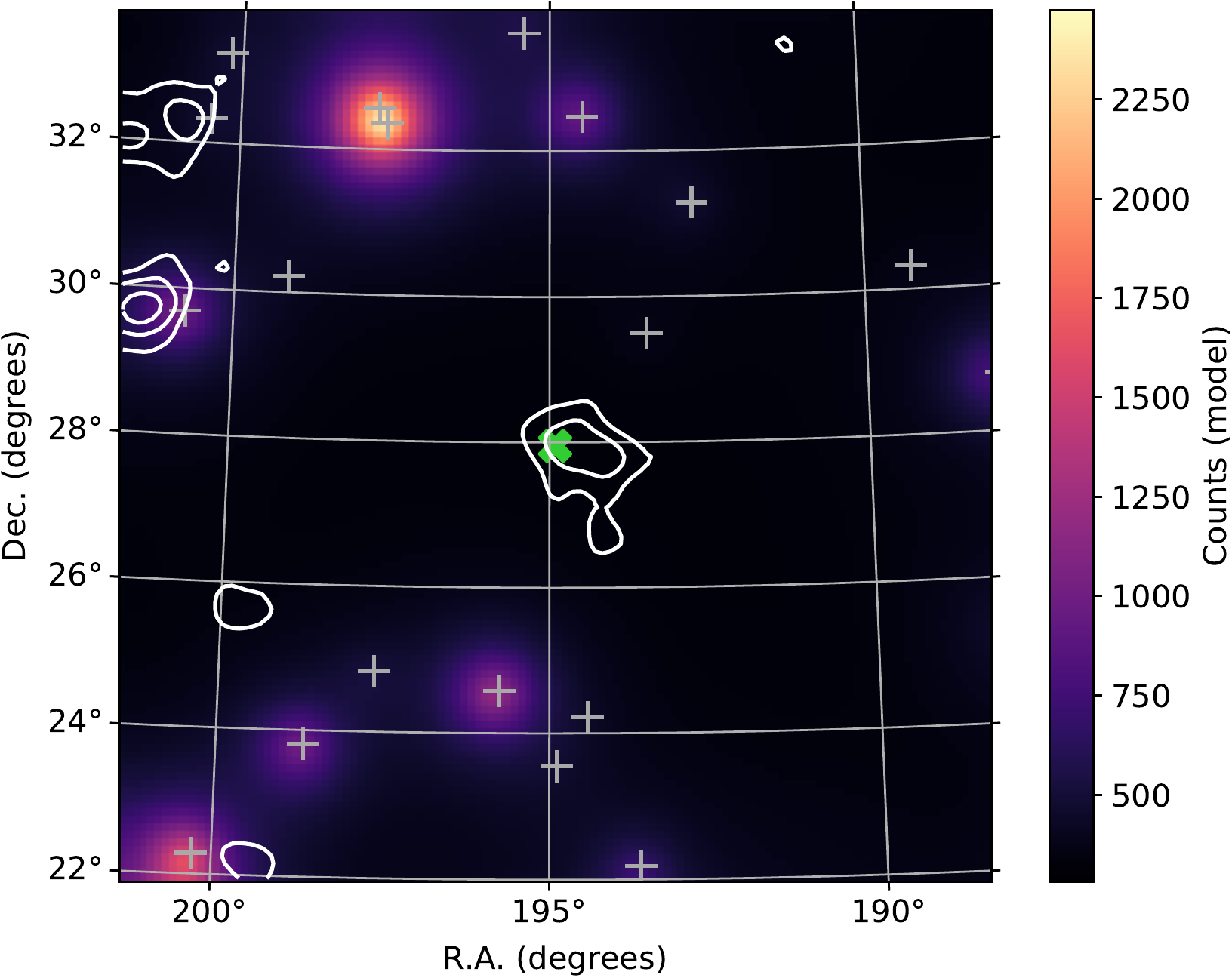}
	\includegraphics[width=0.3\textwidth]{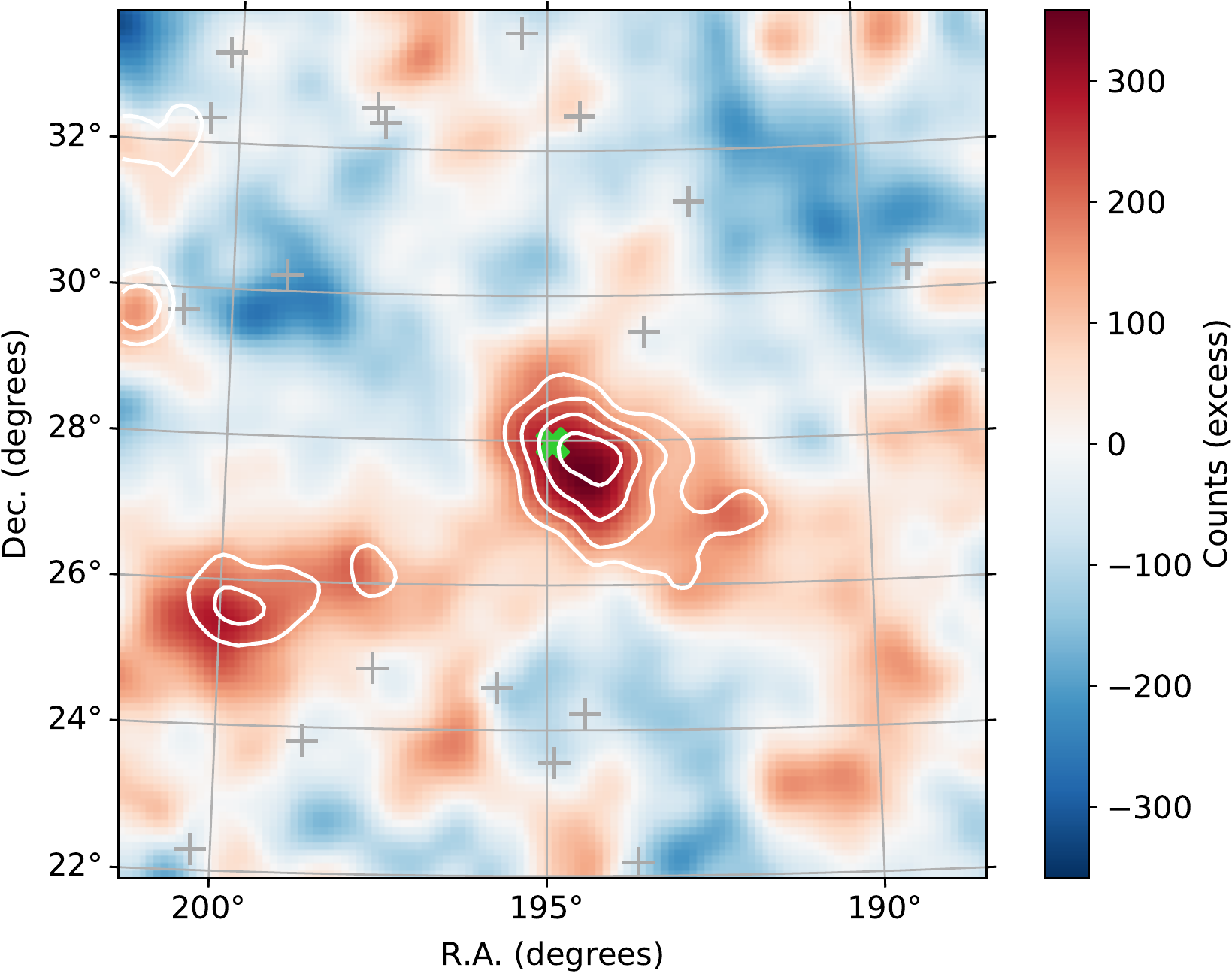}
	\put(-180,30){\rotatebox{90}{\bf \footnotesize Residual (without Coma)}}
	\includegraphics[width=0.3\textwidth]{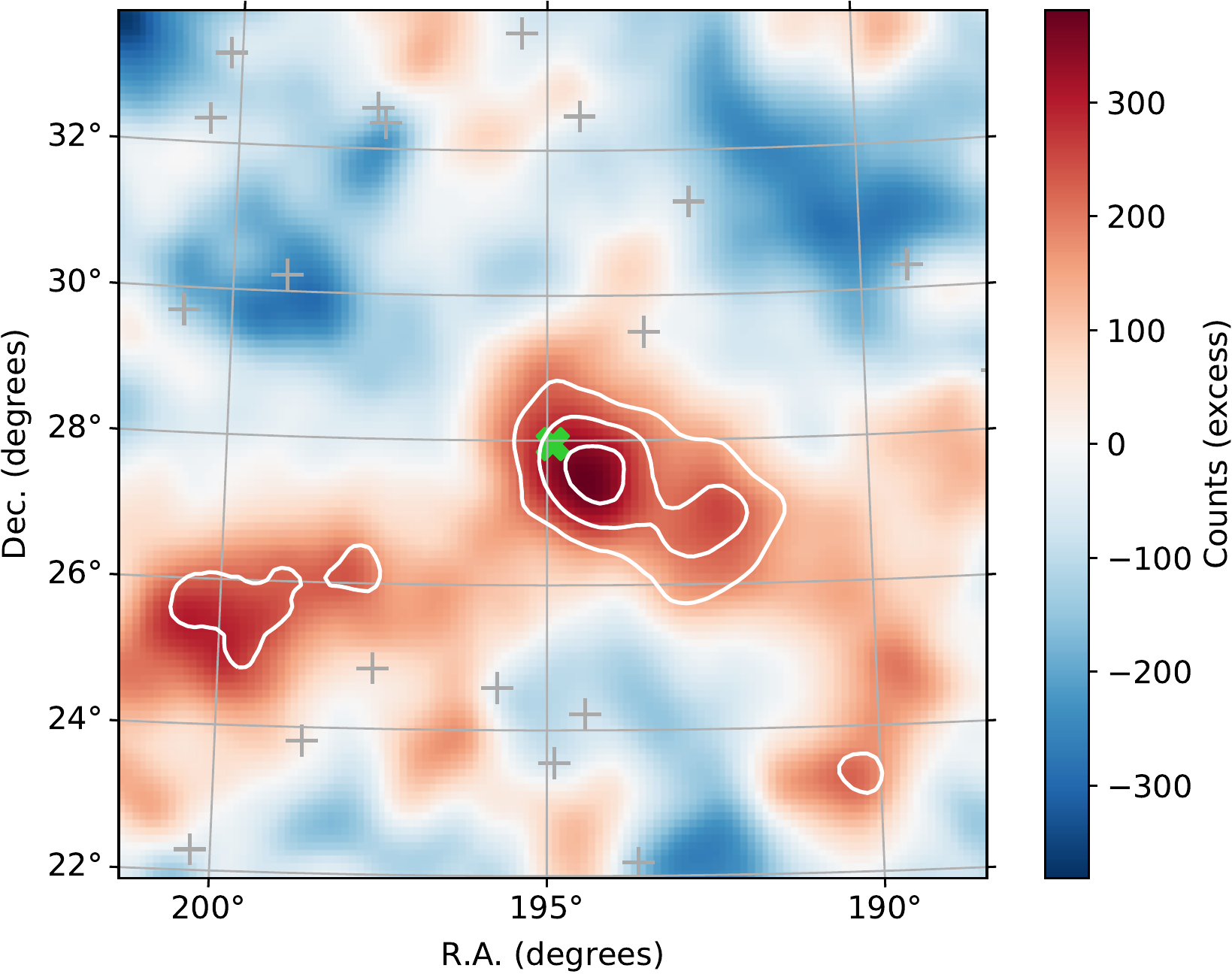}
	\includegraphics[width=0.3\textwidth]{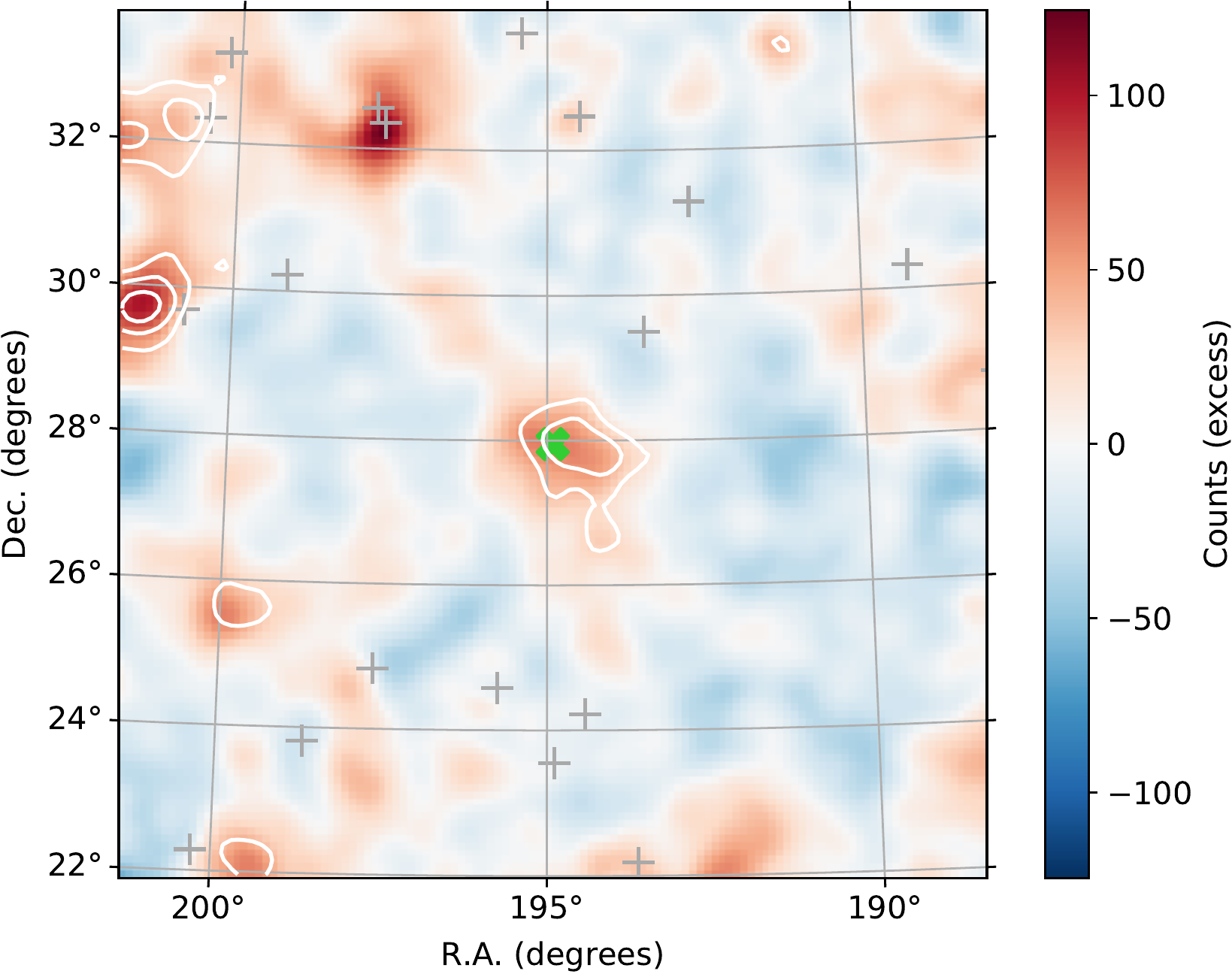}
	\includegraphics[width=0.3\textwidth]{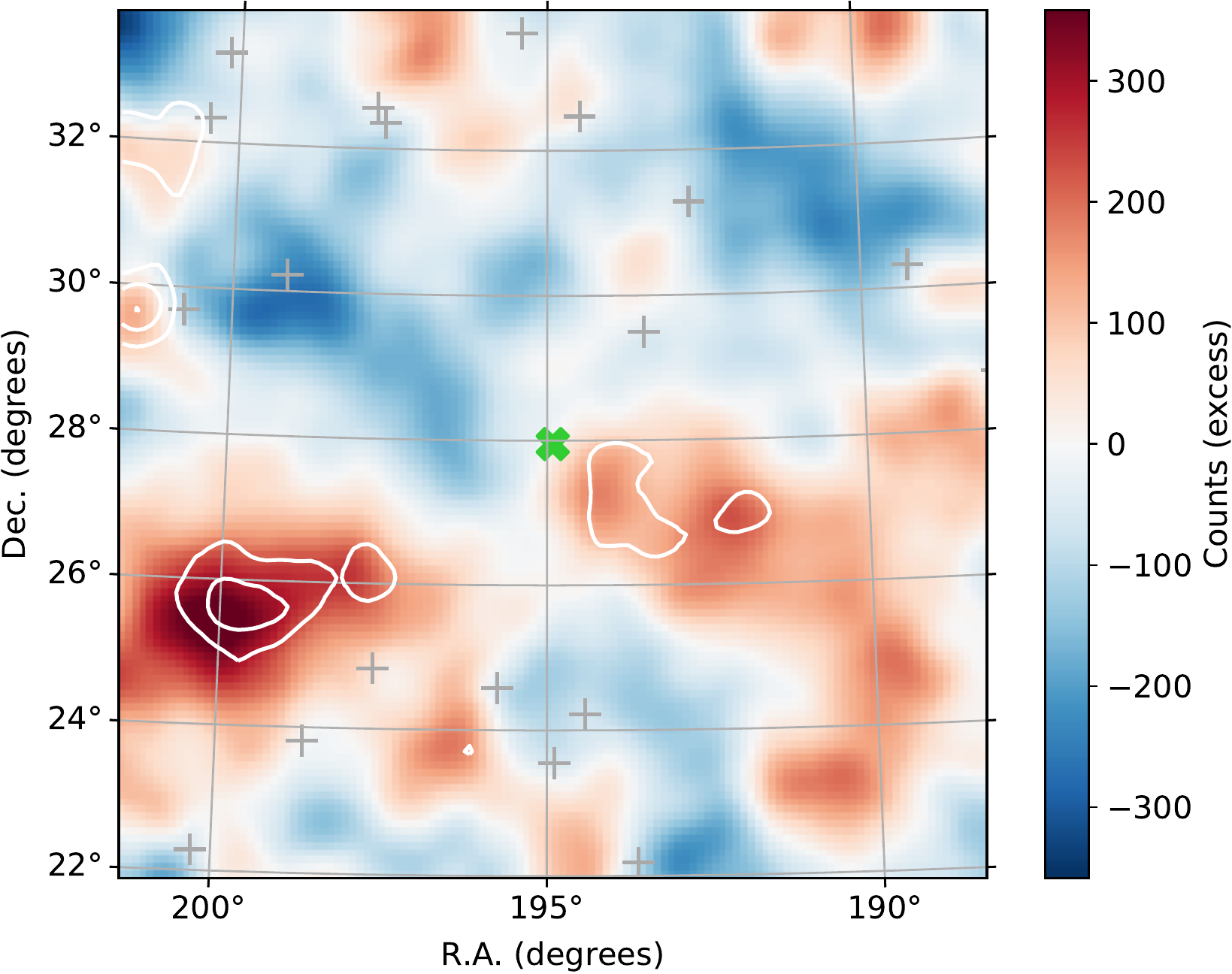}
	\put(-180,50){\rotatebox{90}{\bf \footnotesize Residual (total)}}
	\includegraphics[width=0.3\textwidth]{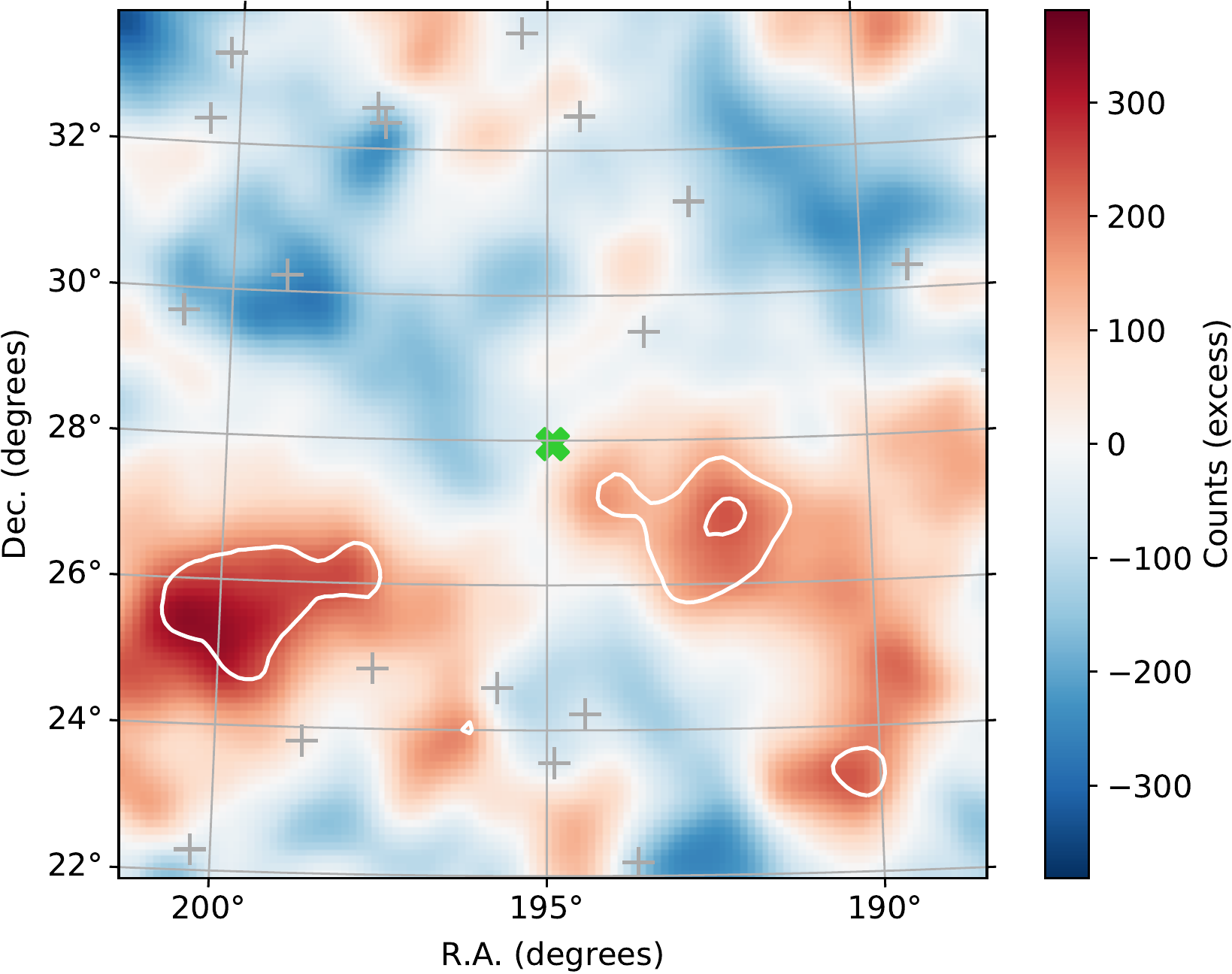}
	\includegraphics[width=0.3\textwidth]{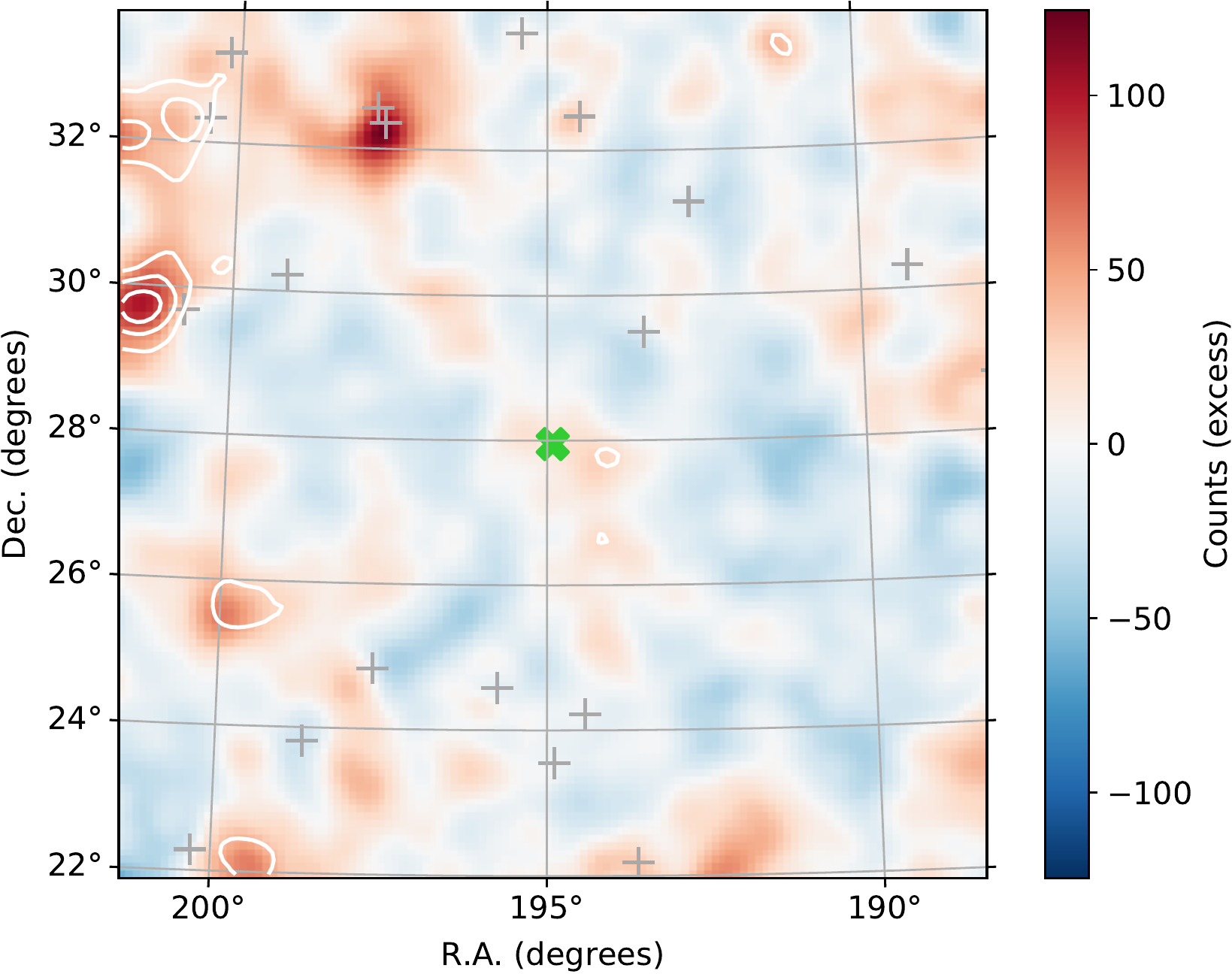}

	\rule{17cm}{0.01cm}
	
	\rule{17cm}{0.01cm}
	\put(-350,2){\footnotesize Point source 4FGL~J1256.9+2736 included (scenario 1)}

	\includegraphics[width=0.3\textwidth]{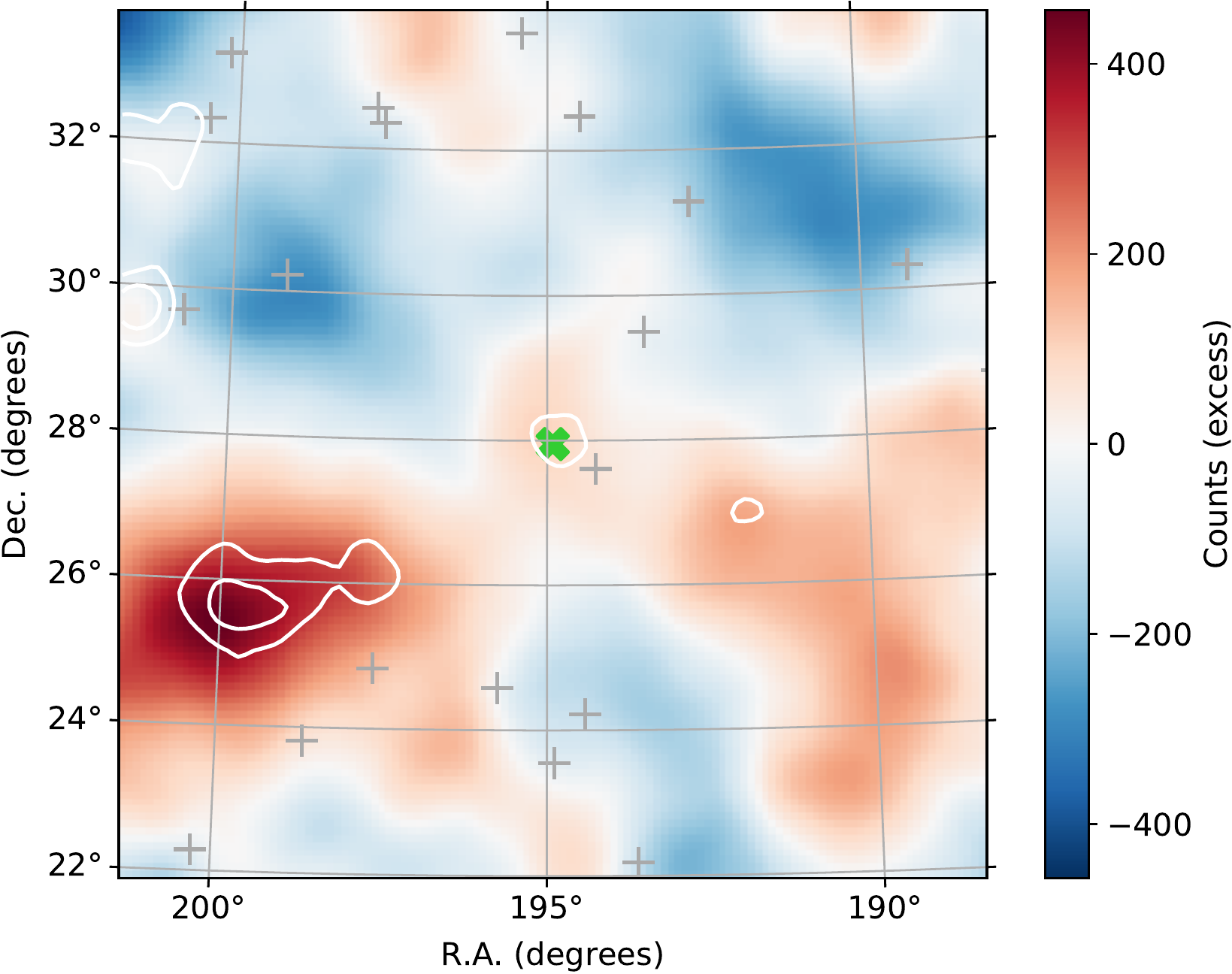}
	\put(-180,30){\rotatebox{90}{\bf \footnotesize Residual (without Coma)}}
	\includegraphics[width=0.3\textwidth]{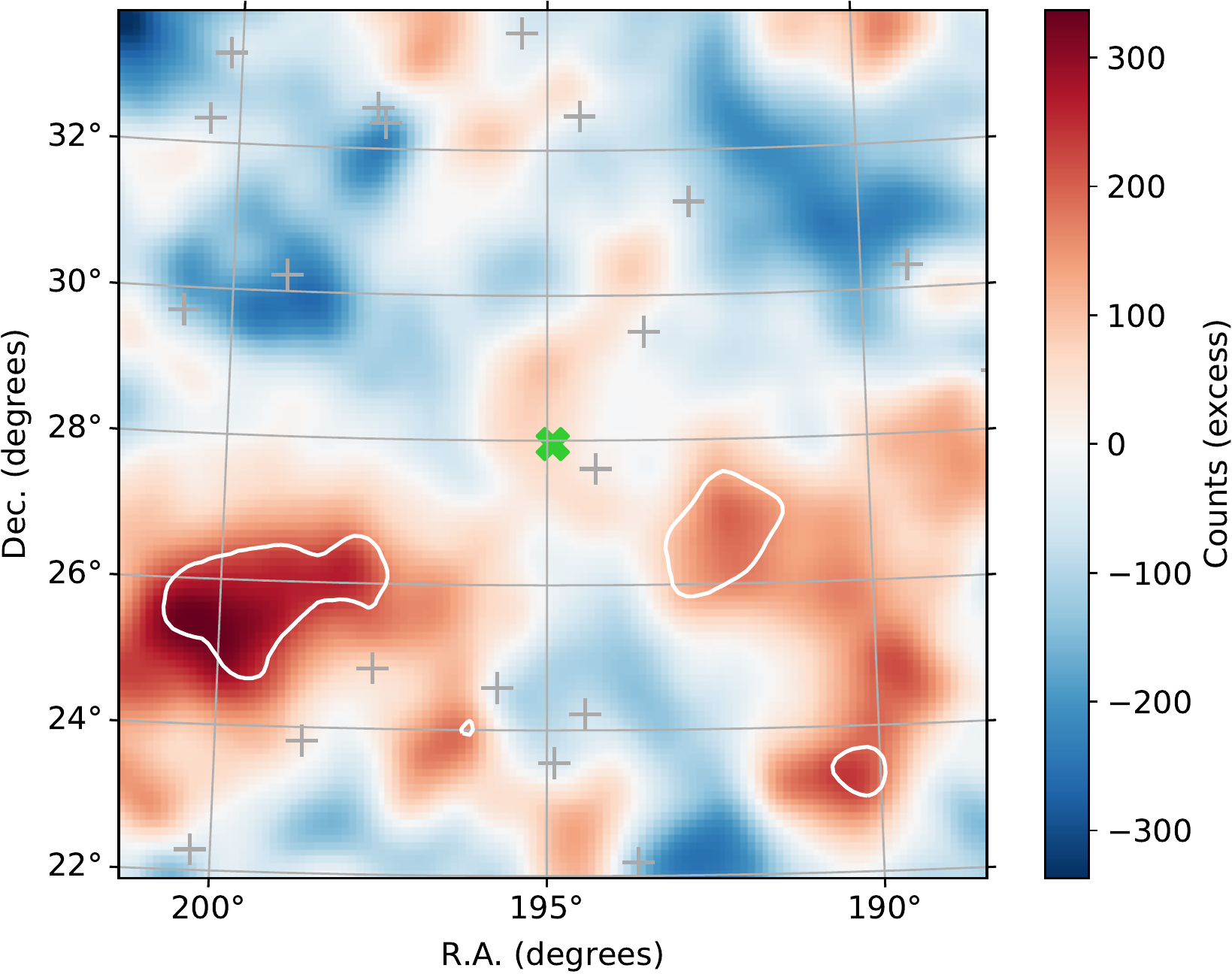}
	\includegraphics[width=0.3\textwidth]{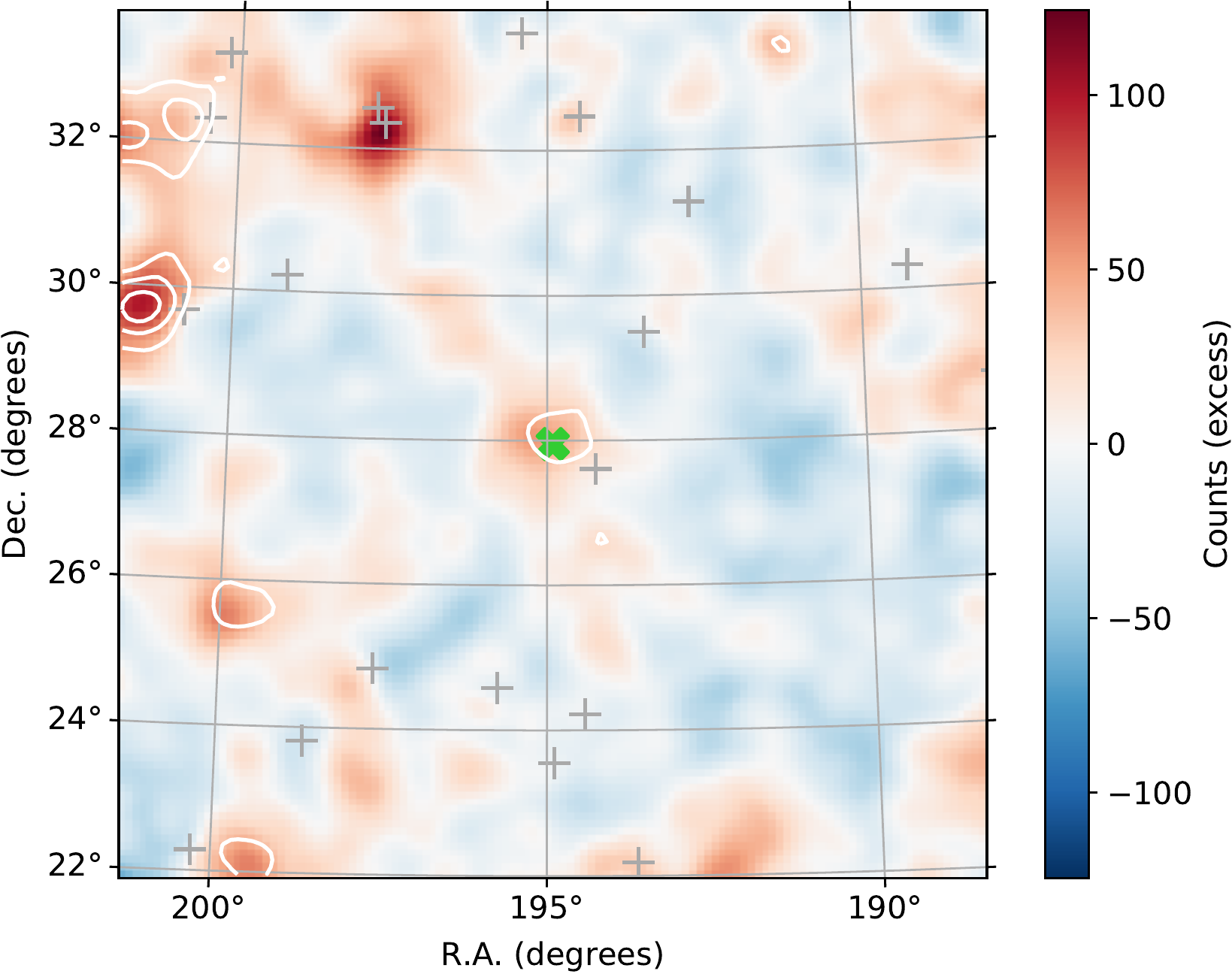}
	\caption{\small \textit{Fermi}-LAT imaging centered on Coma and comparison to the model (baseline model, $n_{\rm CRp} \propto n_e^{1/2}$). We show the \textit{Fermi}-LAT data (first row), the best-fit model (second row), the residual excluding the cluster model (third row), and the residual with respect to the total model (fourth row).  We also show the residual excluding the cluster when accounting for 4FGL~J1256.9+2736 (fifth row). The left, middle and right columns correspond to the total (200 MeV - 300 GeV), to the low (200 MeV - 1 GeV), and high (1 GeV - 300 GeV) energy range, respectively. The white contours correspond to ${\rm TS}=[4,9,16,25$]. The gray crosses provide the location of the 4FGL sources and the green cross the Coma center.}
\label{fig:Fermi_imaging}
\end{figure*}

In order to check how the data compare to the different models (in the different scenarios), we computed maps of the data, model and residual (with and without Coma; with and without 4FGL~J1256.9+2736) in three energy bins: from 200 MeV to 300 GeV (the total considered range), from 200 MeV to 1 GeV, and from 1 GeV to 300 GeV. This is shown in Figure~\ref{fig:Fermi_imaging} in the case of the baseline model. 

When excluding both the diffuse cluster emission and 4FGL~J1256.9+2736, we observe a significant excess near the center of the map (where the green cross indicates the Coma reference center). This excess is also visible independently in the high and low energy bands, although with lower significance (${\rm TS}>16$ and ${\rm TS}>9$, respectively). The peak of the excess is slightly offset in the southwest direction with respect to our reference center. The position angle of this elongation agrees in all energy bins, although the elongation itself is more pronounced at low energy. When including 4FGL~J1256.9+2736 in the sky model, the central excess disappears almost entirely (especially since 4FGL~J1256.9+2736 can absorb most of the signal from Coma, if any, as indicated by the correlation matrix and discussed above). The comparison of the \textit{Fermi}-LAT excess to other data will be discussed further in Section~\ref{sec:multiwavelength_comparison}. We also show in Appendix~\ref{app:mc_realizations} that the offset between 4FGL~J1256.9+2736 and the Coma reference center agrees with 4FGL~J1256.9+2736 corresponding to the peak of the diffuse ICM emission (i.e., scenario 2).

In addition to the central excess, we also see two other excesses near RA,Dec = 200, 25.5 degrees and RA,Dec = 202, 29.5 degrees (${\rm TS} > 9$). Another excess count is seen near RA,Dec = 197.5, 32 degrees (in the high energy bin), but its TS remains low because our cluster model does not match the spatial and/or spectral shape of this signal well. All these other excesses remain relatively small and given their significance, it is not clear whether they correspond to noise fluctuations, the mis-modeling of existing sources, or new sources that are not included in the 4FGL catalog. Given their distances to the central excess, we do not expect that they would lead to any significance bias in our analysis.

\begin{figure}
\centering
\includegraphics[width=0.45\textwidth]{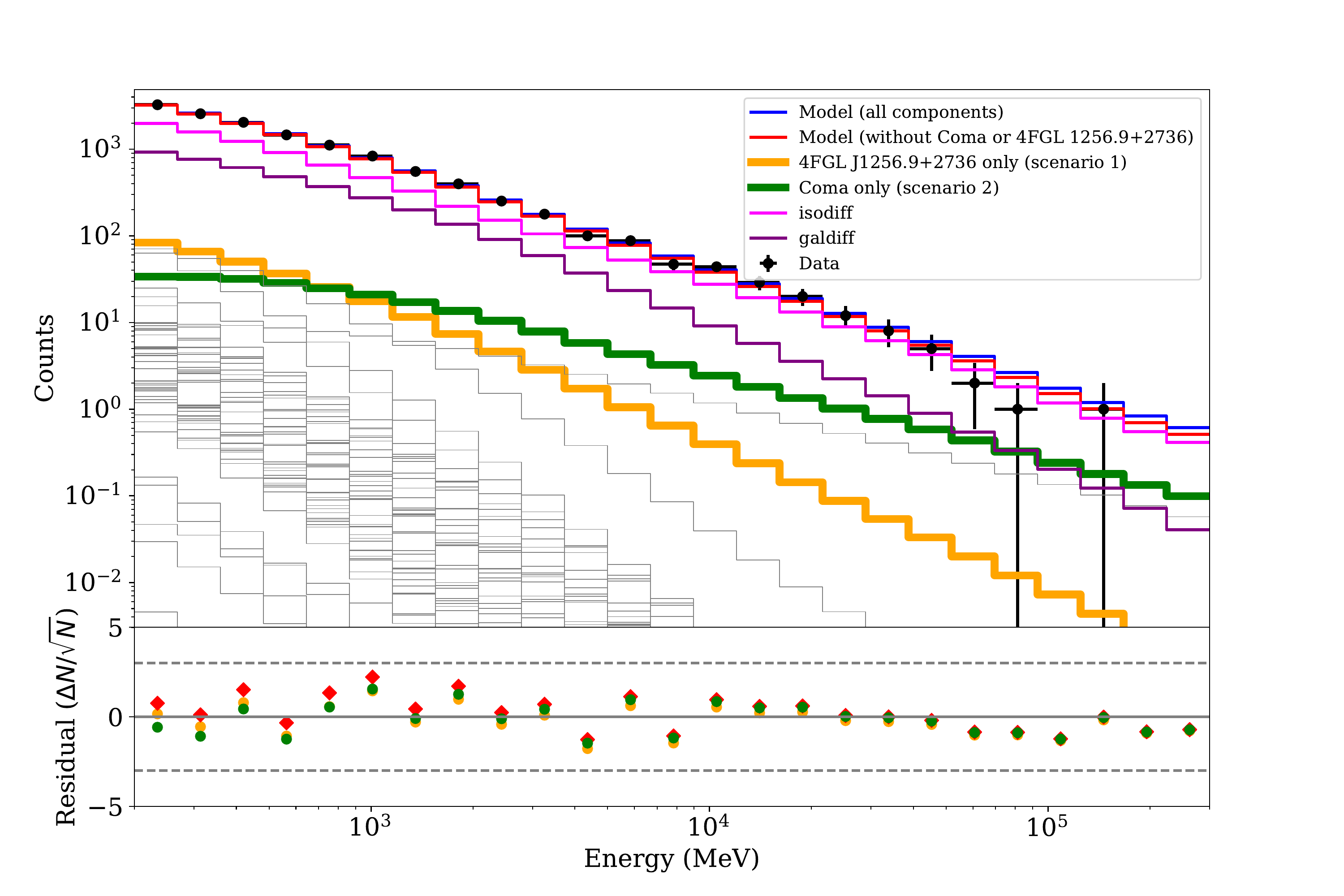}
\caption{\small \textit{Fermi}-LAT counts as a function of energy, computed within $3 \theta_{500}$ from the Coma center. The blue line shows the total model in the case of scenario 2 (but it is not distinguishable from scenario 1 in this figure), the red line shows the model when excluding the Coma cluster diffuse emission or 4FGL~J1256.9+2736, the orange line show the contribution from 4FGL~J1256.9+2736 in the case of scenario 1, and the green line show the Coma cluster diffuse contribution in the case of scenario 2. The contribution from the different sources in the ROI is also indicated as gray lines, except for the isotropic and diffuse backgrounds, given in magenta and purple, respectively. The bottom panel show the residual between the data and the model, in red when both the Coma cluster diffuse emission or 4FGL~J1256.9+2736 are excluded from the model, in orange in the case of scenario 1 and in green in the case of scenario 2.}
\label{fig:Fermi_residual_spectrum}
\end{figure}

We also compute the counts as a function of energy observed within $3 \times \theta_{500}$ from the cluster center and compare it to the different components of the model in Figure~\ref{fig:Fermi_residual_spectrum}. As we can see, the dominant components are from the isotropic and diffuse galactic interstellar emission at almost all energies. The signal from the Coma cluster (in the case of scenario 2) is about an order of magnitude below depending on the energy, but is the dominant compact source except at very low energies where the larger PSF leads to leakage from strong sources within the Coma central region (we note that the source 4FGL~J1253.8+2929, northwest of Coma, falls within $3 \times \theta_{500}$). Similarly in the case of scenario 1, 4FGL~J1256.9+2736 is the dominant compact source at low energy but not at high energy as its spectrum is steeper. Including the Coma cluster or 4FGL~J1256.9+2736 in the model improves the residual, as expected from the result of the likelihood fit, but no significant difference is observed between the two scenarios accounting for the error bars. Given this spectrum, it is very unlikely that other compact sources from the ROI induce the observed excess within $3 \times \theta_{500}$ of the Coma cluster.

\begin{figure}
\centering
\includegraphics[width=0.45\textwidth]{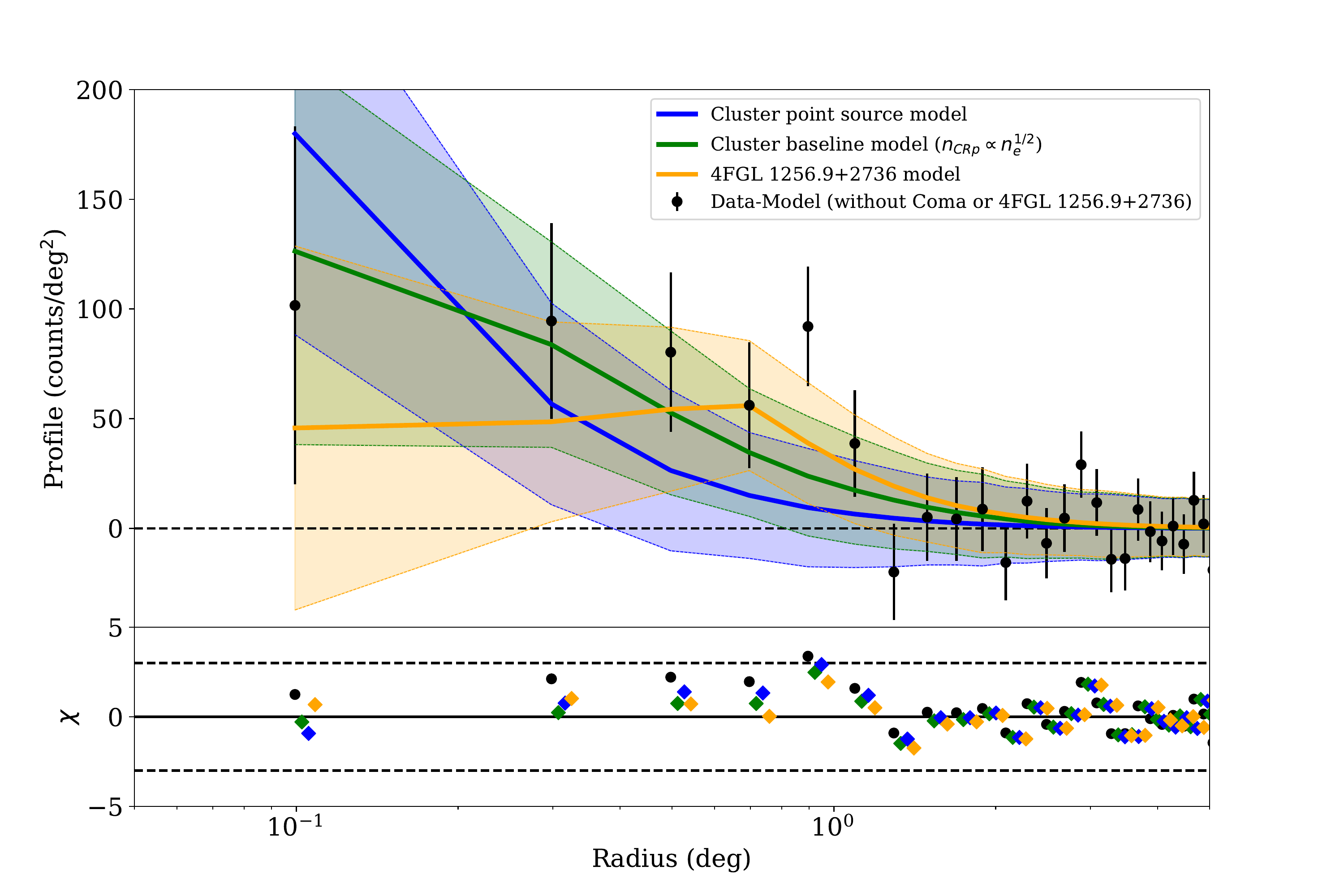}
\caption{\small Radial profile of the \textit{Fermi}-LAT excess count in the total energy band (200 MeV - 300 GeV) for different models. The black data points give the excess count when both 4FGL~J1256.9+2736 or a cluster model is excluded from the model. We show the best-fit model in the case of scenario 1 (4FGL~J1256.9+2736 only, orange line) and in the case of scenario 2 (cluster diffuse model only; baseline in green and point source in blue). The shaded areas correspond to the expected Poisson uncertainties given the respective model. The bottom panel provides the residual normalized by the error bar with a similar color code.}
\label{fig:Fermi_residual_profile}
\end{figure}

In order to address to which extent the \textit{Fermi}-LAT data are sensitive to the shape of our radial model, we compute the profile of the data and model in Figure~\ref{fig:Fermi_residual_profile}. Although each of the 5 first bins (up to 1 degrees extent) are marginally inconsistent with the model when the Coma cluster emission or 4FGL~J1256.9+2736 are not accounted for, they all point to an excess emission at a level of 0.5-3.5 $\sigma$, and thus correspond to a clear overall excess. The baseline cluster model provides a good fit to the data and significantly improves the residual (scenario 2). The point source cluster model, also shown for comparison, is too peaked with respect to the observations. However, it is not in clear disagreement with the data given the error bars. This is in agreement with the results of Table~\ref{tab:table_fermi_analysis}, where the likelihood ratio between the two model is $\Delta {\rm TS} = 9.45$, providing only a hint that the extended model is more appropriate than the point source model. In the case of scenario 1 (4FGL~J1256.9+2736 only), the agreement with the data is also good.

\subsection{Extraction of the cluster spectral energy distribution}\label{sec:sed_extraction}
Having a model for the sky in hand, we used the {\tt sed} function of {\tt Fermipy} to extract the SED of the Coma diffuse emission in the different cases tested in this paper. The {\tt sed} function fits for the flux normalization of a source in independent bins of energy. To do so, we allowed the local photon slope to vary according to the {\tt MINOT} global spectral model and we fixed the background component. However, we note that only minor differences are observed when fixing the slope or leaving the background free. In addition to the flux and error bars in each bin, the {\tt sed} method provides the full likelihood scan for the normalization value in each bin, and we will use this information later in Section~\ref{sec:implication_for_CR_content}.

At this stage, we use the SED results to compute the flux, integrated between 200 MeV and 300 GeV, of the Coma diffuse emission. The measured values are listed in Table~\ref{tab:table_fermi_analysis} for the different models and scenario which we test. We note that in the case of scenario 1 (i.e., no cluster in the model), the flux is by definition equal to zero. The flux ranges from about 13 to $19 \times 10^{-10}$ ph s$^{-1}$ cm$^{-2}$ for the radially symmetric models in scenario 2. The multiwavelength templates lead to similar fluxes, except for the radio relic template for which the flux is only about $9 \times 10^{-10}$ ph s$^{-1}$ cm$^{-2}$. In case of scenario 3, part of the flux is absorbed in the 4FGL~J1256.9+2736 component. This leads to a flux lower by a factor of about two, depending on the cluster model considered.

\subsection{Light curve}
\begin{figure}
\centering
\includegraphics[width=0.45\textwidth]{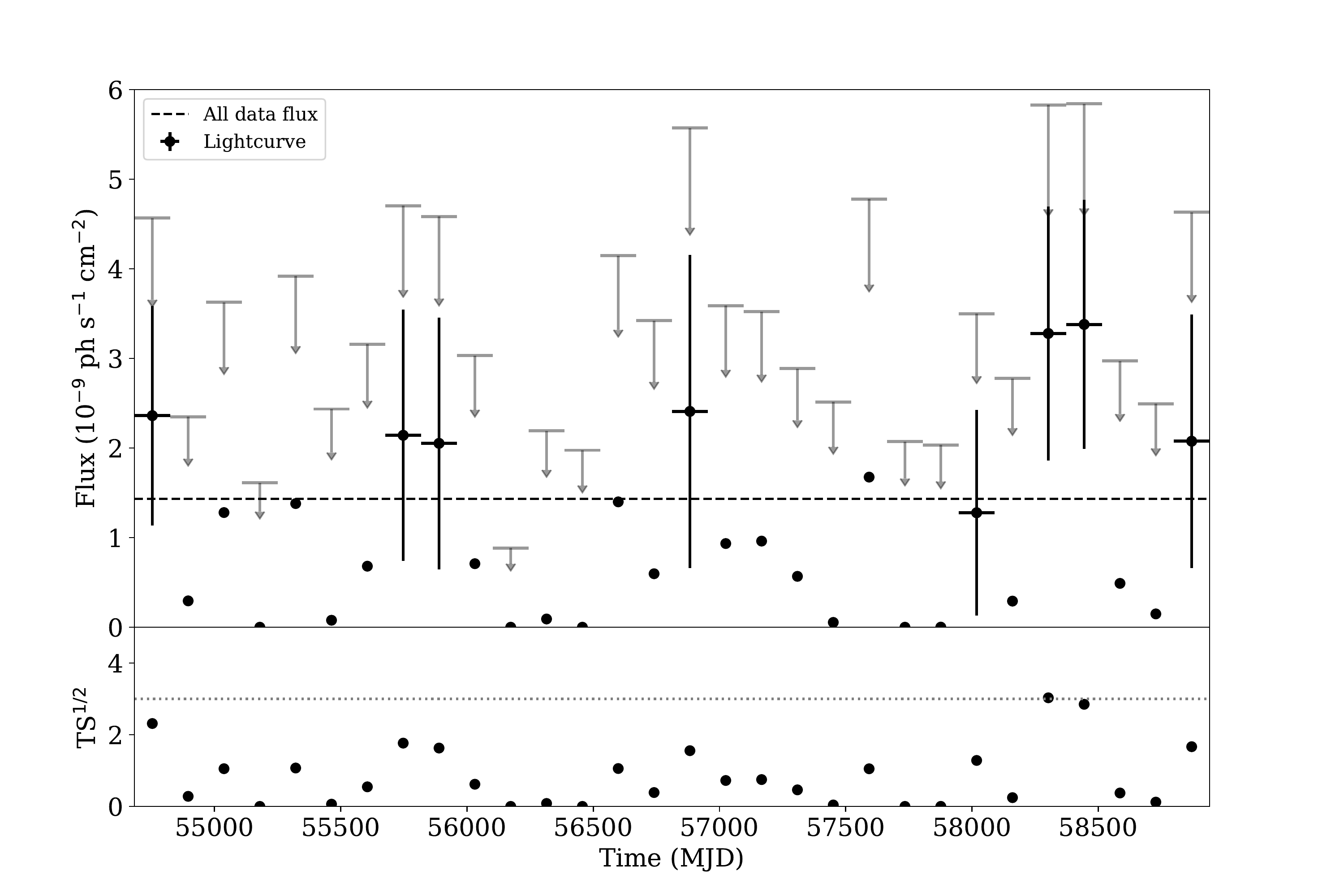}
\caption{\small Light curve associated with the Coma cluster model fit, for the full dataset, in bins of about 4.5 months. for clarity, error bars are shown only for points that are mode than $1\sigma$ away from zero, and upper limits (95\% confidence interval) are also given. The bottom panel presents the square root of the TS associated with the source.}
\label{fig:Fermi_residual_lightcurve}
\end{figure}

In order to check that the observed signal is consistent with diffuse ICM emission, we compute the light curve of the signal. Indeed, while many $\gamma$-ray sources are variable, the $\gamma$-ray emission from galaxy clusters is expected to be steady, at least over the timescale of any human observation, and observing a burst would allow us to rule out the cluster ICM origin of the signal. We thus compute the light curve of the Coma cluster diffuse signal using the {\tt lightcurve} function from {\tt Fermipy}, in the case of scenario 2. We extract the flux within 30 bins, which corresponds to about 4.5 month per bin and report the result in Figure~\ref{fig:Fermi_residual_lightcurve}. Given the significance of the signal for the full dataset, we only expect upper limits in each bin. As we can see, no significant burst is observed. The upper limit excludes the expected flux at 95\% confidence limit in one out of 30 bins (i.e., fewer than 5\%) in agreement with expectation. Therefore, the light curve is consistent with the observed signal being associated with the diffuse Coma cluster ICM emission.

\subsection{Systematic effects}\label{sec:systematic_effects}
\begin{table*}
	\caption{\small Estimation of the systematic effects associated with the \textit{Fermi}-LAT analysis.}
	\begin{center}
	\begin{tabular}{c|ccc}
	\hline
	\hline
	Type & Range explored & Change in $\sqrt{\rm TS}$ & Change in the total flux \\
	\hline
	Diffuse background & alternative models & 20\% & 40\% \\
	Energy threshold$^{(\star)}$ & 100-500 MeV & 6.3\% & 63\% \\
	Energy binning & 5-12 bin per decade & 2.9\% & 5.5\% \\
	Spatial binning & 0.05-0.2 deg/pixel & 1.9\% & 0.9\% \\
	ROI size & 8-15 degrees & 0.9\% & 12.8\% \\
	4FGL source selection size & 20-30 degrees & 0.8\% & 1.8\% \\
	Event selection & alternative selections & 6.3\% & 12.8\% \\
	Rocking angle cut & yes/no & 3.5\% & 3.6\% \\
	\hline
	\end{tabular}
	\end{center}
	{\small {\bf Notes.} $^{(\star)}$ The flux variation is computed by extrapolating the model in the range 200 MeV - 300 GeV. The flux uncertainty is largely driven by uncertainties in the extrapolation.}
	\label{tab:table_fermi_systematics}
\end{table*}

Our \textit{Fermi}-LAT analysis relies on choices that are somewhat arbitrary. To further validate the \textit{Fermi}-LAT results, we thus check the impact of various systematic effects associated with these choices. They are listed below and summarized in Table~\ref{tab:table_fermi_systematics} in terms of the changes on the flux of the cluster emission. We also refer to \cite{Xi2018} and \cite{Ackermann2016} who tested the impact of similar systematic effects in the region around Coma. As discussed below, the uncertainty in the diffuse background is the dominant systematic effect; it is expected to remain below $40$\%.

\paragraph{Diffuse background emission}
Although Coma is located near the galactic north pole (i.e., in a very clean region regarding diffuse galactic emission), the diffuse background is the dominant contaminant (see Figure~\ref{fig:Fermi_residual_spectrum}) and is slightly correlated with the cluster signal (see Figure~\ref{fig:correlation_matrix}). Given the location of Coma, we also note that the galactic emission is nearly isotropic in the ROI, which explains the high degeneracy between the diffuse isotropic component and the diffuse galactic component. Therefore, the diffuse background modeling might lead to a significant systematic effect. The diffuse background related systematic effect was investigated by \cite{Ackermann2016} who showed that above 300 MeV, the systematic effect was less than 22\% (although could reach $\sim 50$\% below 300 MeV). Similarly, \cite{Xi2018} concluded that the uncertainty in the diffuse background was $<30$\% within 0.2-300 GeV.

To test the impact of the diffuse background, we first reproduce our results by fixing the background to its expected model value (i.e., normalization set to 1 and no spectral index variation allowed). In this case, we obtain ${\rm TS} = 60.0$ and the flux of Coma increases by 90\% (scenario 2). Such an increase is expected as the background mis-modeling can be partially absorbed by the cluster template. In fact, we note that in this case the residual image presents a significant positive offset, which is partially absorbed by the cluster model and explains the boost of the signal. Such values for the Coma cluster can thus only be taken as upper limits.

We then reproduce our results using previous background models, namely {\tt gll\_iem\_v06} and {\tt iso\_P8R3\_SOURCE\_V2} for the galactic and isotropic emission, respectively. In scenario 2, the TS value slightly increases, to 31.93, and the flux remains stable within $<4$\%. 

We also consider the 16 alternative background models discussed in \cite{Xi2018}. They are obtained using the GALPROP web interface for computation \citep{Vladimirov2011}\footnote{See \url{https://galprop.stanford.edu/}, and \url{https://galprop.stanford.edu/webrun.php} for the web interface.}, given the parameter definition files provided by \cite{Ackermann2012}. We obtain a set of 16 versions of the diffuse galactic background components (and the corresponding isotropic diffuse emission that was model simultaneously) including bremsstrahlung, inverse Compton and pion decay emission. We first fit the sum of the galactic background components with a single free normalization and spectral slope parameter. In this case, the TS value is systematically higher than in our baseline fit, ranging from 33.19 to 39.26, with a flux from 20 to 40\% higher. Then we consider a free normalization and spectral slope for each individual components separately, but do not consider the subcomponents separately (e.g., inverse Compton from optical, far infrared and CMB scattering individually). The TS values range from 30.98 to 39.42 and the flux is stable within $[-8.5,+21.8]$\%. We also note that the morphology of the residual signal does not change significantly depending on the considered background model.

\paragraph{Energy range} 
We test the impact of our choice of the considered energy range. As the systematic effects are expected to be dominant at low energy, we test changing the nominal minimum energy of 200 MeV to 500 MeV. We also consider the case of 100 MeV as the low energy threshold. We extrapolate the flux assuming our physical model with a CRp spectral index in the range $\alpha_{\rm CRp} = [2.6 - 3.2]$, and also considering power law extrapolation with photon index from 2.25 to 2.65. The statistics is slightly reduced and our results change within 63\%. However, we note that this large number is mainly driven by the uncertainty in the extrapolation when converting the flux to the 200 MeV - 300 GeV band, and the significance of the detection remains stable within 6.3\%.

\paragraph{Energy and spatial binning} 
Our default binning choice is 0.1 degree per pixel, and eight energy bins per decade. We vary the pixel resolution from 0.05 to 0.2 degree and the energy binning from 5 to 12 bins per decade. The changes are less than 6\% overall.

\paragraph{ROI size}
We use a ROI size of 12 by 12 degrees. To make sure that this choice does not introduce any significant bias, e.g., due to the presence of poorly constrained sources at the periphery of the ROI, we reproduce the analysis using a ROI of 8 and 15 degrees. This does not change our results by more than 13\%.

\paragraph{4FGL source selection size}
Similarly, the choice of including sources from the 4FGL catalog that are within a 20 degrees width region from the ROI center is checked by increasing this value by 50\%. The changes in our results are less than 2\%.

\paragraph{Event class}
We reproduce our results by applying the {\tt P8R3\_ULTRACLEANVETO\_V2} selection (event class 1024, i.e., the cleanest Pass 8 event class). We also use the {\tt FRONT} only (event type 1) converting photons. The changes are less than 13\%.

\paragraph{Rocking angle cut}
It is recommended by the \textit{Fermi}-LAT collaboration to check the impact on the rocking angle cut on the results. As a baseline, we apply a cut on rocking angle less than 52 degrees. We reproduce the results presented here without considering this cut. While the statistics slightly increases, the changes are less than 4\% on the results.

\subsection{On the nature of the signal toward the Coma cluster}
In Section~\ref{sec:Fermi_analysis}, we have shown that a significant $\gamma$-ray signal is observed within the characteristic radius $\theta_{500}$ of the Coma cluster. This agrees with the results by \cite{Xi2018} and the 4FGL-DR2 catalog \citep{Abdollahi2020,Ballet2020}. 

The source 4FGL~J1256.9+2736, as modeled in the 4FGL-DR2 catalog (a point source), is strongly degenerate with the diffuse cluster models that we have tested, and could correspond to the peak of the diffuse ICM emission. The comparison between the models based on a single cluster diffuse ICM component and a single point source (4FGL~J1256.9+2736) cannot be strongly discriminated, although diffuse models based on multiwavelength templates provide the best agreement with the data. Models with two components (diffuse ICM plus point source) better match the data, but the improvement is marginal given the additional component involved.

In the next sections, we will explore the consequences for the cosmic ray population in the Coma cluster of the scenarios in which the signal is (at least partly) associated with the diffuse ICM. We will thus consider the scenarios 2 in which the signal is entirely associated with the diffuse ICM, and the scenario 3 in which the diffuse ICM and an independent point source (4FGL~J1256.9+2736) both account for the signal. However, given the data, it is not possible to exclude that the signal arises essentially from a point source independent from the cluster diffuse ICM (scenario 1), but this will not be considered in the following since no CRp are needed in this scenario.

\section{Multiwavelength comparison}\label{sec:multiwavelength_comparison}
\begin{figure}
	\centering
	\includegraphics[width=0.5\textwidth]{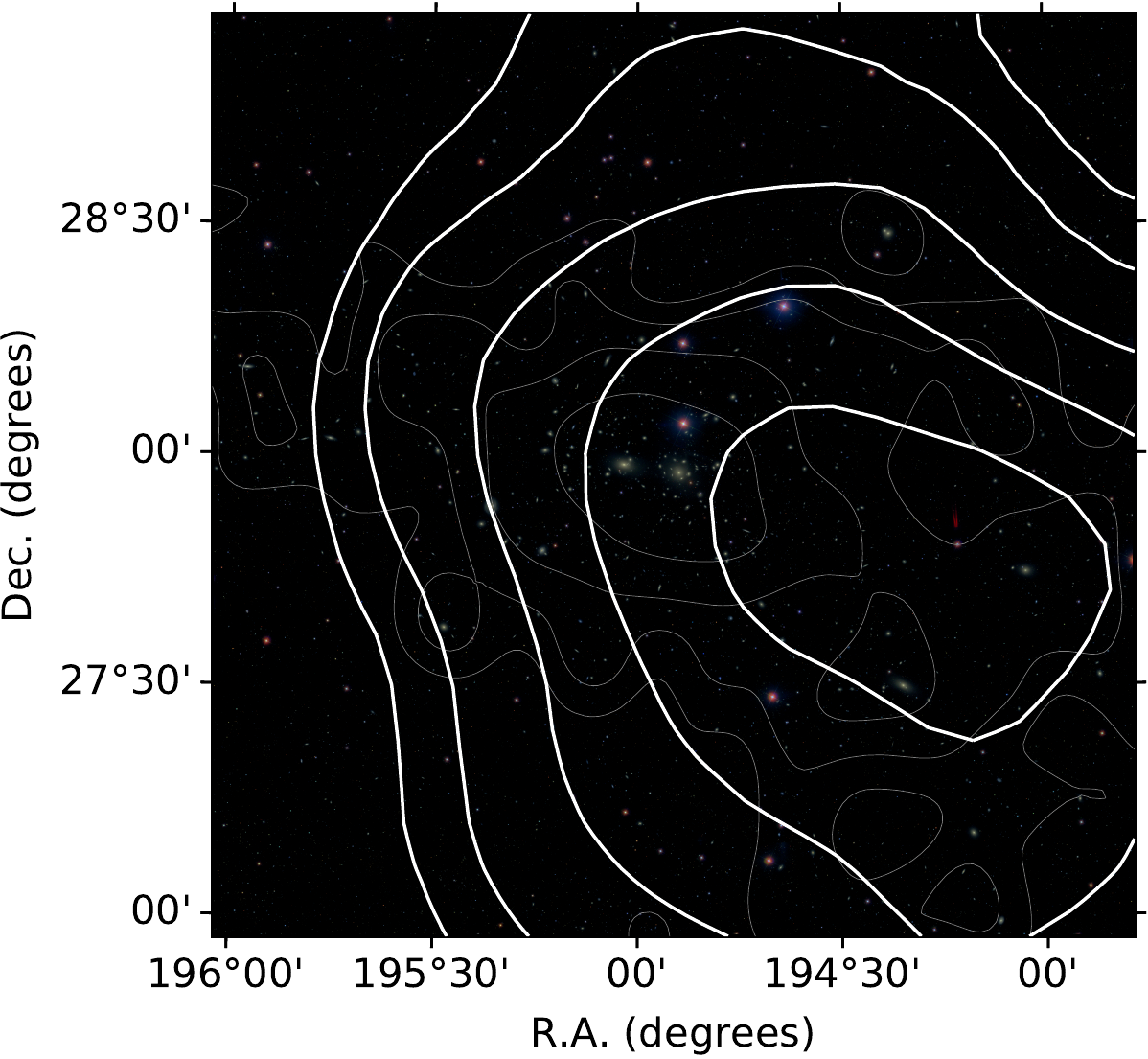}
	\caption{\small SDSS color image of the central $2\times2$ deg$^2$ cluster region. The thin gray contours provide the galaxy density as shown in Figure \ref{fig:multiL_comparison}. The white contours give the TS map levels of 2, 4, 9, 16 and 25. The image was constructed using the $g$, $r$ and $i$ filters of SDSS.}
        \label{fig:sdss_rgb}
\end{figure}

\begin{figure*}
	\centering
	\includegraphics[height=7cm]{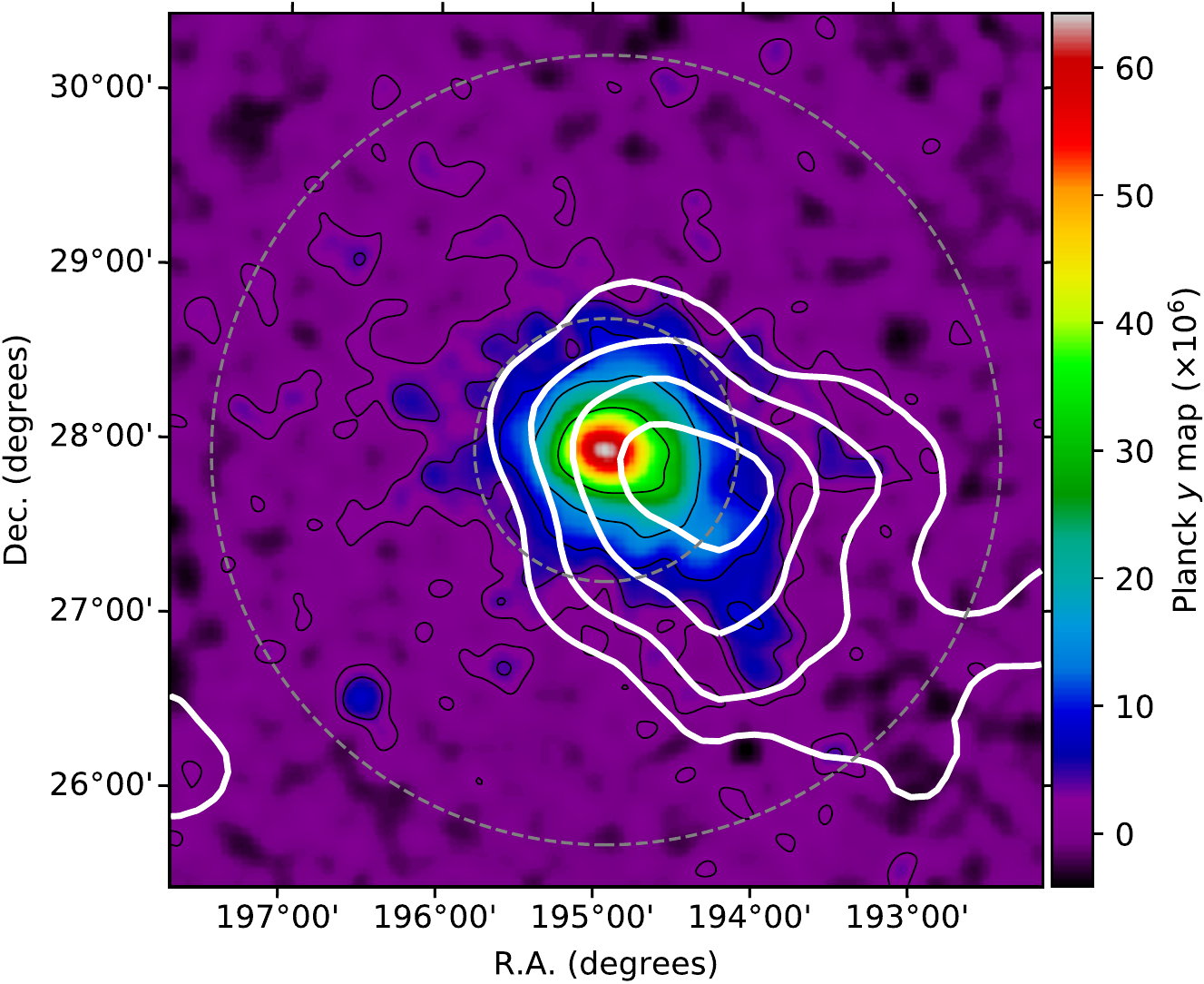}
	\includegraphics[height=7cm]{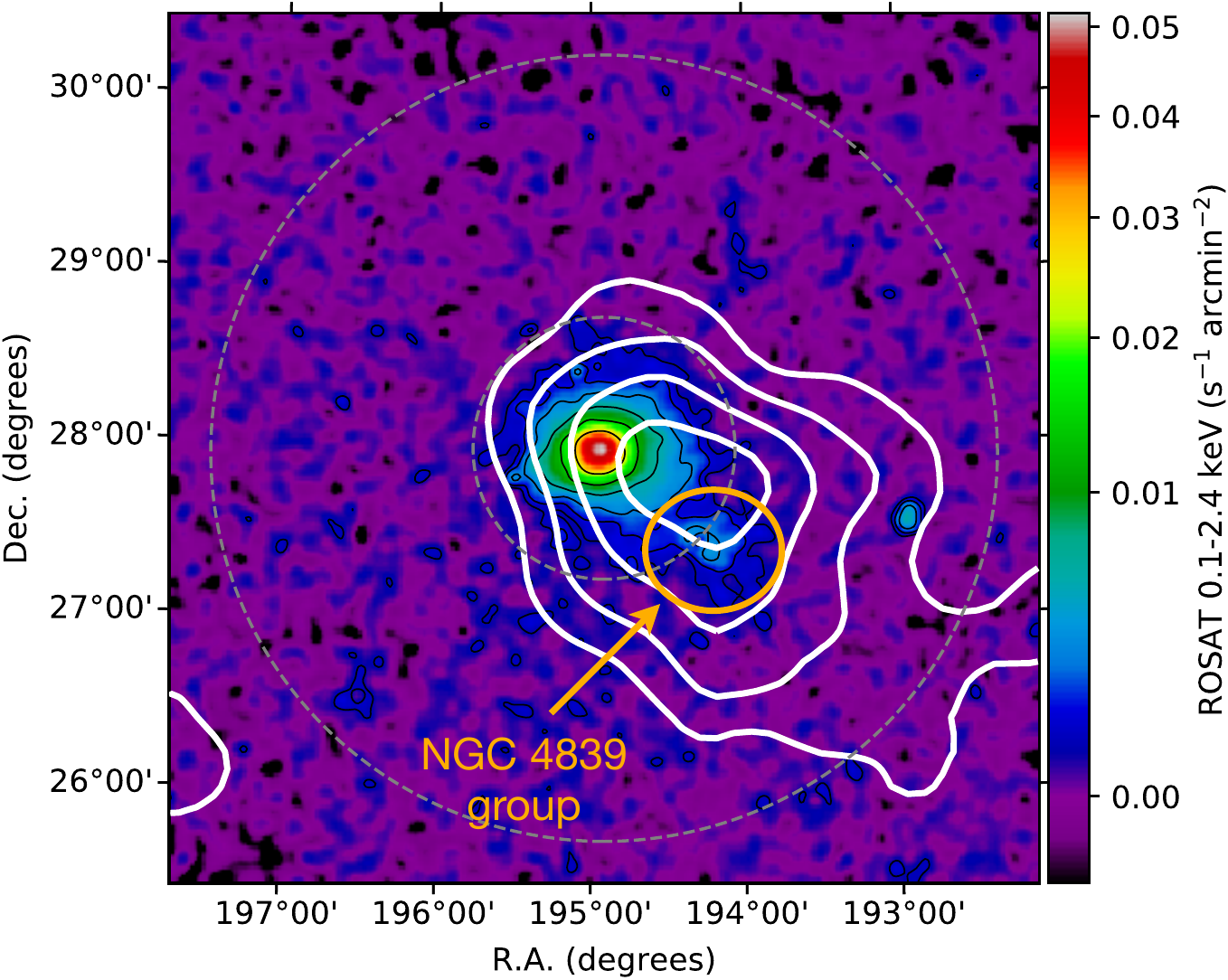}
	\includegraphics[height=7cm]{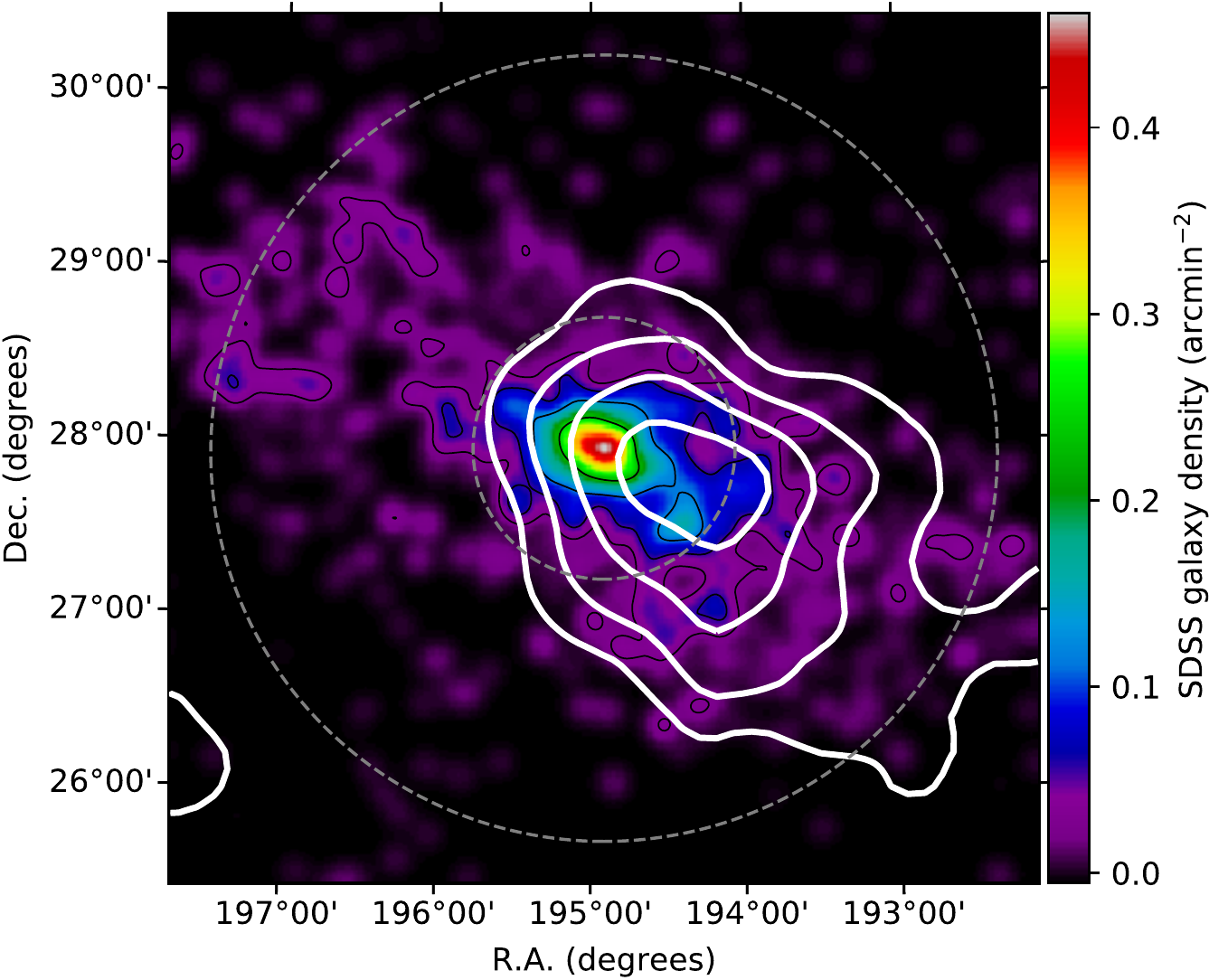}
	\includegraphics[height=7cm]{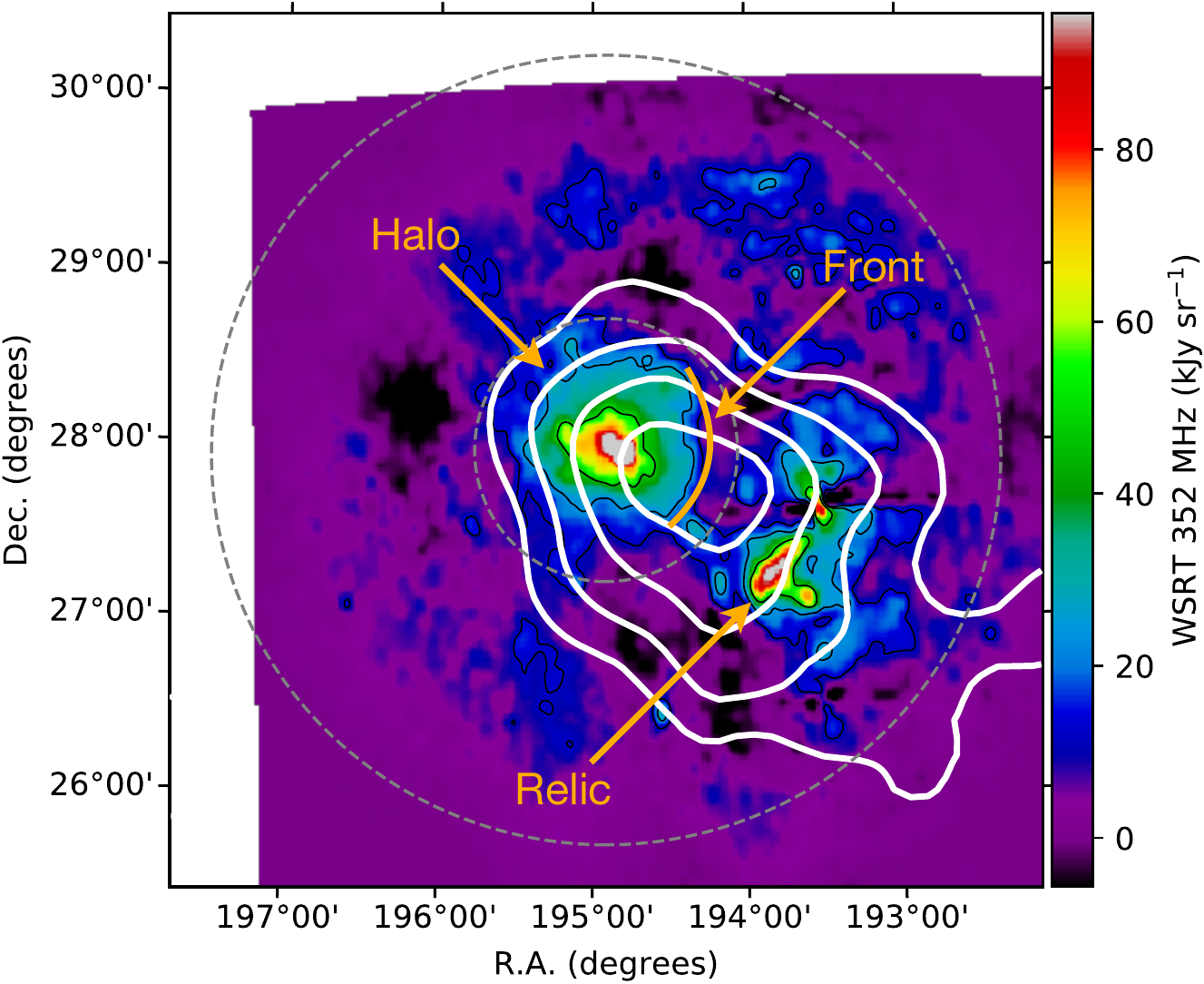}
	\caption{\small Multiwavelength morphological comparison of the Coma cluster signal to the \textit{Fermi}-LAT TS map obtained in our baseline model. {\bf Top left}: \textit{Planck} tSZ. {\bf Top Right}: \textit{ROSAT} X-ray. {\bf Bottom left}: SDSS galaxy density. {\bf Bottom right}: WSRT 352 MHz radio signal. The field of view of all images is $5\times5$ deg$^2$. The white contours give the \textit{Fermi}-LAT TS map (contours at 4, 9, 16 and 25) for the reference {\tt MINOT} model ($n_{\rm CRp} \propto n_e^{1/2}$). For all panels, the black contours correspond to the maximum of the image divided by $2^i$, with $i$ the index of the contours. The dashed gray circle provides the radius $\theta_{500}$ and $3 \times \theta_{500}$. Several relevant features are also indicated in orange. For display purposes, the WSRT image has been apodized at large radii to reduce the larger noise fluctuations present on the edge of the field. As a complementary figure, Figure~\ref{fig:sdss_rgb} provides an optical image of the central region.}
        \label{fig:multiL_comparison}
\end{figure*}

In this Section, we qualitatively compare the \textit{Fermi}-LAT excess map obtained in Section~\ref{sec:Fermi_analysis} to data at other wavelengths, as described in Section~\ref{sec:data}. The interpretation assumes that all the $\gamma$-ray emission arises from the diffuse ICM component.

First, we briefly compare the \textit{Fermi}-LAT TS map (baseline model) to the optical image constructed using SDSS data in Figure~\ref{fig:sdss_rgb}. This provides a visual appreciation of the scales probes by \textit{Fermi}-LAT. The galaxy density contours are also shown for visual purpose. As we can see, the TS peak is located about 10 arcmin north of the NGC 4839 group and about 20-30 arcmin from the Coma center and its two brightest galaxies. However, this offset is small compared to the \textit{Fermi}-LAT angular resolution. The TS value remains larger than 16 (about $4 \sigma$) in most of the region where the bulk of the galaxy is located. However, the \textit{Fermi}-LAT excess also extends further in the southwest direction, as discussed below.

In Figure~\ref{fig:multiL_comparison}, we compare the TS map to images of the tSZ, X-ray, galaxy density, and radio signal. The northwest-southeast elongated morphology of the TS map (and excess counts) matches well what is seen at other wavelengths. The best match is observed for the galaxy distribution and the tSZ map. The later being proportional to the product between the thermal gas density and temperature, this indicates that the CRp distribution matches the temperature well. The temperature being fairly constant up to $\sim$Mpc scales, this would favor relatively flat CRp distributions. Alternatively, it could favor a scenario in which the signal comes from the sum of unresolved sources, as traced by the galaxies within the cluster region, and is not necessarily associated with the diffuse ICM component. The X-ray morphology is more compact than the \textit{Fermi}-LAT excess, although the difference could be largely due to the relatively poor angular resolution of \textit{Fermi}-LAT. The $\gamma$-ray excess matches the radio halo, but also extend toward the relic and could thus provide a good match to the combination of the two. In the case of an association with the relic, it would suggest that a significant fraction of the signal arises from inverse Compton emission toward the relic because very little target thermal gas is available for hadronic interaction in the relic region, but we leave this interpretation for future work \citep[see][for discussions about a possible shock at the relic location and the accretion shock interpretation for its origin]{Brown2011,Ogrean2013,Akamatsu2013}. 

As noted in Figure~\ref{fig:sdss_rgb}, the TS peak is slightly off centered with respect to the cluster center, and better coincides with the southwest extension associated with the merger with the NGC 4839 group (between the halo and the relic). It also coincides well with the location where a radio front was identified \citep{Brown2011}. This front is coincident with a discontinuity seen in the X-ray surface brightness, temperature and entropy \citep{Simionescu2013,Uchida2016,Mirakhor2020} and SZ signal \citep{PlanckX2013}, possibly suggesting that the local CR might have been accelerated by a shock. However, given the significance of the excess and the \textit{Fermi}-LAT angular resolution, this remains difficult to interpret further, as also shown in Section~\ref{sec:Fermi_analysis} and in particular in Figure~\ref{fig:Fermi_residual_profile}.

In conclusion, we find good morphological agreement with data at other wavelengths. Nevertheless, the interpretation remains difficult due to the low significance of the excess and the \textit{Fermi}-LAT angular resolution. This comparison does not allow us to exclude that the signal observed is entirely, if any, associated with the diffuse ICM emission. It could also be the sum of several components.

\section{Implications for the cosmic ray content of the Coma cluster}\label{sec:implication_for_CR_content}
\begin{figure*}
\centering
\includegraphics[width=0.49\textwidth]{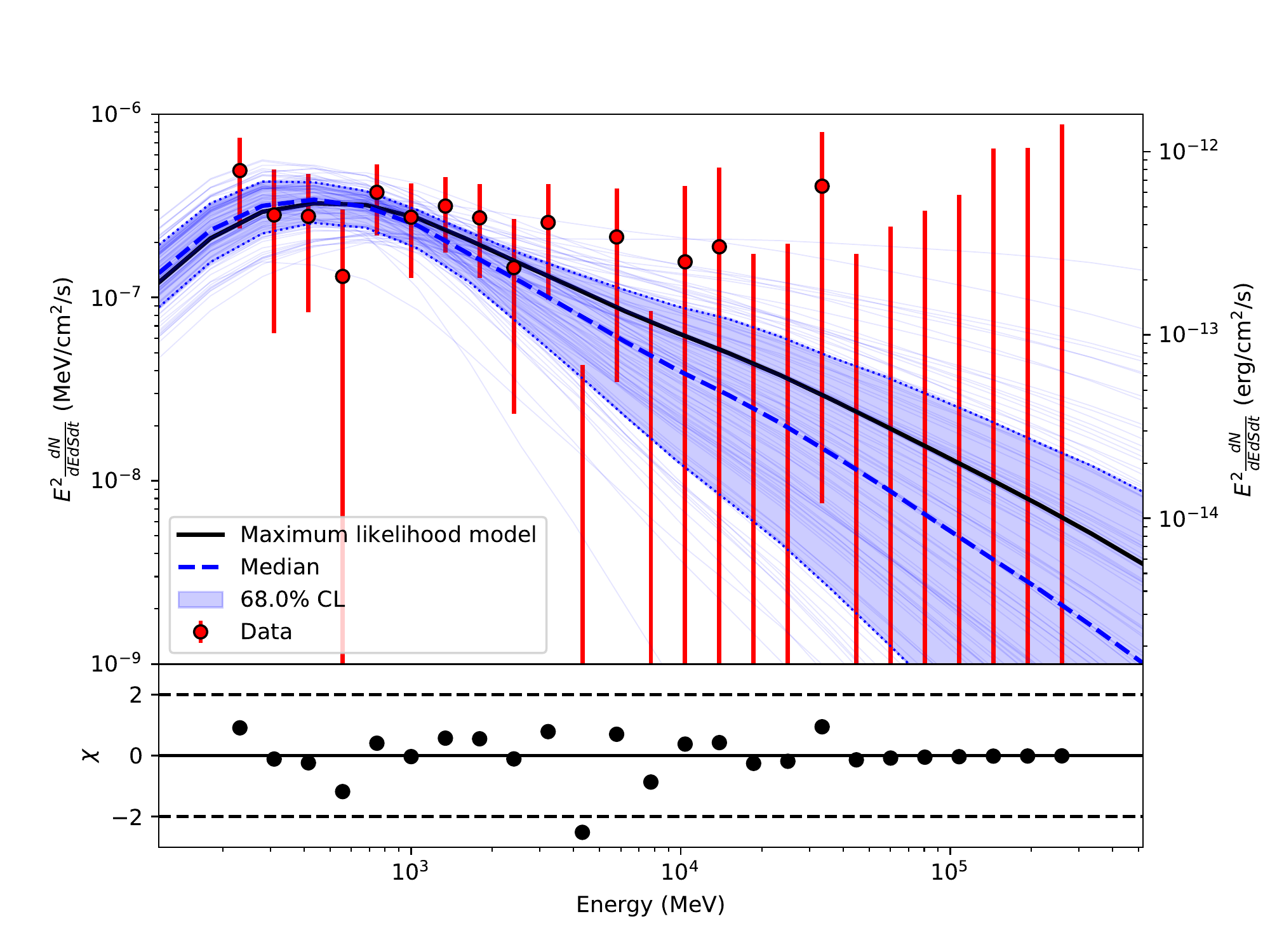}
\includegraphics[width=0.49\textwidth]{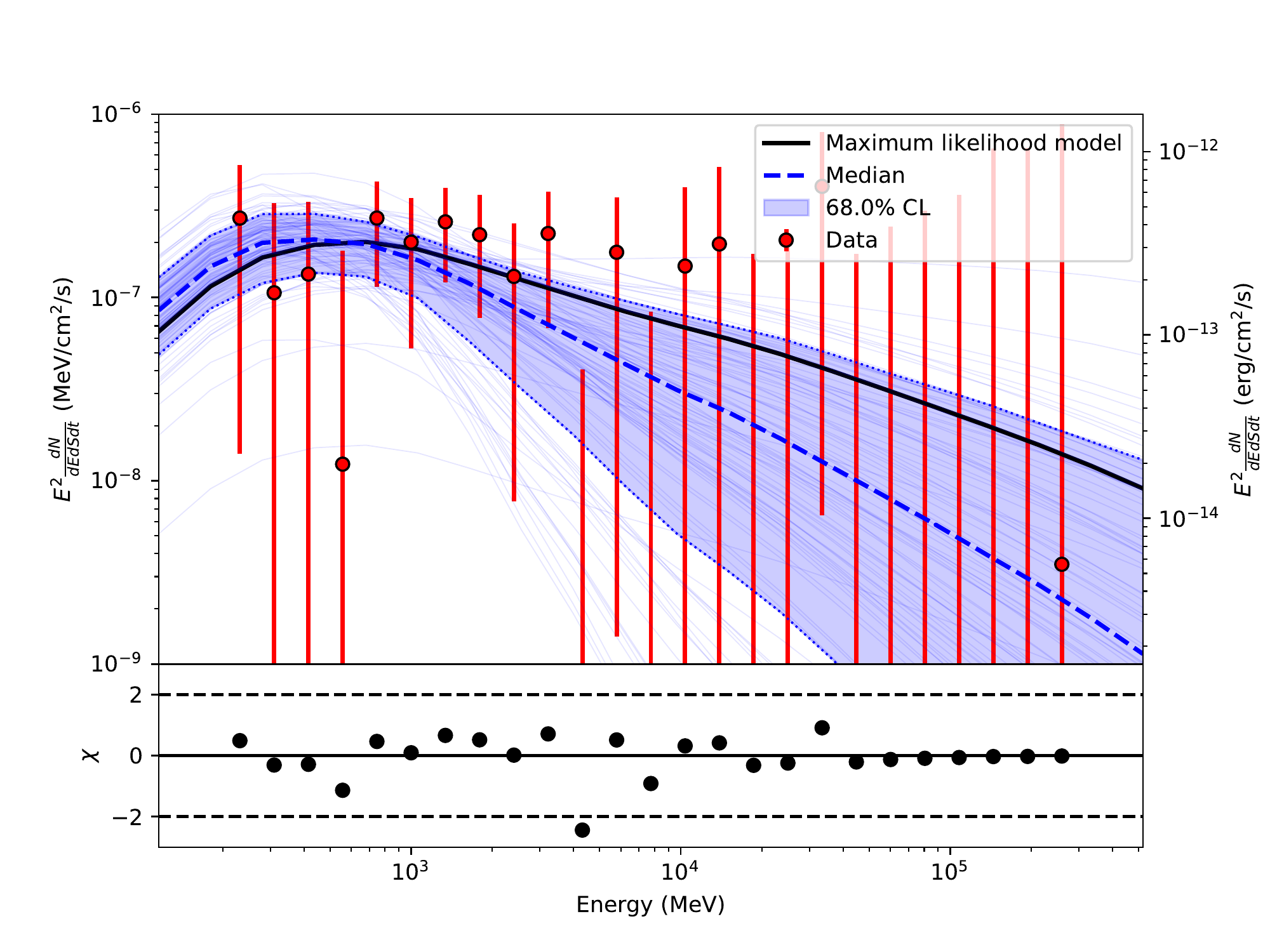}
\caption{\small Total SED recovered from the \textit{Fermi}-LAT and MCMC constraint in the case of the reference model ($n_{\rm CRp} \propto n_e^{1/2}$) when 4FGL~J1256.9+2736 is replaced by the ICM component in the sky model (left) or both are included (right). The best-fit model is shown in black and the 68\% confidence interval is show in blue, together with 100 models randomly sampled from the MCMC parameter chains. The median of these models is also shown as a dashed blue line. The residual provides the difference between the data and the best-fit model normalized by the error bars.}
\label{fig:sed_modeling}
\end{figure*}

\begin{figure*}
\centering
\includegraphics[width=0.75\textwidth]{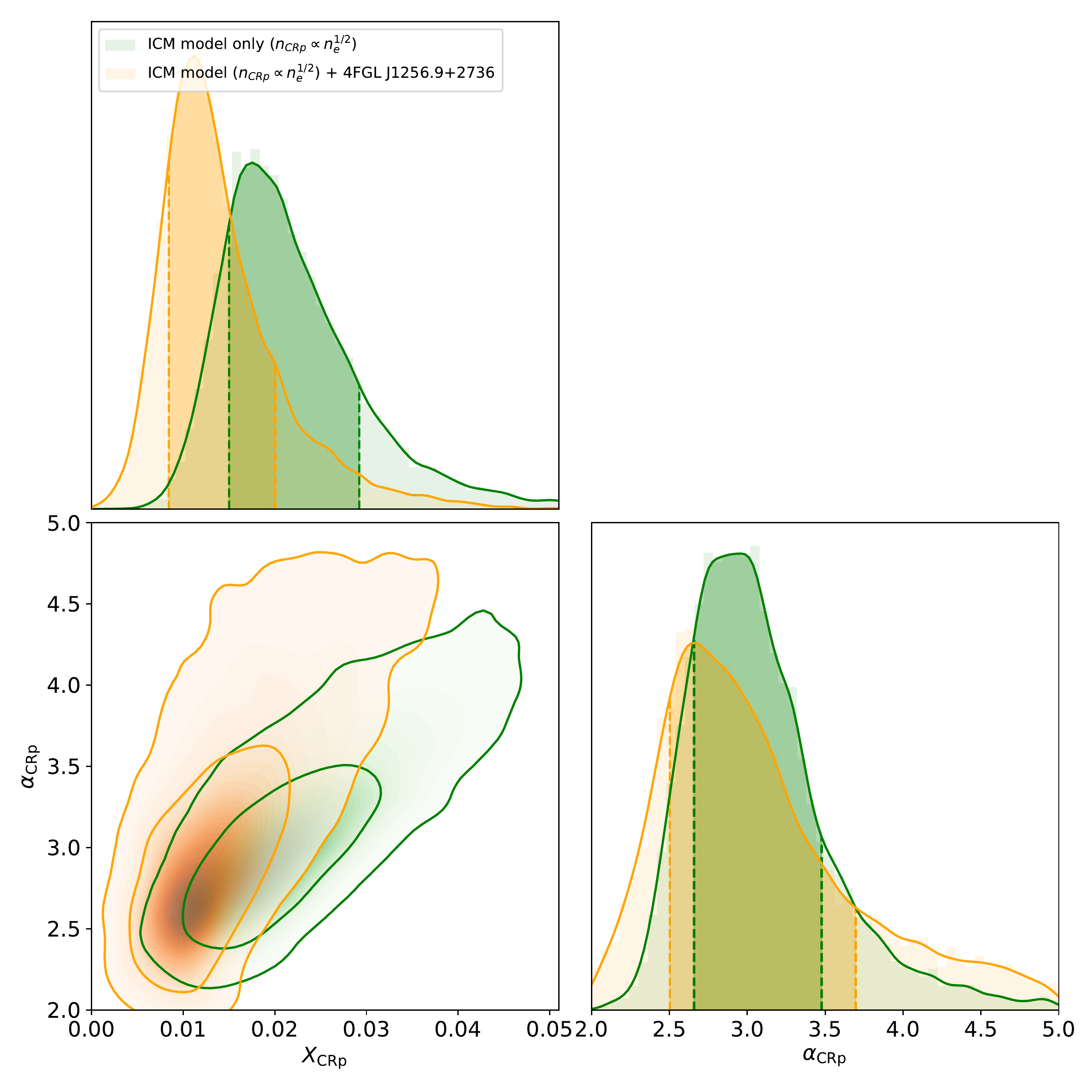}
\caption{\small MCMC constraints on the model parameters, in the case of the reference model ($n_{\rm CRp} \propto n_e^{1/2}$) when 4FGL~J1256.9+2736 is excluded from the sky model (green) or included (orange). The bottom left panel provides the constraint in the plane $X_{\rm CRp}(R_{500}) - \alpha_{\rm CRp}$, where the contours correspond to 68\% and 95\% confidence interval. The marginalized posterior probability distributions are also shown for the parameters $X_{\rm CRp}(R_{500})$ (top) and $\alpha_{\rm CRp}$ (right), where the shaded area provides the 68\% confidence interval.}
\label{fig:mcmc_constraint}
\end{figure*}

\begin{table*}
	\caption{\small Constraints on the CRp population and associated flux and luminosity in the case of the radial models. The quoted values and uncertainties correspond to the maximum likelihood and 68\% confidence interval of the distributions.}
	\begin{center}
	\begin{tabular}{c|cccc}
	\hline
	\hline
	Model & $X_{\rm CRp}$ (\%) & $\alpha_{\rm CRp}$ & Flux ($10^{-10}$ ph s$^{-1}$ cm$^{-2}$) & Luminosity ($10^{41}$ erg s$^{-1}$) \\
	\hline
	\hline
	& \multicolumn{4}{c}{4FGL~J1256.9+2736 replaced by ICM model (scenario 2)} \\
	\hline
	Compact model ($n_{\rm CRp} \propto n_{\rm gas}$) & $1.49_{-0.25}^{+1.21}$ & $2.89_{-0.11}^{+0.82}$ & $11.52_{-2.08}^{+3.98}$ & $14.98_{-3.75}^{+2.43}$ \\
	Extended model ($n_{\rm CRp} \propto n_{\rm gas}^{1/2}$) & $1.79_{-0.29}^{+1.13}$ & $2.79_{-0.13}^{+0.69}$ & $13.61_{-2.71}^{+3.90}$ & $19.15_{-5.35}^{+3.15}$ \\
	Flat model ($n_{\rm CRp} = {\rm constant}$) & $1.45_{-0.19}^{+0.86}$ & $2.71_{-0.11}^{+0.75}$ & $16.58_{-3.12}^{+4.67}$ & $24.90_{-8.14}^{+3.76}$ \\	
	Isobar ($n_{\rm CRp} = P_{\rm gas}$) & $1.49_{-0.29}^{+1.16}$ & $2.92_{-0.15}^{+0.78}$ & $11.32_{-2.35}^{+3.22}$ & $13.58_{-3.08}^{+3.21}$ \\	
	\hline
	\hline
	& \multicolumn{4}{c}{Both 4FGL~J1256.9+2736 and ICM models included (scenario 3)} \\
	\hline
	Compact model ($n_{\rm CRp} \propto n_{\rm gas}$) & $0.75_{-0.11}^{+1.05}$ & $2.56_{-0.06}^{+1.39}$ & $6.20_{-1.75}^{+3.87}$ & $10.89_{-5.38}^{+1.75}$ \\	
	Extended model ($n_{\rm CRp} \propto n_{\rm gas}^{1/2}$) & $1.06_{-0.22}^{+0.96}$ & $2.58_{-0.09}^{+1.12}$ & $8.32_{-3.12}^{+3.46}$ & $14.25_{-7.21}^{+1.66}$ \\
	Flat model ($n_{\rm CRp} = {\rm constant}$) & $0.83_{-0.16}^{+0.85}$ & $2.58_{-0.10}^{+1.38}$ & $9.86_{-3.46}^{+3.61}$ & $13.06_{-5.80}^{+6.52}$ \\
	Isobar ($n_{\rm CRp} = P_{\rm gas}$) & $0.73_{-0.11}^{+0.95}$ & $2.59_{-0.07}^{+1.25}$ & $6.98_{-2.77}^{+2.92}$ & $9.19_{-3.18}^{+3.32}$ \\
	\hline
	\end{tabular}
	\end{center}
	\label{tab:table_mcmc_analysis}
\end{table*}

In this Section, we use the $\gamma$-ray SED extracted in Section~\ref{sec:Fermi_analysis}, together with the model described in Section~\ref{sec:modeling}, in order to constrain the CR population in the Coma cluster. Although they provide the best match to the data, the multiwavelength data templates are not used because they do not allow us to have a three-dimensional physical model of the cluster, which is needed to constrain the CR content.

\subsection{Methodology}
We aim at using the \textit{Fermi}-LAT extracted SED to constrain the CRp population of our model. As discussed in Section~\ref{sec:modeling}, and in more details in \cite{Adam2020}, the hadronically induced $\gamma$-ray emission depends on the thermal gas (modeled via the pressure and density), and the CRp population (spatial and spectral distribution). The thermal gas pressure and density are well constrained from \textit{Planck} and \textit{ROSAT} data, respectively, and are thus kept fixed in this analysis. Since the CRp spatial distribution was kept fixed when extracting the SED, we keep it fixed when constraining the parameters based on the SED fit. However, we consider the different spatial shapes as in Section~\ref{sec:Fermi_analysis} because we have seen that it was not possible to discriminate between the different cases. We also consider the case of scenario 3, where both the diffuse ICM and 4FGL~J1256.9+2736 are included in the sky model. We are left with two parameters to be constrained: 1) the CRp normalization defined as the energy stored in the CRp relative to the thermal energy $X_{\rm CRp}(R_{500})$, which we define at a radius $R_{500}$; 2) the slope of the CRp energy spectrum $\alpha_{\rm CRp}$. These two parameters should be constrained for all the different models considered (compact, extended, flat, and isobar), which will provide an assessment of the systematic effect associated with the spatial model.

We employed a Markov chain Monte Carlo (MCMC) approach in order to constrain the parameters space, using the {\tt emcee} package \citep{Foreman2013}. In brief, the chains move in the parameter space according to a proposal function and the likelihood of the model given the data. We adopted flat priors across the range $X_{\rm CRp}(R_{500}) \in [0, 0.2]$ and $\alpha_{\rm CRp} \in [2, 5]$, which corresponds to the physically acceptable parameter range, but we checked that this limit does not affect our results. This method allows us to efficiently find both the best-fit parameters (as the parameters that maximize the likelihood) and the estimate of the posterior probability distribution.

The likelihood function is defined as
\begin{equation}
	{\rm ln} \mathcal{L}(\vec{\theta} | D) = \sum_i {\rm ln} \mathcal{L}_i (M_i(\vec{\theta})),
\end{equation}
where $i$ runs over each energy bin and the model parameters are $\vec{\theta} \equiv [X_{\rm CRp}(R_{500}), \alpha_{\rm CRp}]$. The value $\mathcal{L}_i$ is the likelihood of measuring a given flux normalization in the energy bin $i$, depending on the model flux $M_i(\vec{\theta})$. It is obtained by interpolating the likelihood scan, as provided by {\tt Fermipy}, when extracting the SED in Section~\ref{sec:sed_extraction}.

Once the chains have converged, and after removing the burn-in phase, the two-dimensional histogram in the plane $X_{\rm CRp}(R_{500}) - \alpha_{\rm CRp}$ can be integrated to provide the constraints up to a given confidence interval. The individual chain histograms also provide the marginalized posterior probability distribution of each parameter. The integrated posterior probability distribution up to 68\% probability gives the uncertainties. In addition to the model parameters themselves we obtain the constraint of the spectrum. To do so, we compute the model SED for each set of parameters and calculate the envelope of all the models as the 68\% confidence limit measured from the model histogram in each energy bin. The same procedure is applied to measure the $\gamma$-ray flux (and luminosity) between 200 MeV and 300 GeV according to the MCMC sampling.

\subsection{Constraints on the cosmic ray proton population}
The SED measured in the case of our baseline model is shown in Figure~\ref{fig:sed_modeling} for both scenario 2 and 3 (see subsection~\ref{sec:ROI_modeling} for the detailed definition). Error bars are the $1 \sigma$ error on the SED as evaluated from the likelihood scan curvature. The maximum likelihood model is also shown in black, as well as 100 models randomly sampled from the MCMC chains, their median and the associated 68\% confidence interval. We can observe that the best-fit model is in good agreement with the data in both cases, as also highlighted by the residual. The model is relatively well constrained around 300 MeV - 1 GeV, but the uncertainties remain large at larger energies. The peak SED, around 500 MeV, reaches about $3 \times 10^{-7}$ MeV cm$^{-2}$ s$^{-1}$. In the case of scenario 3 (right panel, both the ICM and point source in the model), we can see that the spectrum amplitude is reduced and that error contours are significantly larger.

The corresponding constraint on the posterior likelihood of the parameters $X_{\rm CRp}(R_{500})$ and $\alpha_{\rm CRp}$ is shown in Figure~\ref{fig:mcmc_constraint}. The constraints on the model SED lead to a relatively tight constraint on the normalization, but the constraint on the slope remains fairly loose. The two parameters are degenerate as increasing the normalization and the slope simultaneously does not strongly change the flux at high energies (see also Figure~\ref{fig:template_minot}, bottom right panel). The constraint on the CRp to thermal energy is about 1.8\% and the slope about 2.8, as also shown on the marginalized distributions.
When both the ICM component and 4FGL~J1256.9+2736 are included in the sky model, the constraint on the normalization shifts toward zero and the posterior only excludes $X_{\rm CRp}(R_{500})=0$ at about $2\sigma$. The uncertainty on the slope increases and the best-fit slightly decreases to about $\alpha_{\rm CRp} \simeq 2.6$.

In Table~\ref{tab:table_mcmc_analysis} we provide the MCMC constraints on these parameters in the case of all tested spatial models. The fluxes and corresponding luminosities are also constrained according to the MCMC fit of the model. Given the uncertainties, all models are in agreement and the associated systematic shift on the parameter is about 20\% on the CRp to thermal energy ratio, and about 7\% on the slope. When including the source 4FGL~J1256.9+2736 and the cluster simultaneously in the sky model, the slope is slightly reduced (albeit with increased uncertainty) and the normalization is reduced by about 70\%.

\subsection{Implications for the cosmic ray electrons}\label{sec:implication_for_CR_electrons}
\begin{figure*}
\centering
\includegraphics[width=0.45\textwidth]{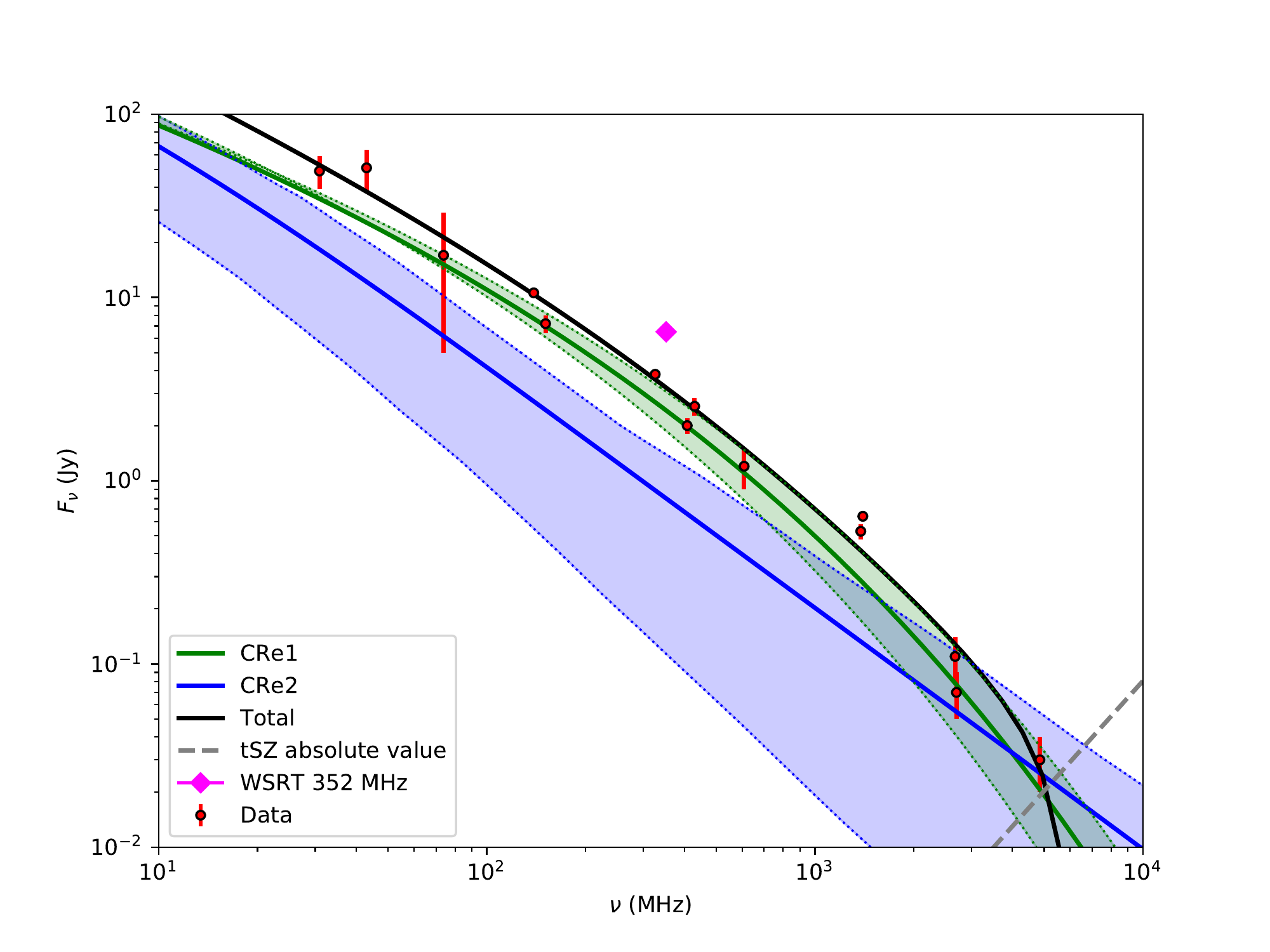}
\includegraphics[width=0.45\textwidth]{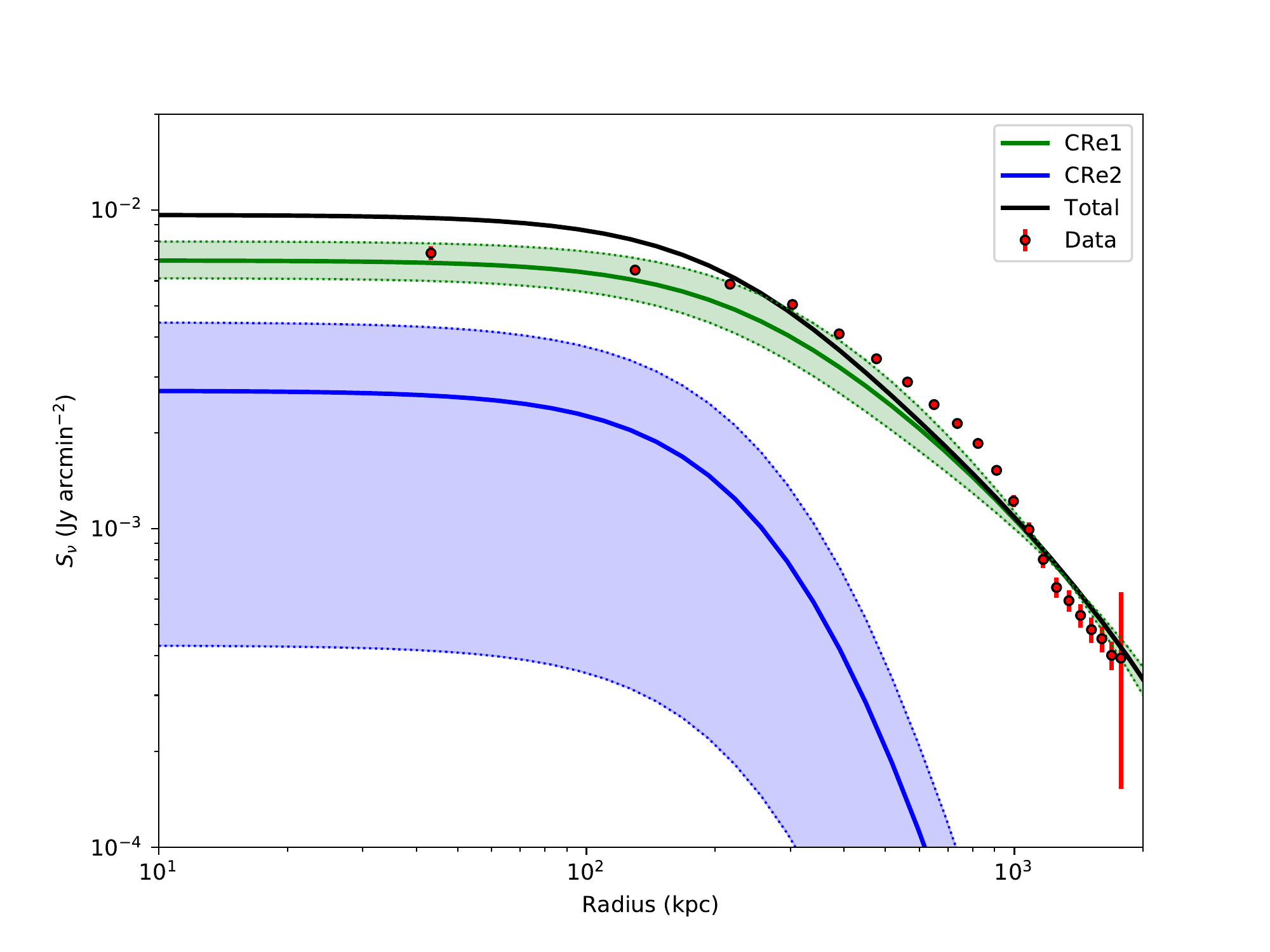}
\includegraphics[width=0.45\textwidth]{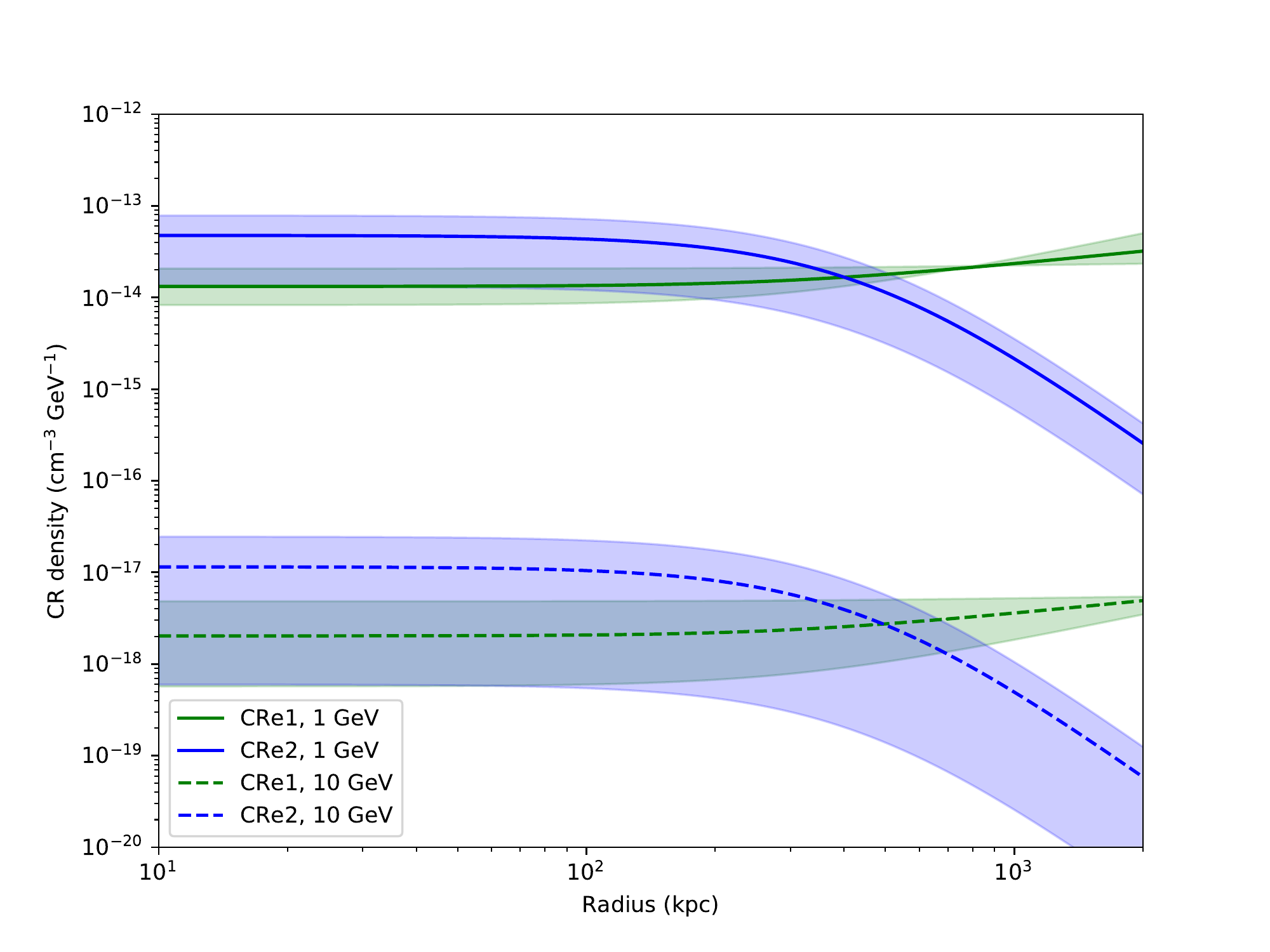}
\includegraphics[width=0.45\textwidth]{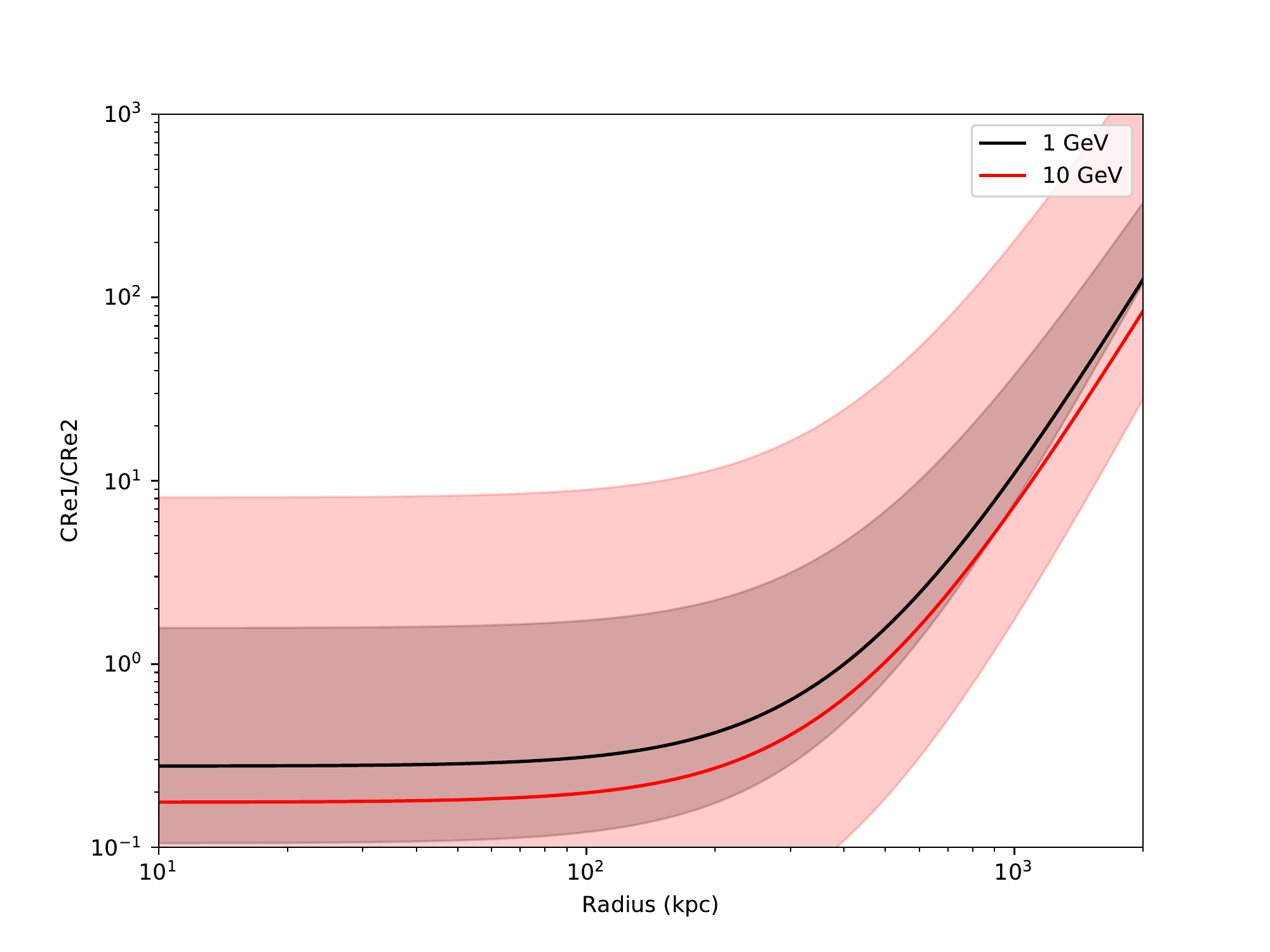}
\caption{\small Constraint on the CRe populations in the case of scenario 1 (distinct CRe$_1$ and CRe$_2$ populations, no reacceleration).
{\bf Top left}: Radio spectrum of Coma, as compiled from \cite{Pizzo2010} and constraint from the reference CRp spatial model ($n_{\rm CRp} \propto n_e^{1/2}$) and the reference CRe$_1$ spectral model ({\tt ExponentialCutoffPowerLaw}). The measurement from the \cite{Brown2011} data is also shown as the magenta diamond. The contribution from the CRe$_2$ is shown in blue together with its 68\% confidence interval, and the remaining contribution from CRe$_1$ is shown in green. The sum of the two is given as the black line. The dashed gray line provide the expected amplitude of the tSZ signal. All fluxes are computed using cylindrical integration within $R = 0.48 \times R_{500} = 629 \ {\rm kpc} \equiv 0.36$ deg.
{\bf Top right}: Radio profile measured from the WSRT map at 352 GHz and comparison to the reference model. The contributions from CRe$_2$ and CRe$_1$ are as in the left panel.
{\bf Bottom left}: Absolute number density distributions of CRe$_1$ and CRe$_2$ taken at 1 GeV and 10 GeV.
{\bf Bottom right}: Ratio between the CRe$_1$ and CRe$_2$ number populations.
We note that in the case of this figure, the confidence limits where computed using a resampling of only 100 parameters, and are thus not very accurate. We also stress that this figure depends on the magnetic field modeling (see Section~\ref{sec:discussions_cr_physics} for discussions).}
\label{fig:radio_implication_observable}
\end{figure*}

\begin{figure*}
\centering
\includegraphics[width=0.45\textwidth]{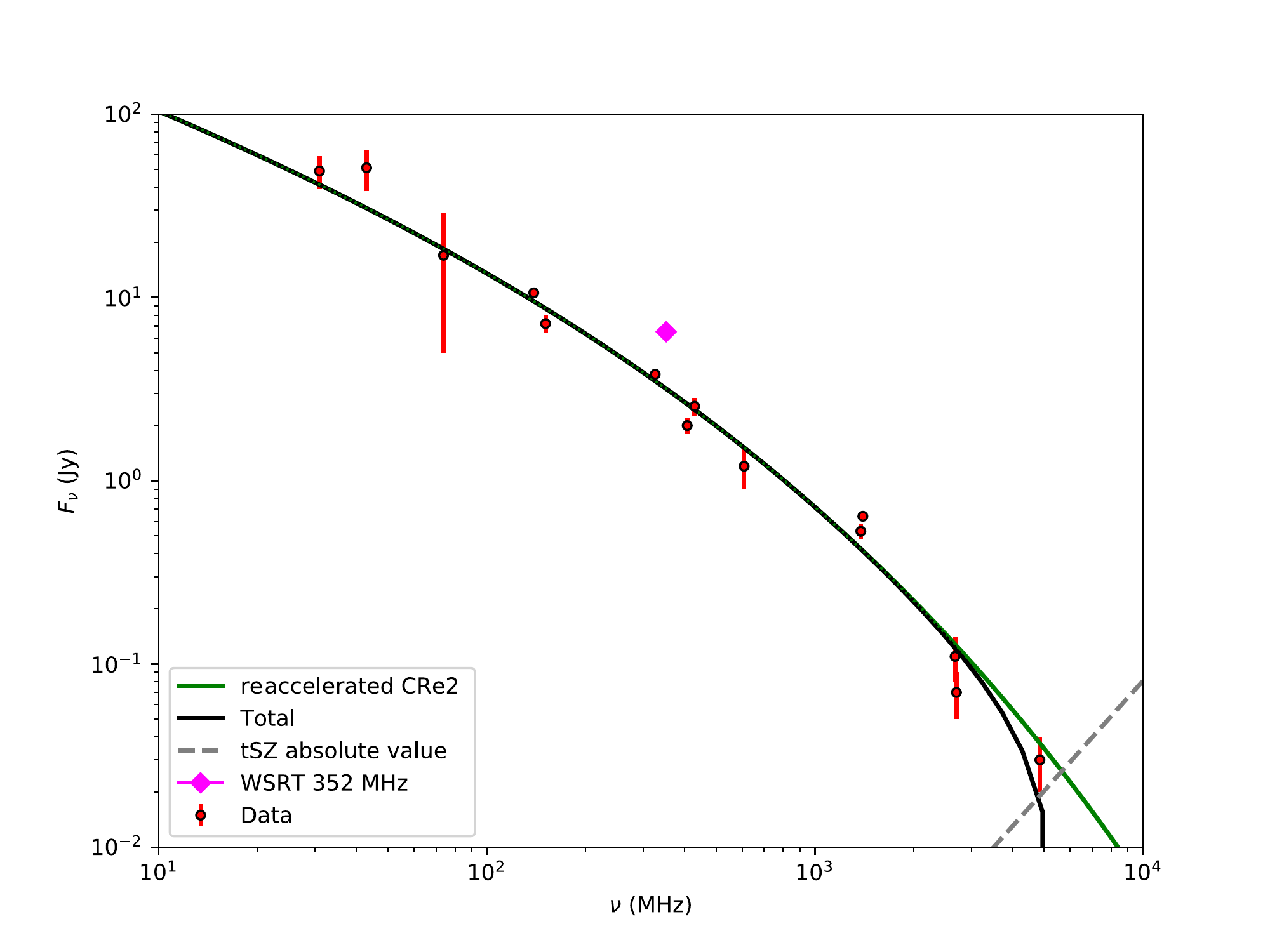}
\includegraphics[width=0.45\textwidth]{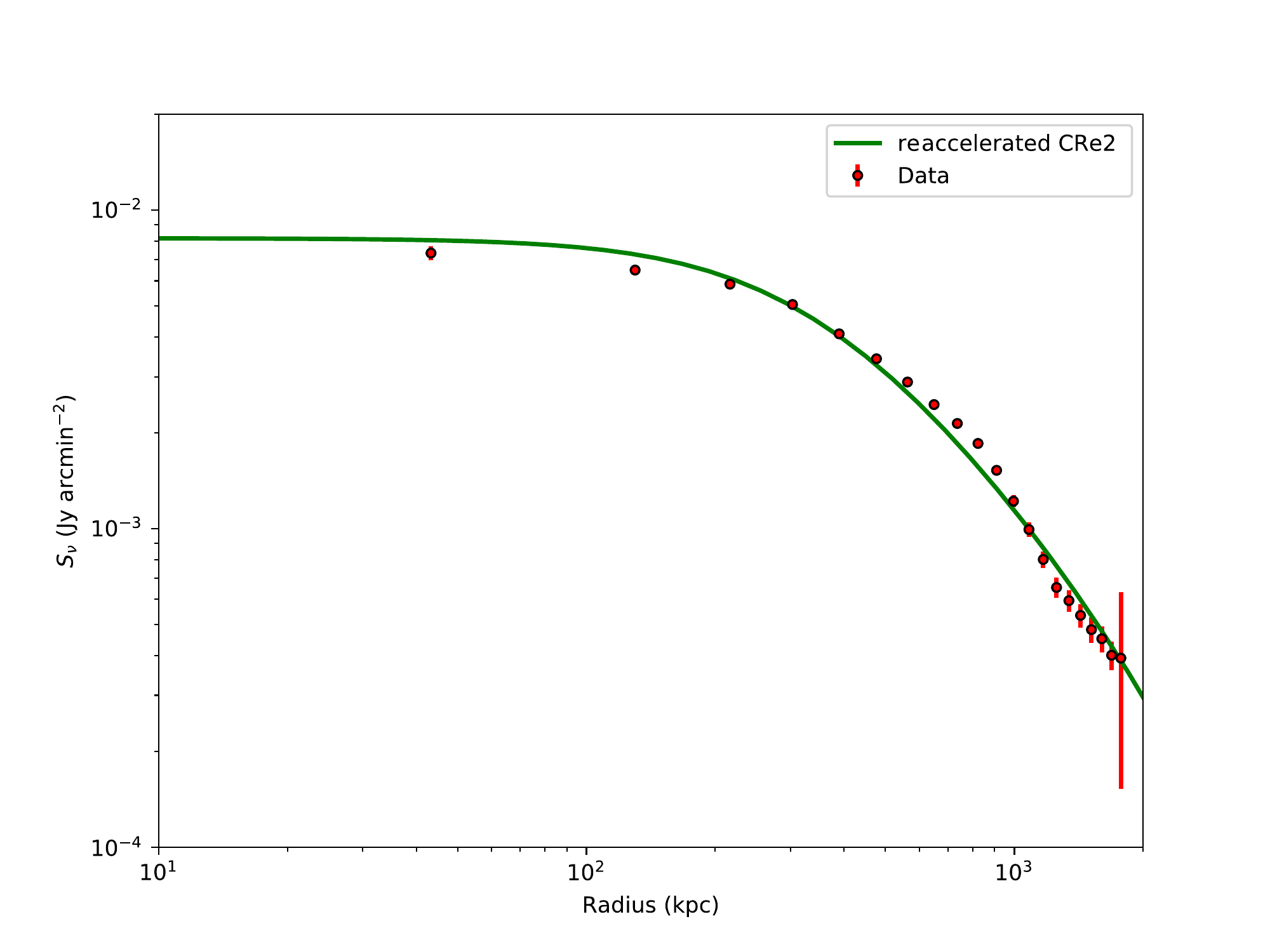}
\includegraphics[width=0.45\textwidth]{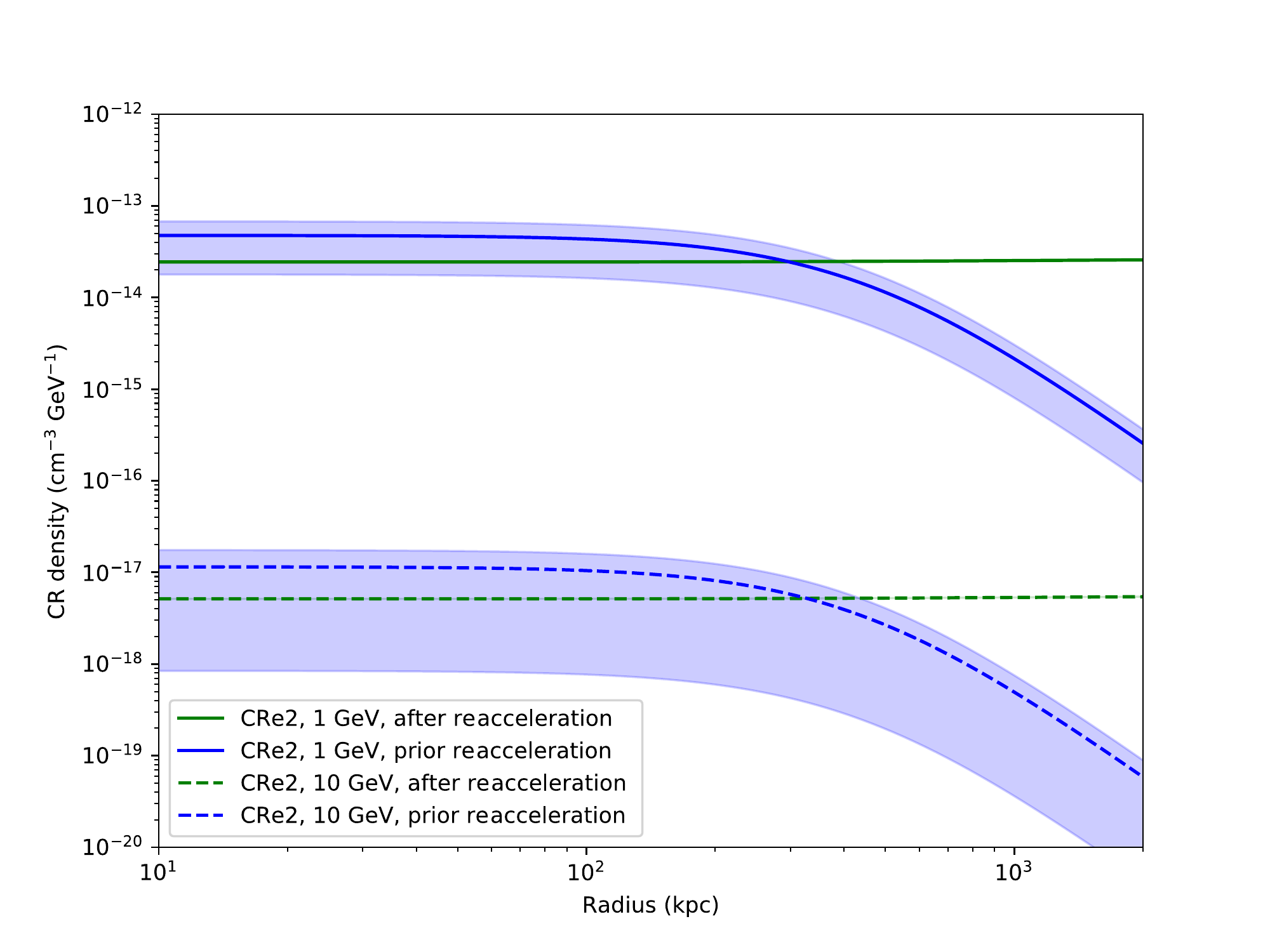}
\includegraphics[width=0.45\textwidth]{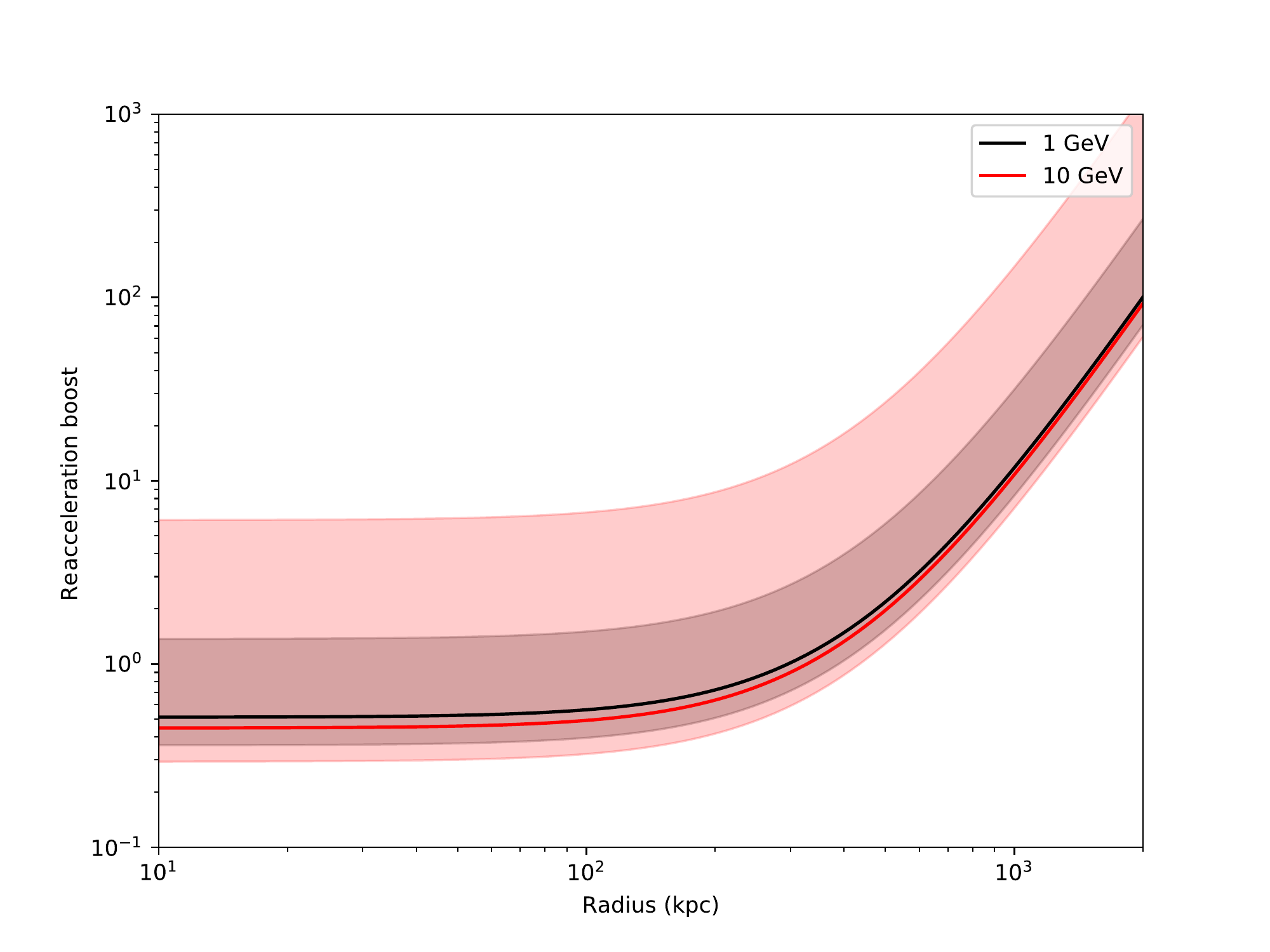}
\caption{\small Same as Figure~\ref{fig:radio_implication_observable} in the case of scenario 2 (CRe$_1$ interpreted as the reaccelerated CRe$_2$ population). The CRe$_2$ prior reacceleration refer to the steady state hadronic model. The reacceleration boost is given relative to the steady state hadronic model.}
\label{fig:radio_implication_observable2}
\end{figure*}

The presence of CRp, as traced by the hadronic $\gamma$-ray emission, implies the production of secondary CRe, which should contribute to the observed radio emission. We compute both the profile and the spectrum associated with this population given the modeling discussed in Section \ref{sec:modeling}. We stress that this is done assuming a steady state scenario. This calculation is done for the \textit{Fermi}-LAT SED best-fit model, as well as for the set of parameters sampled in the MCMC to obtain the 68\% confidence limit on the radio emission. 

As discussed in Section~\ref{sec:modeling}, the primary CRe, i.e., CRe$_1$, are also included in our model and we consider two cases for interpreting their origin. 1) Since our model does not include explicitly any reacceleration \citep[see][for reacceleration models]{Brunetti2007,Brunetti2011}, the population of CRe$_1$ that we constrain is supposed to be independent of the CRp and the CRe$_2$. It would correspond, for instance, to a CRe$_1$ population arising from star formation activity spread over the cluster volume, or the direct shock acceleration in the ICM. In this case, the CRe$_1$ and CRe$_2$ populations coexist and the radio emission accounts for the sum of the two. 2) We could also interpret the CRe$_1$ population in our model as the reaccelerated CRe$_2$ population. In this case, the CRe$_2$ would be the seed for the CRe$_1$. The CRe$_2$ population should thus not contribute directly to the radio emission. Instead the ratio CRe$_1$/CRe$_2$ would measure the amount of reacceleration needed to explain the emission with respect to the purely hadronic steady state scenario, as a function of energy and radius. We stress that in this second interpretation, we only provide a constraint relative to the hadronic steady state scenario since this is an assumption made when computing the CRe$_2$ population. In contrast, reacceleration models do not assume steady state, and both CRp and CRe populations evolve together according to turbulent reacceleration. Nevertheless, we consider this second case as it is still instructive regarding reacceleration physics.

The CRe$_1$ are parametrized using the {\tt ExponentialCutoffPowerLaw} model (our baseline), the {\tt InitialInjection} model, or the {\tt ContinuousInjection} model, as described in Section~\ref{sec:CRe1model}. All models include a normalization $X_{\rm CRe_1}(R_{500})$, a spectral slope $\alpha_{\rm CRe_1}$, a break energy ($E_{\rm cut, CRe_1}$), and a spatial scaling relative to the thermal density ($\eta_{\rm CRe_1}$). We fit for these parameters using simultaneously the radio spectrum and the 352 GHz radio profile data (see Section~\ref{sec:data}). Regarding the spectrum, we compute our model using a cylindrical integration within $R = 0.48 \times R_{500}$ (629 kpc, 0.36 deg) as it corresponds to the extent of the radio halo \citep{Brunetti2013}. We note, however, that the radio spectrum data are not strictly homogeneous in terms of aperture radius used for flux measurement and our model value only provides an effective radius. Because the profile is extracted from a single instrument (WSRT), which arguably could be the best one in terms of capturing the diffuse emission, but which does not necessarily perfectly agree with other measurements (as seen in Figure~\ref{fig:radio_implication_observable}), we also allow for a cross-calibration of the profile measurement by adding an extra normalization which we fit simultaneously. The fit is performed with the least square function {\tt curve\_fit} from the {\tt scipy} {\tt python} package. Depending on the considered case, the radio model includes both the CRe$_1$ and CRe$_2$ contributions (case 1, no reacceleration), or only the CRe$_1$ as the reaccelerated CRe$_2$ (case 2, pure reacceleration).  We compute the error contour on the CRe$_1$ fitted population by reproducing the fit in the case of the lower and upper bounds for the CRe$_2$ population. We thus assume that the uncertainties from the CRe$_2$ population (given by the $\gamma$-ray) are dominant over the uncertainties associated with the radio data.

In Figure~\ref{fig:radio_implication_observable}, we show the constraint on the CRe population from the radio synchrotron spectrum and profile fit in the first case using the {\tt ExponentialCutoffPowerLaw} model. We note that the Figure is provided for our baseline CRp radial model ($n_{\rm CRp} \propto n_{\rm gas}^{1/2}$) in the scenario in which 4FGL~J1256.9+2736 was replace by the ICM component (scenario 2). In Appendix~\ref{app:CRe_alternative_constraints}, we also provide these constraints in the case of the alternative models considered. First, we note that the tSZ signal, included as a dashed gray line given our pressure model, is not negligible for the highest frequency data point, but we correct for it \citep[see also][for a dedicated analysis]{Brunetti2013}. Our model provides a reasonable fit to the data for both the spectrum and the profile (this is also the case for the other considered models, see Appendix~\ref{app:CRe_alternative_constraints}, and our results do not depend significantly on the considered CRe$_1$ spectral model). The slope of the synchrotron emission from CRe$_2$ is similar to the one from the CRe$_1$, but it is significantly less curved and no high energy cutoff is present in the CRe$_2$ distribution. We can see on the spectrum that the radio emission within $0.48 \times R_{500}$ (629 kpc, 0.36 deg) from CRe$_2$ is overall a factor of about four lower than the total emission (except at high frequency where it reaches similar values). On the profile, the CRe$_2$ emission is significantly more concentrated than the total radio signal in the case of this CRp spatial model (although it is also true for all CRp model, see Appendix~\ref{app:CRe_alternative_constraints}) and lead to slightly over-fitting the total synchrotron emission in the core when added to the CRe$_1$ contribution. This high concentration is expected because the CRe$_2$ profile arises as the product of the gas density and the CRp density. The synchrotron profile, which arises from the product of the magnetic field profile and the total CRe density, is so flat that it requires a nearly flat CRe distribution given the fixed magnetic field profile \citep[as also noted in][]{Zandanel2014b}. Thus, this could be achieved for the CRe$_2$ only at the cost of an inverted CRp profile (rising with radius). The number density of CRe is at a level of about $10^{-14}$ and $10^{-17}$ cm$^{-3}$ GeV$^{-1}$ at 1 GeV and 10 GeV, respectively, for both populations, but with opposite radial dependences. These constraints on the CRe populations translate into a ratio CRe$_1$ to CRe$_2$ that increases with radius and that does not depend much on energy up to 10 GeV (before the cutoff; best-fit $E_{\rm cut, CRe_1} = 17$ GeV). Given this CRp spatial model, the CRe$_1$/CRe$_2$ ratio is slightly below unity in the core, and strongly rises to reach about 100 at 2 Mpc.

In Figure~\ref{fig:radio_implication_observable2}, we show the constraint on the CRe population in the second case (CRe$_2$ are the seed to the CRe$_1$, pure reacceleration). As for Figure~\ref{fig:radio_implication_observable}, the alternative models considered in this paper are shown in Appendix~\ref{app:CRe_alternative_constraints}. The model also provides a good fit to the data as shown in the top panels, where only the CRe$_1$ population (as the reaccelerated CRe$_2$ population) is included. On the bottom panels, the interpretation is now different as the amount of CRe$_1$ now correspond to the CRe$_2$ population after reacceleration. As can be observed, the best-fit CRe$_1$ profile is nearly flat, while the original seed population is more concentrated in the core. The amount of reacceleration relative to the hadronic steady state case is thus relatively low in the core (in agreement with no reacceleration there, or even favoring a boost lower than one for the most compact CRp profile), but strongly increases in the outskirt reaching about a factor of 100 at 2 Mpc. As in the previous case, this radial dependence depends on the considered model for the CRp distribution and we show in Appendix~\ref{app:CRe_alternative_constraints} that flatter is the CRp distribution, flatter will be the reacceleration boost profile. Nevertheless, in all the considered cases, the reacceleration boost (relative to the steady state hadronic model) increases with radius. The energy dependence of the boost, comparing the values for 1 GeV and 10 GeV electrons, is not much affected by the choice of the spectral model for the CRe$_1$ population.

To summarize how the radio data were used, we used the WSRT data to establish the radio profile, but for the spectral dependence, we were forced to use only the data within $0.48 \times R_{500}$ where measurements at other frequencies were available. The larger halo size and flux from the WSRT data, which are due to its increased sensitivity, are discussed in detail by \cite{Brunetti2013}. Those authors showed (in their Figure 2, right) that the observed size of the halo was a strong function of the signal to noise ratio, and that the WSRT data were therefore the most reliable. The magenta diamond in Figure~\ref{fig:radio_implication_observable} and~\ref{fig:radio_implication_observable2} shows how the WSRT flux is above  the rest of the spectral points. In order to examine the effects of increasing the entire spectrum by a factor of two to match the WSRT point, we added an extra normalization parameter, as discussed earlier in this Section. We find that the ratio between the total radio emission and that arising from CRe$_2$ would increase from 4 to 8. However, the effect of changing such normalization would not impact the conclusions of this paper.

In Appendix~\ref{app:CRe_alternative_constraints}, we also provide the results when including both the ICM component and 4FGL~J1256.9+2736 in the \textit{Fermi}-LAT sky model (scenario 3). While the CRe$_2$ synchrotron spectrum is slightly steeper and the amplitude of the CRe$_2$ component is reduced, the shapes of the CRe$_2$/CRe$_1$ profile (case 1), or reacceleration boost profile (case 2) are not significantly changed given the large uncertainties.

\begin{figure*}
\centering
\includegraphics[width=0.45\textwidth]{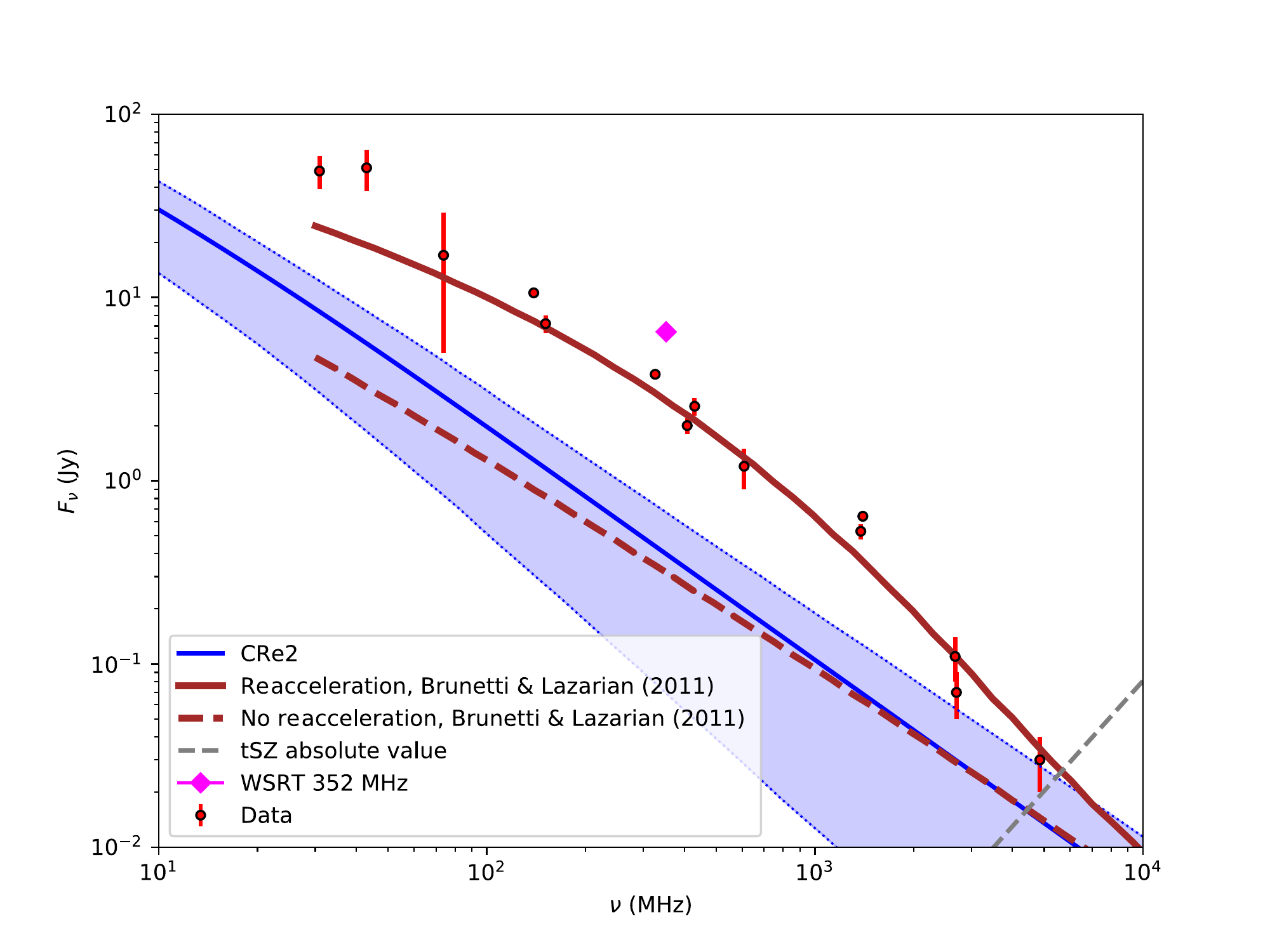}
\includegraphics[width=0.45\textwidth]{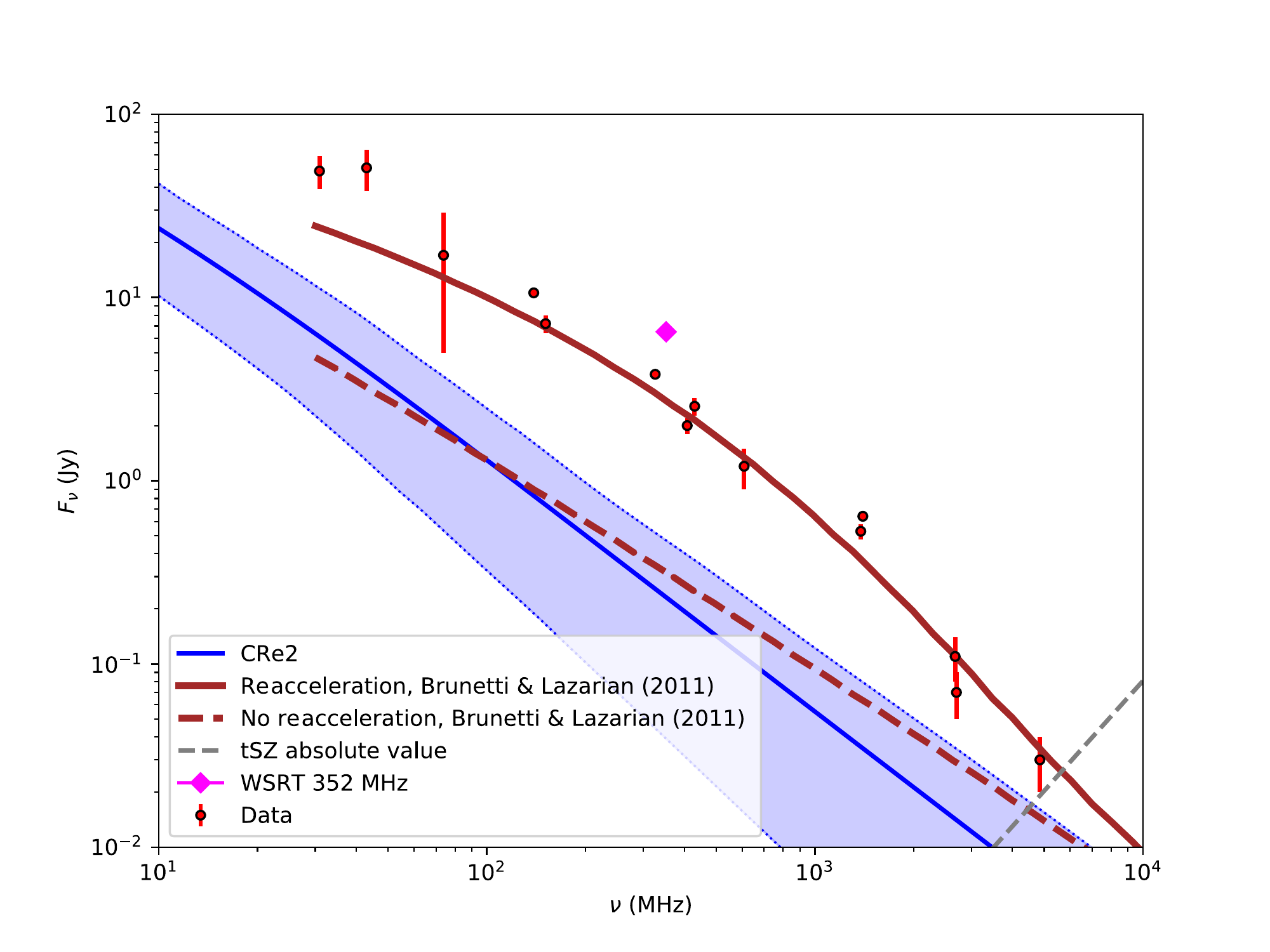}
\caption{\small Comparison between the CRe$_2$ induced synchrotron spectrum to the reacceleration model developed by \cite{Brunetti2011}, in the case of a flat CRp population. The solid brown line corresponds to the full reacceleration model, while the dashed brown line corresponds to the case where reacceleration is switched off \citep[see][for more details]{Brunetti2011}. {\bf Left}: Replacing 4FGL~J1256.9+2736 by the cluster diffuse component (scenario 2). {\bf Right}: Including both 4FGL~J1256.9+2736 and the cluster diffuse component in the sky model (scenario 3).}
\label{fig:radio_reacceleration_comparison}
\end{figure*}

Finally, in order to compare the distribution of CRp that we measure to that expected in reacceleration models to explain the radio emission, we compare our CRe$_2$ induced synchrotron spectrum to that of the model developed by \cite{Brunetti2011} in Figure~\ref{fig:radio_reacceleration_comparison}. To best match the model developed by \cite{Brunetti2011}, we use the ICM model in which the CRp radial profile is flat, but we note that the two models are not strictly equal. For instance, \cite{Brunetti2011} use an isothermal $\beta$-model, while we set the thermodynamic profiles to X-ray and tSZ data \citep[see Section 4.3.1 from][for more details]{Brunetti2011}. As the reacceleration model was calibrated on the Coma cluster, the total radio synchrotron (solid brown line) matches well the data, as expected. When the reacceleration is switched off (dashed brown line), only the emission from secondaries directly produced from hadronic collisions remains, which compares very well to the prediction from our model. This shows that the distribution of CRp that we measure provides an excellent match to what is needed in the reacceleration model developed by \cite{Brunetti2011} to explain the overall radio spectrum, when including reacceleration.

\section{Discussions}\label{sec:Discussions}
\subsection{Comparison to previous analysis}
Constraints on the $\gamma$-ray emission of the Coma cluster have been obtained in earlier work using \textit{Fermi}-LAT data \citep[see][]{Arlen2012,Prokhorov2014,Zandanel2014,Ackermann2016,Keshet2018,Xi2018}. In general, the signal is modeled using spatial templates together with power law distribution for the photon spectrum. Our results push forward such analysis by directly using a physical model for the CRp and thermal gas. Thus we are able to directly constrain the physics of CR by comparing model and data, and push the analysis by investigating the implications of $\gamma$-ray emission for the CRe population. Our results are in broad agreement with previous searches and limits. Nevertheless, we stress that the reported CRp to thermal energy (or pressure) are generally obtained under the assumption of a given spectral distribution (harder the spectrum and more stringent the limit). Spectral index values used in the literature correspond to spectra that are generally much harder ($\sim 2.1 - 2.3$) than what we obtain here (see also below for further discussions).

The \textit{Fermi}-LAT collaboration observed $\gamma$-ray excess within the virial radius of Coma using six years of data \citep{Ackermann2016}. However, the signal was too faint for detailed investigation and they published upper limits on the signal for various templates. Our results are consistent with that of \cite{Ackermann2016}, although using twice the amount of data and an analysis that differs in various aspects. They provide an upper limit of $5.2 \times 10^{-9}$ ph cm$^{-2}$ s$^{-1}$ ($E>100$ MeV). Extrapolating our best-fit model in the same energy range, we obtain a compatible total flux of $2.1 \times 10^{-9}$ ph cm$^{-2}$ s$^{-1}$ (in scenario 2, to match \citealt{Ackermann2016}). Among the main differences, we note that they use a power law emission model with a spectral index of 2.3 and spatial distribution based on WSRT, while our baseline model is directly connected to the underlying CR population and the spectrum that we measure is significantly steeper than their model. Given this framework, they obtain a TS value of 13, compared to about 27 in our case.

The detection of $\gamma$-ray emission toward Coma was first claimed by \cite{Xi2018}, who used an unbinned likelihood approach (reaching TS values of about 40-50 depending on the model). While the morphology of the signal and the spectral slope of the emission that we observe agree with their results, we obtain fluxes that are significantly lower (about a factor of two), depending on the exact model used. For instance, they obtain fluxes of about $2.3 - 3.1 \times 10^{-9}$ ph cm$^{-2}$ s$^{-1}$ for disk, core, or radio and X-ray based templates ($E = [200,300]$ GeV). For the same energy range and similar models, our fluxes are $1.3-1.6 \times 10^{-9}$ ph cm$^{-2}$ s$^{-1}$. We note that their fluxes would also be excluded by the \cite{Ackermann2016} limit when extrapolating down to 100 MeV using a photon power law index of 2.7-2.8, and would nearly reach the limit when extrapolating with our best-fit model.

\cite{Keshet2018} claimed the detection of a ring-like signal at a position that correspond to the expected location of the accretion shock \citep[see also][for the detection of such an accretion shock with \textit{Planck} in A2319]{Hurier2019}. In contrast, our results do not show any ring-like structure, especially when looking at the excess profile around 3 Mpc radius (about 2 degrees).

\subsection{Cosmic ray physics}\label{sec:discussions_cr_physics}
The amount of CRp and their spectral and spatial distributions has been predicted from numerical simulation \citep[e.g.,][]{Pfrommer2007,Pinzke2010}. The amplitude of the CRp that we obtain, relative to the thermal energy, is in line with predictions from such simulations. We obtain a ratio of about 1.5\% (about 0.8\% when including 4FGL~J1256.9+2736), while simulations suggest a few percent. For instance, in \cite{Ackermann2014}, based on the work by \cite{Pinzke2010}, the CRp to thermal energy ratio of Coma was expected to be $X_{\rm CRp}(R_{200 }= 2.2 \ {\rm Mpc}) = 2.4 \times 10^{-2}$ (we account for the fact that the pressure ratio that they use is half the energy ration that we use). This value is just a factor of $\sim 2$ above the one that we constrain, although it was obtained for a different CRp index and may also depend on radius so that the comparison depends on the exact CRp spatial distribution. This CRp to thermal energy ratio is related to the CRp injection efficiency for the diffusive shock acceleration process, which increases with the Mach number. Thus, it gives access, in principle, to the microphysics of CRp acceleration. We refer to \cite{Ackermann2014} for more discussions.

As also found by \cite{Xi2018}, the spectral index of the CRp that we constrain is significantly larger than what is generally assumed \citep[$\alpha_{\rm CRp} \sim 2.5 - 3.2$ for our constraints versus $2.1 - 2.3$; see e.g.,][]{Arlen2012,Zandanel2014,Pinzke2010}. For a given shock, the CRp index is related to the Mach number: higher the Mach number and harder the spectrum \citep[see][for discussions]{Pfrommer2006}. Therefore, our results could point to shock acceleration for which the Mach numbers are overall smaller than usually expected.

The shape of the CRp is often assumed to follow the shape of the thermal density profile, with possible diffusion sometimes included \citep[e.g.,][]{Zandanel2014b}. While our results favor models with intermediate scaling ($n_{\rm CRp} \propto n_e^{\sim 1/2}$), the data are not sufficient to firmly constrain the shape of the CRp distribution. In the context of the reacceleration of CRe$_2$, a compact CRp profile would imply that reacceleration strongly increases with increasing radius, as already highlighted in \cite{Pinzke2017}. However, even flat CRp profiles lead to an increasing reacceleration boost with radius.

Our results are in agreement with earlier work that have shown that pure hadronic models were excluded given the magnetic field strength inferred from Faraday rotation measures \citep{Brunetti2012,Zandanel2014,Zandanel2014b,Brunetti2017}. Nevertheless, the hadronically induced CRe are only a factor of a few lower than the amount required, and could thus provide seeds for turbulent reacceleration models \citep{Brunetti2007,Brunetti2011}.

Throughout this work, we have fixed the magnetic field strength model to the best-fit result obtained by \cite{Bonafede2010}, despite the fact that there are relatively large uncertainties in the constraint. We also refer to the recent work by \cite{Johnson2020} who showed that even under ideal conditions, the central magnetic field cannot be determined to better than a range of 3, with corresponding uncertainties in the scaling parameter $\eta$. These uncertainties affect the results presented in Section~\ref{sec:implication_for_CR_electrons}. In our work, increasing (decreasing) the magnetic field would imply more (fewer) synchrotron emission for a given CRe population. Thus it would lead to fewer (more) CRe$_1$ for case 1, or fewer (more) reacceleration boost in case 2. This effect also applies to the radial dependence. Given the uncertainties in the magnetic field measure, these should contribute significantly to our constraints. The uncertainties in the CRe$_2$ population are nonetheless expected to dominate, and we leave the joint investigation of the CRe population and magnetic field distribution aside.

\subsection{Contamination from discrete sources}
In this paper, we have generally assumed that the $\gamma$-ray emission observed in the direction of the Coma was originating from hadronic interactions in the ICM. We have also considered the case where 4FGL~J1256.9+2736 was included in the overall model as a point source contaminant in addition to the diffuse emission, but still modeling the diffuse component as from hadronic interactions.

In fact, cluster member and radio galaxies may also contribute significantly to the total signal. In \cite{Ackermann2016}, a minimum flux was estimated for the two dominant central radio galaxies NGC 4869 and NGC 4874 assuming that the electrons responsible for the radio emission also generate inverse Compton emission in the \textit{Fermi}-LAT band and assuming simple scaling relations for the calculation. They obtained luminosities of about $6 \times 10^{40}$ and $2 \times 10^{40}$ erg s$^{-1}$ ($0.1 < E < 10$ GeV), which is a factor of $\sim 20$ lower than what we measure for the hadronic emission at $E>0.2$ GeV (thus even lower in our energy range for usual spectra).

However, other sources may also contribute and might lead to an unresolved diffuse component. In particular star forming galaxies could lead to $\gamma$-ray emission for which the associated flux is very uncertain and within the range of our constraints \citep[in the range $3 \times 10^{40} - 3 \times 10^{42}$ erg s$^{-1}$ for energies in 0.1-100 GeV; see][]{Storm2012}.

The cluster member radio and star forming galaxies are also expected to generate CR that diffuse in the ICM and would contribute to the population which we model in this paper. \cite{Rephaeli2016} calculated that they might account for a significant amount of the radio diffuse emission, as well as in the $\gamma$-rays.

\section{Summary and conclusions}\label{sec:Summary_and_conclusions}
This paper presented the analysis of nearly 12 years of \textit{Fermi}-LAT data toward the Coma cluster, together with multiwavelength data and using the {\tt MINOT} software in order to model the signal. Different scenarios were considered to model the signal: no cluster diffuse emission, replacing the source 4FGL~J1256.9+2736 by a diffuse cluster model, or accounting for a combination of both. Assuming that the diffuse emission was associated with the hadronically induced $\gamma$-rays in the ICM, we investigated the implications for the CR physics, for both CRp and CRe.

The signal was modeled assuming that the $\gamma$-ray emission arises from hadronic interactions between CRp and the thermal gas. The thermal gas model was set using \textit{ROSAT} X-ray and \textit{Planck} tSZ data. In this model secondary CRe are also produced, and we fixed the magnetic field to that obtained from Faraday rotation measurements in order to compute the associated radio synchrotron emission. The CRp spatial model was calibrated assuming a scaling relative to the thermal gas profiles. The CRp normalization and spectrum were defined relative to the thermal gas energy and using a power law for the spectrum, respectively. In addition to spherically symmetric models, we built two-dimentional spatial templates based on X-ray, tSZ, radio and galaxy density images to fit the \textit{Fermi}-LAT data.

We detect $\gamma$-ray emission in the direction of the Coma cluster. The detection level depends on the model considered, in the range ${\rm TS} \sim 24 - 34$, corresponding to a significance of about $4.9 - 5.8 \sigma$. While extended models provide a better description of the data, it is not possible to strictly exclude that the signal is associated with a point source, or a combination of the two, and we include this possibility in our analysis. The morphology of the signal is elongated in the northeast to southwest direction, in agreement with other wavelengths. The peak of the emission is about 0.5 degree offset in the southwest direction with respect to the X-ray peak, and coincides with the well-know merger extension.

Using an MCMC approach, we constrained the amplitude and the spectral index of the CRp population in the cluster assuming that at least part of the signal was associated with the diffuse ICM hadronic interactions. We find that the energy stored in the CRp is about 1.5\% of the thermal energy of the Coma cluster (0.8\% when including 4FGL~J1256.9+2736 and the cluster in the model). The slope is larger than what is usually assumed, around 2.8, although with large error bars. In the framework of diffuse shock acceleration, this could implies that CRp acceleration arises in weaker shocks than what is usually assumed.

Secondary CRe are also expected from hadronic interactions. Their population was computed in a steady state scenario, leading to a radio synchrotron emission that is about 4 times lower than what is observed. While a pure hadronic origin of the radio emission is ruled out, these secondary CRe could serve as the seeds for turbulent reacceleration. In this model, the reacceleration should increase with radius, depending on the exact CRp spatial distribution. Alternatively, an independent CRe population could be at the origin of the remaining radio emission, but it would require a nearly flat radial distribution.

Our results show that after almost 12 years of observations, the diffuse $\gamma$-ray emission from galaxy clusters might now become accessible with the \textit{Fermi}-LAT. Since the hadronic emission is expected to be a universal property of galaxy clusters, our results might be reproducible for other clusters, renewing the interest of such analysis, although Coma might be the best target for such searches. In the scenario in which the signal indeed arises from diffuse ICM hadronic interactions, if the large value of the CRp spectral slope is confirmed and if the CRp content of clusters is, as expected, a universal property, the very high energy $\gamma$-ray emission ($\sim$TeV) could therefore be much lower than usually assumed. In this context, it would be challenging for future ground-based $\gamma$-ray observatories \citep[e.g., the Cherenkov Telescope Array,][]{CTA2019} to detect the diffuse emission associated with hadronic interactions in the ICM of galaxy clusters.

\begin{acknowledgements}
We are thankful to the anonymous referee for useful comments that helped improve the quality of the paper.
We thank Gianfranco Brunetti for useful discussions and for providing us with the reacceleration model used in Figure~\ref{fig:radio_reacceleration_comparison}.
Partial support for LR comes from U.S. National Science Foundation grant  AST 17-14205 to the University of Minnesota. 
This research made use of Astropy, a community-developed core Python package for Astronomy \citep{Astropy2013}, in addition to NumPy \citep{VanDerWalt2011}, SciPy \citep{Jones2001} and Ipython \citep{Perez2007}. Figures were generated using Matplotlib \citep{Hunter2007} and seaborn \citep{Waskom2020}.
\end{acknowledgements}

\bibliography{coma_biblio}

\begin{thebibliography}{97}
\expandafter\ifx\csname natexlab\endcsname\relax\def\natexlab#1{#1}\fi

\bibitem[{{Abdollahi} {et~al.}(2020){Abdollahi}, {Acero}, {Ackermann},
  {Ajello}, {Atwood}, {Axelsson}, {Baldini}, {Ballet}, {Barbiellini},
  {Bastieri}, {Becerra Gonzalez}, {Bellazzini}, {Berretta}, {Bissaldi}, {Bland
  ford}, {Bloom}, {Bonino}, {Bottacini}, {Brandt}, {Bregeon}, {Bruel},
  {Buehler}, {Burnett}, {Buson}, {Cameron}, {Caputo}, {Caraveo}, {Casandjian},
  {Castro}, {Cavazzuti}, {Charles}, {Chaty}, {Chen}, {Cheung}, {Chiaro},
  {Ciprini}, {Cohen-Tanugi}, {Cominsky}, {Coronado-Bl{\'a}zquez}, {Costantin},
  {Cuoco}, {Cutini}, {D'Ammando}, {DeKlotz}, {Torre Luque}, {de Palma},
  {Desai}, {Digel}, {Lalla}, {Mauro}, {Venere}, {Dom{\'\i}nguez}, {Dumora},
  {Dirirsa}, {Fegan}, {Ferrara}, {Franckowiak}, {Fukazawa}, {Funk}, {Fusco},
  {Gargano}, {Gasparrini}, {Giglietto}, {Giommi}, {Giordano}, {Giroletti},
  {Glanzman}, {Green}, {Grenier}, {Griffin}, {Grondin}, {Grove}, {Guiriec},
  {Harding}, {Hayashi}, {Hays}, {Hewitt}, {Horan}, {J{\'o}hannesson},
  {Johnson}, {Kamae}, {Kerr}, {Kocevski}, {Kovac'evic'}, {Kuss}, {Landriu},
  {Larsson}, {Latronico}, {Lemoine-Goumard}, {Li}, {Liodakis}, {Longo},
  {Loparco}, {Lott}, {Lovellette}, {Lubrano}, {Madejski}, {Maldera},
  {Malyshev}, {Manfreda}, {Marchesini}, {Marcotulli}, {Mart{\'\i}-Devesa},
  {Martin}, {Massaro}, {Mazziotta}, {McEnery}, {Mereu}, {Meyer}, {Michelson},
  {Mirabal}, {Mizuno}, {Monzani}, {Morselli}, {Moskalenko}, {Negro}, {Nuss},
  {Ojha}, {Omodei}, {Orienti}, {Orlando}, {Ormes}, {Palatiello}, {Paliya},
  {Paneque}, {Pei}, {Pe{\~n}a-Herazo}, {Perkins}, {Persic}, {Pesce-Rollins},
  {Petrosian}, {Petrov}, {Piron}, {Poon}, {Porter}, {Principe}, {Rain{\`o}},
  {Rando}, {Razzano}, {Razzaque}, {Reimer}, {Reimer}, {Remy}, {Reposeur},
  {Romani}, {Parkinson}, {Schinzel}, {Serini}, {Sgr{\`o}}, {Siskind}, {Smith},
  {Spandre}, {Spinelli}, {Strong}, {Suson}, {Tajima}, {Takahashi}, {Tak},
  {Thayer}, {Thompson}, {Tibaldo}, {Torres}, {Torresi}, {Valverde}, {Klaveren},
  {Zyl}, {Wood}, {Yassine}, \& {Zaharijas}}]{Abdollahi2020}
{Abdollahi}, S. {et~al.} 2020, \apjs, 247, 33, 2005.11208

\bibitem[{{Ackermann} {et~al.}(2014){Ackermann}, {Ajello}, {Albert},
  {Allafort}, {Atwood}, {Baldini}, {Ballet}, {Barbiellini}, {Bastieri},
  {Bechtol}, {Bellazzini}, {Bloom}, {Bonamente}, {Bottacini}, {Brandt},
  {Bregeon}, {Brigida}, {Bruel}, {Buehler}, {Buson}, {Caliandro}, {Cameron},
  {Caraveo}, {Cavazzuti}, {Chaves}, {Chiang}, {Chiaro}, {Ciprini}, {Claus},
  {Cohen-Tanugi}, {Conrad}, {D'Ammando}, {de Angelis}, {de Palma}, {Dermer},
  {Digel}, {Drell}, {Drlica-Wagner}, {Favuzzi}, {Franckowiak}, {Funk}, {Fusco},
  {Gargano}, {Gasparrini}, {Germani}, {Giglietto}, {Giordano}, {Giroletti},
  {Godfrey}, {Gomez-Vargas}, {Grenier}, {Guiriec}, {Gustafsson}, {Hadasch},
  {Hayashida}, {Hewitt}, {Hughes}, {Jeltema}, {J{\'o}hannesson}, {Johnson},
  {Kamae}, {Kataoka}, {Kn{\"o}dlseder}, {Kuss}, {Lande}, {Larsson},
  {Latronico}, {Llena Garde}, {Longo}, {Loparco}, {Lovellette}, {Lubrano},
  {Mayer}, {Mazziotta}, {McEnery}, {Michelson}, {Mitthumsiri}, {Mizuno},
  {Monzani}, {Morselli}, {Moskalenko}, {Murgia}, {Nemmen}, {Nuss}, {Ohsugi},
  {Orienti}, {Orlando}, {Ormes}, {Perkins}, {Pesce-Rollins}, {Piron}, {Pivato},
  {Rain{\`o}}, {Rando}, {Razzano}, {Razzaque}, {Reimer}, {Reimer}, {Ruan},
  {S{\'a}nchez-Conde}, {Schulz}, {Sgr{\`o}}, {Siskind}, {Spandre}, {Spinelli},
  {Storm}, {Strong}, {Suson}, {Takahashi}, {Thayer}, {Thayer}, {Thompson},
  {Tibaldo}, {Tinivella}, {Torres}, {Troja}, {Uchiyama}, {Usher},
  {Vandenbroucke}, {Vianello}, {Vitale}, {Winer}, {Wood}, {Zimmer}, {Fermi-LAT
  Collaboration}, {Pinzke}, \& {Pfrommer}}]{Ackermann2014}
{Ackermann}, M. {et~al.} 2014, \apj, 787, 18, 1308.5654

\bibitem[{{Ackermann} {et~al.}(2016){Ackermann}, {Ajello}, {Albert}, {Atwood},
  {Baldini}, {Ballet}, {Barbiellini}, {Bastieri}, {Bechtol}, {Bellazzini},
  {Bissaldi}, {Blandford}, {Bloom}, {Bonino}, {Bottacini}, {Bregeon}, {Bruel},
  {Buehler}, {Caliandro}, {Cameron}, {Caragiulo}, {Caraveo}, {Casandjian},
  {Cavazzuti}, {Cecchi}, {Charles}, {Chekhtman}, {Chiaro}, {Ciprini},
  {Cohen-Tanugi}, {Conrad}, {Cutini}, {D'Ammando}, {de Angelis}, {de Palma},
  {Desiante}, {Digel}, {Di Venere}, {Drell}, {Favuzzi}, {Fegan}, {Fukazawa},
  {Funk}, {Fusco}, {Gargano}, {Gasparrini}, {Giglietto}, {Giordano},
  {Giroletti}, {Godfrey}, {Green}, {Grenier}, {Guiriec}, {Hays}, {Hewitt},
  {Horan}, {J{\'o}hannesson}, {Kuss}, {Larsson}, {Latronico}, {Li}, {Li},
  {Longo}, {Loparco}, {Lovellette}, {Lubrano}, {Madejski}, {Maldera},
  {Manfreda}, {Mayer}, {Mazziotta}, {Michelson}, {Mitthumsiri}, {Mizuno},
  {Monzani}, {Morselli}, {Moskalenko}, {Murgia}, {Nuss}, {Ohsugi}, {Orienti},
  {Orlando}, {Ormes}, {Paneque}, {Pesce-Rollins}, {Petrosian}, {Piron},
  {Pivato}, {Porter}, {Rain{\`o}}, {Rando}, {Razzano}, {Reimer}, {Reimer},
  {S{\'a}nchez-Conde}, {Sgr{\`o}}, {Siskind}, {Spada}, {Spandre}, {Spinelli},
  {Tajima}, {Takahashi}, {Thayer}, {Tibaldo}, {Torres}, {Tosti}, {Troja},
  {Vianello}, {Wood}, {Zimmer}, {Fermi-LAT Collaboration}, \&
  {Rephaeli}}]{Ackermann2016}
------. 2016, \apj, 819, 149, 1507.08995

\bibitem[{{Ackermann} {et~al.}(2010){Ackermann}, {Ajello}, {Allafort},
  {Baldini}, {Ballet}, {Barbiellini}, {Bastieri}, {Bechtol}, {Bellazzini},
  {Blandford}, {Blasi}, {Bloom}, {Bonamente}, {Borgland }, {Bouvier}, {Brandt},
  {Bregeon}, {Brigida}, {Bruel}, {Buehler}, {Buson}, {Caliandro}, {Cameron},
  {Caraveo}, {Carrigan}, {Casandjian}, {Cavazzuti}, {Cecchi}, {{\c{C}}elik},
  {Charles}, {Chekhtman}, {Cheung}, {Chiang}, {Ciprini}, {Claus},
  {Cohen-Tanugi}, {Colafrancesco}, {Cominsky}, {Conrad}, {Dermer}, {de Palma},
  {Silva}, {Drell}, {Dubois}, {Dumora}, {Edmonds}, {Farnier}, {Favuzzi},
  {Frailis}, {Fukazawa}, {Funk}, {Fusco}, {Gargano}, {Gasparrini}, {Gehrels},
  {Germani}, {Giglietto}, {Giordano}, {Giroletti}, {Glanzman}, {Godfrey},
  {Grenier}, {Grondin}, {Guiriec}, {Hadasch}, {Harding}, {Hayashida}, {Hays},
  {Horan}, {Hughes}, {Jeltema}, {J{\'o}hannesson}, {Johnson}, {Johnson},
  {Johnson}, {Kamae}, {Katagiri}, {Kataoka}, {Kerr}, {Kn{\"o}dlseder}, {Kuss},
  {Lande}, {Latronico}, {Lee}, {Lemoine-Goumard}, {Longo}, {Loparco}, {Lott},
  {Lovellette}, {Lubrano}, {Madejski}, {Makeev}, {Mazziotta}, {Michelson},
  {Mitthumsiri}, {Mizuno}, {Moiseev}, {Monte}, {Monzani}, {Morselli},
  {Moskalenko}, {Murgia}, {Naumann-Godo}, {Nolan}, {Norris}, {Nuss}, {Ohsugi},
  {Omodei}, {Orlando}, {Ormes}, {Ozaki}, {Paneque}, {Panetta}, {Pepe},
  {Pesce-Rollins}, {Petrosian}, {Pfrommer}, {Piron}, {Porter}, {Profumo},
  {Rain{\`o}}, {Rando}, {Razzano}, {Reimer}, {Reimer}, {Reposeur}, {Ripken},
  {Ritz}, {Rodriguez}, {Romani}, {Roth}, {Sadrozinski}, {Sander}, {Saz
  Parkinson}, {Scargle}, {Sgr{\`o}}, {Siskind}, {Smith}, {Spandre}, {Spinelli},
  {Starck}, {Stawarz}, {Strickman}, {Strong}, {Suson}, {Tajima}, {Takahashi},
  {Takahashi}, {Tanaka}, {Thayer}, {Thayer}, {Tibaldo}, {Tibolla}, {Torres},
  {Tosti}, {Tramacere}, {Uchiyama}, {Usher}, {Vandenbroucke}, {Vasileiou},
  {Vilchez}, {Vitale}, {Waite}, {Wang}, {Winer}, {Wood}, {Yang}, {Ylinen}, \&
  {Ziegler}}]{Ackermann2010}
------. 2010, \apjl, 717, L71, 1006.0748

\bibitem[{{Ackermann} {et~al.}(2012){Ackermann}, {Ajello}, {Atwood}, {Baldini},
  {Ballet}, {Barbiellini}, {Bastieri}, {Bechtol}, {Bellazzini}, {Berenji},
  {Bland ford}, {Bloom}, {Bonamente}, {Borgland }, {Brandt}, {Bregeon},
  {Brigida}, {Bruel}, {Buehler}, {Buson}, {Caliandro}, {Cameron}, {Caraveo},
  {Cavazzuti}, {Cecchi}, {Charles}, {Chekhtman}, {Chiang}, {Ciprini}, {Claus},
  {Cohen-Tanugi}, {Conrad}, {Cutini}, {de Angelis}, {de Palma}, {Dermer},
  {Digel}, {Silva}, {Drell}, {Drlica-Wagner}, {Falletti}, {Favuzzi}, {Fegan},
  {Ferrara}, {Focke}, {Fortin}, {Fukazawa}, {Funk}, {Fusco}, {Gaggero},
  {Gargano}, {Germani}, {Giglietto}, {Giordano}, {Giroletti}, {Glanzman},
  {Godfrey}, {Grove}, {Guiriec}, {Gustafsson}, {Hadasch}, {Hanabata},
  {Harding}, {Hayashida}, {Hays}, {Horan}, {Hou}, {Hughes}, {J{\'o}hannesson},
  {Johnson}, {Johnson}, {Kamae}, {Katagiri}, {Kataoka}, {Kn{\"o}dlseder},
  {Kuss}, {Lande}, {Latronico}, {Lee}, {Lemoine-Goumard}, {Longo}, {Loparco},
  {Lott}, {Lovellette}, {Lubrano}, {Mazziotta}, {McEnery}, {Michelson},
  {Mitthumsiri}, {Mizuno}, {Monte}, {Monzani}, {Morselli}, {Moskalenko},
  {Murgia}, {Naumann-Godo}, {Norris}, {Nuss}, {Ohsugi}, {Okumura}, {Omodei},
  {Orlando}, {Ormes}, {Paneque}, {Panetta}, {Parent}, {Pesce-Rollins},
  {Pierbattista}, {Piron}, {Pivato}, {Porter}, {Rain{\`o}}, {Rando}, {Razzano},
  {Razzaque}, {Reimer}, {Reimer}, {Sadrozinski}, {Sgr{\`o}}, {Siskind},
  {Spandre}, {Spinelli}, {Strong}, {Suson}, {Takahashi}, {Tanaka}, {Thayer},
  {Thayer}, {Thompson}, {Tibaldo}, {Tinivella}, {Torres}, {Tosti}, {Troja},
  {Usher}, {Vandenbroucke}, {Vasileiou}, {Vianello}, {Vitale}, {Waite}, {Wang},
  {Winer}, {Wood}, {Wood}, {Yang}, {Ziegler}, \& {Zimmer}}]{Ackermann2012}
------. 2012, \apj, 750, 3, 1202.4039

\bibitem[{{Adam} {et~al.}(2020){Adam}, {Goksu}, {Leing{\"a}rtner-Goth},
  {Ettori}, {Gnatyk}, {Hnatyk}, {H{\"u}tten}, {P{\'e}rez-Romero},
  {S{\'a}nchez-Conde}, \& {Sergijenko}}]{Adam2020}
{Adam}, R. {et~al.} 2020, arXiv e-prints, arXiv:2009.05373, 2009.05373

\bibitem[{{Aharonian} {et~al.}(2009){Aharonian}, {Akhperjanian}, {Anton},
  {Barres de Almeida}, {Bazer-Bachi}, {Becherini}, {Behera}, {Bernl{\"o}hr},
  {Boisson}, {Bochow}, {Borrel}, {Brion}, {Brucker}, {Brun}, {B{\"u}hler},
  {Bulik}, {B{\"u}sching}, {Boutelier}, {Chadwick}, {Charbonnier}, {Chaves},
  {Cheesebrough}, {Chounet}, {Clapson}, {Coignet}, {Dalton}, {Daniel},
  {Davids}, {Degrange}, {Deil}, {Dickinson}, {Djannati-Ata{\"\i}}, {Domainko},
  {O'C. Drury}, {Dubois}, {Dubus}, {Dyks}, {Dyrda}, {Egberts},
  {Emmanoulopoulos}, {Espigat}, {Farnier}, {Feinstein}, {Fiasson},
  {F{\"o}rster}, {Fontaine}, {F{\"u}{\ss}ling}, {Gabici}, {Gallant},
  {G{\'e}rard}, {Giebels}, {Glicenstein}, {Gl{\"u}ck}, {Goret}, {G{\"o}hring},
  {Hauser}, {Hauser}, {Heinz}, {Heinzelmann}, {Henri}, {Hermann}, {Hinton},
  {Hoffmann}, {Hofmann}, {Holleran}, {Hoppe}, {Horns}, {Inoue}, {Jacholkowska},
  {de Jager}, {Jahn}, {Jung}, {Katarzy{\'n}ski}, {Katz}, {Kaufmann},
  {Kendziorra}, {Kerschhaggl}, {Khangulyan}, {Kh{\'e}lifi}, {Keogh},
  {Klu{\'z}niak}, {Kneiske}, {Komin}, {Kosack}, {Lamanna}, {Lenain}, {Lohse},
  {Marandon}, {Martin}, {Martineau-Huynh}, {Marcowith}, {Maurin}, {McComb},
  {Medina}, {Moderski}, {Moulin}, {Naumann-Godo}, {de Naurois}, {Nedbal},
  {Nekrassov}, {Niemiec}, {Nolan}, {Ohm}, {Olive}, {de O{\~n}a Wilhelmi},
  {Orford}, {Ostrowski}, {Panter}, {Paz Arribas}, {Pedaletti}, {Pelletier},
  {Petrucci}, {Pita}, {P{\"u}hlhofer}, {Punch}, {Quirrenbach}, {Raubenheimer},
  {Raue}, {Rayner}, {Renaud}, {Reimer}, {Rieger}, {Ripken}, {Rob},
  {Rosier-Lees}, {Rowell}, {Rudak}, {Rulten}, {Ruppel}, {Sahakian},
  {Santangelo}, {Schlickeiser}, {Sch{\"o}ck}, {Schr{\"o}der}, {Schwanke},
  {Schwarzburg}, {Schwemmer}, {Shalchi}, {Sikora}, {Skilton}, {Sol},
  {Spanglfoer}, {Stawarz}, {Steenkamp}, {Stegmann}, {Superina}, {Szostek},
  {Tam}, {Tavernet}, {Terrier}, {Tibolla}, {Tluczykont}, {van Eldik},
  {Vasileiadis}, {Venter}, {Venter}, {Vialle}, {Vincent}, {Vivier}, {V{\"o}lk},
  {Volpe}, {Wagner}, {Ward}, {Zdziarski}, \& {Zech}}]{Aharonian2009}
{Aharonian}, F. {et~al.} 2009, \aap, 502, 437, 0907.0727

\bibitem[{{Aharonian} {et~al.}(2010){Aharonian}, {Kelner}, \&
  {Prosekin}}]{Aharonian2010}
{Aharonian}, F.~A., {Kelner}, S.~R., \& {Prosekin}, A.~Y. 2010, \prd, 82,
  043002, 1006.1045

\bibitem[{{Akamatsu} {et~al.}(2013){Akamatsu}, {Inoue}, {Sato}, {Matsusita},
  {Ishisaki}, \& {Sarazin}}]{Akamatsu2013}
{Akamatsu}, H., {Inoue}, S., {Sato}, T., {Matsusita}, K., {Ishisaki}, Y., \&
  {Sarazin}, C.~L. 2013, \pasj, 65, 89, 1302.2907

\bibitem[{{Aleksi{\'c}} {et~al.}(2010){Aleksi{\'c}}, {Antonelli}, {Antoranz},
  {Backes}, {Baixeras}, {Balestra}, {Barrio}, {Bastieri}, {Becerra
  Gonz{\'a}lez}, {Bednarek}, {Berdyugin}, {Berger}, {Bernardini}, {Biland},
  {Bock}, {Bonnoli}, {Bordas}, {Borla Tridon}, {Bosch-Ramon}, {Bose}, {Braun},
  {Bretz}, {Britzger}, {Camara}, {Carmona}, {Carosi}, {Colin}, {Commichau},
  {Contreras}, {Cortina}, {Costado}, {Covino}, {Dazzi}, {De Angelis}, {De Cea
  del Pozo}, {De los Reyes}, {De Lotto}, {De Maria}, {De Sabata}, {Delgado
  Mendez}, {Doert}, {Dom{\'\i}nguez}, {Dominis Prester}, {Dorner}, {Doro},
  {Elsaesser}, {Errando}, {Ferenc}, {Fonseca}, {Font}, {Galante}, {Garc{\'\i}a
  L{\'o}pez}, {Garczarczyk}, {Gaug}, {Godinovic}, {Hadasch}, {Herrero},
  {Hildebrand }, {H{\"o}hne-M{\"o}nch}, {Hose}, {Hrupec}, {Hsu}, {Jogler},
  {Klepser}, {Kr{\"a}henb{\"u}hl}, {Kranich}, {La Barbera}, {Laille},
  {Leonardo}, {Lindfors}, {Lombardi}, {Longo}, {L{\'o}pez}, {Lorenz},
  {Majumdar}, {Maneva}, {Mankuzhiyil}, {Mannheim}, {Maraschi}, {Mariotti},
  {Mart{\'\i}nez}, {Mazin}, {Meucci}, {Miranda}, {Mirzoyan}, {Miyamoto},
  {Mold{\'o}n}, {Moles}, {Moralejo}, {Nieto}, {Nilsson}, {Ninkovic}, {Orito},
  {Oya}, {Paiano}, {Paoletti}, {Paredes}, {Partini}, {Pasanen}, {Pascoli},
  {Pauss}, {Pegna}, {Perez-Torres}, {Persic}, {Peruzzo}, {Prada}, {Prandini},
  {Puchades}, {Puljak}, {Reichardt}, {Rhode}, {Rib{\'o}}, {Rico}, {Rissi},
  {R{\"u}gamer}, {Saggion}, {Saito}, {Salvati}, {S{\'a}nchez-Conde},
  {Satalecka}, {Scalzotto}, {Scapin}, {Schultz}, {Schweizer}, {Shayduk},
  {Shore}, {Sierpowska-Bartosik}, {Sillanp{\"a}{\"a}}, {Sitarek}, {Sobczynska},
  {Spanier}, {Spiro}, {Stamerra}, {Steinke}, {Struebig}, {Suric}, {Takalo},
  {Tavecchio}, {Temnikov}, {Terzic}, {Tescaro}, {Teshima}, {Torres}, {Vankov},
  {Wagner}, {Zabalza}, {Zandanel}, {Zanin}, {Zapatero}, {Pfrommer}, {Pinzke},
  {En{\ss}lin}, {Inoue}, {Ghisellini}, \& {MAGIC Collaboration}}]{Aleksic2010}
{Aleksi{\'c}}, J. {et~al.} 2010, \apj, 710, 634, 0909.3267

\bibitem[{{Arlen} {et~al.}(2012){Arlen}, {Aune}, {Beilicke}, {Benbow},
  {Bouvier}, {Buckley}, {Bugaev}, {Byrum}, {Cannon}, {Cesarini}, {Ciupik},
  {Collins-Hughes}, {Connolly}, {Cui}, {Dickherber}, {Dumm}, {Falcone},
  {Federici}, {Feng}, {Finley}, {Finnegan}, {Fortson}, {Furniss}, {Galante},
  {Gall}, {Godambe}, {Griffin}, {Grube}, {Gyuk}, {Holder}, {Huan}, {Hughes},
  {Humensky}, {Imran}, {Kaaret}, {Karlsson}, {Kertzman}, {Khassen}, {Kieda},
  {Krawczynski}, {Krennrich}, {Lee}, {Madhavan}, {Maier}, {Majumdar},
  {McArthur}, {McCann}, {Moriarty}, {Mukherjee}, {Nelson}, {O'Faol{\'a}in de
  Bhr{\'o}ithe}, {Ong}, {Orr}, {Otte}, {Park}, {Perkins}, {Pohl}, {Prokoph},
  {Quinn}, {Ragan}, {Reyes}, {Reynolds}, {Roache}, {Ruppel}, {Saxon},
  {Schroedter}, {Sembroski}, {Skole}, {Smith}, {Telezhinsky},
  {Te{\v{s}}i{\'c}}, {Theiling}, {Thibadeau}, {Tsurusaki}, {Varlotta},
  {Vivier}, {Wakely}, {Ward}, {Weinstein}, {Welsing}, {Williams}, {Zitzer},
  {Pfrommer}, \& {Pinzke}}]{Arlen2012}
{Arlen}, T. {et~al.} 2012, \apj, 757, 123, 1208.0676

\bibitem[{{Astropy Collaboration} {et~al.}(2013){Astropy Collaboration},
  {Robitaille}, {Tollerud}, {Greenfield}, {Droettboom}, {Bray}, {Aldcroft},
  {Davis}, {Ginsburg}, {Price-Whelan}, {Kerzendorf}, {Conley}, {Crighton},
  {Barbary}, {Muna}, {Ferguson}, {Grollier}, {Parikh}, {Nair}, {Unther},
  {Deil}, {Woillez}, {Conseil}, {Kramer}, {Turner}, {Singer}, {Fox}, {Weaver},
  {Zabalza}, {Edwards}, {Azalee Bostroem}, {Burke}, {Casey}, {Crawford},
  {Dencheva}, {Ely}, {Jenness}, {Labrie}, {Lim}, {Pierfederici}, {Pontzen},
  {Ptak}, {Refsdal}, {Servillat}, \& {Streicher}}]{Astropy2013}
{Astropy Collaboration} {et~al.} 2013, \aap, 558, A33, 1307.6212

\bibitem[{{Atwood} {et~al.}(2009){Atwood}, {Abdo}, {Ackermann}, {Althouse},
  {Anderson}, {Axelsson}, {Baldini}, {Ballet}, {Band}, {Barbiellini},
  {Bartelt}, {Bastieri}, {Baughman}, {Bechtol}, {B{\'e}d{\'e}r{\`e}de},
  {Bellardi}, {Bellazzini}, {Berenji}, {Bignami}, {Bisello}, {Bissaldi},
  {Blandford}, {Bloom}, {Bogart}, {Bonamente}, {Bonnell}, {Borgland },
  {Bouvier}, {Bregeon}, {Brez}, {Brigida}, {Bruel}, {Burnett}, {Busetto},
  {Caliandro}, {Cameron}, {Caraveo}, {Carius}, {Carlson}, {Casandjian},
  {Cavazzuti}, {Ceccanti}, {Cecchi}, {Charles}, {Chekhtman}, {Cheung},
  {Chiang}, {Chipaux}, {Cillis}, {Ciprini}, {Claus}, {Cohen-Tanugi},
  {Condamoor}, {Conrad}, {Corbet}, {Corucci}, {Costamante}, {Cutini}, {Davis},
  {Decotigny}, {DeKlotz}, {Dermer}, {de Angelis}, {Digel}, {do Couto e Silva},
  {Drell}, {Dubois}, {Dumora}, {Edmonds}, {Fabiani}, {Farnier}, {Favuzzi},
  {Flath}, {Fleury}, {Focke}, {Funk}, {Fusco}, {Gargano}, {Gasparrini},
  {Gehrels}, {Gentit}, {Germani}, {Giebels}, {Giglietto}, {Giommi}, {Giordano},
  {Glanzman}, {Godfrey}, {Grenier}, {Grondin}, {Grove}, {Guillemot}, {Guiriec},
  {Haller}, {Harding}, {Hart}, {Hays}, {Healey}, {Hirayama}, {Hjalmarsdotter},
  {Horn}, {Hughes}, {J{\'o}hannesson}, {Johansson}, {Johnson}, {Johnson},
  {Johnson}, {Johnson}, {Kamae}, {Katagiri}, {Kataoka}, {Kavelaars}, {Kawai},
  {Kelly}, {Kerr}, {Klamra}, {Kn{\"o}dlseder}, {Kocian}, {Komin}, {Kuehn},
  {Kuss}, {Landriu}, {Latronico}, {Lee}, {Lee}, {Lemoine-Goumard}, {Lionetto},
  {Longo}, {Loparco}, {Lott}, {Lovellette}, {Lubrano}, {Madejski}, {Makeev},
  {Marangelli}, {Massai}, {Mazziotta}, {McEnery}, {Menon}, {Meurer},
  {Michelson}, {Minuti}, {Mirizzi}, {Mitthumsiri}, {Mizuno}, {Moiseev},
  {Monte}, {Monzani}, {Moretti}, {Morselli}, {Moskalenko}, {Murgia},
  {Nakamori}, {Nishino}, {Nolan}, {Norris}, {Nuss}, {Ohno}, {Ohsugi}, {Omodei},
  {Orlando}, {Ormes}, {Paccagnella}, {Paneque}, {Panetta}, {Parent}, {Pearce},
  {Pepe}, {Perazzo}, {Pesce-Rollins}, {Picozza}, {Pieri}, {Pinchera}, {Piron},
  {Porter}, {Poupard}, {Rain{\`o}}, {Rando}, {Rapposelli}, {Razzano}, {Reimer},
  {Reimer}, {Reposeur}, {Reyes}, {Ritz}, {Rochester}, {Rodriguez}, {Romani},
  {Roth}, {Russell}, {Ryde}, {Sabatini}, {Sadrozinski}, {Sanchez}, {Sand er},
  {Sapozhnikov}, {Parkinson}, {Scargle}, {Schalk}, {Scolieri}, {Sgr{\`o}},
  {Share}, {Shaw}, {Shimokawabe}, {Shrader}, {Sierpowska-Bartosik}, {Siskind},
  {Smith}, {Smith}, {Spandre}, {Spinelli}, {Starck}, {Stephens}, {Strickman},
  {Strong}, {Suson}, {Tajima}, {Takahashi}, {Takahashi}, {Tanaka}, {Tenze},
  {Tether}, {Thayer}, {Thayer}, {Thompson}, {Tibaldo}, {Tibolla}, {Torres},
  {Tosti}, {Tramacere}, {Turri}, {Usher}, {Vilchez}, {Vitale}, {Wang},
  {Watters}, {Winer}, {Wood}, {Ylinen}, \& {Ziegler}}]{Atwood2009}
{Atwood}, W.~B. {et~al.} 2009, \apj, 697, 1071, 0902.1089

\bibitem[{{Ballet} {et~al.}(2020){Ballet}, {Burnett}, {Digel}, \&
  {Lott}}]{Ballet2020}
{Ballet}, J., {Burnett}, T.~H., {Digel}, S.~W., \& {Lott}, B. 2020, arXiv
  e-prints, arXiv:2005.11208, 2005.11208

\bibitem[{{Birkinshaw}(1999)}]{Birkinshaw1999}
{Birkinshaw}, M. 1999, \physrep, 310, 97, arXiv:astro-ph/9808050

\bibitem[{{Blasi} \& {Colafrancesco}(1999)}]{Blasi1999}
{Blasi}, P., \& {Colafrancesco}, S. 1999, Astroparticle Physics, 12, 169,
  astro-ph/9905122

\bibitem[{{B{\"o}hringer} \& {Werner}(2010)}]{Bohringer2010}
{B{\"o}hringer}, H., \& {Werner}, N. 2010, \aapr, 18, 127

\bibitem[{{Bonafede} {et~al.}(2020){Bonafede}, {Brunetti}, {Vazza},
  {Simionescu}, {Giovannini}, {Bonnassieux}, {Shimwell}, {Br{\"u}ggen}, {van
  Weeren}, {Botteon}, {Brienza}, {Cassano}, {Drabent}, {Feretti}, {de
  Gasperin}, {Gastaldello}, {di Gennaro}, {Rossetti}, {Rottgering}, {Stuardi},
  \& {Venturi}}]{Bonafede2020}
{Bonafede}, A. {et~al.} 2020, arXiv e-prints, arXiv:2011.08856, 2011.08856

\bibitem[{{Bonafede} {et~al.}(2010){Bonafede}, {Feretti}, {Murgia}, {Govoni},
  {Giovannini}, {Dallacasa}, {Dolag}, \& {Taylor}}]{Bonafede2010}
{Bonafede}, A., {Feretti}, L., {Murgia}, M., {Govoni}, F., {Giovannini}, G.,
  {Dallacasa}, D., {Dolag}, K., \& {Taylor}, G.~B. 2010, \aap, 513, A30,
  1002.0594

\bibitem[{{Bonafede} {et~al.}(2014{\natexlab{a}}){Bonafede}, {Intema},
  {Br{\"u}ggen}, {Girardi}, {Nonino}, {Kantharia}, {van Weeren}, \&
  {R{\"o}ttgering}}]{Bonafede2014}
{Bonafede}, A., {Intema}, H.~T., {Br{\"u}ggen}, M., {Girardi}, M., {Nonino},
  M., {Kantharia}, N., {van Weeren}, R.~J., \& {R{\"o}ttgering}, H.~J.~A.
  2014{\natexlab{a}}, \apj, 785, 1, 1402.1492

\bibitem[{{Bonafede} {et~al.}(2014{\natexlab{b}}){Bonafede}, {Intema},
  {Bruggen}, {Russell}, {Ogrean}, {Basu}, {Sommer}, {van Weeren}, {Cassano},
  {Fabian}, \& {Rottgering}}]{Bonafede2014b}
{Bonafede}, A. {et~al.} 2014{\natexlab{b}}, \mnras, 444, L44, 1407.4801

\bibitem[{{Branchini} {et~al.}(2017){Branchini}, {Camera}, {Cuoco}, {Fornengo},
  {Regis}, {Viel}, \& {Xia}}]{Branchini2017}
{Branchini}, E., {Camera}, S., {Cuoco}, A., {Fornengo}, N., {Regis}, M.,
  {Viel}, M., \& {Xia}, J.-Q. 2017, \apjs, 228, 8, 1612.05788

\bibitem[{{Briel} {et~al.}(1992){Briel}, {Henry}, \& {Boehringer}}]{Briel1992}
{Briel}, U.~G., {Henry}, J.~P., \& {Boehringer}, H. 1992, \aap, 259, L31

\bibitem[{{Brown} \& {Rudnick}(2011)}]{Brown2011}
{Brown}, S., \& {Rudnick}, L. 2011, \mnras, 412, 2, 1009.4258

\bibitem[{{Brunetti} {et~al.}(2012){Brunetti}, {Blasi}, {Reimer}, {Rudnick},
  {Bonafede}, \& {Brown}}]{Brunetti2012}
{Brunetti}, G., {Blasi}, P., {Reimer}, O., {Rudnick}, L., {Bonafede}, A., \&
  {Brown}, S. 2012, \mnras, 426, 956, 1207.3025

\bibitem[{{Brunetti} \& {Jones}(2014)}]{Brunetti2014}
{Brunetti}, G., \& {Jones}, T.~W. 2014, International Journal of Modern Physics
  D, 23, 1430007, 1401.7519

\bibitem[{{Brunetti} \& {Lazarian}(2007)}]{Brunetti2007}
{Brunetti}, G., \& {Lazarian}, A. 2007, \mnras, 378, 245, astro-ph/0703591

\bibitem[{{Brunetti} \& {Lazarian}(2011)}]{Brunetti2011}
------. 2011, \mnras, 410, 127, 1008.0184

\bibitem[{{Brunetti} {et~al.}(2013){Brunetti}, {Rudnick}, {Cassano},
  {Mazzotta}, {Donnert}, \& {Dolag}}]{Brunetti2013}
{Brunetti}, G., {Rudnick}, L., {Cassano}, R., {Mazzotta}, P., {Donnert}, J., \&
  {Dolag}, K. 2013, \aap, 558, A52, 1309.1820

\bibitem[{{Brunetti} {et~al.}(2017){Brunetti}, {Zimmer}, \&
  {Zandanel}}]{Brunetti2017}
{Brunetti}, G., {Zimmer}, S., \& {Zandanel}, F. 2017, \mnras, 472, 1506,
  1707.02085

\bibitem[{{Cassano} {et~al.}(2010){Cassano}, {Ettori}, {Giacintucci},
  {Brunetti}, {Markevitch}, {Venturi}, \& {Gitti}}]{Cassano2010}
{Cassano}, R., {Ettori}, S., {Giacintucci}, S., {Brunetti}, G., {Markevitch},
  M., {Venturi}, T., \& {Gitti}, M. 2010, \apjl, 721, L82, 1008.3624

\bibitem[{{Cavaliere} \& {Fusco-Femiano}(1978)}]{Cavaliere1978}
{Cavaliere}, A., \& {Fusco-Femiano}, R. 1978, \aap, 70, 677

\bibitem[{{Cherenkov Telescope Array Consortium} {et~al.}(2019){Cherenkov
  Telescope Array Consortium}, {Acharya}, {Agudo}, {Al Samarai}, {Alfaro},
  {Alfaro}, {Alispach}, {Alves Batista}, {Amans}, {Amato}, {Ambrosi},
  {Antolini}, {Antonelli}, {Aramo}, {Araya}, {Armstrong}, {Arqueros},
  {Arrabito}, {Asano}, {Ashley}, {Backes}, {Balazs}, {Balbo}, {Ballester},
  {Ballet}, {Bamba}, {Barkov}, {Barres de Almeida}, {Barrio}, {Bastieri},
  {Becherini}, {Belfiore}, {Benbow}, {Berge}, {Bernardini}, {Bernardini},
  {Bernardos}, {Bernl{\"o}hr}, {Bertucci}, {Biasuzzi}, {Bigongiari}, {Biland},
  {Bissaldi}, {Biteau}, {Blanch}, {Blazek}, {Boisson}, {Bolmont}, {Bonanno},
  {Bonardi}, {Bonavolont{\`a}}, {Bonnoli}, {Bosnjak}, {B{\"o}ttcher},
  {Braiding}, {Bregeon}, {Brill}, {Brown}, {Brun}, {Brunetti}, {Buanes},
  {Buckley}, {Bugaev}, {B{\"u}hler}, {Bulgarelli}, {Bulik}, {Burton},
  {Burtovoi}, {Busetto}, {Canestrari}, {Capalbi}, {Capitanio}, {Caproni},
  {Caraveo}, {C{\'a}rdenas}, {Carlile}, {Carosi}, {Carqu{\'\i}n}, {Carr},
  {Casanova}, {Cascone}, {Catalani}, {Catalano}, {Cauz}, {Cerruti}, {Chadwick},
  {Chaty}, {Chaves}, {Chen}, {Chen}, {Chernyakova}, {Chikawa}, {Christov},
  {Chudoba}, {Cie{\'s}lar}, {Coco}, {Colafrancesco}, {Colin}, {Conforti},
  {Connaughton}, {Conrad}, {Contreras}, {Cortina}, {Costa}, {Costantini},
  {Cotter}, {Covino}, {Crocker}, {Cuadra}, {Cuevas}, {Cumani}, {D'A{\`\i}},
  {D'Ammando}, {D'Avanzo}, {D'Urso}, {Daniel}, {Davids}, {Dawson}, {Dazzi}, {De
  Angelis}, {de C{\'a}ssia dos Anjos}, {De Cesare}, {De Franco}, {de Gouveia
  Dal Pino}, {de la Calle}, {de los Reyes Lopez}, {De Lotto}, {De Luca}, {De
  Lucia}, {de Naurois}, {de O{\~n}a Wilhelmi}, {De Palma}, {De Persio}, {de
  Souza}, {Deil}, {Del Santo}, {Delgado}, {della Volpe}, {Di Girolamo}, {Di
  Pierro}, {Di Venere}, {D{\'\i}az}, {Dib}, {Diebold}, {Djannati-Ata{\"\i}},
  {Dom{\'\i}nguez}, {Dominis Prester}, {Dorner}, {Doro}, {Drass}, {Dravins},
  {Dubus}, {Dwarkadas}, {Ebr}, {Eckner}, {Egberts}, {Einecke}, {Ekoume},
  {Els{\"a}sser}, {Ernenwein}, {Espinoza}, {Evoli}, {Fairbairn},
  {Falceta-Goncalves}, {Falcone}, {Farnier}, {Fasola}, {Fedorova}, {Fegan},
  {Fernand ez-Alonso}, {Fern{\'a}ndez-Barral}, {Ferrand}, {Fesquet},
  {Filipovic}, {Fioretti}, {Fontaine}, {Fornasa}, {Fortson}, {Freixas
  Coromina}, {Fruck}, {Fujita}, {Fukazawa}, {Funk}, {F{\"u}{\ss}ling},
  {Gabici}, {Gadola}, {Gallant}, {Garcia}, {Garcia L{\'o}pez}, {Garczarczyk},
  {Gaskins}, {Gasparetto}, {Gaug}, {Gerard}, {Giavitto}, {Giglietto}, {Giommi},
  {Giordano}, {Giro}, {Giroletti}, {Giuliani}, {Glicenstein}, {Gnatyk},
  {Godinovic}, {Goldoni}, {G{\'o}mez-Vargas}, {Gonz{\'a}lez}, {Gonz{\'a}lez},
  {G{\"o}tz}, {Graham}, {Grand i}, {Granot}, {Green}, {Greenshaw}, {Griffiths},
  {Gunji}, {Hadasch}, {Hara}, {Hardcastle}, {Hassan}, {Hayashi}, {Hayashida},
  {Heller}, {Helo}, {Hermann}, {Hinton}, {Hnatyk}, {Hofmann}, {Holder},
  {Horan}, {H{\"o}randel}, {Horns}, {Horvath}, {Hovatta}, {Hrabovsky},
  {Hrupec}, {Humensky}, {H{\"u}tten}, {Iarlori}, {Inada}, {Inome}, {Inoue},
  {Inoue}, {Inoue}, {Iocco}, {Ioka}, {Iori}, {Ishio}, {Iwamura}, {Jamrozy},
  {Janecek}, {Jankowsky}, {Jean}, {Jung-Richardt}, {Jurysek}, {Kaaret},
  {Karkar}, {Katagiri}, {Katz}, {Kawanaka}, {Kazanas}, {Kh{\'e}lifi}, {Kieda},
  {Kimeswenger}, {Kimura}, {Kisaka}, {Knapp}, {Kn{\"o}dlseder}, {Koch},
  {Kohri}, {Komin}, {Kosack}, {Kraus}, {Krause}, {Krau{\ss}}, {Kubo}, {Kukec
  Mezek}, {Kuroda}, {Kushida}, {La Palombara}, {Lamanna}, {Lang}, {Lapington},
  {Le Blanc}, {Leach}, {Lees}, {Lefaucheur}, {Leigui de Oliveira}, {Lenain},
  {Lico}, {Limon}, {Lindfors}, {Lohse}, {Lombardi}, {Longo}, {L{\'o}pez},
  {L{\'o}pez-Coto}, {Lu}, {Lucarelli}, {Luque-Escamilla}, {Lyard}, {Maccarone},
  {Maier}, {Majumdar}, {Malaguti}, {Mandat}, {Maneva}, {Manganaro}, {Mangano},
  {Marcowith}, {Mar{\'\i}n}, {Markoff}, {Mart{\'\i}}, {Martin},
  {Mart{\'\i}nez}, {Mart{\'\i}nez}, {Masetti}, {Masuda}, {Maurin}, {Maxted},
  {Mazin}, {Medina}, {Melandri}, {Mereghetti}, {Meyer}, {Minaya}, {Mirabal},
  {Mirzoyan}, {Mitchell}, {Mizuno}, {Moderski}, {Mohammed}, {Mohrmann},
  {Montaruli}, {Moralejo}, {Morcuende-Parrilla}, {Mori}, {Morlino}, {Morris},
  {Morselli}, {Moulin}, {Mukherjee}, {Mundell}, {Murach}, {Muraishi}, {Murase},
  {Nagai}, {Nagataki}, {Nagayoshi}, {Naito}, {Nakamori}, {Nakamura}, {Niemiec},
  {Nieto}, {Niko{\l}ajuk}, {Nishijima}, {Noda}, {Nosek}, {Novosyadlyj},
  {Nozaki}, {O'Brien}, {Oakes}, {Ohira}, {Ohishi}, {Ohm}, {Okazaki}, {Okumura},
  {Ong}, {Orienti}, {Orito}, {Osborne}, {Ostrowski}, {Otte}, {Oya}, {Padovani},
  {Paizis}, {Palatiello}, {Palatka}, {Paoletti}, {Paredes}, {Pareschi},
  {Parsons}, {Pe'er}, {Pech}, {Pedaletti}, {Perri}, {Persic}, {Petrashyk},
  {Petrucci}, {Petruk}, {Peyaud}, {Pfeifer}, {Piano}, {Pisarski}, {Pita},
  {Pohl}, {Polo}, {Pozo}, {Prandini}, {Prast}, {Principe}, {Prokhorov},
  {Prokoph}, {Prouza}, {P{\"u}hlhofer}, {Punch}, {P{\"u}rckhauer}, {Queiroz},
  {Quirrenbach}, {Rain{\`o}}, {Razzaque}, {Reimer}, {Reimer}, {Reisenegger},
  {Renaud}, {Rezaeian}, {Rhode}, {Ribeiro}, {Rib{\'o}}, {Richtler}, {Rico},
  {Rieger}, {Riquelme}, {Rivoire}, {Rizi}, {Rodriguez}, {Rodriguez Fernandez},
  {Rodr{\'\i}guez V{\'a}zquez}, {Rojas}, {Romano}, {Romeo}, {Rosado}, {Rovero},
  {Rowell}, {Rudak}, {Rugliancich}, {Rulten}, {Sadeh}, {Safi-Harb}, {Saito},
  {Sakaki}, {Sakurai}, {Salina}, {S{\'a}nchez-Conde}, {Sandaker}, {Sandoval},
  {Sangiorgi}, {Sanguillon}, {Sano}, {Santand er}, {Sarkar}, {Satalecka},
  {Saturni}, {Schioppa}, {Schlenstedt}, {Schneider}, {Schoorlemmer},
  {Schovanek}, {Schulz}, {Schussler}, {Schwanke}, {Sciacca}, {Scuderi},
  {Seitenzahl}, {Semikoz}, {Sergijenko}, {Servillat}, {Shalchi}, {Shellard},
  {Sidoli}, {Siejkowski}, {Sillanp{\"a}{\"a}}, {Sironi}, {Sitarek}, {Sliusar},
  {Slowikowska}, {Sol}, {Stamerra}, {Stani{\v{c}}}, {Starling}, {Stawarz},
  {Stefanik}, {Stephan}, {Stolarczyk}, {Stratta}, {Straumann}, {Suomijarvi},
  {Supanitsky}, {Tagliaferri}, {Tajima}, {Tavani}, {Tavecchio}, {Tavernet},
  {Tayabaly}, {Tejedor}, {Temnikov}, {Terada}, {Terrier}, {Terzic}, {Teshima},
  {Testa}, {Thoudam}, {Tian}, {Tibaldo}, {Tluczykont}, {Todero Peixoto},
  {Tokanai}, {Tomastik}, {Tonev}, {Tornikoski}, {Torres}, {Torresi}, {Tosti},
  {Tothill}, {Tovmassian}, {Travnicek}, {Trichard}, {Trifoglio}, {Troyano
  Pujadas}, {Tsujimoto}, {Umana}, {Vagelli}, {Vagnetti}, {Valentino},
  {Vallania}, {Valore}, {van Eldik}, {Vand enbroucke}, {Varner}, {Vasileiadis},
  {Vassiliev}, {V{\'a}zquez Acosta}, {Vecchi}, {Vega}, {Vercellone}, {Veres},
  {Vergani}, {Verzi}, {Vettolani}, {Viana}, {Vigorito}, {Villanueva}, {Voelk},
  {Vollhardt}, {Vorobiov}, {Vrastil}, {Vuillaume}, {Wagner}, {Wagner},
  {Walter}, {Ward}, {Warren}, {Watson}, {Werner}, {White}, {White},
  {Wierzcholska}, {Wilcox}, {Will}, {Williams}, {Wischnewski}, {Wood},
  {Yamamoto}, {Yamazaki}, {Yanagita}, {Yang}, {Yoshida}, {Yoshiike},
  {Yoshikoshi}, {Zacharias}, {Zaharijas}, {Zampieri}, {Zand anel}, {Zanin},
  {Zavrtanik}, {Zavrtanik}, {Zdziarski}, {Zech}, {Zechlin}, {Zhdanov},
  {Ziegler}, \& {Zorn}}]{CTA2019}
{Cherenkov Telescope Array Consortium} {et~al.} 2019, {Science with the
  Cherenkov Telescope Array}

\bibitem[{{Colavincenzo} {et~al.}(2020){Colavincenzo}, {Tan}, {Ammazzalorso},
  {Camera}, {Regis}, {Xia}, \& {Fornengo}}]{Colavincenzo2020}
{Colavincenzo}, M., {Tan}, X., {Ammazzalorso}, S., {Camera}, S., {Regis}, M.,
  {Xia}, J.-Q., \& {Fornengo}, N. 2020, \mnras, 491, 3225, 1907.05264

\bibitem[{{Dennison}(1980)}]{Dennison1980}
{Dennison}, B. 1980, \apjl, 239, L93

\bibitem[{{Dolag} \& {En{\ss}lin}(2000)}]{Dolag2000}
{Dolag}, K., \& {En{\ss}lin}, T.~A. 2000, \aap, 362, 151, astro-ph/0008333

\bibitem[{{Dutson} {et~al.}(2013){Dutson}, {White}, {Edge}, {Hinton}, \&
  {Hogan}}]{Dutson2013}
{Dutson}, K.~L., {White}, R.~J., {Edge}, A.~C., {Hinton}, J.~A., \& {Hogan},
  M.~T. 2013, \mnras, 429, 2069, 1211.6344

\bibitem[{{Ferrari} {et~al.}(2011){Ferrari}, {Intema}, {Orr{\`u}}, {Govoni},
  {Murgia}, {Mason}, {Bourdin}, {Asad}, {Mazzotta}, {Wise}, {Mroczkowski}, \&
  {Croston}}]{Ferrari2011}
{Ferrari}, C. {et~al.} 2011, \aap, 534, L12, 1107.5945

\bibitem[{{Foreman-Mackey} {et~al.}(2013){Foreman-Mackey}, {Hogg}, {Lang}, \&
  {Goodman}}]{Foreman2013}
{Foreman-Mackey}, D., {Hogg}, D.~W., {Lang}, D., \& {Goodman}, J. 2013, \pasp,
  125, 306, 1202.3665

\bibitem[{{Gavazzi} {et~al.}(2009){Gavazzi}, {Adami}, {Durret}, {Cuillandre},
  {Ilbert}, {Mazure}, {Pell{\'o}}, \& {Ulmer}}]{Gavazzi2009}
{Gavazzi}, R., {Adami}, C., {Durret}, F., {Cuillandre}, J.~C., {Ilbert}, O.,
  {Mazure}, A., {Pell{\'o}}, R., \& {Ulmer}, M.~P. 2009, \aap, 498, L33,
  0904.0220

\bibitem[{{Giacintucci} {et~al.}(2017){Giacintucci}, {Markevitch}, {Cassano},
  {Venturi}, {Clarke}, \& {Brunetti}}]{Giacintucci2017}
{Giacintucci}, S., {Markevitch}, M., {Cassano}, R., {Venturi}, T., {Clarke},
  T.~E., \& {Brunetti}, G. 2017, \apj, 841, 71, 1701.01364

\bibitem[{{Giovannini} {et~al.}(1993){Giovannini}, {Feretti}, {Venturi}, {Kim},
  \& {Kronberg}}]{Giovannini1993}
{Giovannini}, G., {Feretti}, L., {Venturi}, T., {Kim}, K.~T., \& {Kronberg},
  P.~P. 1993, \apj, 406, 399

\bibitem[{{Griffin} {et~al.}(2014){Griffin}, {Dai}, \&
  {Kochanek}}]{Griffin2014}
{Griffin}, R.~D., {Dai}, X., \& {Kochanek}, C.~S. 2014, \apjl, 795, L21,
  1405.7047

\bibitem[{{Huber} {et~al.}(2013){Huber}, {Tchernin}, {Eckert}, {Farnier},
  {Manalaysay}, {Straumann}, \& {Walter}}]{Huber2013}
{Huber}, B., {Tchernin}, C., {Eckert}, D., {Farnier}, C., {Manalaysay}, A.,
  {Straumann}, U., \& {Walter}, R. 2013, \aap, 560, A64, 1308.6278

\bibitem[{Hunter(2007)}]{Hunter2007}
Hunter, J.~D. 2007, Computing In Science \& Engineering, 9, 90

\bibitem[{{Hurier} {et~al.}(2019){Hurier}, {Adam}, \& {Keshet}}]{Hurier2019}
{Hurier}, G., {Adam}, R., \& {Keshet}, U. 2019, \aap, 622, A136

\bibitem[{{Hurier} {et~al.}(2013){Hurier}, {Mac{\'\i}as-P{\'e}rez}, \&
  {Hildebrandt}}]{Hurier2013}
{Hurier}, G., {Mac{\'\i}as-P{\'e}rez}, J.~F., \& {Hildebrandt}, S. 2013, \aap,
  558, A118, 1007.1149

\bibitem[{{Johnson} {et~al.}(2020){Johnson}, {Rudnick}, {Jones}, {Mendygral},
  \& {Dolag}}]{Johnson2020}
{Johnson}, A.~R., {Rudnick}, L., {Jones}, T.~W., {Mendygral}, P.~J., \&
  {Dolag}, K. 2020, \apj, 888, 101, 2001.00903

\bibitem[{Jones {et~al.}(2001)Jones, Oliphant, Peterson, {et~al.}}]{Jones2001}
Jones, E., Oliphant, T., Peterson, P., {et~al.} 2001, {SciPy}: Open source
  scientific tools for {Python}, http://www.scipy.org/

\bibitem[{{Kafexhiu} {et~al.}(2014){Kafexhiu}, {Aharonian}, {Taylor}, \&
  {Vila}}]{Kafexhiu2014}
{Kafexhiu}, E., {Aharonian}, F., {Taylor}, A.~M., \& {Vila}, G.~S. 2014, \prd,
  90, 123014, 1406.7369

\bibitem[{{Kelner} {et~al.}(2006){Kelner}, {Aharonian}, \&
  {Bugayov}}]{Kelner2006}
{Kelner}, S.~R., {Aharonian}, F.~A., \& {Bugayov}, V.~V. 2006, \prd, 74,
  034018, astro-ph/0606058

\bibitem[{{Kent} \& {Gunn}(1982)}]{Kent1982}
{Kent}, S.~M., \& {Gunn}, J.~E. 1982, \aj, 87, 945

\bibitem[{{Keshet} \& {Reiss}(2018)}]{Keshet2018}
{Keshet}, U., \& {Reiss}, I. 2018, \apj, 869, 53

\bibitem[{{Kravtsov} \& {Borgani}(2012)}]{Kravtsov2012}
{Kravtsov}, A.~V., \& {Borgani}, S. 2012, \araa, 50, 353, 1205.5556

\bibitem[{{Liang} {et~al.}(2018){Liang}, {Xia}, {Xi}, {Li}, {Shen}, \&
  {Fan}}]{Liang2018}
{Liang}, Y.-F., {Xia}, Z.-Q., {Xi}, S.-Q., {Li}, S., {Shen}, Z.-Q., \& {Fan},
  Y.-Z. 2018, arXiv e-prints, arXiv:1801.01645, 1801.01645

\bibitem[{{Mattox} {et~al.}(1996){Mattox}, {Bertsch}, {Chiang}, {Dingus},
  {Digel}, {Esposito}, {Fierro}, {Hartman}, {Hunter}, {Kanbach}, {Kniffen},
  {Lin}, {Macomb}, {Mayer-Hasselwander}, {Michelson}, {von Montigny},
  {Mukherjee}, {Nolan}, {Ramanamurthy}, {Schneid}, {Sreekumar}, {Thompson}, \&
  {Willis}}]{Mattox1996}
{Mattox}, J.~R. {et~al.} 1996, \apj, 461, 396

\bibitem[{{Mirakhor} \& {Walker}(2020)}]{Mirakhor2020}
{Mirakhor}, M.~S., \& {Walker}, S.~A. 2020, \mnras, 497, 3204, 2007.12194

\bibitem[{{Mroczkowski} {et~al.}(2019){Mroczkowski}, {Nagai}, {Basu}, {Chluba},
  {Sayers}, {Adam}, {Churazov}, {Crites}, {Di Mascolo}, {Eckert},
  {Macias-Perez}, {Mayet}, {Perotto}, {Pointecouteau}, {Romero}, {Ruppin},
  {Scannapieco}, \& {ZuHone}}]{Mroczkowski2019}
{Mroczkowski}, T. {et~al.} 2019, \ssr, 215, 17, 1811.02310

\bibitem[{{Nagai} {et~al.}(2007){Nagai}, {Vikhlinin}, \&
  {Kravtsov}}]{Nagai2007}
{Nagai}, D., {Vikhlinin}, A., \& {Kravtsov}, A.~V. 2007, \apj, 655, 98,
  arXiv:astro-ph/0609247

\bibitem[{{Ogrean} \& {Br{\"u}ggen}(2013)}]{Ogrean2013}
{Ogrean}, G.~A., \& {Br{\"u}ggen}, M. 2013, \mnras, 433, 1701, 1211.3419

\bibitem[{P\'erez \& Granger(2007)}]{Perez2007}
P\'erez, F., \& Granger, B.~E. 2007, Computing in Science and Engineering, 9,
  21

\bibitem[{{Perkins}(2008)}]{Perkins2008}
{Perkins}, J.~S. 2008, in American Institute of Physics Conference Series, Vol.
  1085, American Institute of Physics Conference Series, ed. F.~A. {Aharonian},
  W.~{Hofmann}, \& F.~{Rieger}, 569--572, 0810.0302

\bibitem[{{Pfrommer} {et~al.}(2007){Pfrommer}, {En{\ss}lin}, {Springel},
  {Jubelgas}, \& {Dolag}}]{Pfrommer2007}
{Pfrommer}, C., {En{\ss}lin}, T.~A., {Springel}, V., {Jubelgas}, M., \&
  {Dolag}, K. 2007, \mnras, 378, 385, astro-ph/0611037

\bibitem[{{Pfrommer} {et~al.}(2006){Pfrommer}, {Springel}, {En{\ss}lin}, \&
  {Jubelgas}}]{Pfrommer2006}
{Pfrommer}, C., {Springel}, V., {En{\ss}lin}, T.~A., \& {Jubelgas}, M. 2006,
  \mnras, 367, 113, astro-ph/0603483

\bibitem[{{Pinzke} {et~al.}(2017){Pinzke}, {Oh}, \& {Pfrommer}}]{Pinzke2017}
{Pinzke}, A., {Oh}, S.~P., \& {Pfrommer}, C. 2017, \mnras, 465, 4800,
  1611.07533

\bibitem[{{Pinzke} \& {Pfrommer}(2010)}]{Pinzke2010}
{Pinzke}, A., \& {Pfrommer}, C. 2010, \mnras, 409, 449, 1001.5023

\bibitem[{{Pizzo}(2010)}]{Pizzo2010}
{Pizzo}, R.~F. 2010, PhD thesis, University of Groningen

\bibitem[{{Planck Collaboration} {et~al.}(2013){Planck Collaboration}, {Ade},
  {Aghanim}, {Arnaud}, {Ashdown}, {Atrio-Barandela}, {Aumont}, {Baccigalupi},
  {Balbi}, {Banday}, {Barreiro}, {Bartlett}, {Battaner}, {Benabed},
  {Beno{\^\i}t}, {Bernard}, {Bersanelli}, {Bikmaev}, {B{\"o}hringer},
  {Bonaldi}, {Bond}, {Borrill}, {Bouchet}, {Bourdin}, {Brown}, {Brown},
  {Burenin}, {Burigana}, {Cabella}, {Cardoso}, {Carvalho}, {Catalano},
  {Cay{\'o}n}, {Chiang}, {Chon}, {Christensen}, {Churazov}, {Clements},
  {Colafrancesco}, {Colombo}, {Coulais}, {Crill}, {Cuttaia}, {Da Silva},
  {Dahle}, {Danese}, {Davis}, {de Bernardis}, {de Gasperis}, {de Rosa}, {de
  Zotti}, {Delabrouille}, {D{\'e}mocl{\`e}s}, {D{\'e}sert}, {Dickinson},
  {Diego}, {Dolag}, {Dole}, {Donzelli}, {Dor{\'e}}, {D{\"o}rl}, {Douspis},
  {Dupac}, {En{\ss}lin}, {Eriksen}, {Finelli}, {Flores-Cacho}, {Forni},
  {Frailis}, {Franceschi}, {Frommert}, {Galeotta}, {Ganga},
  {G{\'e}nova-Santos}, {Giard}, {Gilfanov}, {Gonz{\'a}lez-Nuevo}, {G{\'o}rski},
  {Gregorio}, {Gruppuso}, {Hansen}, {Harrison}, {Henrot-Versill{\'e}},
  {Hern{\'a}ndez-Monteagudo}, {Hildebrandt}, {Hivon}, {Hobson}, {Holmes},
  {Hornstrup}, {Hovest}, {Huffenberger}, {Hurier}, {Jaffe}, {Jagemann},
  {Jones}, {Juvela}, {Keih{\"a}nen}, {Khamitov}, {Kneissl}, {Knoche}, {Knox},
  {Kunz}, {Kurki-Suonio}, {Lagache}, {L{\"a}hteenm{\"a}ki}, {Lamarre},
  {Lasenby}, {Lawrence}, {Le Jeune}, {Leonardi}, {Lilje}, {Linden-V{\o}rnle},
  {L{\'o}pez-Caniego}, {Lubin}, {Mac{\'\i}as-P{\'e}rez}, {Maffei}, {Maino},
  {Mand olesi}, {Maris}, {Marleau}, {Mart{\'\i}nez-Gonz{\'a}lez}, {Masi},
  {Massardi}, {Matarrese}, {Matthai}, {Mazzotta}, {Mei}, {Melchiorri}, {Melin},
  {Mendes}, {Mennella}, {Mitra}, {Miville-Desch{\^e}nes}, {Moneti}, {Montier},
  {Morgante}, {Munshi}, {Murphy}, {Naselsky}, {Natoli}, {N{\o}rgaard-Nielsen},
  {Noviello}, {Novikov}, {Novikov}, {Osborne}, {Pajot}, {Paoletti},
  {Perdereau}, {Perrotta}, {Piacentini}, {Piat}, {Pierpaoli}, {Piffaretti},
  {Plaszczynski}, {Pointecouteau}, {Polenta}, {Ponthieu}, {Popa}, {Poutanen},
  {Pratt}, {Prunet}, {Puget}, {Rachen}, {Rebolo}, {Reinecke}, {Remazeilles},
  {Renault}, {Ricciardi}, {Riller}, {Ristorcelli}, {Rocha}, {Roman}, {Rosset},
  {Rossetti}, {Rubi{\~n}o-Mart{\'\i}n}, {Rudnick}, {Rusholme}, {Sandri},
  {Savini}, {Schaefer}, {Scott}, {Smoot}, {Stivoli}, {Sudiwala}, {Sunyaev},
  {Sutton}, {Suur-Uski}, {Sygnet}, {Tauber}, {Terenzi}, {Toffolatti}, {Tomasi},
  {Tristram}, {Tuovinen}, {T{\"u}rler}, {Umana}, {Valenziano}, {Van Tent},
  {Varis}, {Vielva}, {Villa}, {Vittorio}, {Wade}, {Wandelt}, {Welikala},
  {White}, {Yvon}, {Zacchei}, {Zaroubi}, \& {Zonca}}]{PlanckX2013}
{Planck Collaboration} {et~al.} 2013, \aap, 554, A140, 1208.3611

\bibitem[{{Planck Collaboration} {et~al.}(2016{\natexlab{a}}){Planck
  Collaboration}, {Ade}, {Aghanim}, {Arnaud}, {Ashdown}, {Aumont},
  {Baccigalupi}, {Banday}, {Barreiro}, {Barrena}, {Bartlett}, {Bartolo},
  {Battaner}, {Battye}, {Benabed}, {Beno{\^\i}t}, {Benoit-L{\'e}vy}, {Bernard},
  {Bersanelli}, {Bielewicz}, {Bikmaev}, {B{\"o}hringer}, {Bonaldi}, {Bonavera},
  {Bond}, {Borrill}, {Bouchet}, {Bucher}, {Burenin}, {Burigana}, {Butler},
  {Calabrese}, {Cardoso}, {Carvalho}, {Catalano}, {Challinor}, {Chamballu},
  {Chary}, {Chiang}, {Chon}, {Christensen}, {Clements}, {Colombi}, {Colombo},
  {Combet}, {Comis}, {Couchot}, {Coulais}, {Crill}, {Curto}, {Cuttaia},
  {Dahle}, {Danese}, {Davies}, {Davis}, {de Bernardis}, {de Rosa}, {de Zotti},
  {Delabrouille}, {D{\'e}sert}, {Dickinson}, {Diego}, {Dolag}, {Dole},
  {Donzelli}, {Dor{\'e}}, {Douspis}, {Ducout}, {Dupac}, {Efstathiou},
  {Eisenhardt}, {Elsner}, {En{\ss}lin}, {Eriksen}, {Falgarone}, {Fergusson},
  {Feroz}, {Ferragamo}, {Finelli}, {Forni}, {Frailis}, {Fraisse}, {Franceschi},
  {Frejsel}, {Galeotta}, {Galli}, {Ganga}, {G{\'e}nova-Santos}, {Giard},
  {Giraud-H{\'e}raud}, {Gjerl{\o}w}, {Gonz{\'a}lez-Nuevo}, {G{\'o}rski},
  {Grainge}, {Gratton}, {Gregorio}, {Gruppuso}, {Gudmundsson}, {Hansen},
  {Hanson}, {Harrison}, {Hempel}, {Henrot-Versill{\'e}},
  {Hern{\'a}ndez-Monteagudo}, {Herranz}, {Hildebrandt}, {Hivon}, {Hobson},
  {Holmes}, {Hornstrup}, {Hovest}, {Huffenberger}, {Hurier}, {Jaffe}, {Jaffe},
  {Jin}, {Jones}, {Juvela}, {Keih{\"a}nen}, {Keskitalo}, {Khamitov}, {Kisner},
  {Kneissl}, {Knoche}, {Kunz}, {Kurki-Suonio}, {Lagache}, {Lamarre}, {Lasenby},
  {Lattanzi}, {Lawrence}, {Leonardi}, {Lesgourgues}, {Levrier}, {Liguori},
  {Lilje}, {Linden-V{\o}rnle}, {L{\'o}pez-Caniego}, {Lubin},
  {Mac{\'\i}as-P{\'e}rez}, {Maggio}, {Maino}, {Mak}, {Mandolesi}, {Mangilli},
  {Martin}, {Mart{\'\i}nez-Gonz{\'a}lez}, {Masi}, {Matarrese}, {Mazzotta},
  {McGehee}, {Mei}, {Melchiorri}, {Melin}, {Mendes}, {Mennella}, {Migliaccio},
  {Mitra}, {Miville-Desch{\^e}nes}, {Moneti}, {Montier}, {Morgante},
  {Mortlock}, {Moss}, {Munshi}, {Murphy}, {Naselsky}, {Nastasi}, {Nati},
  {Natoli}, {Netterfield}, {N{\o}rgaard-Nielsen}, {Noviello}, {Novikov},
  {Novikov}, {Olamaie}, {Oxborrow}, {Paci}, {Pagano}, {Pajot}, {Paoletti},
  {Pasian}, {Patanchon}, {Pearson}, {Perdereau}, {Perotto}, {Perrott},
  {Perrotta}, {Pettorino}, {Piacentini}, {Piat}, {Pierpaoli}, {Pietrobon},
  {Plaszczynski}, {Pointecouteau}, {Polenta}, {Pratt}, {Pr{\'e}zeau}, {Prunet},
  {Puget}, {Rachen}, {Reach}, {Rebolo}, {Reinecke}, {Remazeilles}, {Renault},
  {Renzi}, {Ristorcelli}, {Rocha}, {Rosset}, {Rossetti}, {Roudier}, {Rozo},
  {Rubi{\~n}o-Mart{\'\i}n}, {Rumsey}, {Rusholme}, {Rykoff}, {Sandri}, {Santos},
  {Saunders}, {Savelainen}, {Savini}, {Schammel}, {Scott}, {Seiffert},
  {Shellard}, {Shimwell}, {Spencer}, {Stanford}, {Stern}, {Stolyarov},
  {Stompor}, {Streblyanska}, {Sudiwala}, {Sunyaev}, {Sutton}, {Suur-Uski},
  {Sygnet}, {Tauber}, {Terenzi}, {Toffolatti}, {Tomasi}, {Tramonte},
  {Tristram}, {Tucci}, {Tuovinen}, {Umana}, {Valenziano}, {Valiviita}, {Van
  Tent}, {Vielva}, {Villa}, {Wade}, {Wandelt}, {Wehus}, {White}, {Wright},
  {Yvon}, {Zacchei}, \& {Zonca}}]{Planck2016XXVII}
------. 2016{\natexlab{a}}, \aap, 594, A27, 1502.01598

\bibitem[{{Planck Collaboration} {et~al.}(2016{\natexlab{b}}){Planck
  Collaboration}, {Ade}, {Aghanim}, {Arnaud}, {Ashdown}, {Aumont},
  {Baccigalupi}, {Banday}, {Barreiro}, {Bartlett}, \& et~al.}]{Planck2016XIII}
------. 2016{\natexlab{b}}, \aap, 594, A13, 1502.01589

\bibitem[{{Planck Collaboration} {et~al.}(2016{\natexlab{c}}){Planck
  Collaboration}, {Aghanim}, {Arnaud}, {Ashdown}, {Aumont}, {Baccigalupi},
  {Band ay}, {Barreiro}, {Bartlett}, {Bartolo}, {Battaner}, {Battye},
  {Benabed}, {Beno{\^\i}t}, {Benoit-L{\'e}vy}, {Bernard}, {Bersanelli},
  {Bielewicz}, {Bock}, {Bonaldi}, {Bonavera}, {Bond}, {Borrill}, {Bouchet},
  {Burigana}, {Butler}, {Calabrese}, {Cardoso}, {Catalano}, {Challinor},
  {Chiang}, {Christensen}, {Churazov}, {Clements}, {Colombo}, {Combet},
  {Comis}, {Coulais}, {Crill}, {Curto}, {Cuttaia}, {Danese}, {Davies}, {Davis},
  {de Bernardis}, {de Rosa}, {de Zotti}, {Delabrouille}, {D{\'e}sert},
  {Dickinson}, {Diego}, {Dolag}, {Dole}, {Donzelli}, {Dor{\'e}}, {Douspis},
  {Ducout}, {Dupac}, {Efstathiou}, {Elsner}, {En{\ss}lin}, {Eriksen},
  {Fergusson}, {Finelli}, {Forni}, {Frailis}, {Fraisse}, {Franceschi},
  {Frejsel}, {Galeotta}, {Galli}, {Ganga}, {G{\'e}nova-Santos}, {Giard},
  {Gonz{\'a}lez-Nuevo}, {G{\'o}rski}, {Gregorio}, {Gruppuso}, {Gudmundsson},
  {Hansen}, {Harrison}, {Henrot-Versill{\'e}}, {Hern{\'a}ndez-Monteagudo},
  {Herranz}, {Hildebrand t}, {Hivon}, {Holmes}, {Hornstrup}, {Huffenberger},
  {Hurier}, {Jaffe}, {Jones}, {Juvela}, {Keih{\"a}nen}, {Keskitalo}, {Kneissl},
  {Knoche}, {Kunz}, {Kurki-Suonio}, {Lacasa}, {Lagache}, {L{\"a}hteenm{\"a}ki},
  {Lamarre}, {Lasenby}, {Lattanzi}, {Leonardi}, {Lesgourgues}, {Levrier},
  {Liguori}, {Lilje}, {Linden-V{\o}rnle}, {L{\'o}pez-Caniego},
  {Mac{\'\i}as-P{\'e}rez}, {Maffei}, {Maggio}, {Maino}, {Mandolesi},
  {Mangilli}, {Maris}, {Martin}, {Mart{\'\i}nez-Gonz{\'a}lez}, {Masi},
  {Matarrese}, {Melchiorri}, {Melin}, {Migliaccio}, {Miville-Desch{\^e}nes},
  {Moneti}, {Montier}, {Morgante}, {Mortlock}, {Munshi}, {Murphy}, {Naselsky},
  {Nati}, {Natoli}, {Noviello}, {Novikov}, {Novikov}, {Paci}, {Pagano},
  {Pajot}, {Paoletti}, {Pasian}, {Patanchon}, {Perdereau}, {Perotto},
  {Pettorino}, {Piacentini}, {Piat}, {Pierpaoli}, {Pietrobon}, {Plaszczynski},
  {Pointecouteau}, {Polenta}, {Ponthieu}, {Pratt}, {Prunet}, {Puget}, {Rachen},
  {Reinecke}, {Remazeilles}, {Renault}, {Renzi}, {Ristorcelli}, {Rocha},
  {Rossetti}, {Roudier}, {Rubi{\~n}o-Mart{\'\i}n}, {Rusholme}, {Sandri},
  {Santos}, {Sauv{\'e}}, {Savelainen}, {Savini}, {Scott}, {Spencer},
  {Stolyarov}, {Stompor}, {Sunyaev}, {Sutton}, {Suur-Uski}, {Sygnet}, {Tauber},
  {Terenzi}, {Toffolatti}, {Tomasi}, {Tramonte}, {Tristram}, {Tucci},
  {Tuovinen}, {Valenziano}, {Valiviita}, {Van Tent}, {Vielva}, {Villa}, {Wade},
  {Wandelt}, {Wehus}, {Yvon}, {Zacchei}, \& {Zonca}}]{PlanckXXII2016}
------. 2016{\natexlab{c}}, \aap, 594, A22, 1502.01596

\bibitem[{{Prokhorov}(2014)}]{Prokhorov2014}
{Prokhorov}, D.~A. 2014, \mnras, 441, 2309

\bibitem[{{Reimer} {et~al.}(2003){Reimer}, {Pohl}, {Sreekumar}, \&
  {Mattox}}]{Reimer2003}
{Reimer}, O., {Pohl}, M., {Sreekumar}, P., \& {Mattox}, J.~R. 2003, \apj, 588,
  155, astro-ph/0301362

\bibitem[{{Rephaeli} \& {Sadeh}(2016)}]{Rephaeli2016}
{Rephaeli}, Y., \& {Sadeh}, S. 2016, \prd, 93, 101301, 1605.04461

\bibitem[{{Rudnick}(2002)}]{Rudnick2002}
{Rudnick}, L. 2002, \pasp, 114, 427

\bibitem[{{Sarazin}(1999)}]{Sarazin1999}
{Sarazin}, C.~L. 1999, \apj, 520, 529, astro-ph/9901061

\bibitem[{{Savini} {et~al.}(2019){Savini}, {Bonafede}, {Br{\"u}ggen},
  {Rafferty}, {Shimwell}, {Botteon}, {Brunetti}, {Intema}, {Wilber}, {Cassano},
  {Vazza}, {van Weeren}, {Cuciti}, {De Gasperin}, {R{\"o}ttgering}, {Sommer},
  {B{\^\i}rzan}, \& {Drabent}}]{Savini2019}
{Savini}, F. {et~al.} 2019, \aap, 622, A24, 1811.08410

\bibitem[{{Simionescu} {et~al.}(2013){Simionescu}, {Werner}, {Urban}, {Allen},
  {Fabian}, {Mantz}, {Matsushita}, {Nulsen}, {Sanders}, {Sasaki}, {Sato},
  {Takei}, \& {Walker}}]{Simionescu2013}
{Simionescu}, A. {et~al.} 2013, \apj, 775, 4, 1302.4140

\bibitem[{{Storm} {et~al.}(2012){Storm}, {Jeltema}, \& {Profumo}}]{Storm2012}
{Storm}, E.~M., {Jeltema}, T.~E., \& {Profumo}, S. 2012, \apj, 755, 117,
  1206.1676

\bibitem[{{Sunyaev} \& {Zeldovich}(1972)}]{Sunyaev1972}
{Sunyaev}, R.~A., \& {Zeldovich}, Y.~B. 1972, Comments on Astrophysics and
  Space Physics, 4, 173

\bibitem[{{Thierbach} {et~al.}(2003){Thierbach}, {Klein}, \&
  {Wielebinski}}]{Thierbach2003}
{Thierbach}, M., {Klein}, U., \& {Wielebinski}, R. 2003, \aap, 397, 53,
  astro-ph/0210147

\bibitem[{{Truemper}(1993)}]{Truemper1993}
{Truemper}, J. 1993, Science, 260, 1769

\bibitem[{{Uchida} {et~al.}(2016){Uchida}, {Simionescu}, {Takahashi}, {Werner},
  {Ichinohe}, {Allen}, {Urban}, \& {Matsushita}}]{Uchida2016}
{Uchida}, Y., {Simionescu}, A., {Takahashi}, T., {Werner}, N., {Ichinohe}, Y.,
  {Allen}, S.~W., {Urban}, O., \& {Matsushita}, K. 2016, \pasj, 68, S20,
  1509.01901

\bibitem[{{van der Walt} {et~al.}(2011){van der Walt}, {Colbert}, \&
  {Varoquaux}}]{VanDerWalt2011}
{van der Walt}, S., {Colbert}, S.~C., \& {Varoquaux}, G. 2011, Computing in
  Science and Engineering, 13, 22, 1102.1523

\bibitem[{{van Weeren} {et~al.}(2019){van Weeren}, {de Gasperin}, {Akamatsu},
  {Br{\"u}ggen}, {Feretti}, {Kang}, {Stroe}, \& {Zandanel}}]{vanWeeren2019}
{van Weeren}, R.~J., {de Gasperin}, F., {Akamatsu}, H., {Br{\"u}ggen}, M.,
  {Feretti}, L., {Kang}, H., {Stroe}, A., \& {Zandanel}, F. 2019, \ssr, 215,
  16, 1901.04496

\bibitem[{{van Weeren} {et~al.}(2010){van Weeren}, {R{\"o}ttgering},
  {Br{\"u}ggen}, \& {Hoeft}}]{vanWeeren2010}
{van Weeren}, R.~J., {R{\"o}ttgering}, H. J.~A., {Br{\"u}ggen}, M., \& {Hoeft},
  M. 2010, Science, 330, 347, 1010.4306

\bibitem[{{Vladimirov} {et~al.}(2011){Vladimirov}, {Digel}, {J{\'o}hannesson},
  {Michelson}, {Moskalenko}, {Nolan}, {Orland o}, {Porter}, \&
  {Strong}}]{Vladimirov2011}
{Vladimirov}, A.~E. {et~al.} 2011, Computer Physics Communications, 182, 1156,
  1008.3642

\bibitem[{{Voges} {et~al.}(1999){Voges}, {Aschenbach}, {Boller},
  {Br{\"a}uninger}, {Briel}, {Burkert}, {Dennerl}, {Englhauser}, {Gruber},
  {Haberl}, {Hartner}, {Hasinger}, {K{\"u}rster}, {Pfeffermann}, {Pietsch},
  {Predehl}, {Rosso}, {Schmitt}, {Tr{\"u}mper}, \& {Zimmermann}}]{Voges1999}
{Voges}, W. {et~al.} 1999, \aap, 349, 389, astro-ph/9909315

\bibitem[{{Voges} {et~al.}(2000){Voges}, {Aschenbach}, {Boller}, {Brauninger},
  {Briel}, {Burkert}, {Dennerl}, {Englhauser}, {Gruber}, {Haberl}, {Hartner},
  {Hasinger}, {Pfeffermann}, {Pietsch}, {Predehl}, {Schmitt}, {Trumper}, \&
  {Zimmermann}}]{Voges2000}
------. 2000, \iaucirc, 7432, 3

\bibitem[{Waskom \& the seaborn~development team(2020)}]{Waskom2020}
Waskom, M., \& the seaborn~development team. 2020, mwaskom/seaborn

\bibitem[{{Willson}(1970)}]{Willson1970}
{Willson}, M.~A.~G. 1970, \mnras, 151, 1

\bibitem[{{Wood} {et~al.}(2017){Wood}, {Caputo}, {Charles}, {Di Mauro},
  {Magill}, {Perkins}, \& {Fermi-LAT Collaboration}}]{Wood2017}
{Wood}, M., {Caputo}, R., {Charles}, E., {Di Mauro}, M., {Magill}, J.,
  {Perkins}, J.~S., \& {Fermi-LAT Collaboration}. 2017, in International Cosmic
  Ray Conference, Vol. 301, 35th International Cosmic Ray Conference
  (ICRC2017), 824, 1707.09551

\bibitem[{{Xi} {et~al.}(2018){Xi}, {Wang}, {Liang}, {Peng}, {Yang}, \&
  {Liu}}]{Xi2018}
{Xi}, S.-Q., {Wang}, X.-Y., {Liang}, Y.-F., {Peng}, F.-K., {Yang}, R.-Z., \&
  {Liu}, R.-Y. 2018, \prd, 98, 063006, 1709.08319

\bibitem[{{York} {et~al.}(2000){York}, {Adelman}, {Anderson}, {Anderson},
  {Annis}, {Bahcall}, {Bakken}, {Barkhouser}, {Bastian}, {Berman}, {Boroski},
  {Bracker}, {Briegel}, {Briggs}, {Brinkmann}, {Brunner}, {Burles}, {Carey},
  {Carr}, {Castander}, {Chen}, {Colestock}, {Connolly}, {Crocker}, {Csabai},
  {Czarapata}, {Davis}, {Doi}, {Dombeck}, {Eisenstein}, {Ellman}, {Elms},
  {Evans}, {Fan}, {Federwitz}, {Fiscelli}, {Friedman}, {Frieman}, {Fukugita},
  {Gillespie}, {Gunn}, {Gurbani}, {de Haas}, {Haldeman}, {Harris}, {Hayes},
  {Heckman}, {Hennessy}, {Hindsley}, {Holm}, {Holmgren}, {Huang}, {Hull},
  {Husby}, {Ichikawa}, {Ichikawa}, {Ivezi{\'c}}, {Kent}, {Kim}, {Kinney},
  {Klaene}, {Kleinman}, {Kleinman}, {Knapp}, {Korienek}, {Kron}, {Kunszt},
  {Lamb}, {Lee}, {Leger}, {Limmongkol}, {Lindenmeyer}, {Long}, {Loomis},
  {Loveday}, {Lucinio}, {Lupton}, {MacKinnon}, {Mannery}, {Mantsch}, {Margon},
  {McGehee}, {McKay}, {Meiksin}, {Merelli}, {Monet}, {Munn}, {Narayanan},
  {Nash}, {Neilsen}, {Neswold}, {Newberg}, {Nichol}, {Nicinski}, {Nonino},
  {Okada}, {Okamura}, {Ostriker}, {Owen}, {Pauls}, {Peoples}, {Peterson},
  {Petravick}, {Pier}, {Pope}, {Pordes}, {Prosapio}, {Rechenmacher}, {Quinn},
  {Richards}, {Richmond}, {Rivetta}, {Rockosi}, {Ruthmansdorfer}, {Sand ford},
  {Schlegel}, {Schneider}, {Sekiguchi}, {Sergey}, {Shimasaku}, {Siegmund},
  {Smee}, {Smith}, {Snedden}, {Stone}, {Stoughton}, {Strauss}, {Stubbs},
  {SubbaRao}, {Szalay}, {Szapudi}, {Szokoly}, {Thakar}, {Tremonti}, {Tucker},
  {Uomoto}, {Vanden Berk}, {Vogeley}, {Waddell}, {Wang}, {Watanabe},
  {Weinberg}, {Yanny}, {Yasuda}, \& {SDSS Collaboration}}]{York2000}
{York}, D.~G. {et~al.} 2000, \aj, 120, 1579, astro-ph/0006396

\bibitem[{{Zabalza}(2015)}]{Zabalza2015}
{Zabalza}, V. 2015, Proc.~of International Cosmic Ray Conference 2015, 922,
  1509.03319

\bibitem[{{Zandanel} \& {Ando}(2014)}]{Zandanel2014}
{Zandanel}, F., \& {Ando}, S. 2014, \mnras, 440, 663, 1312.1493

\bibitem[{{Zandanel} {et~al.}(2014){Zandanel}, {Pfrommer}, \&
  {Prada}}]{Zandanel2014b}
{Zandanel}, F., {Pfrommer}, C., \& {Prada}, F. 2014, \mnras, 438, 124,
  1311.4795

\end{thebibliography}

\appendix

\section{Inverse Compton emission}\label{app:neglecting_ic}
\begin{figure*}
\centering
\includegraphics[width=0.45\textwidth]{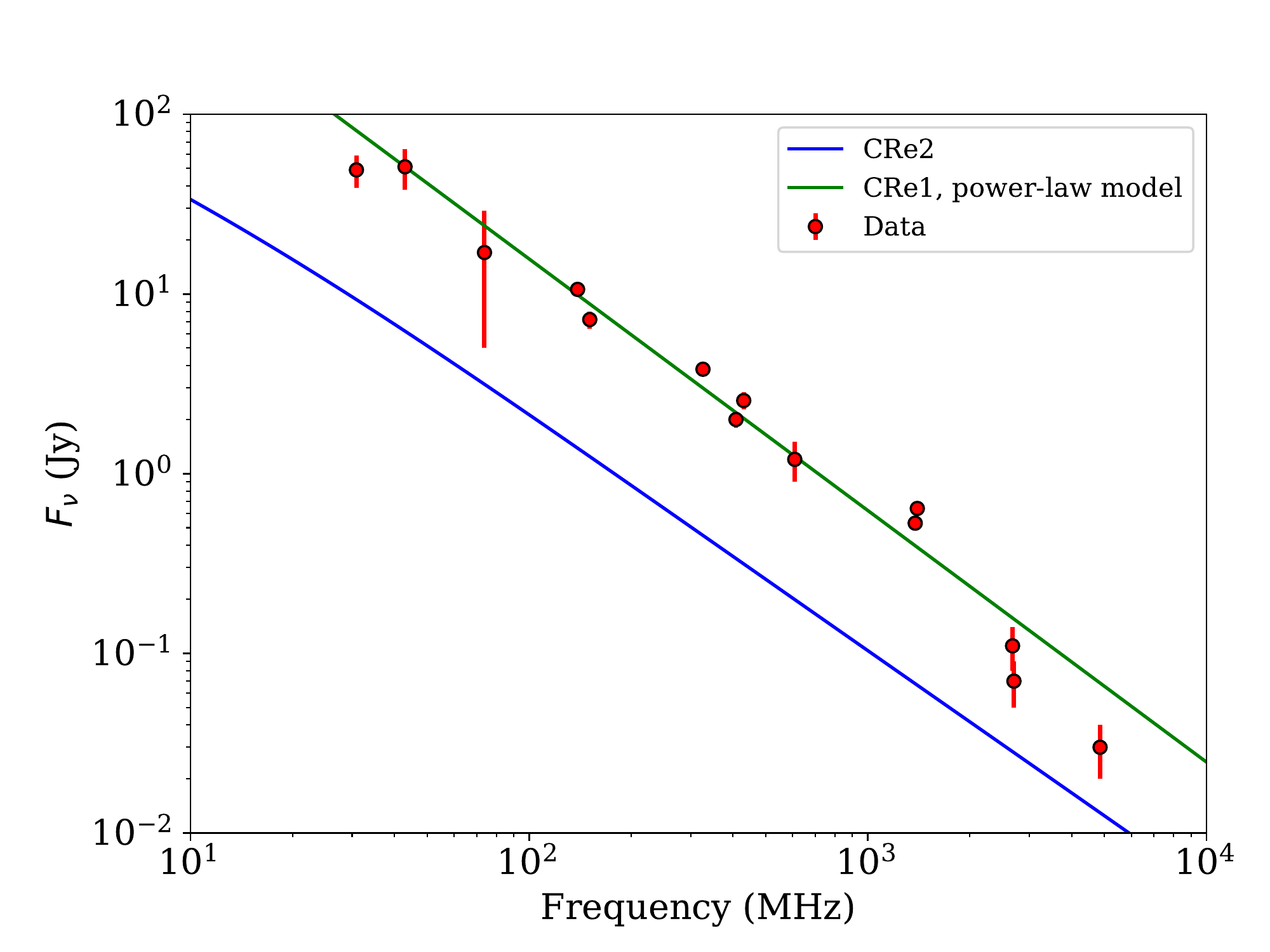}
\includegraphics[width=0.45\textwidth]{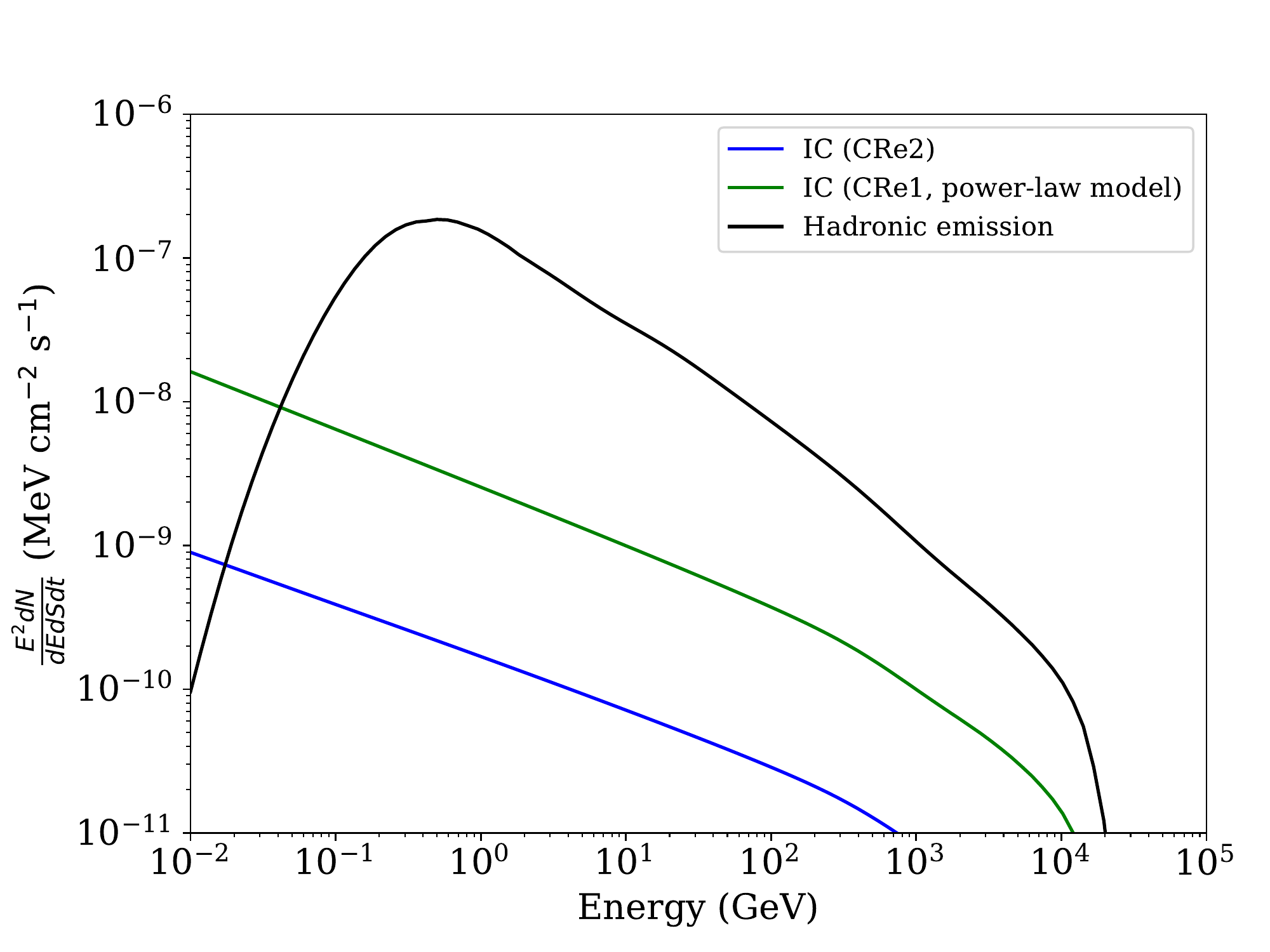}
\caption{\small Estimation of the inverse Compton emission to the $\gamma$-ray signal. {\bf Left}: Radio synchrotron spectrum data to which the CRe population can be matched. {\bf Right}: Associated emission in the $\gamma$-ray energy range.}
\label{fig:IC_checks}
\end{figure*}

This appendix provides an estimate of the inverse Compton emission associated with CRe$_2$ and CRe$_1$.

Firstly, we assume a CRp population with $X_{\rm CRp}(R_{500}) = 10^{-2}$ and $\alpha_{\rm CRp} = 2.8$, which is consistent with the values found in this paper for Coma, and use our baseline scaling $n_{\rm CRp} \propto n_e^{1/2}$. We compute the associated CRe$_2$ population in the steady state approximation, as in Section~\ref{sec:implication_for_CR_content}. We finally compute the inverse Compton emission as explained in \cite{Adam2020}. This provides an estimate of the necessary inverse Compton emission associated with the hadronic collisions.

Secondly, we assume that the CRe$_1$ population is described by a power law, which we match to radio data and extrapolate to very high energies. As strong losses are expected at high energy, this provides an upper limit on the CRe$_1$ population for energies that are large enough to induce inverse Compton emission in the \textit{Fermi}-LAT energy range (energies in the range of about 100 GeV - 100 TeV, while the radio emission probe CRe$_1$ up to a few tens of GeV at GHz frequencies). We match the slope and spectrum of the power law to the radio spectrum and compute the associated inverse Compton emission.

These two cases (CRe$_1$ and CRe$_2$) are reported in Figure~\ref{fig:IC_checks}. On the left panel, we see the contribution of the CRe$_1$ and CRe$_2$ to the radio signal. On the right panel, we see their contribution to the inverse Compton emission. Even in the case of boosting the CRe$_2$ population (i.e., the turbulent reacceleration model) by a factor of five to ten to match the radio emission, this would remain more than an order of magnitude below the hadronically induced $\gamma$-rays. An energy-dependent boost, strongly rising with the energy, could lead to the inverse Compton signal being significant, but this is not expected. In the case of CRe$_1$, the upper limit provided by the power law model remains below the hadronically induced $\gamma$-rays at energies above 100 MeV. As losses are expected to strongly reduce the signal as energy increases, we expect the emission to be negligible, especially since the Coma radio spectrum already present a curvature at GHz frequencies.

While these results may slightly vary depending on the exact shape of the spatial and spectral distribution of the CR, we do not expect any major change in the comparison. We conclude that the Inverse Compton emission from both CRe$_1$ and CRe$_2$ should not contribute significantly to the observed signal at \textit{Fermi}-LAT energies, and it is neglected when modeling the $\gamma$-ray emission.

\section{Comparison to Monte Carlo realizations}\label{app:mc_realizations}
\begin{figure*}
\centering
\includegraphics[width=0.45\textwidth]{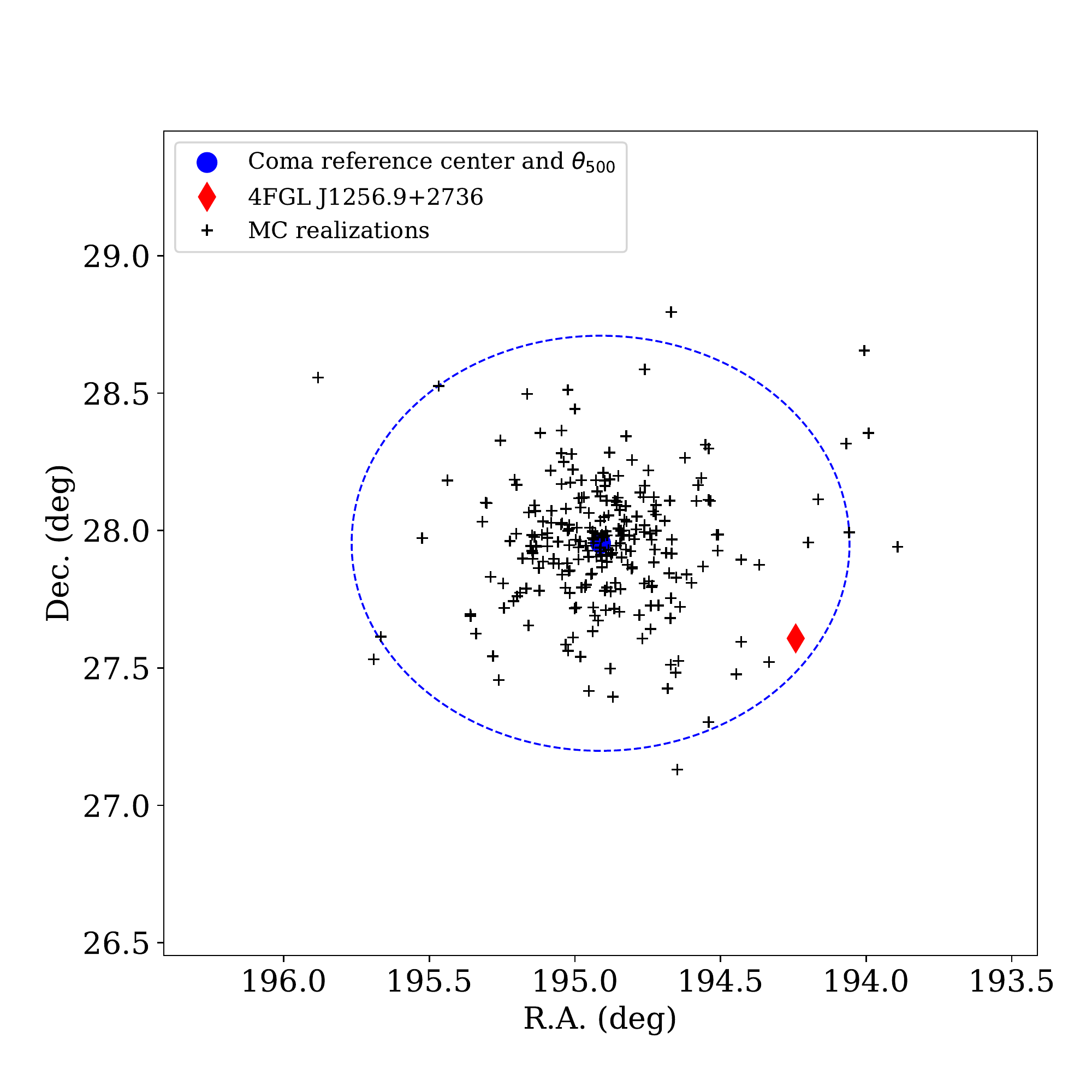}
\includegraphics[width=0.45\textwidth]{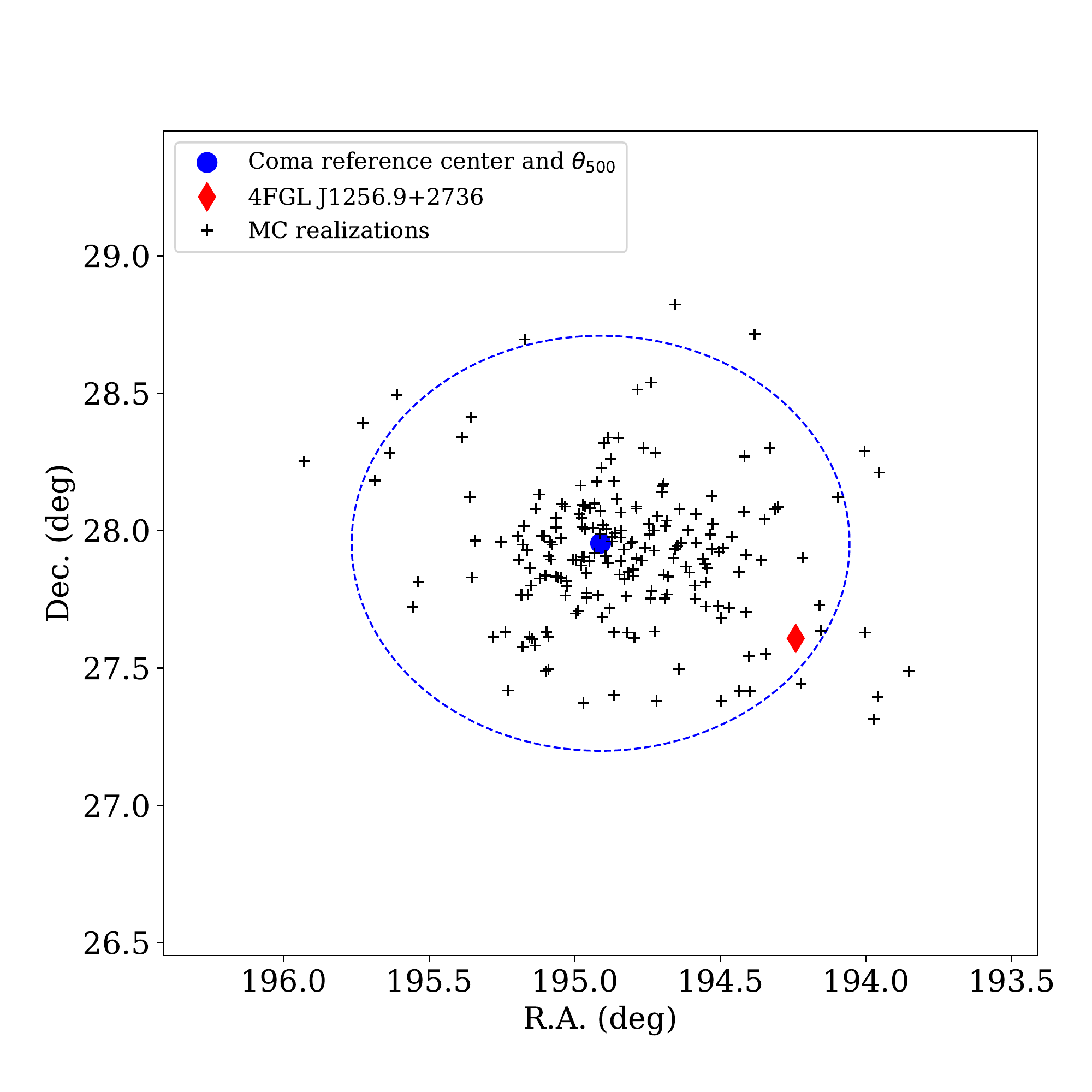}
\includegraphics[width=0.45\textwidth]{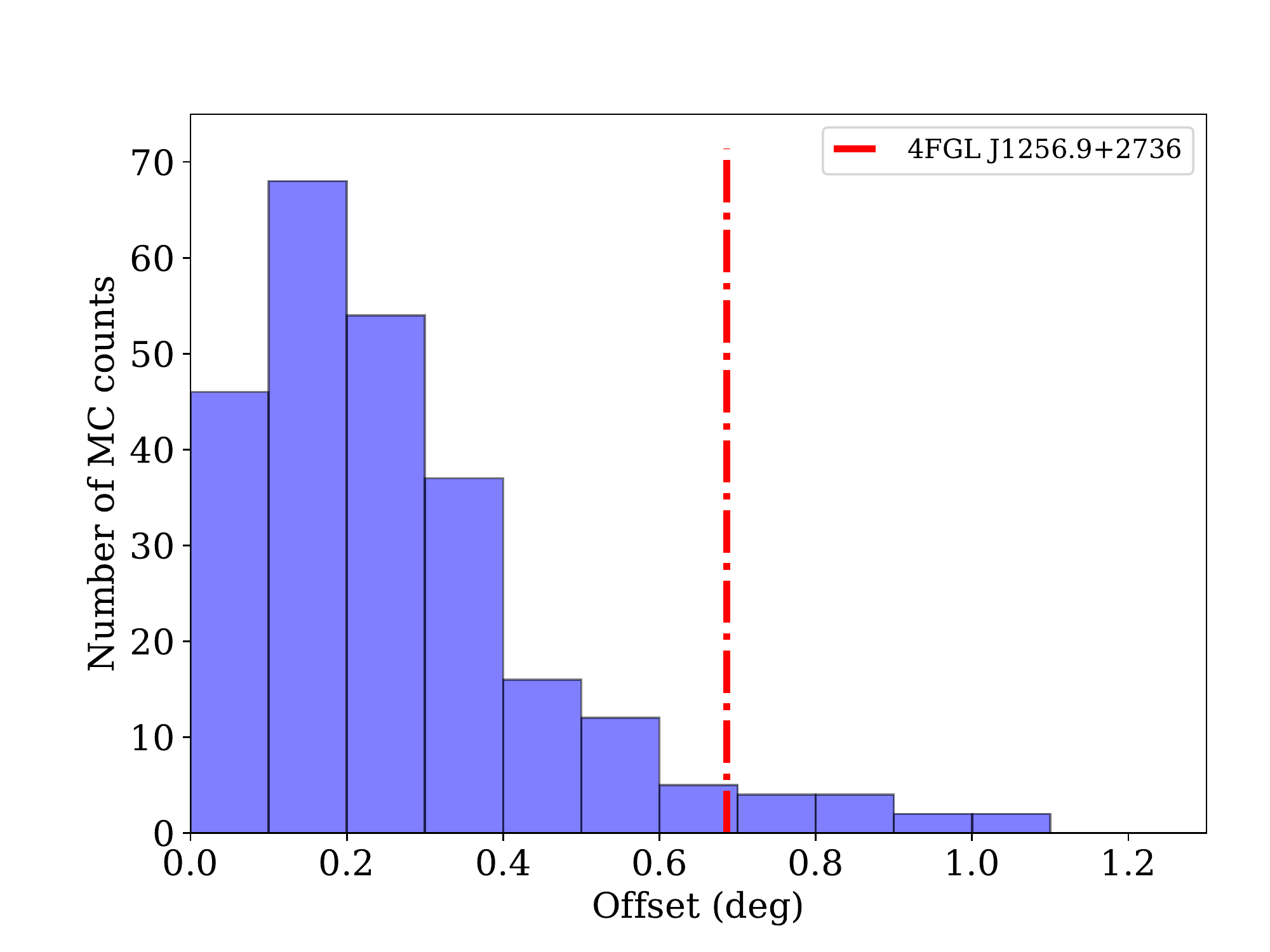}
\includegraphics[width=0.45\textwidth]{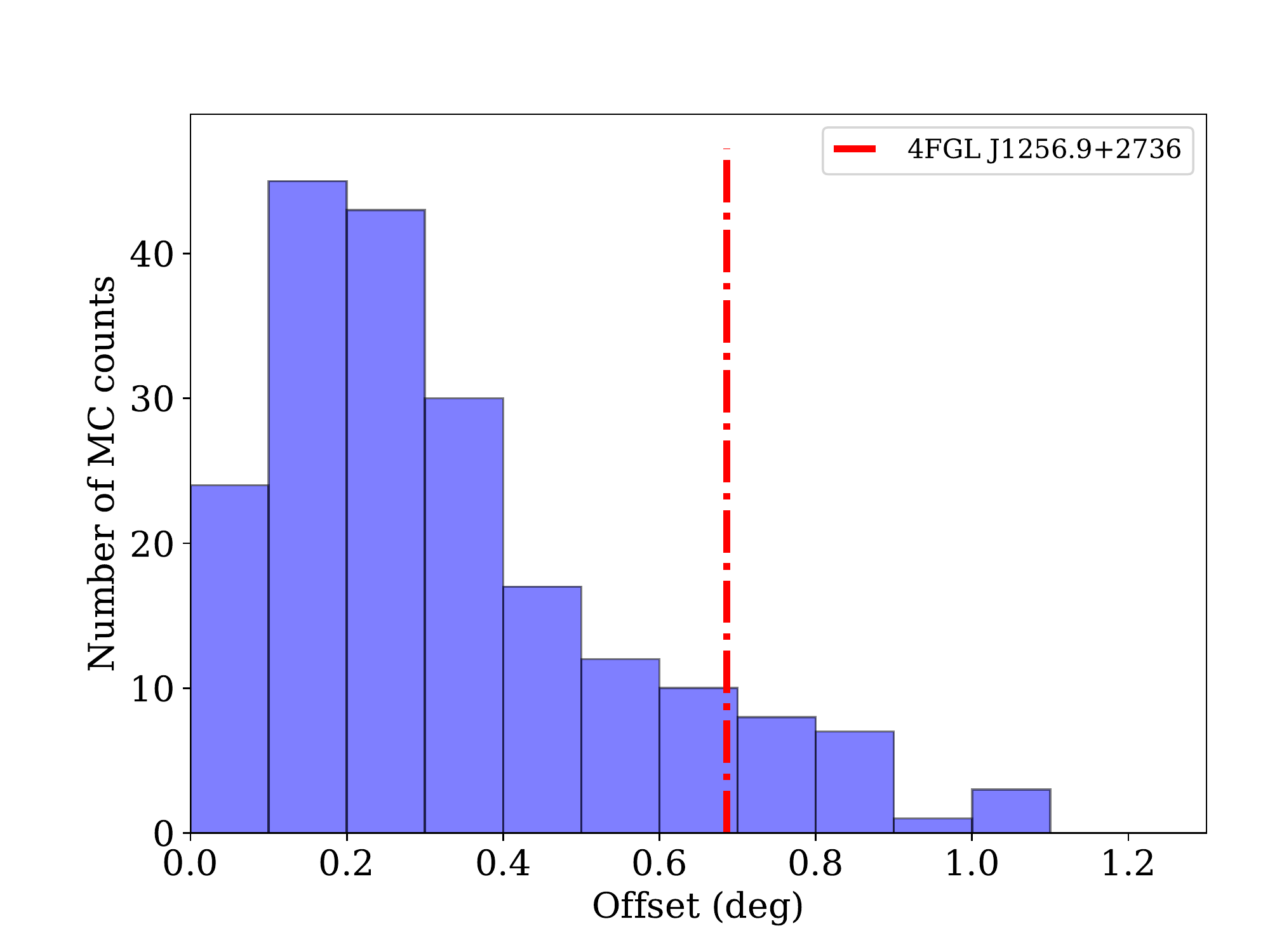}
\caption{\small {\bf Top}: sky distribution of the test source coordinates recovered for each Monte Carlo realization and comparison to the Coma reference center and the 4FGL~J1256.9+2736 location. {\bf Bottom}: distribution of the offset from the Coma reference center of the Monte Carlo realizations. The left panels correspond to the extended ICM model and the right panel to the tSZ template model.}
\label{fig:MC_testsource}
\end{figure*}

In this Appendix, we quantify the likelihood of finding a point source 0.68 degrees away from the Coma cluster center (as 4FGL~J1256.9+2736), in the hypothesis in which the point source corresponds in fact to the peak of the diffuse cluster ICM emission (i.e., scenario 2). To do so we use an approach that is based on Monte Carlo realizations, as follows. First, we select the best-fit sky model of scenario 2 (with 4FGL~J1256.9+2736 replaced by the cluster), compute the number count prediction model, and use it to perform Monte Carlo realizations of the data using poisson sampling. For each realization, we fit the model as done for the real data when excluding the cluster ICM from the sky model (see Section~\ref{sec:ROI_modeling} and Section~\ref{sec:ROI_fitting}). We then include a point source in the sky model, starting at the coordinates of the Coma reference center, and use the {\tt localize} method from \textit{Fermipy} to search for the location which provides the best match for such a test point source within 1.5 degree. We repeat the operation 200 times and compute the probability to find such a source at distance $\theta$ from the cluster.

The results are provided in Figure~\ref{fig:MC_testsource}, where we show the best-fit coordinates of the test point source for all the Monte Carlo realizations and compare it to the position of the Coma reference center and 4FGL~J1256.9+2736. We also display the histogram of the offsets between the recovered test point source and the Coma reference center. The results are shown in the case of our baseline extended ICM model and in the case the spatial ICM template is based on the tSZ map, for comparison. The results for the other cases are expected to slightly vary based on the extension of the spatial template, but are not shown here. As we can see, although it is located 0.68 degrees away from the Coma reference center, 4FGL~J1256.9+2736 does not appear as an outlier compared to our Monte Carlo. We find that 4.8\% of the recovered sources are located further than 0.68 degrees from the Coma center in the extended model, and 9.5\% in the more diffuse tSZ template case. This is due to the combination of the low S/N, the limited Fermi-LAT angular resolution and the extension of the ICM emission model. Indeed, the Fermi-LAT PSF varies with energy and around 1 GeV, where the signal is peaking, the 68\% containment angle is about 1 deg (about 3 deg for 95\% containment). This can be compared to the offset of 0.68 deg between 4FGL~J1256.9+2736 and the cluster center. Given the low S/N and the fact the ICM is an extended source (e.g., surface brightness drops by a factor of $\sim 2$ at $\theta=0.2$ degrees, in the extended case, see Figure~\ref{fig:template_minot}), this agrees with the offset being not significant.

Although a diffuse emission plus point source model provides the best match to the data (see Section~\ref{sec:comparison_data_model}), these results confirm that the scenario 2 agrees with the data. This also agrees with the results presented in the main paper, in which no significant residual is observed around Coma once the best-fit ICM model is subtracted from the data, and where the comparison of the TS given in Table~\ref{tab:table_fermi_analysis} does not point toward the need of including 4FGL~J1256.9+2736 in the sky model if we include a diffuse ICM component. See also Section~\ref{sec:comparison_data_model} for discussions.

\section{ROI modeling and fitting: A null test in the direction of point sources}\label{app:null_test}
In this Appendix, we present a null test, which we use to validate our ROI modeling and fitting procedure. As discussed in the main text, the source 4FGL~J1256.9+2736 can be better described by an extended ICM model (Table~\ref{tab:table_fermi_analysis}) that is centered 0.68 degrees away from the source, leading us to consider the scenario in which 4FGL~J1256.9+2736 corresponds in fact to the peak of the diffuse ICM emission (scenario 2). Here, we check that this is not a general feature of our procedure and that it does not apply to other sources.
 
We select the following sources from our field because their local background is very close to the Coma region (no close by source in the 4FGL catalog). However, we note that their spectra are generally steeper and their TS may differ: 4FGL J1253.8+2929 (${\rm TS}=27.57$, 2 degrees northwest from the Coma center, power law index of 1.9), 4FGL J1250.8+3117 (${\rm TS}=137.29$, 3.8 degrees northwest from the Coma center, power law index of 2.16), 4FGL J1316.5+3013 (${\rm TS}=20.90$, 4.3 degrees northeast from Coma, power law index of 2.1). To match what is done for Coma, we define the cluster center 0.68 degrees away from these sources and test four directions for the offset (north, east, south, west) to increase statistics. We also perform the test with the cluster centered on the source, but we note that this does not match the case of Coma. Because the spectra of these sources are not the same as 4FGL~J1256.9+2736, we model the cluster with a power law of free index instead of a fixed physical spectrum. However, we note that a flatter spectrum will lead to an improved angular resolution since the PSF decreases with increasing energy. We then apply the same fitting procedure as the one done for Coma, i.e., we add the cluster template in the sky model either by replacing the point source with the cluster, or including both the point source and the cluster. We use the baseline extended cluster model for the test performed here.

We report the different TS values that we obtain for the different test cases in Table~\ref{tab:table_null_test}. We can see that for all tests, the TS is much smaller for the cluster than the point source when both are included, even when the two are co-aligned. In the cases where the point source is replaced by the cluster, the TS of the cluster increases, but remains much smaller than for the point source except when the cluster is centered on the source (although it always remains lower than for the point source only). 

We conclude that while the point source 4FGL~J1256.9+2736 can be better explained by an extended ICM model (even when both cluster and point source are included, see Section~\ref{sec:ROI_fitting} and Table~\ref{tab:table_fermi_analysis}), this is not the case for the other sources 4FGL~J1253.8+2929, 4FGL~J1250.8+3117 and 4FGL~J1316.5+3013. This provides us with a null test that allows us to strengthen the modeling and fitting procedure described in the main text.

\begin{table*}
	\caption{\small Recovered TS for the different null tests and comparison to the initial TS of the considered sources.}
	\begin{center}
	\begin{tabular}{c|ccc}
	\hline
	\hline
	Test case & Initial single point source & Point source + cluster & Single cluster  \\
	 & TS$_{\rm point \ source}$ & (TS$_{\rm point \ source}$, TS$_{\rm cluster}$) & TS$_{\rm cluster}$ \\
	\hline
	4FGL J1253.8+2929 -- North & 27.57 & (25.94, 1.57) & 3.82 \\
	4FGL J1253.8+2929 -- South & 27.57 & (27.11, 1.04) & 1.83 \\
	4FGL J1253.8+2929 -- East & 27.57 & (26.38, 0.18) & 3.47 \\
	4FGL J1253.8+2929 -- West & 27.57 & (25.48, 0.49) & 4.72 \\
	4FGL J1253.8+2929 -- Center & 27.57 & (10.75, 3.82) & 25.42 \\
	\hline
	4FGL J1250.8+3117 -- North & 137.29 & (130.92, 5.05) & 23.14 \\
	4FGL J1250.8+3117 -- South & 137.29 & (126.57, 2.47) & 29.25 \\
	4FGL J1250.8+3117 -- East & 137.29 & (131.44, 2.71) & 20.35 \\
	4FGL J1250.8+3117 -- West & 137.29 & (134.05, 0.43) & 19.78 \\
	4FGL J1250.8+3117 -- Center & 137.29 & (133.23, 0.01) & 104.18 \\
	\hline
	4FGL J1316.5+3013 -- North & 20.90 & (20.87, 0.02) & 0.05 \\
	4FGL J1316.5+3013 -- South & 20.90 & (20.87, 0.03) & 0.07 \\
	4FGL J1316.5+3013 -- East & 20.90 & (17.82, 1.74) & 6.53 \\
	4FGL J1316.5+3013 -- West & 20.90 & (20.86, 0.01) & 0.06 \\
	4FGL J1316.5+3013 -- Center & 20.90 & (20.74, 0.03) & 8.90 \\
	\hline
	\end{tabular}
	\end{center}
	\label{tab:table_null_test}
\end{table*}

\section{Constraints on the cosmic ray electron populations with alternative models}\label{app:CRe_alternative_constraints}
\begin{figure}
\centering
	\rule{8cm}{0.01cm}
	\put(-220,5){\footnotesize {\tt InitialInjection}}
	\put(-100,5){\footnotesize {\tt ContinuousInjection}}
	
	\includegraphics[width=0.24\textwidth]{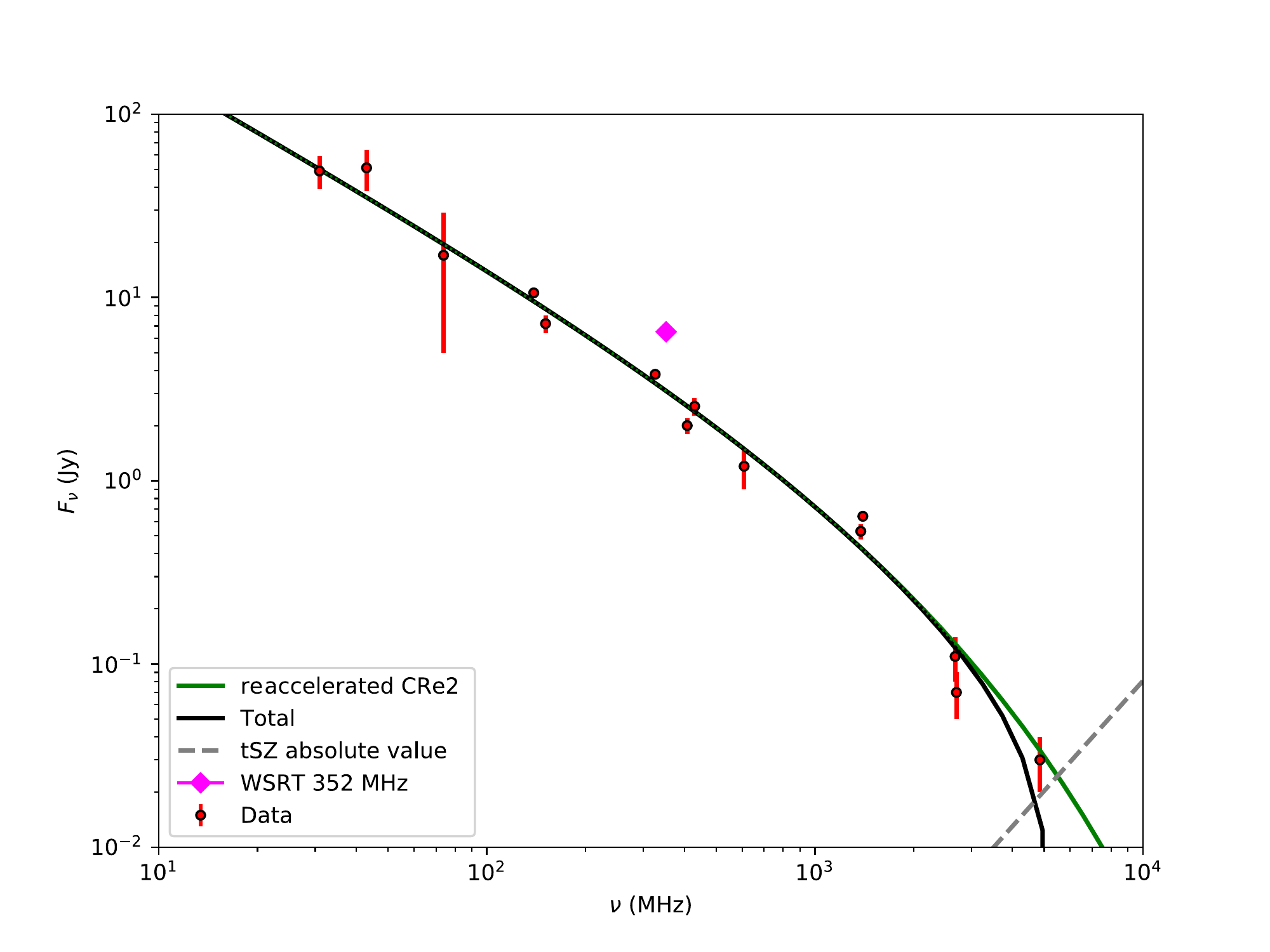}
	\includegraphics[width=0.24\textwidth]{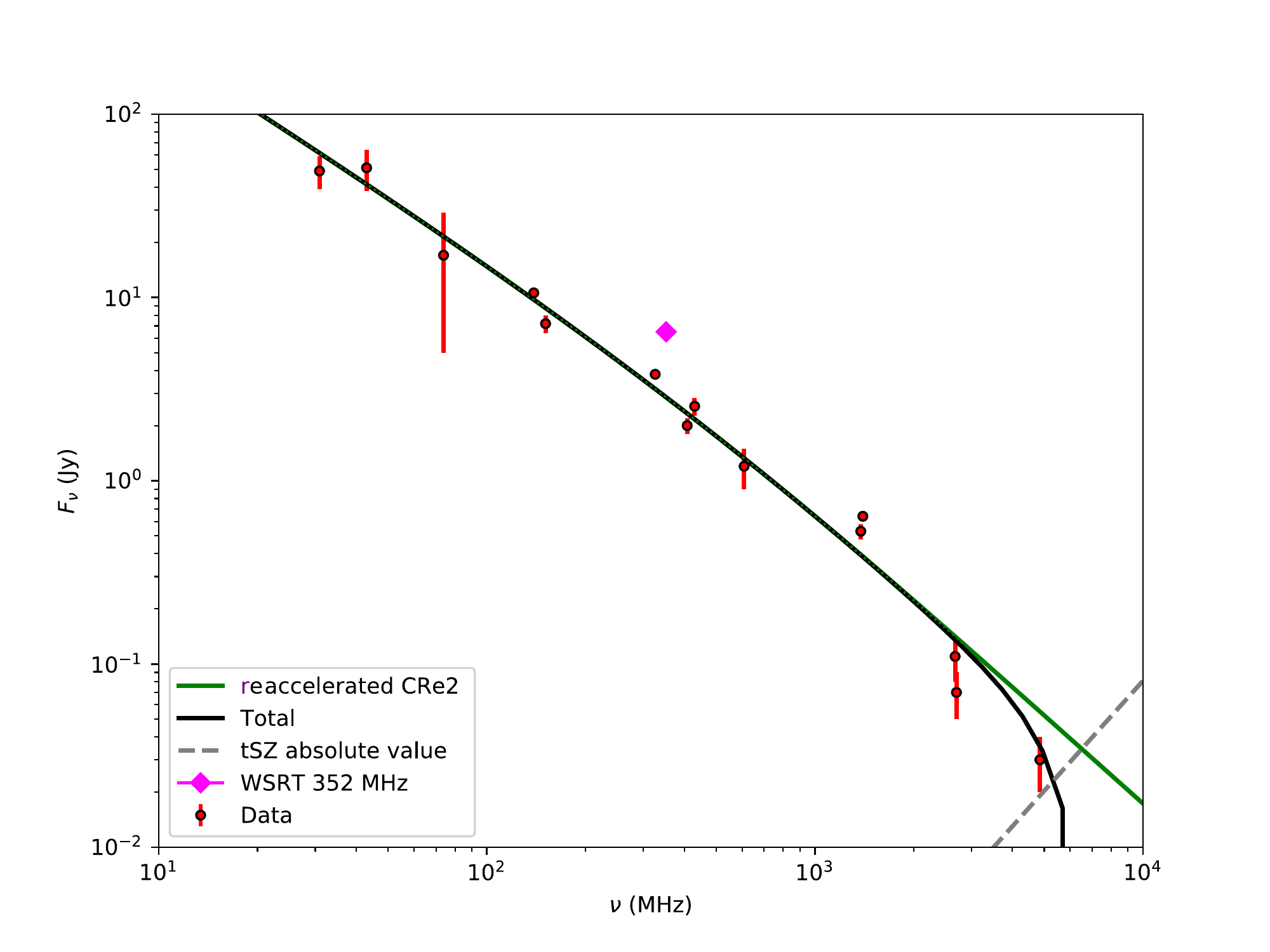}
	\includegraphics[width=0.24\textwidth]{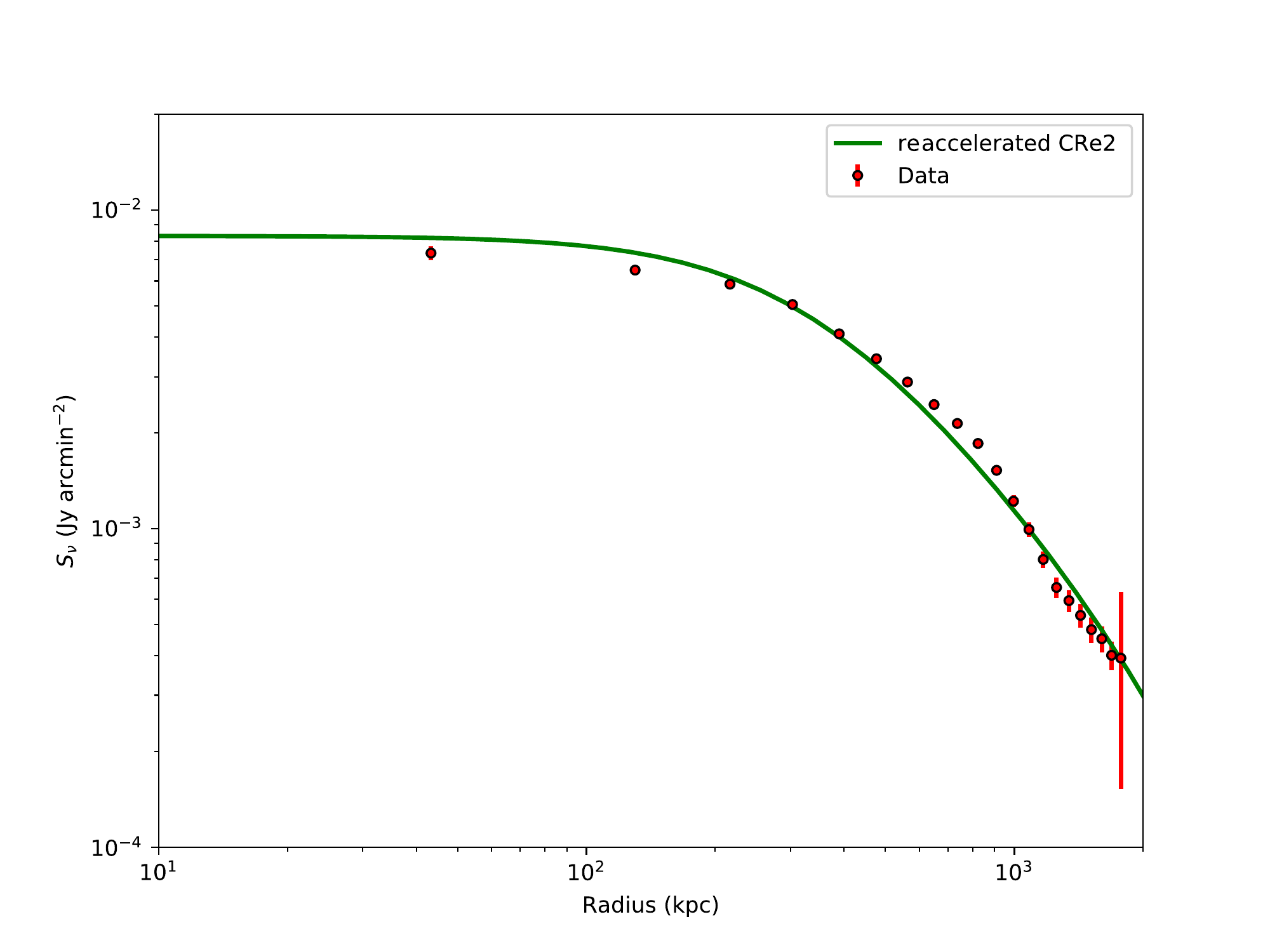}
	\includegraphics[width=0.24\textwidth]{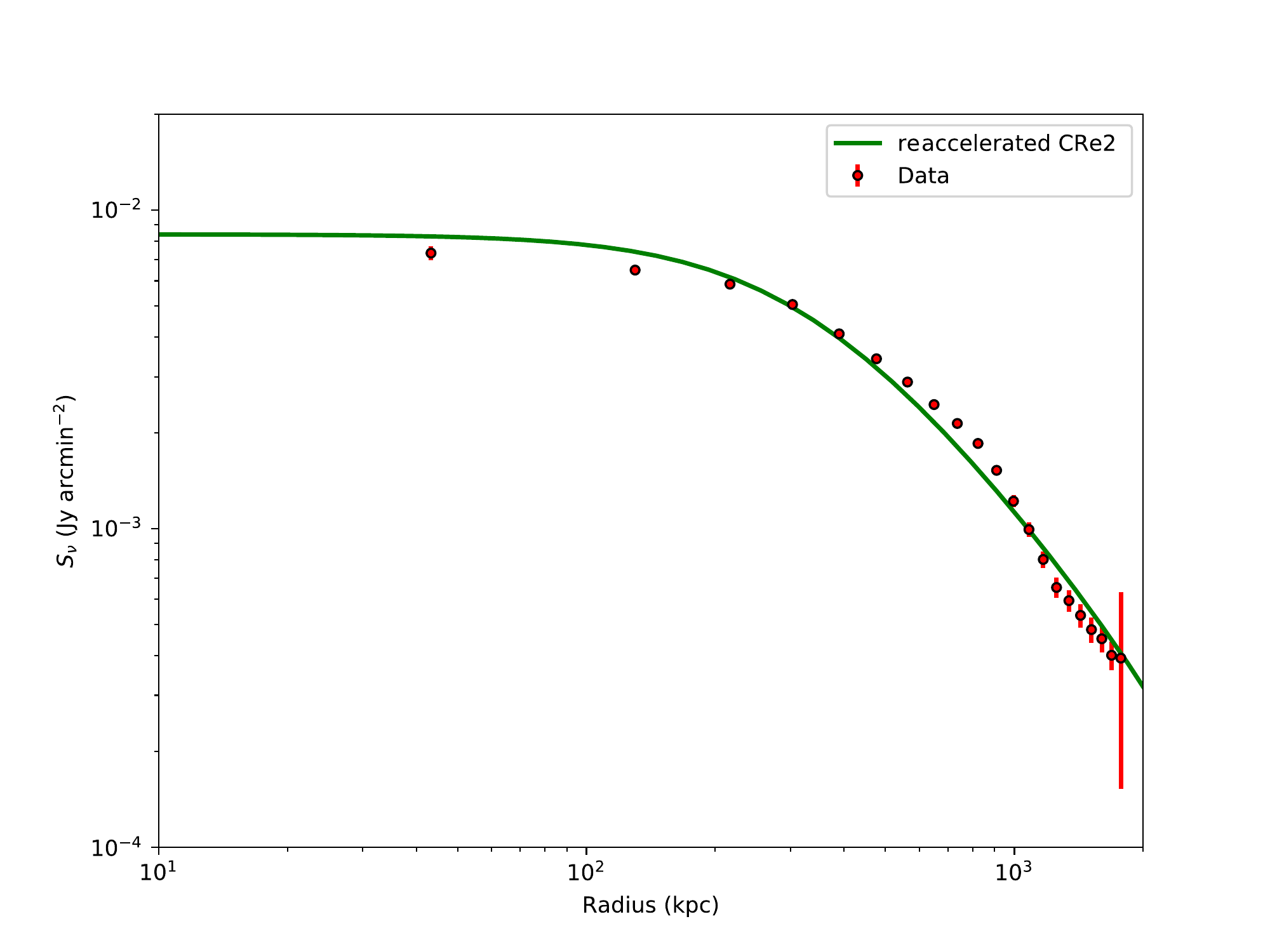}
	\includegraphics[width=0.24\textwidth]{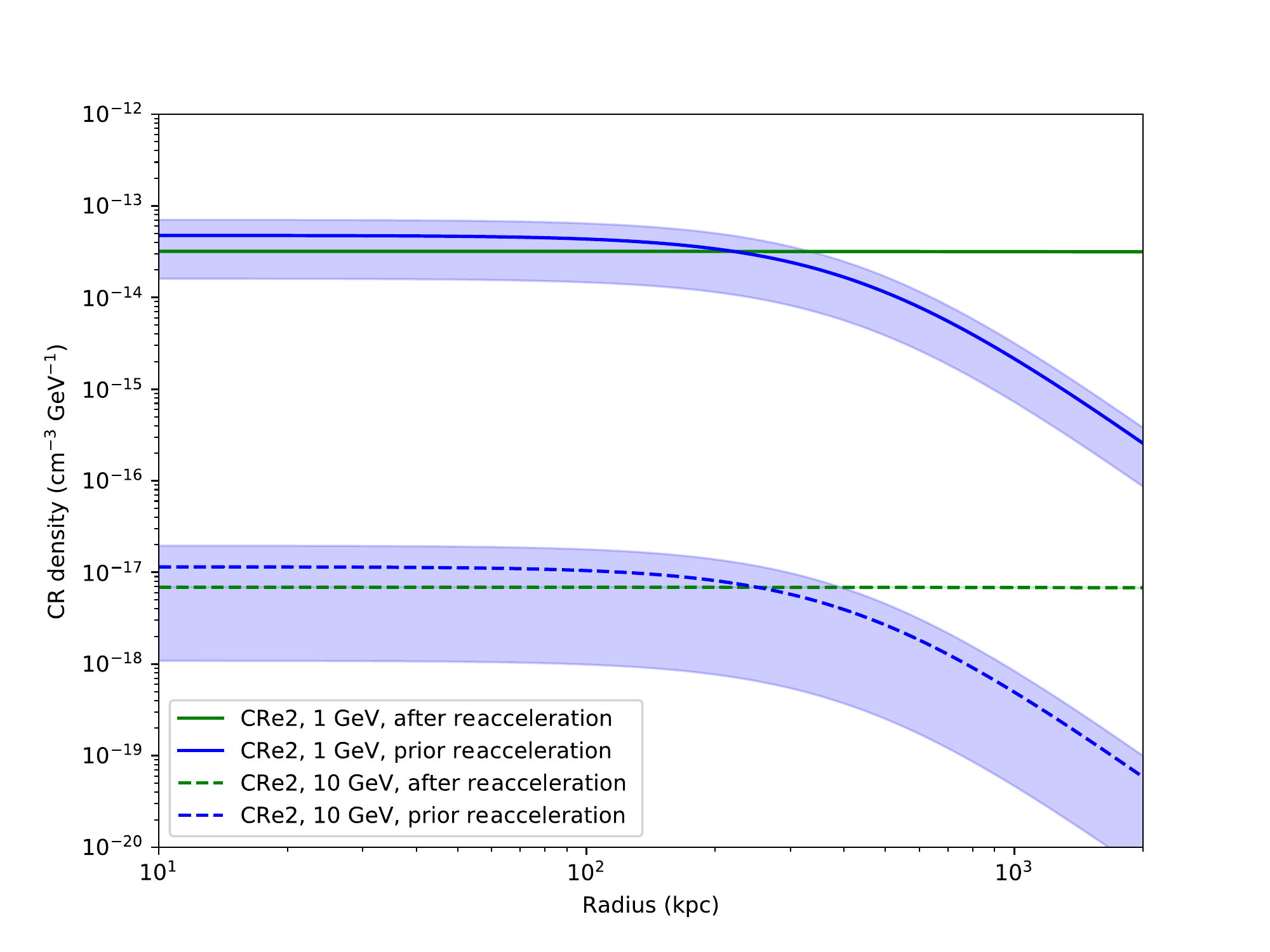}
	\includegraphics[width=0.24\textwidth]{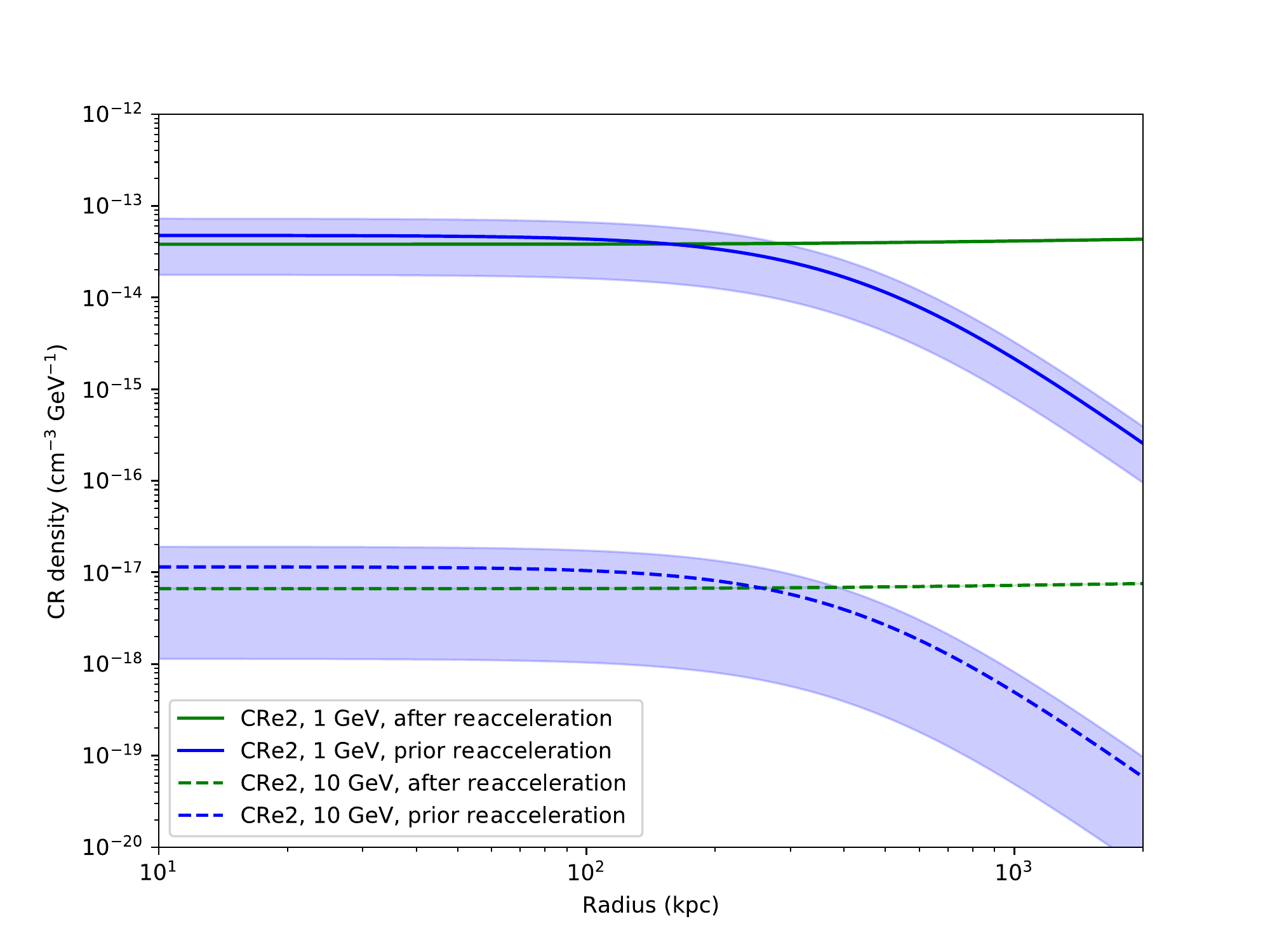}
	\includegraphics[width=0.24\textwidth]{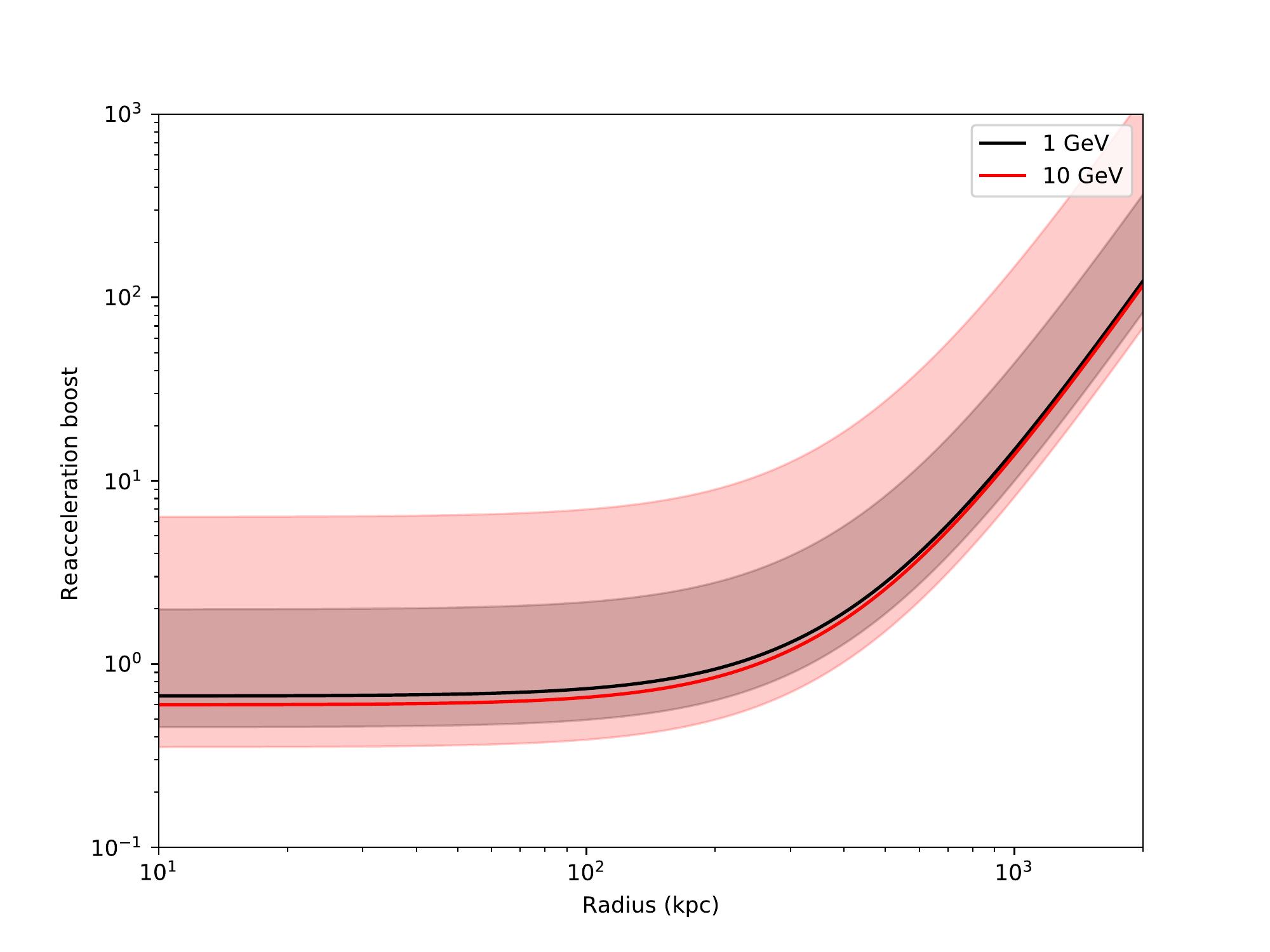}
	\includegraphics[width=0.24\textwidth]{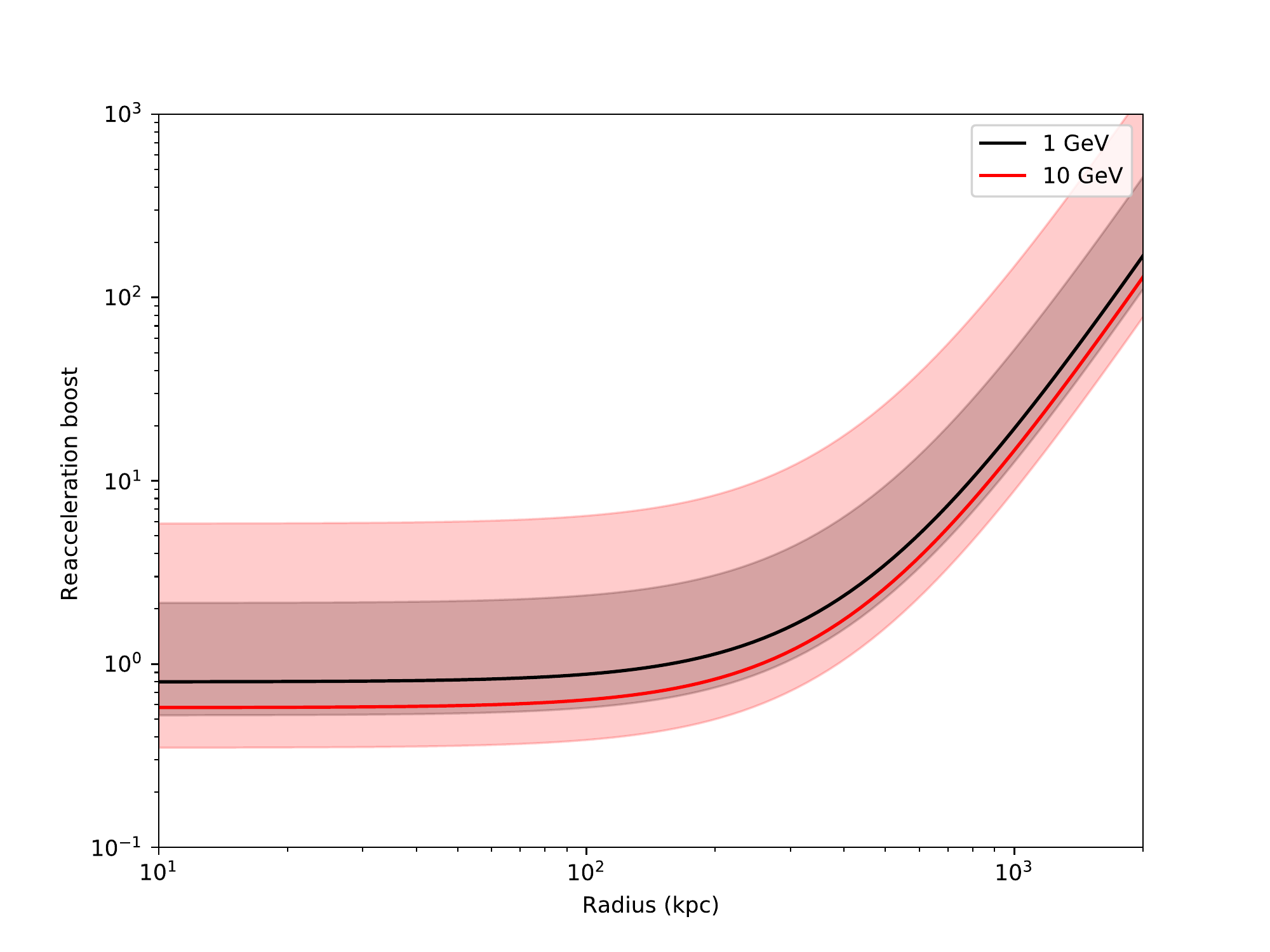}
	\caption{\small Same as Figure~\ref{fig:radio_implication_observable2} for the {\tt InitialInjection} model (left) and the {\tt ContinuousInjection} model (right).}
	\label{fig:radio_implication_observable2_app2}
\end{figure}

\begin{figure*}
\centering
	\rule{17cm}{0.01cm}
	\put(-460,5){\footnotesize $n_{\rm CRp} \propto n_{\rm gas}$}
	\put(-340,5){\footnotesize $n_{\rm CRp} = {\rm constant}$}
	\put(-210,5){\footnotesize $n_{\rm CRp} \propto P_{\rm gas}$}
	\put(-120,5){\footnotesize $n_{\rm CRp} \propto n_{\rm gas}^{1/2}$ with 4FGL~J1256.9+2736}
	
	\rule{17cm}{0.01cm}
	\put(-300,5){\footnotesize CRe$_1$ + CRe$_2$ interpretation}

	\includegraphics[width=0.24\textwidth]{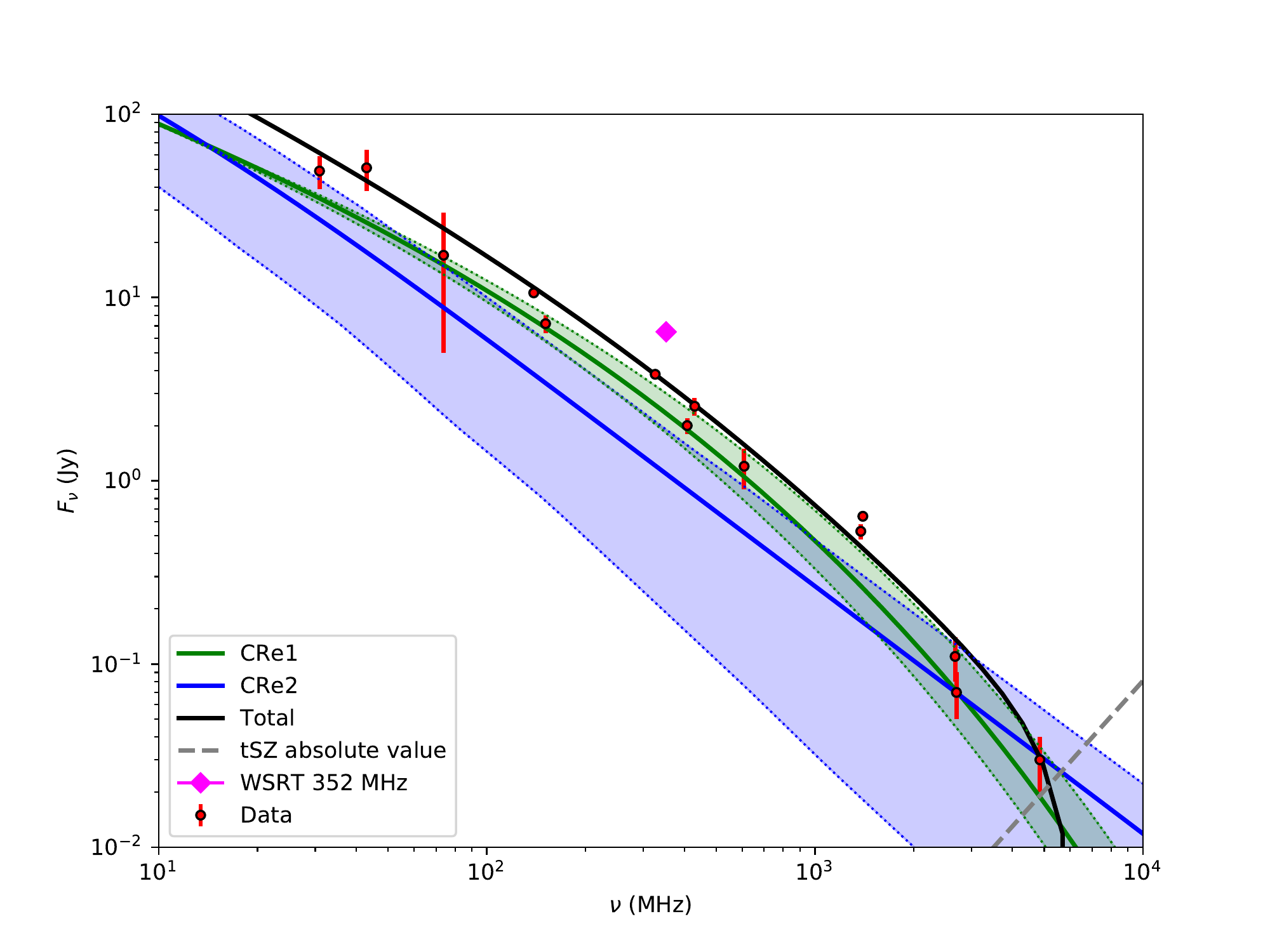}
	\includegraphics[width=0.24\textwidth]{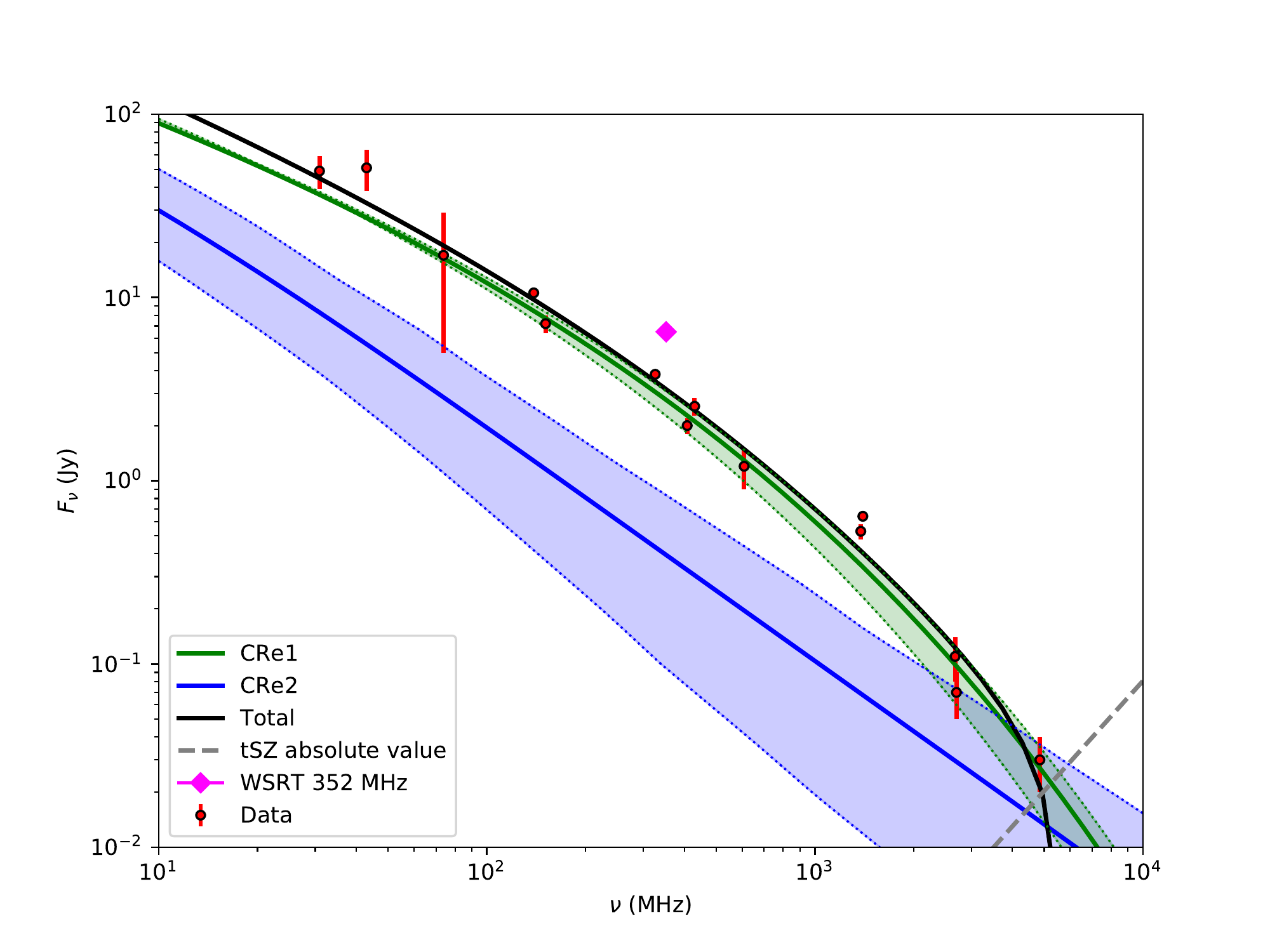}
	\includegraphics[width=0.24\textwidth]{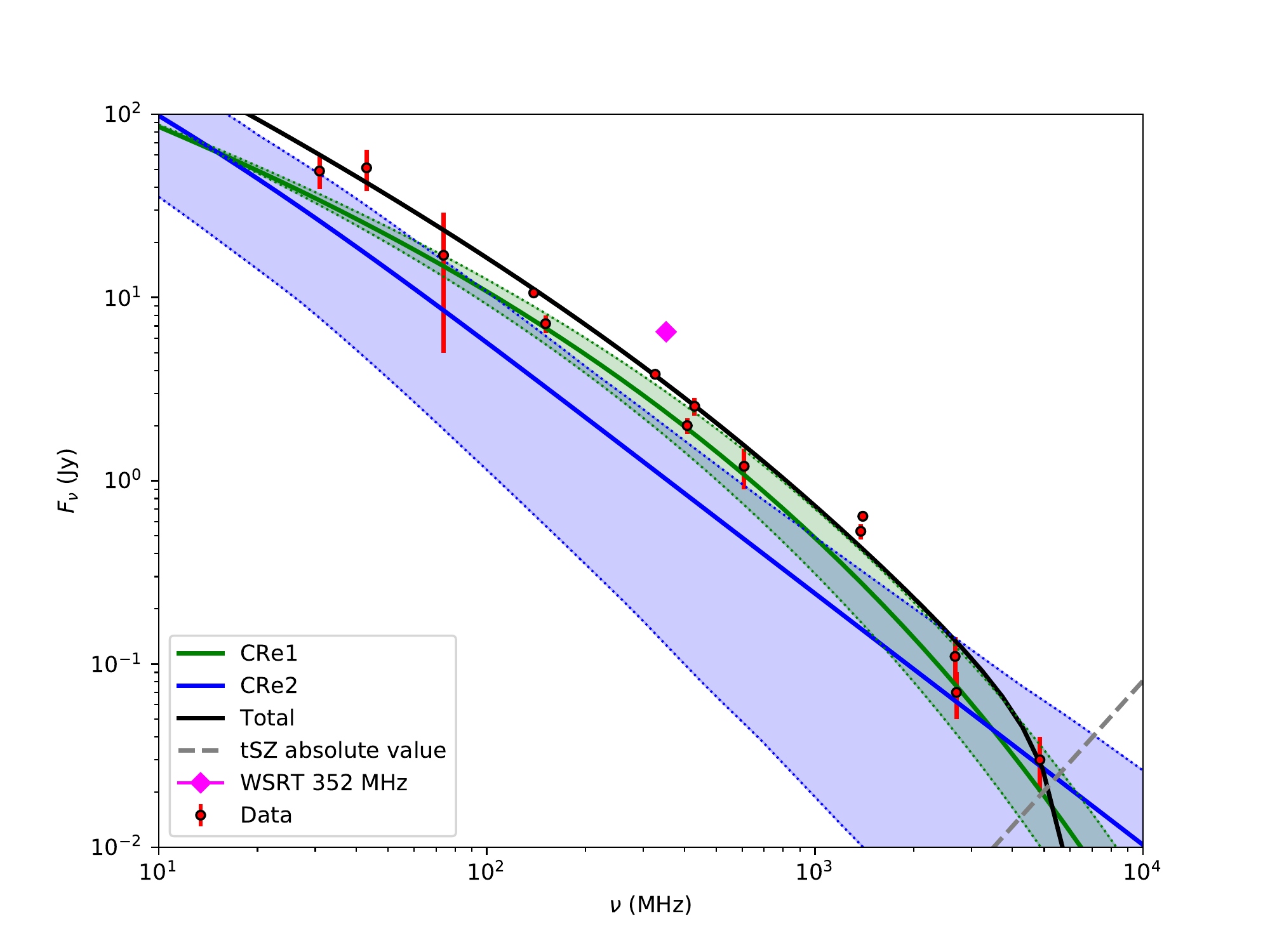}
	\includegraphics[width=0.24\textwidth]{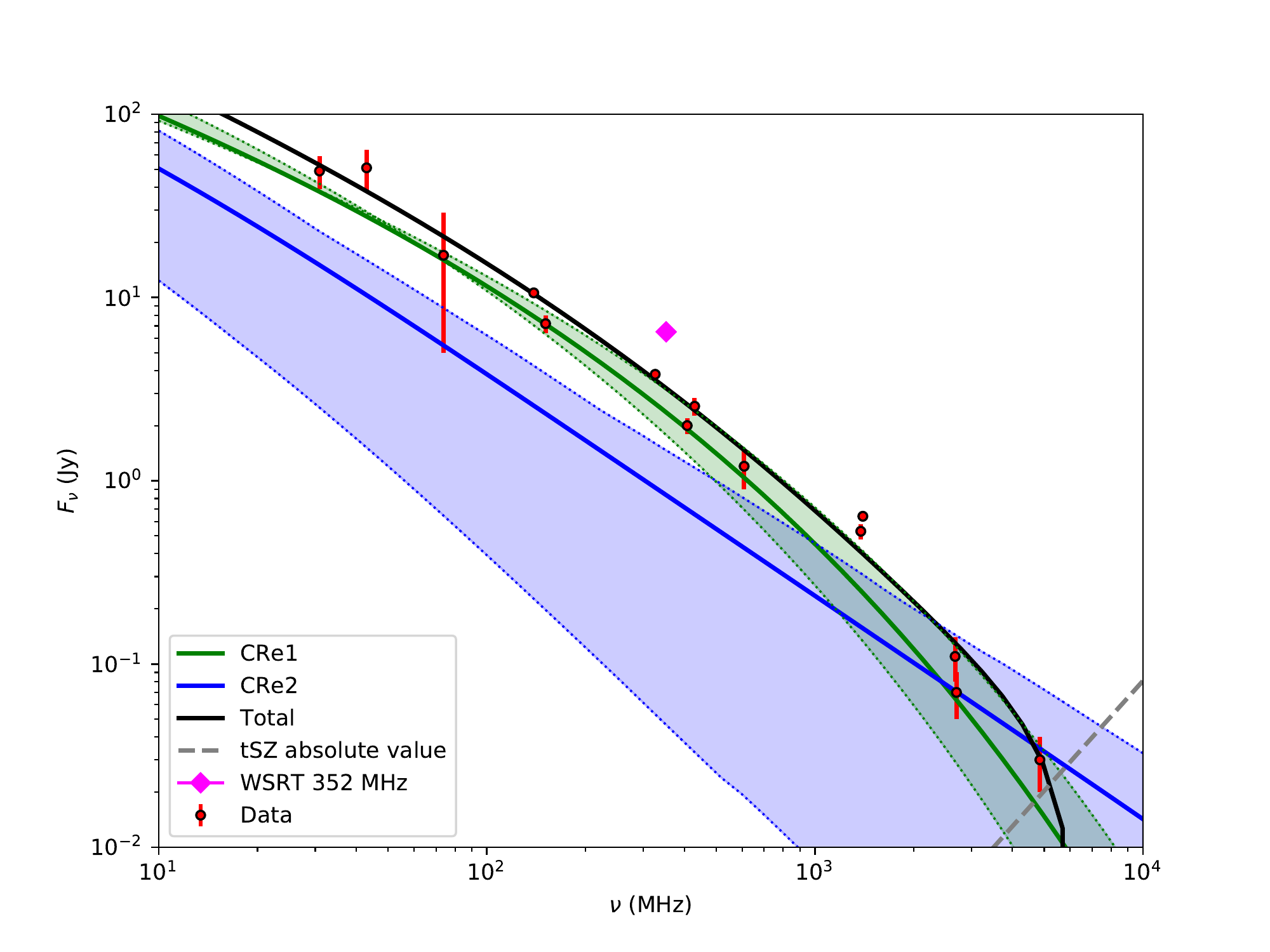}
	\includegraphics[width=0.24\textwidth]{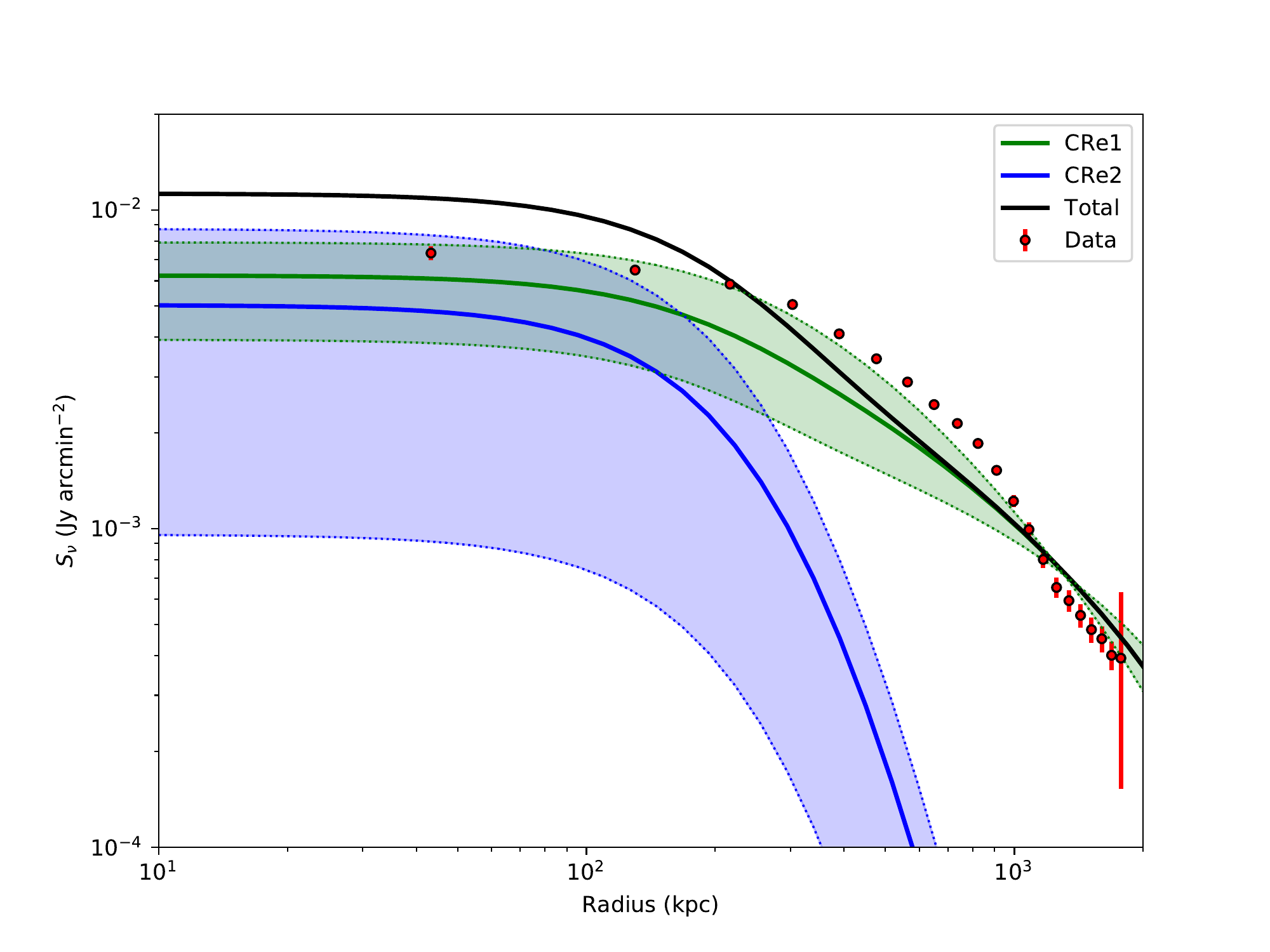}
	\includegraphics[width=0.24\textwidth]{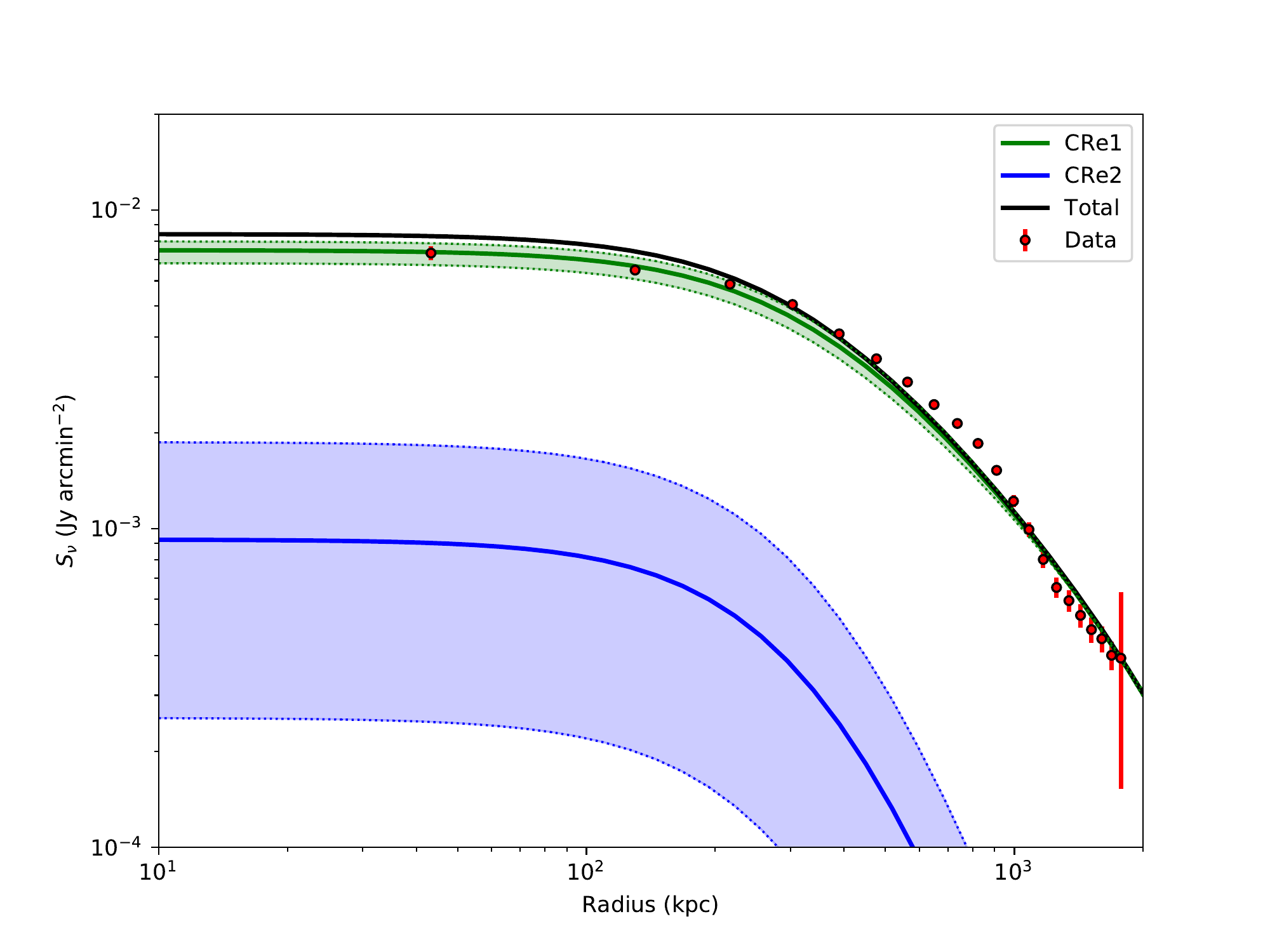}
	\includegraphics[width=0.24\textwidth]{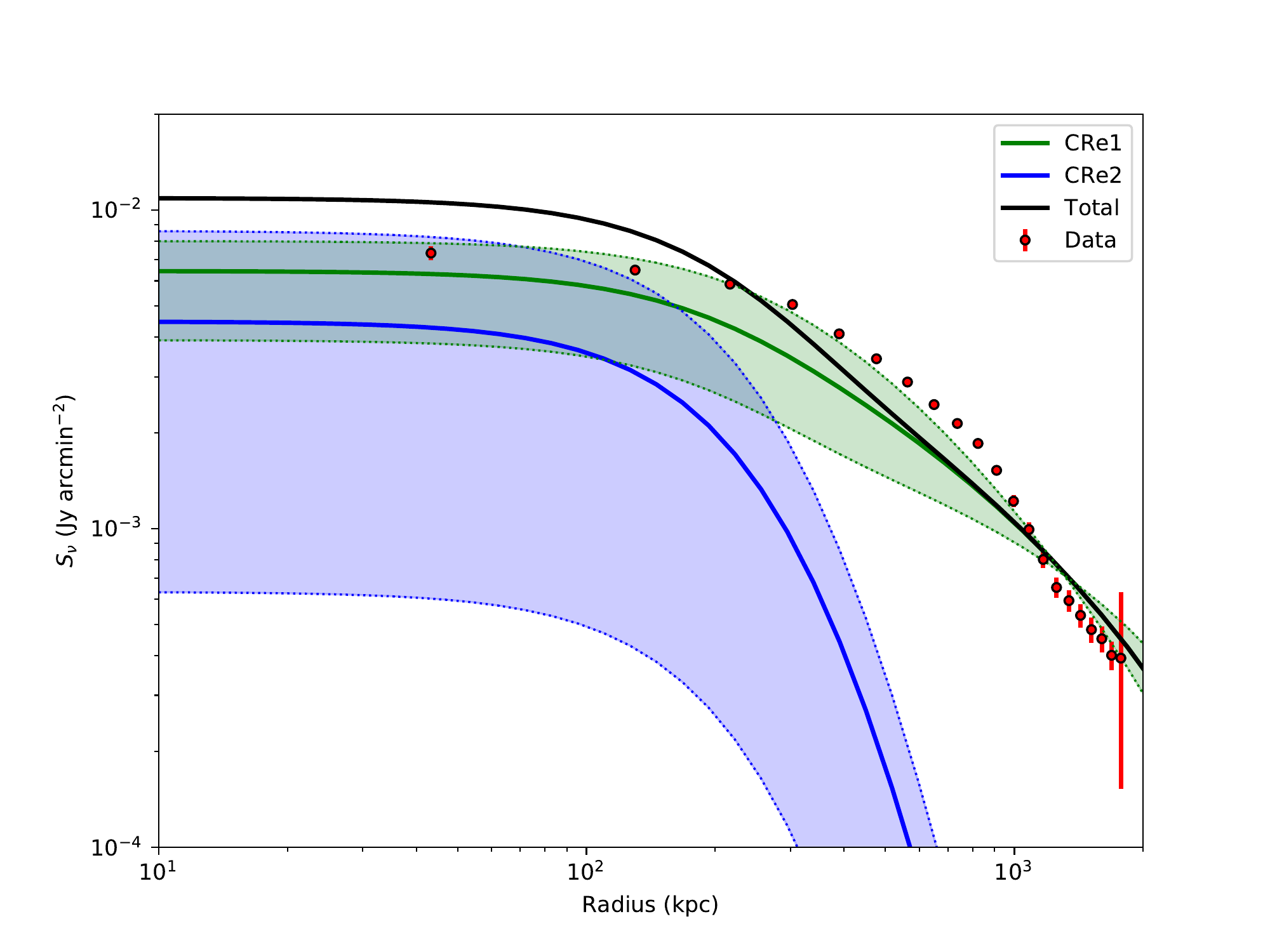}
	\includegraphics[width=0.24\textwidth]{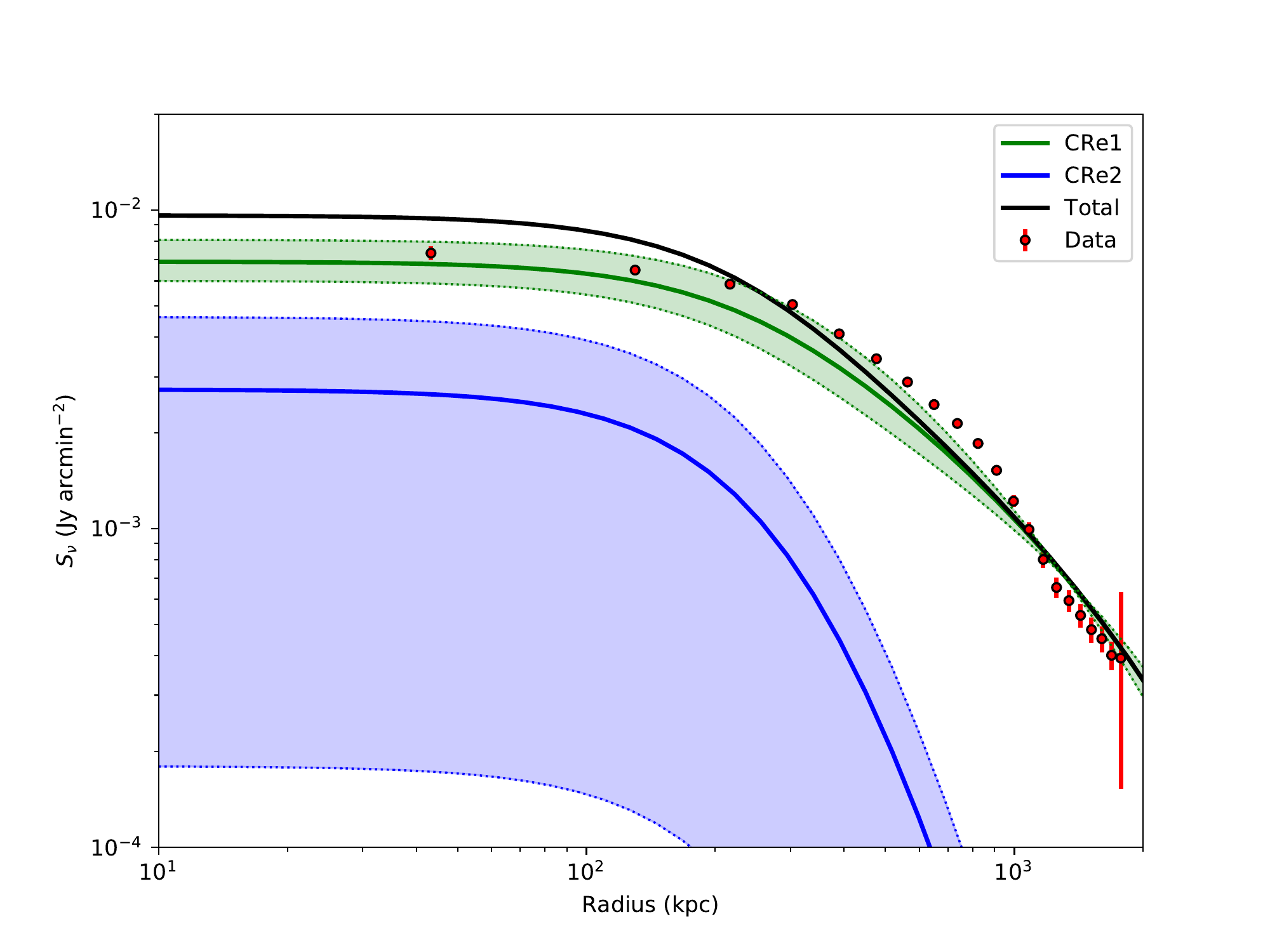}
	\includegraphics[width=0.24\textwidth]{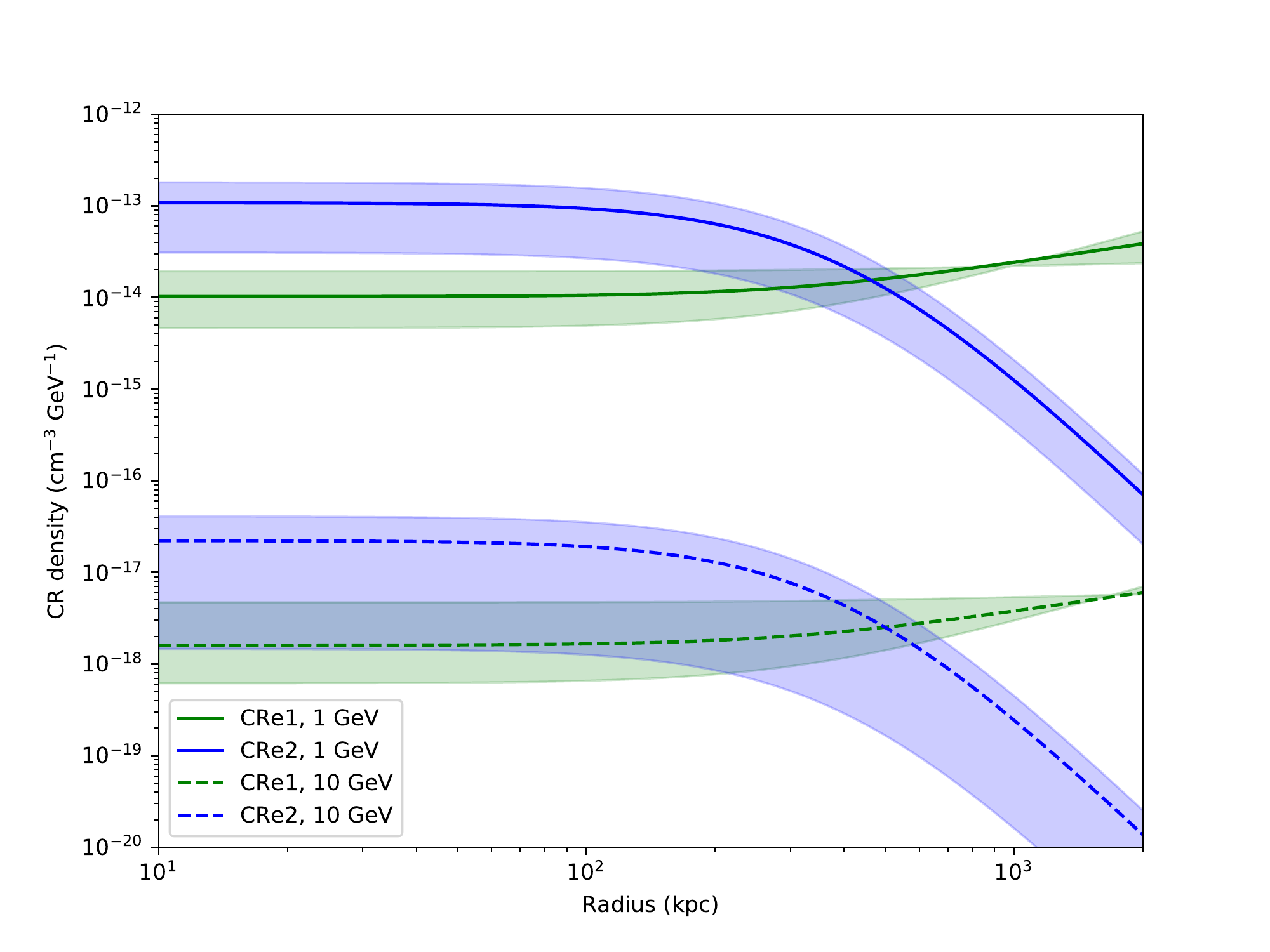}
	\includegraphics[width=0.24\textwidth]{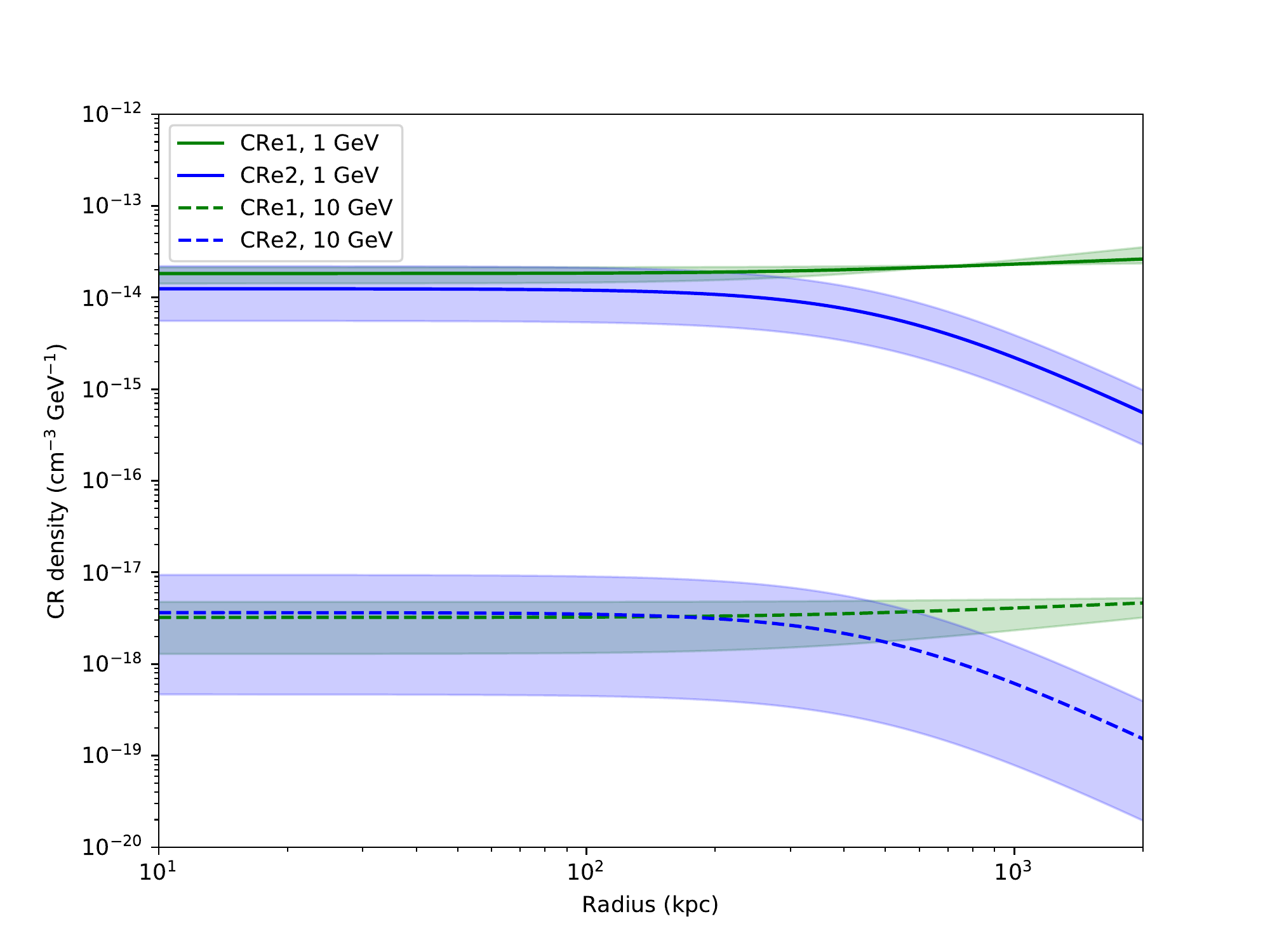}
	\includegraphics[width=0.24\textwidth]{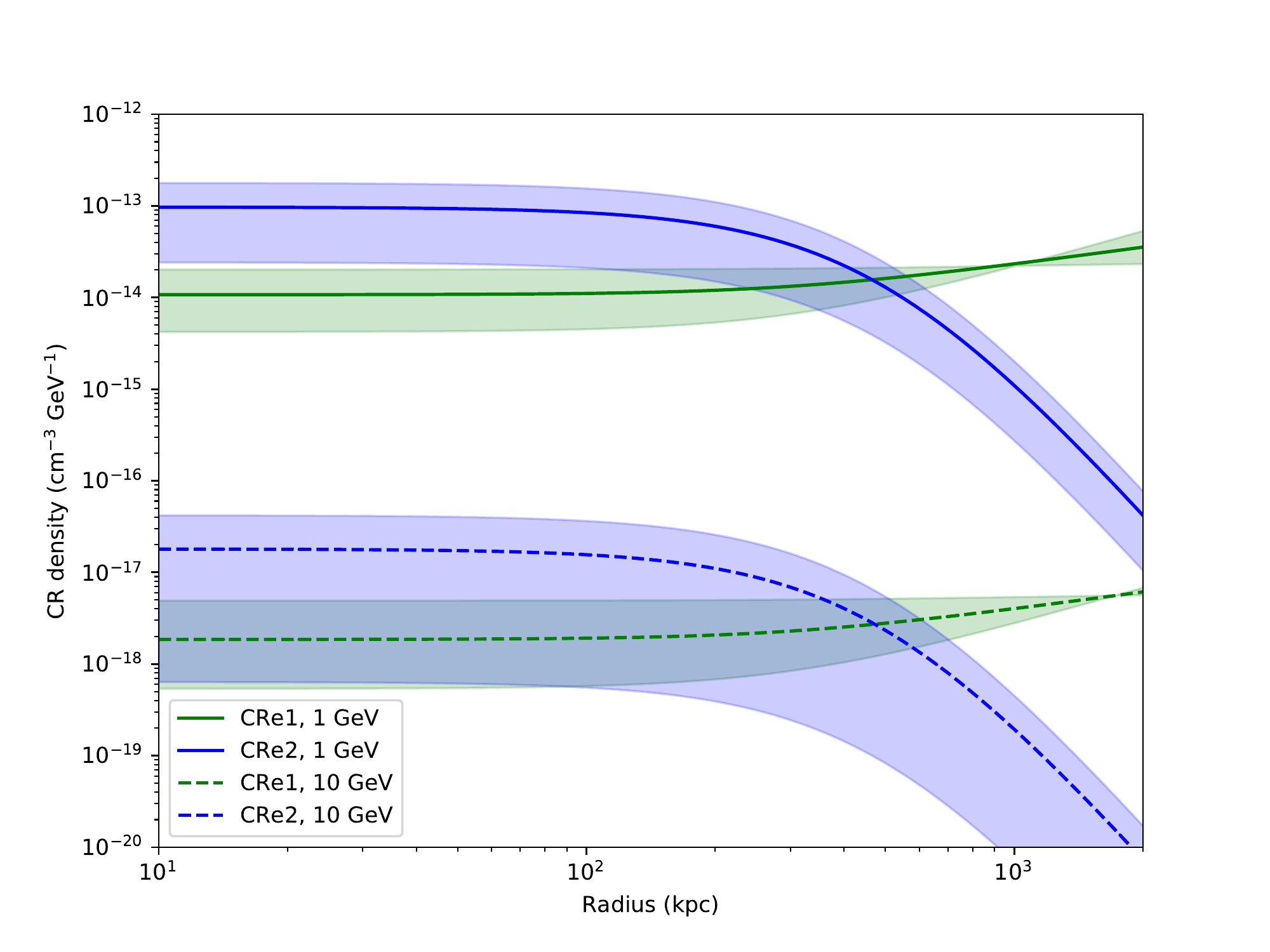}
	\includegraphics[width=0.24\textwidth]{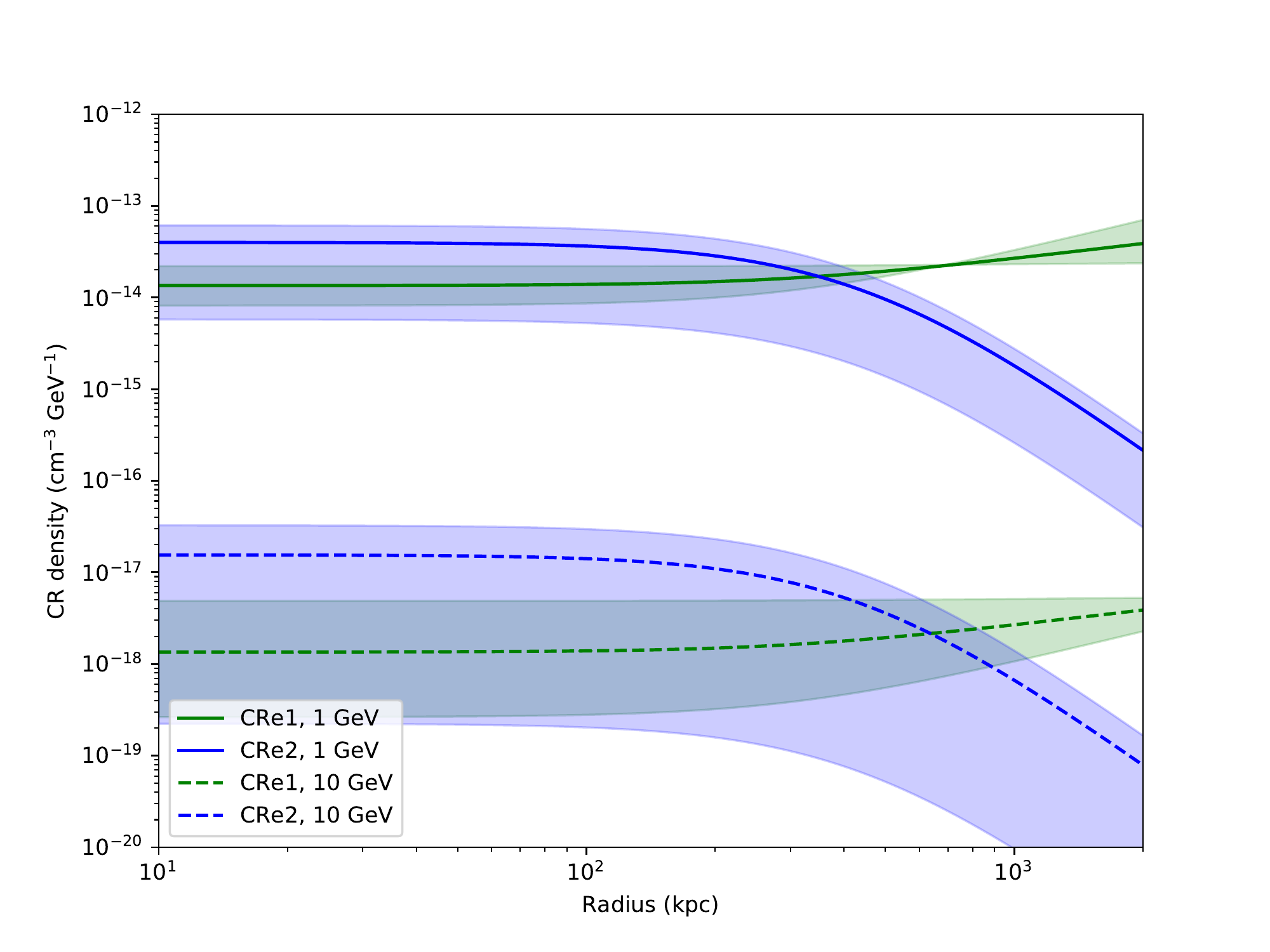}
	\includegraphics[width=0.24\textwidth]{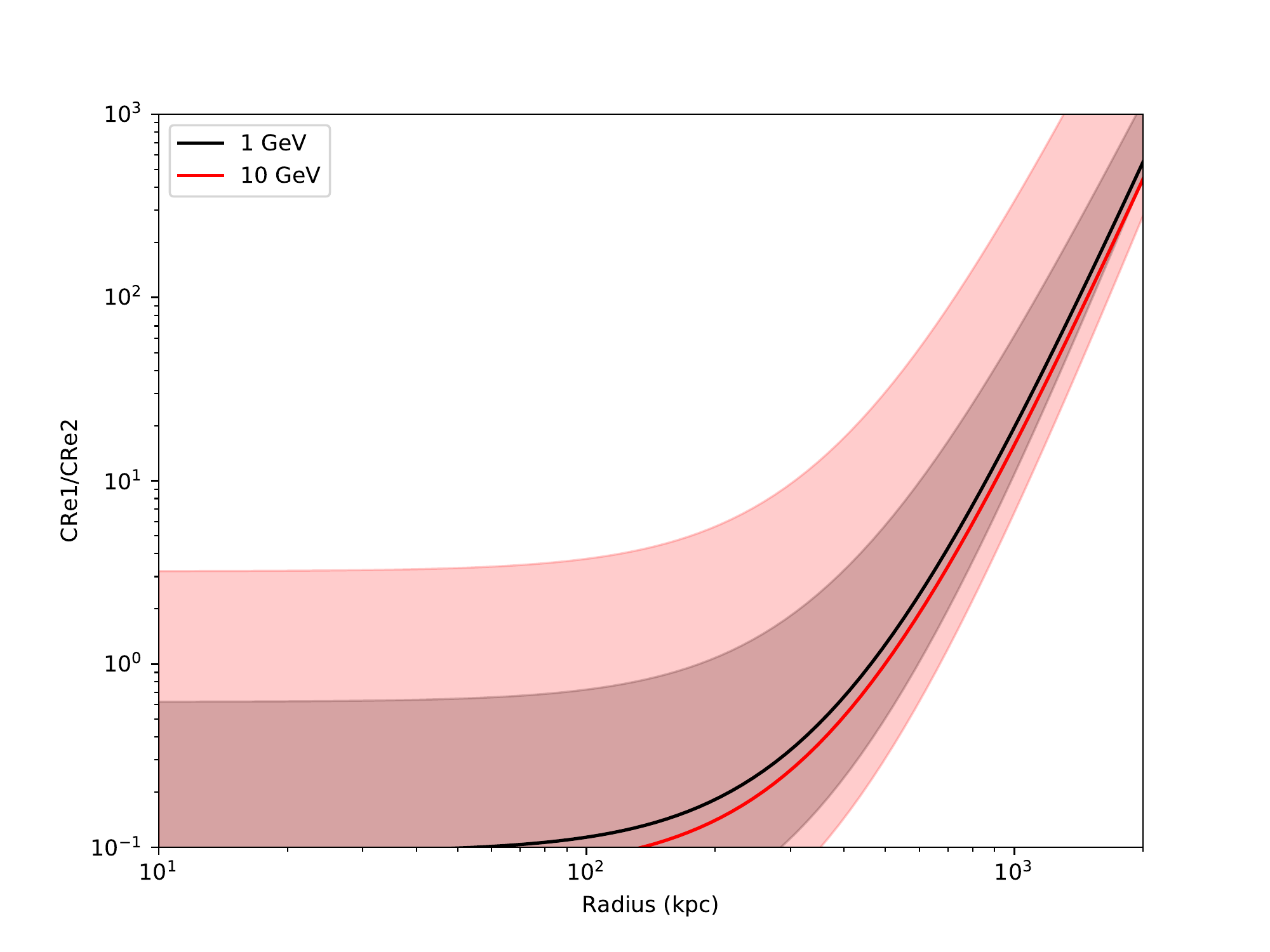}
	\includegraphics[width=0.24\textwidth]{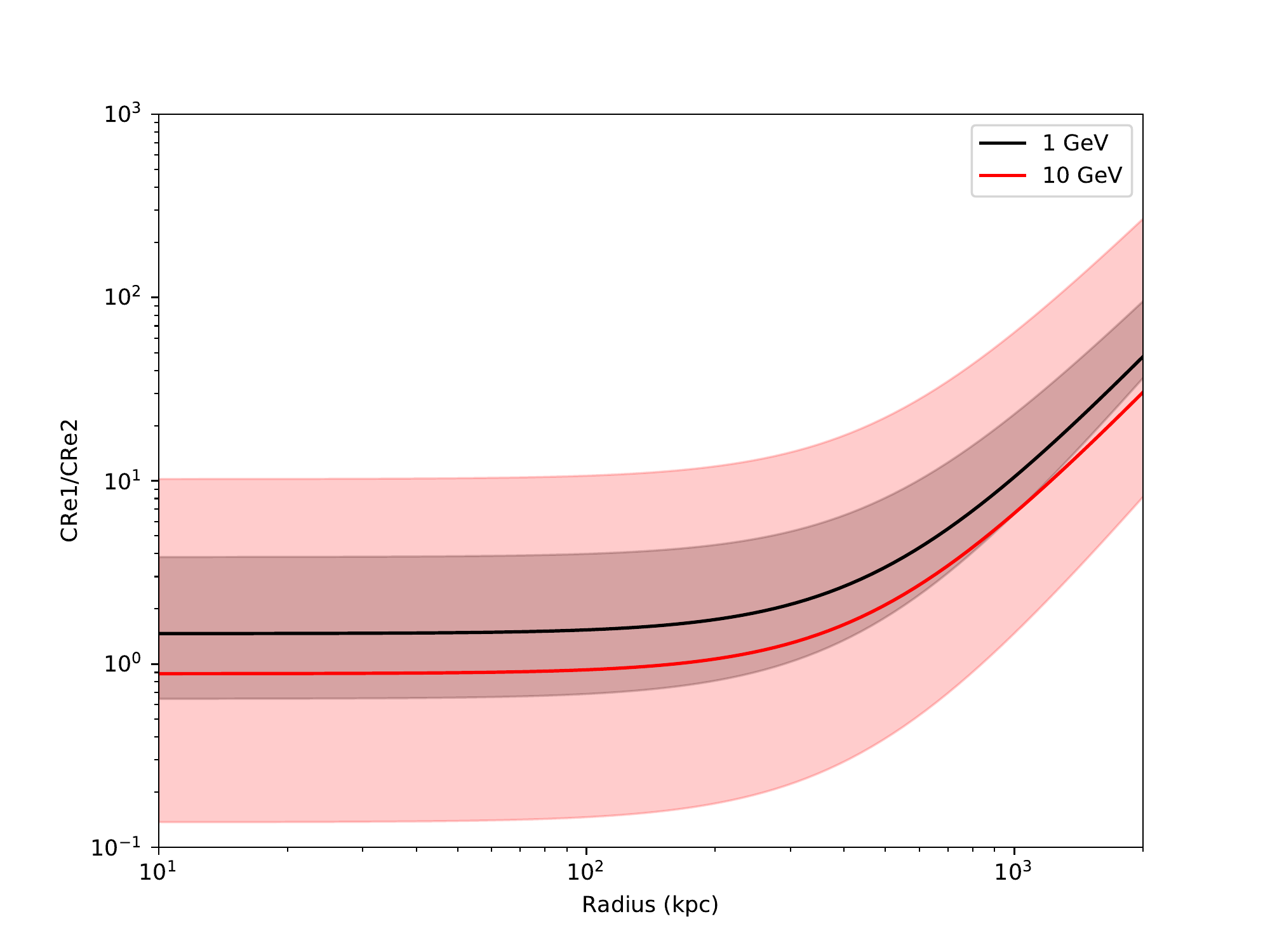}
	\includegraphics[width=0.24\textwidth]{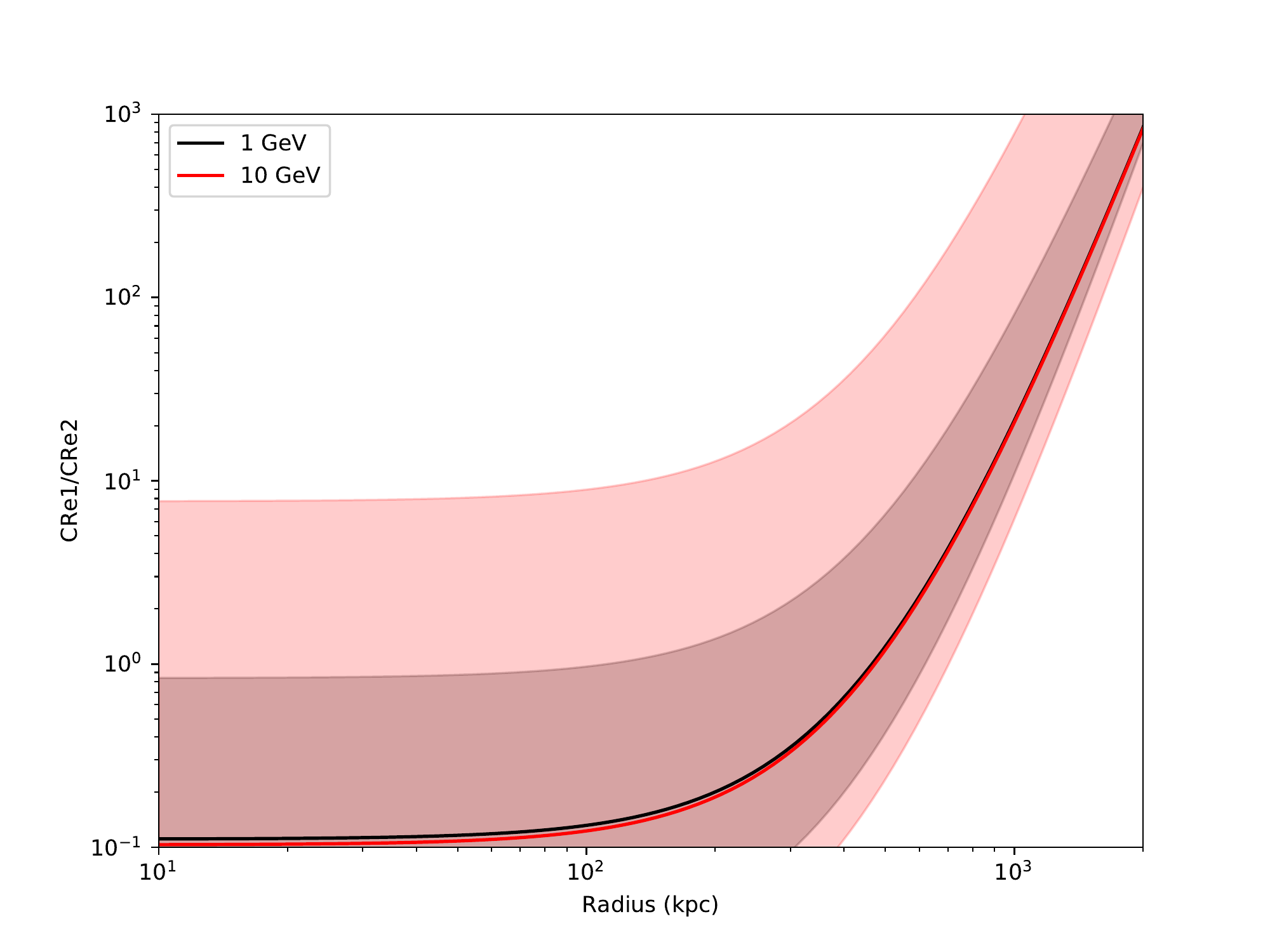}
	\includegraphics[width=0.24\textwidth]{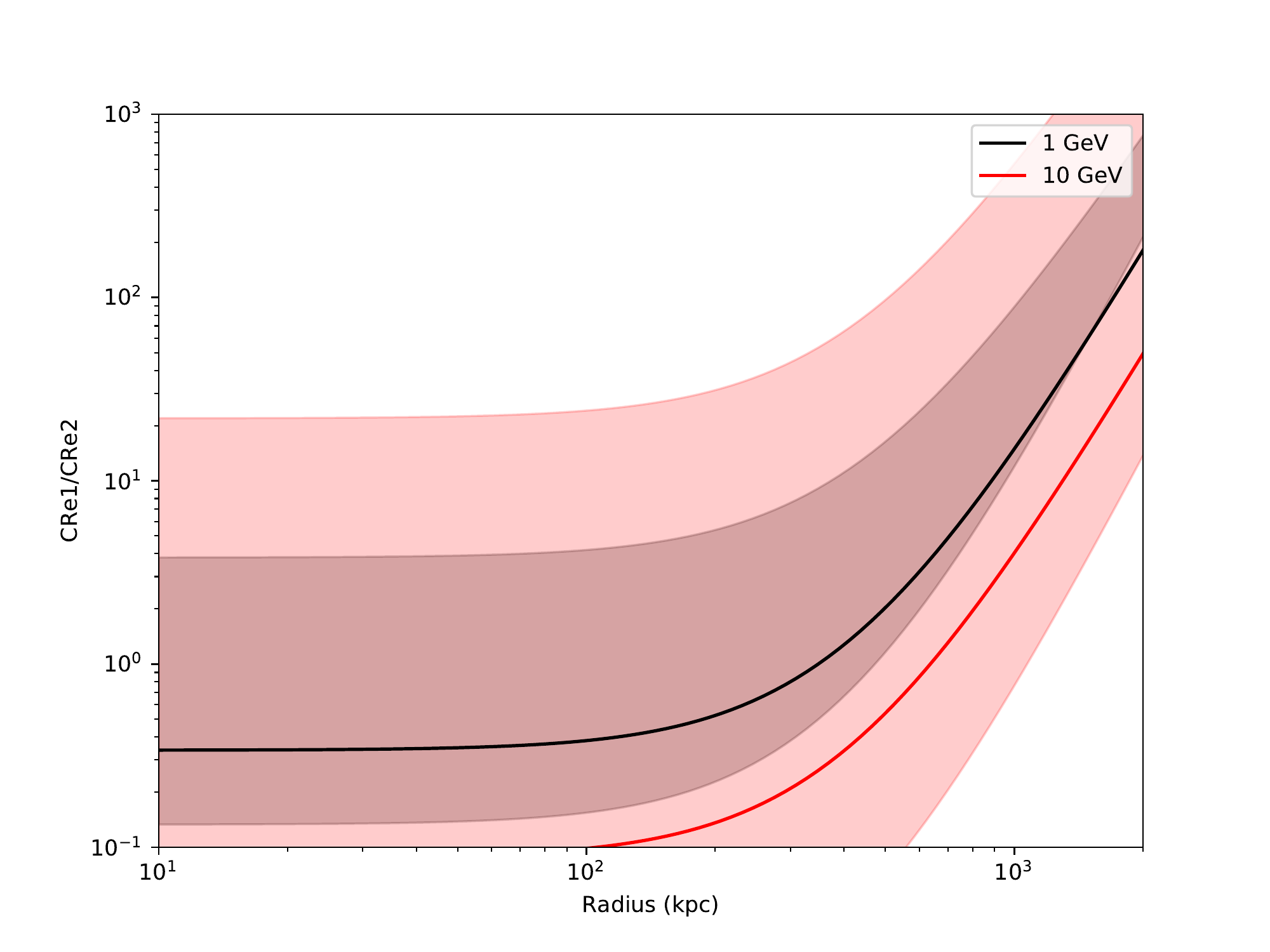}

	\rule{17cm}{0.01cm}
	
	\rule{17cm}{0.01cm}
	\put(-300,2){\footnotesize CRe$_2$ + reacceleration interpretation}
	
	\includegraphics[width=0.24\textwidth]{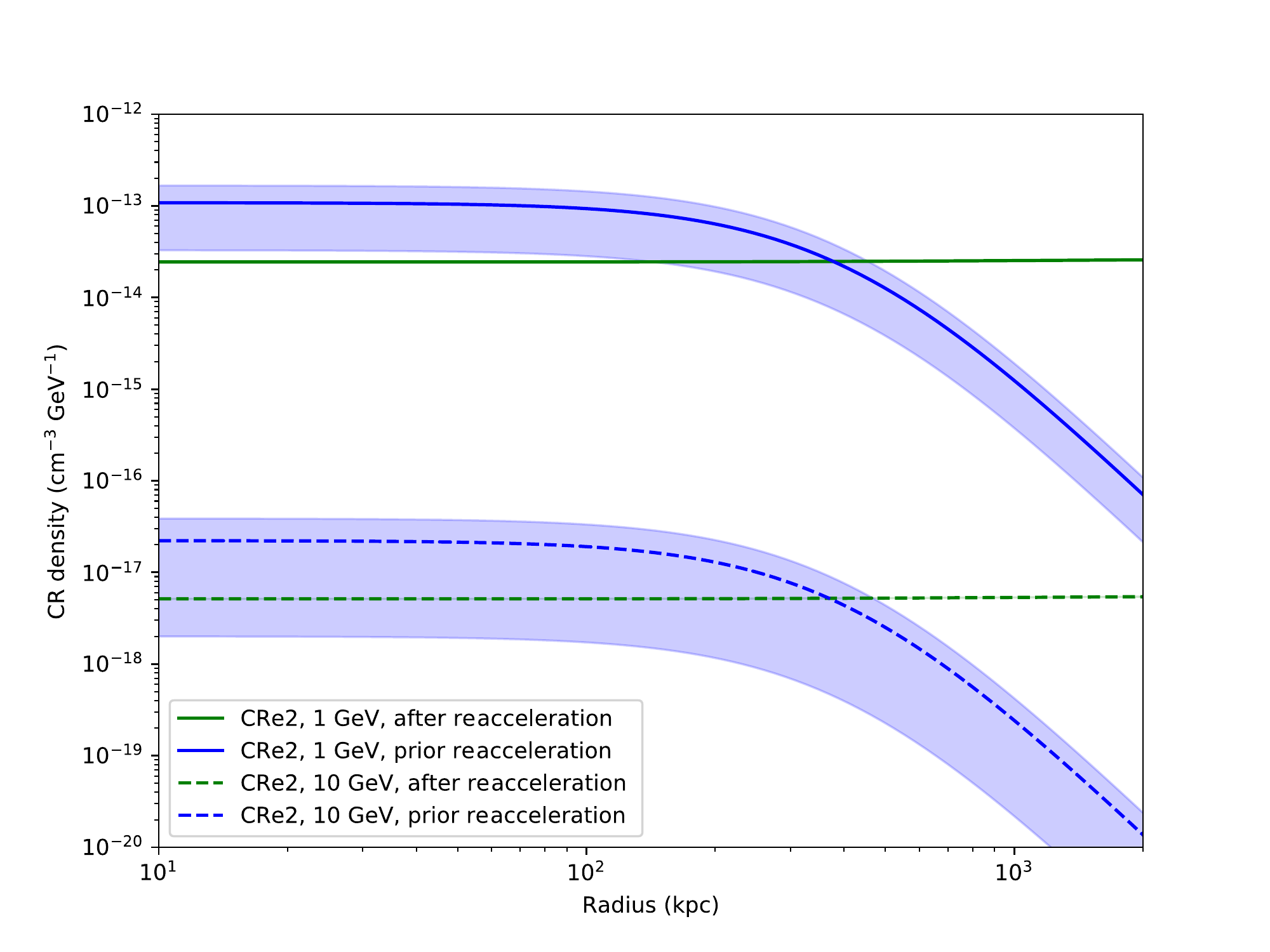}
	\includegraphics[width=0.24\textwidth]{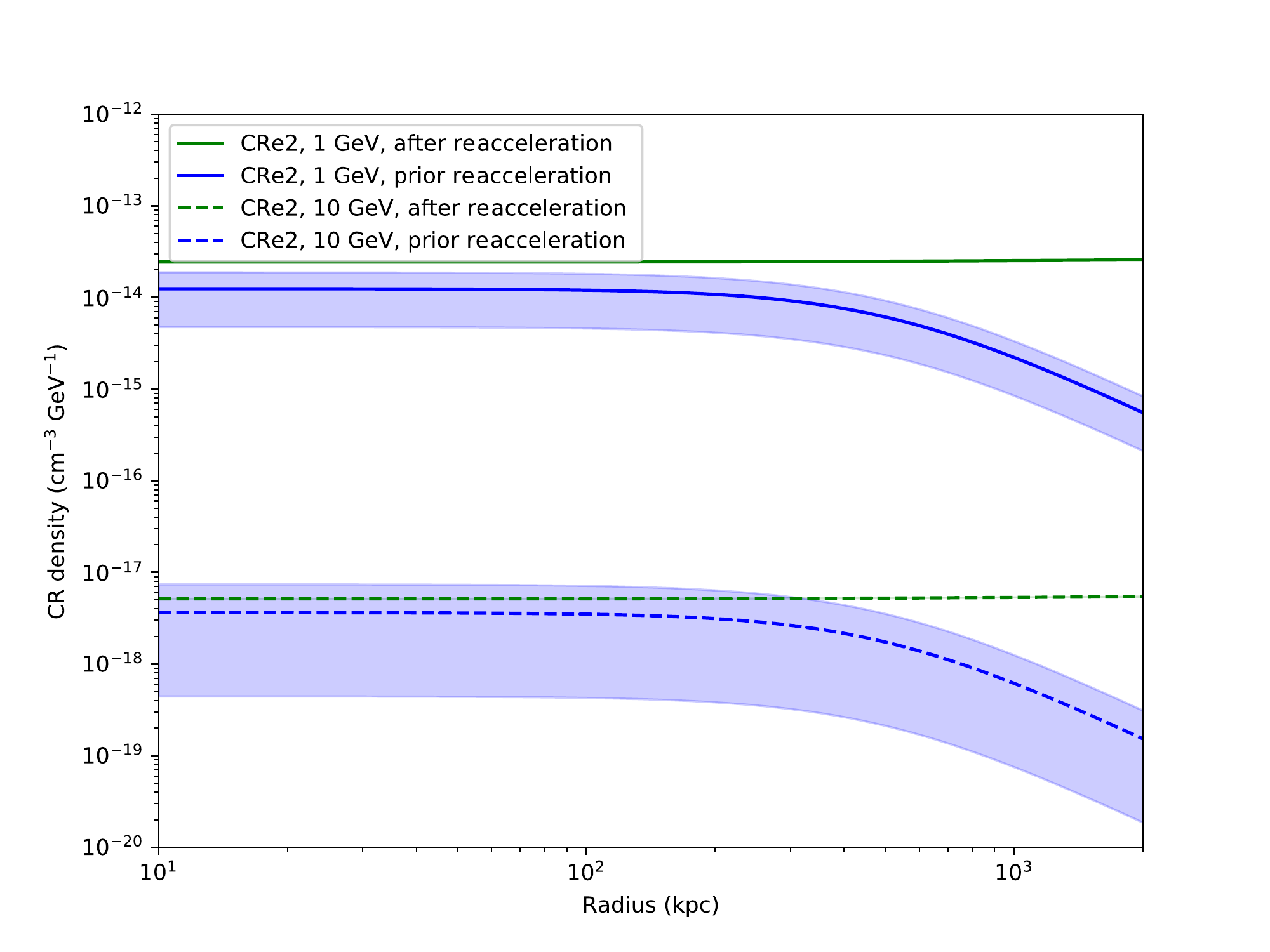}
	\includegraphics[width=0.24\textwidth]{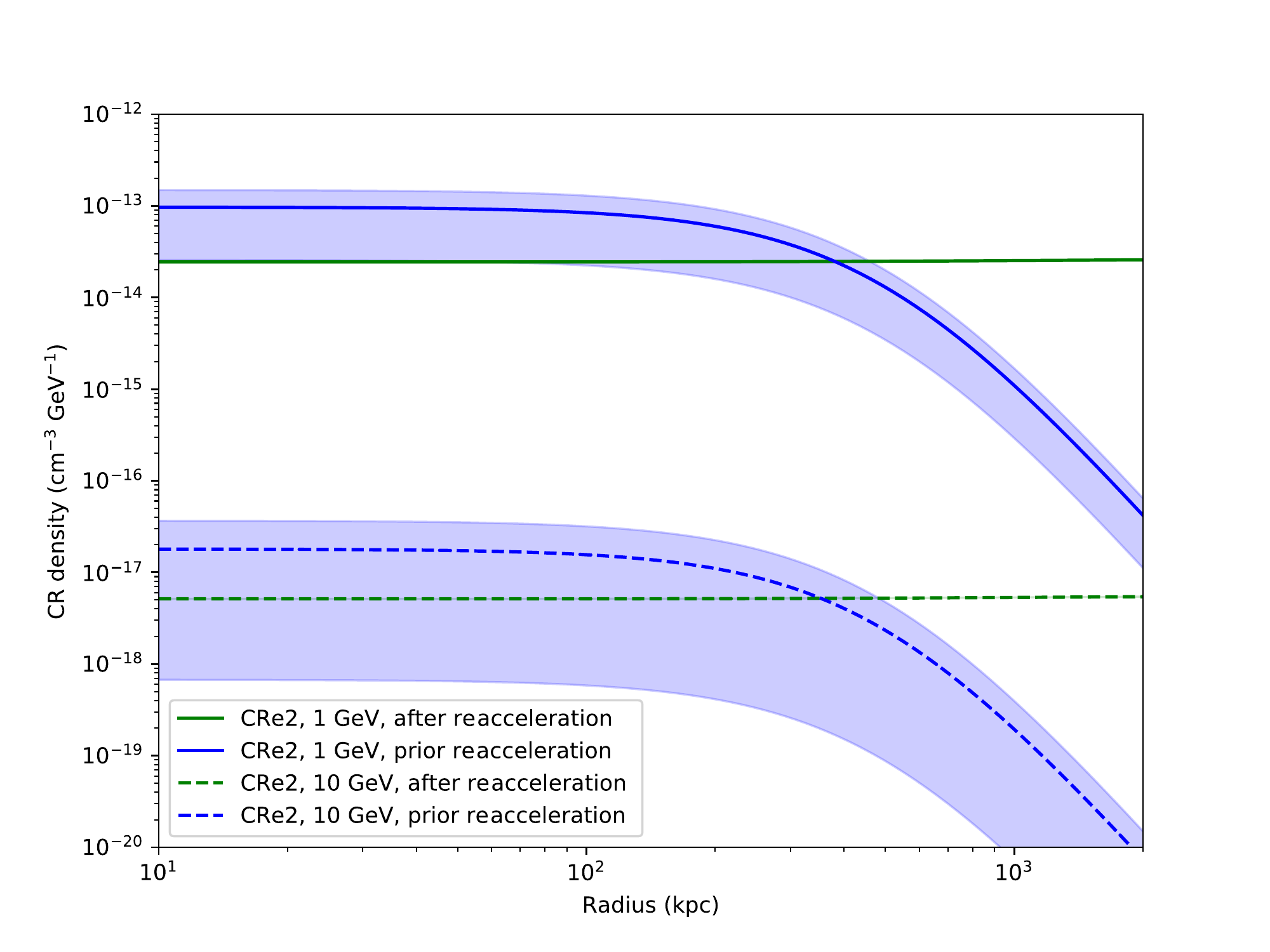}
	\includegraphics[width=0.24\textwidth]{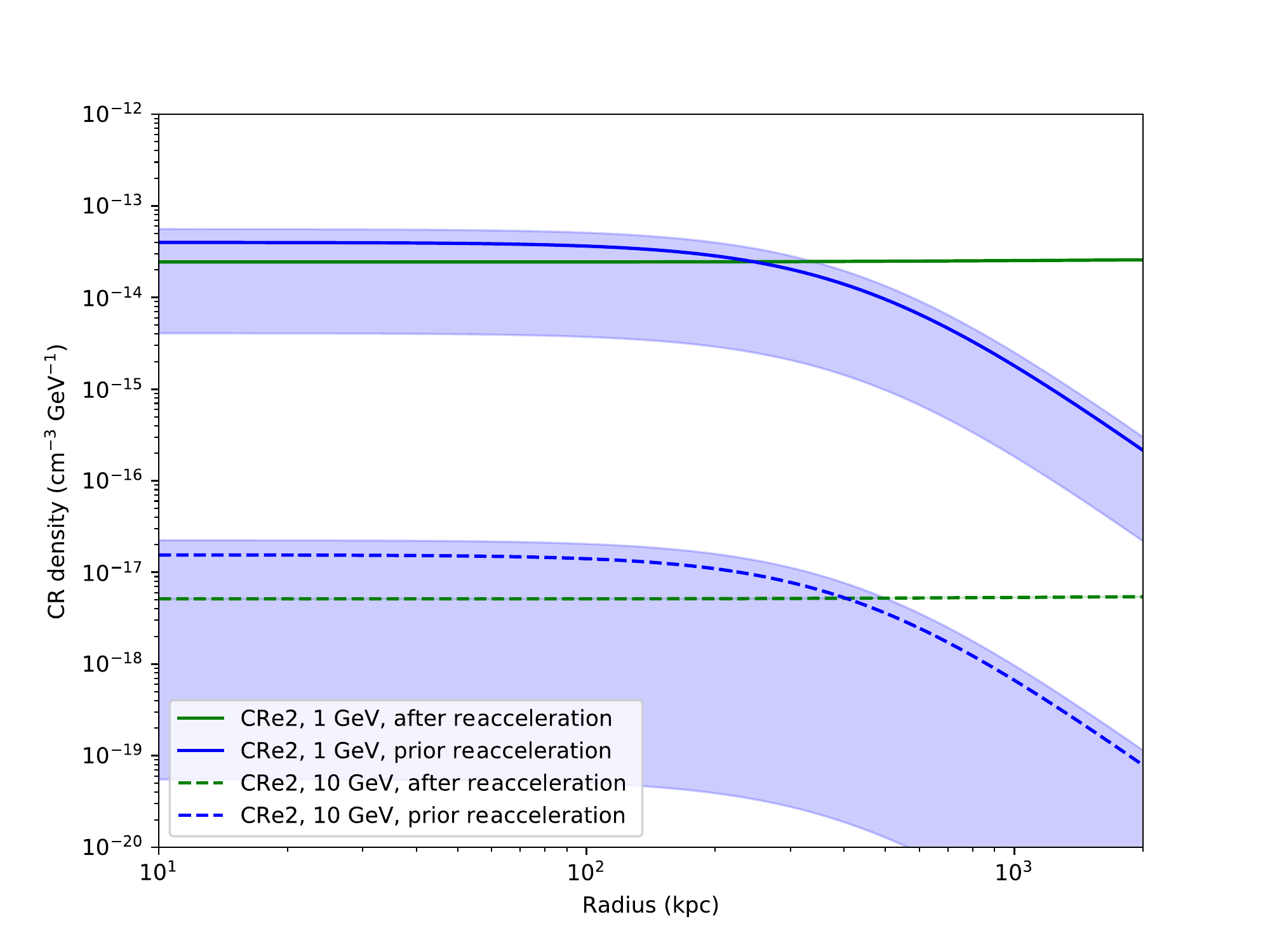}
	\includegraphics[width=0.24\textwidth]{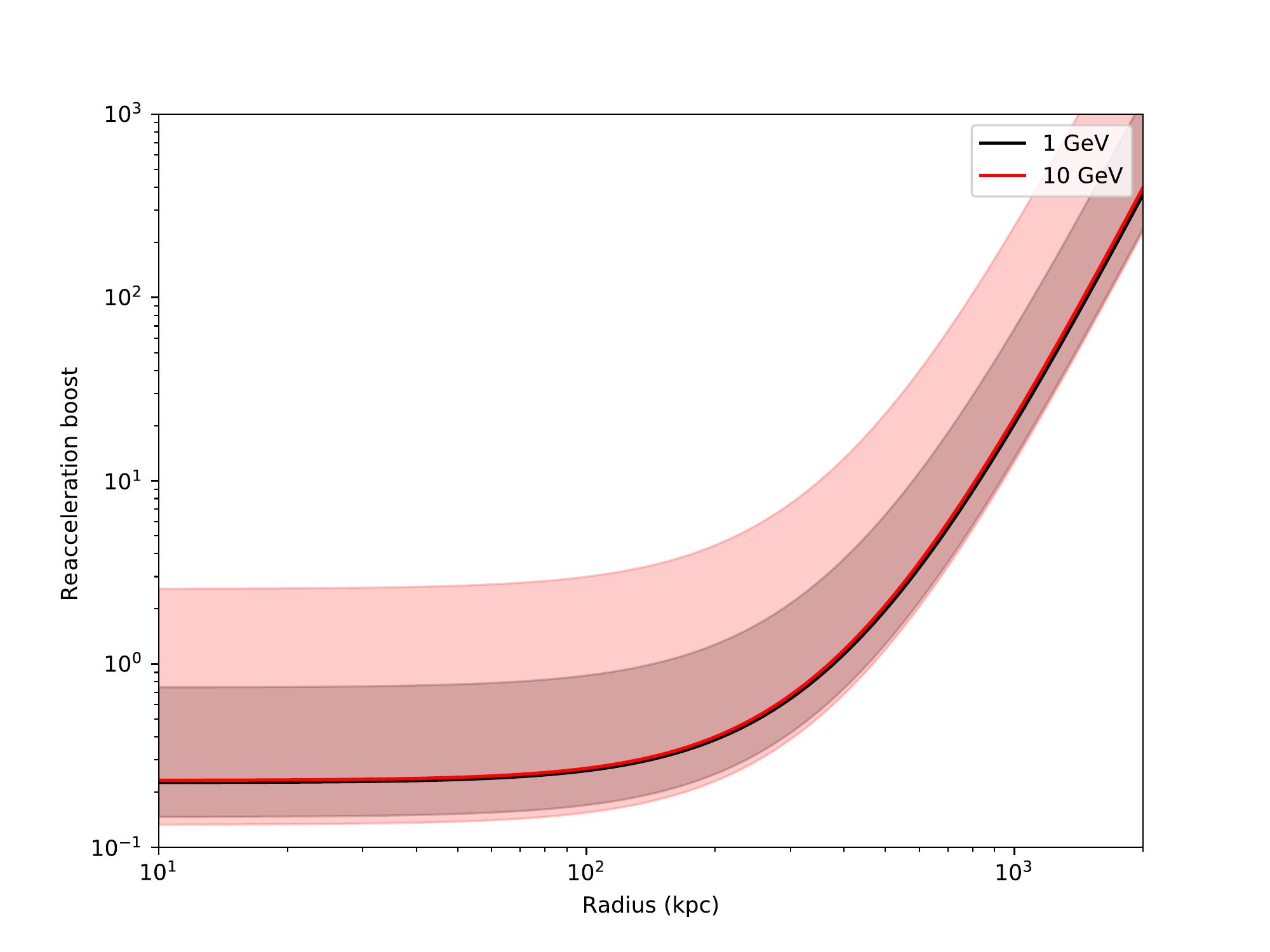}
	\includegraphics[width=0.24\textwidth]{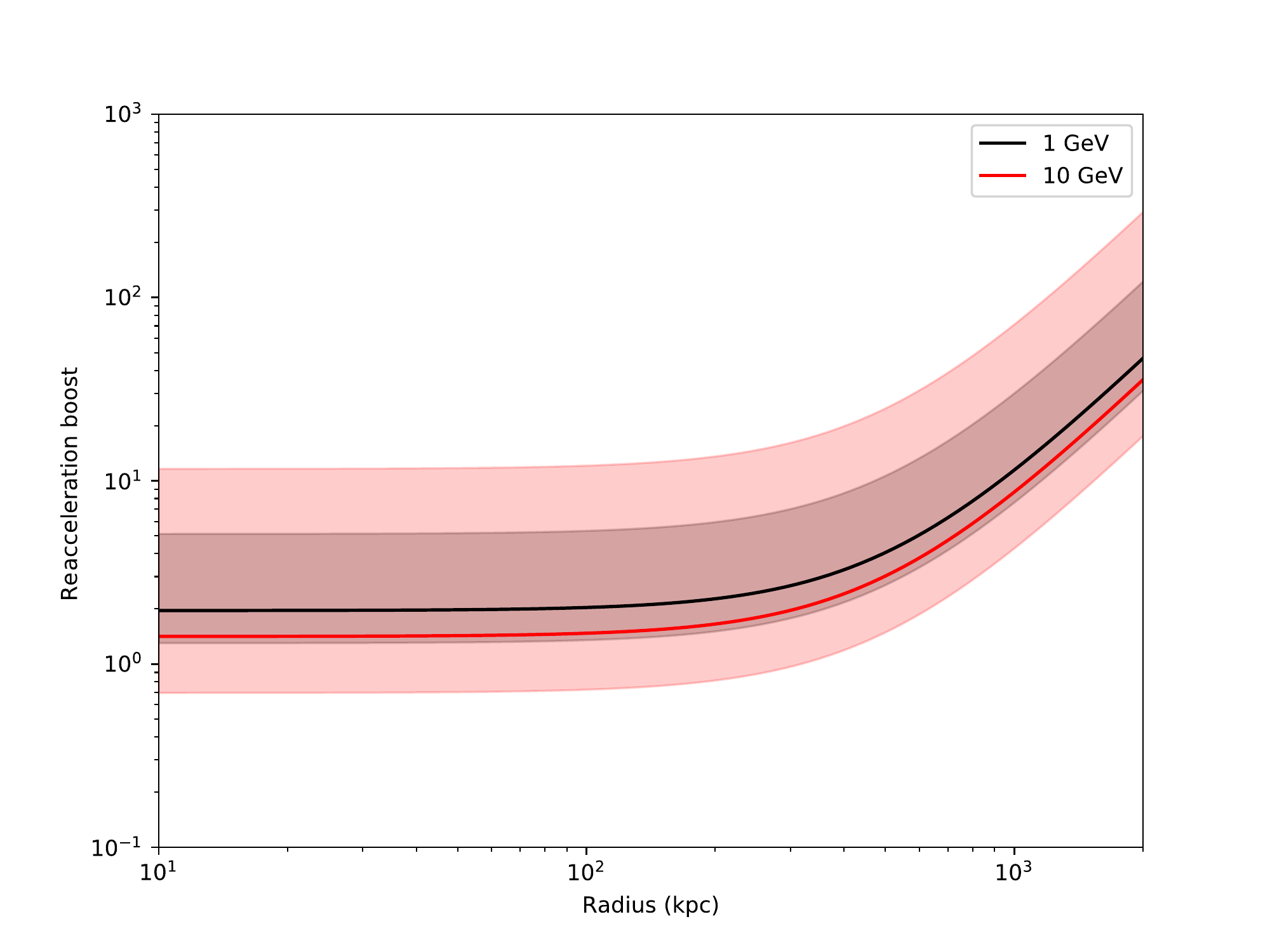}
	\includegraphics[width=0.24\textwidth]{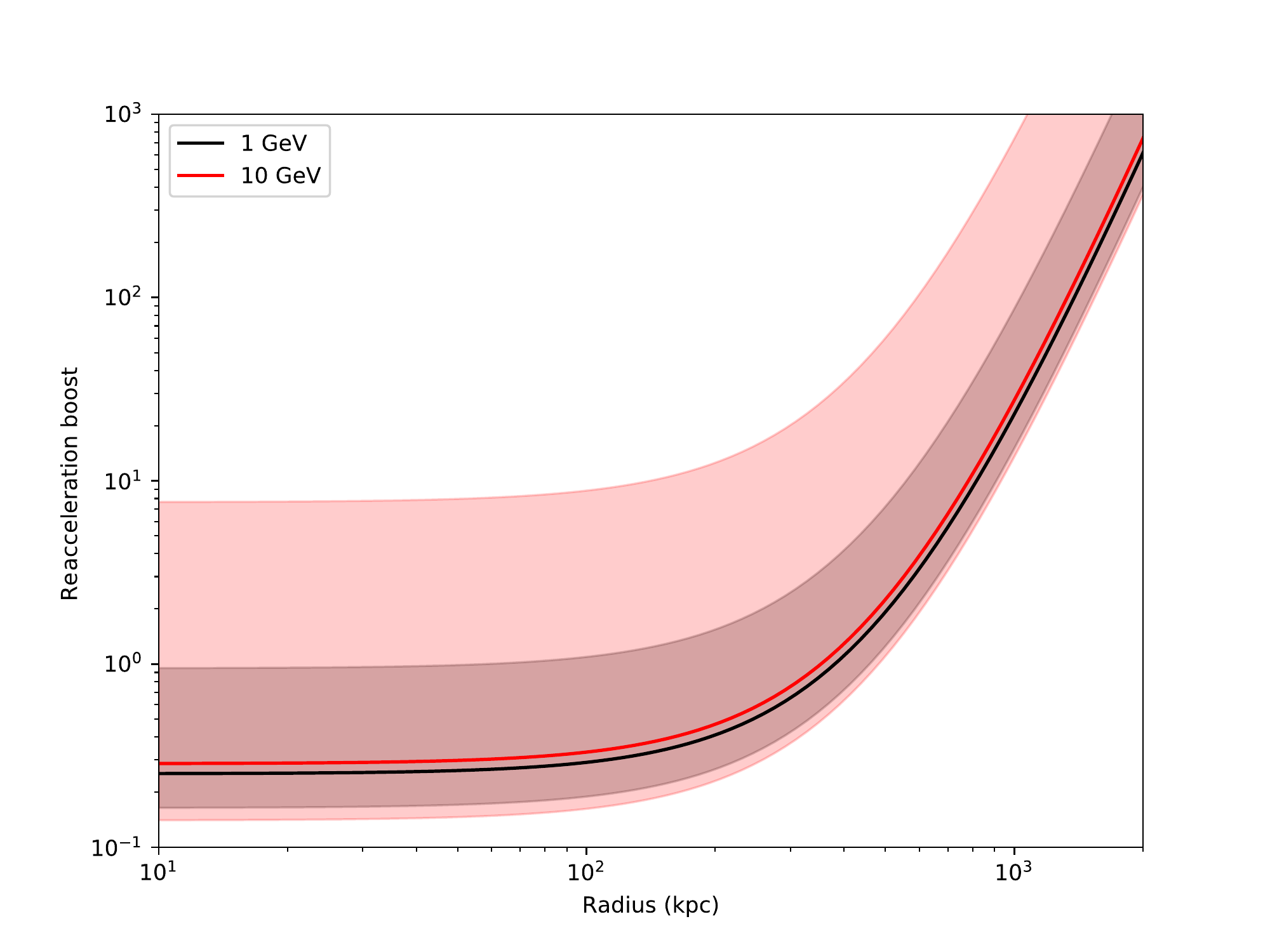}
	\includegraphics[width=0.24\textwidth]{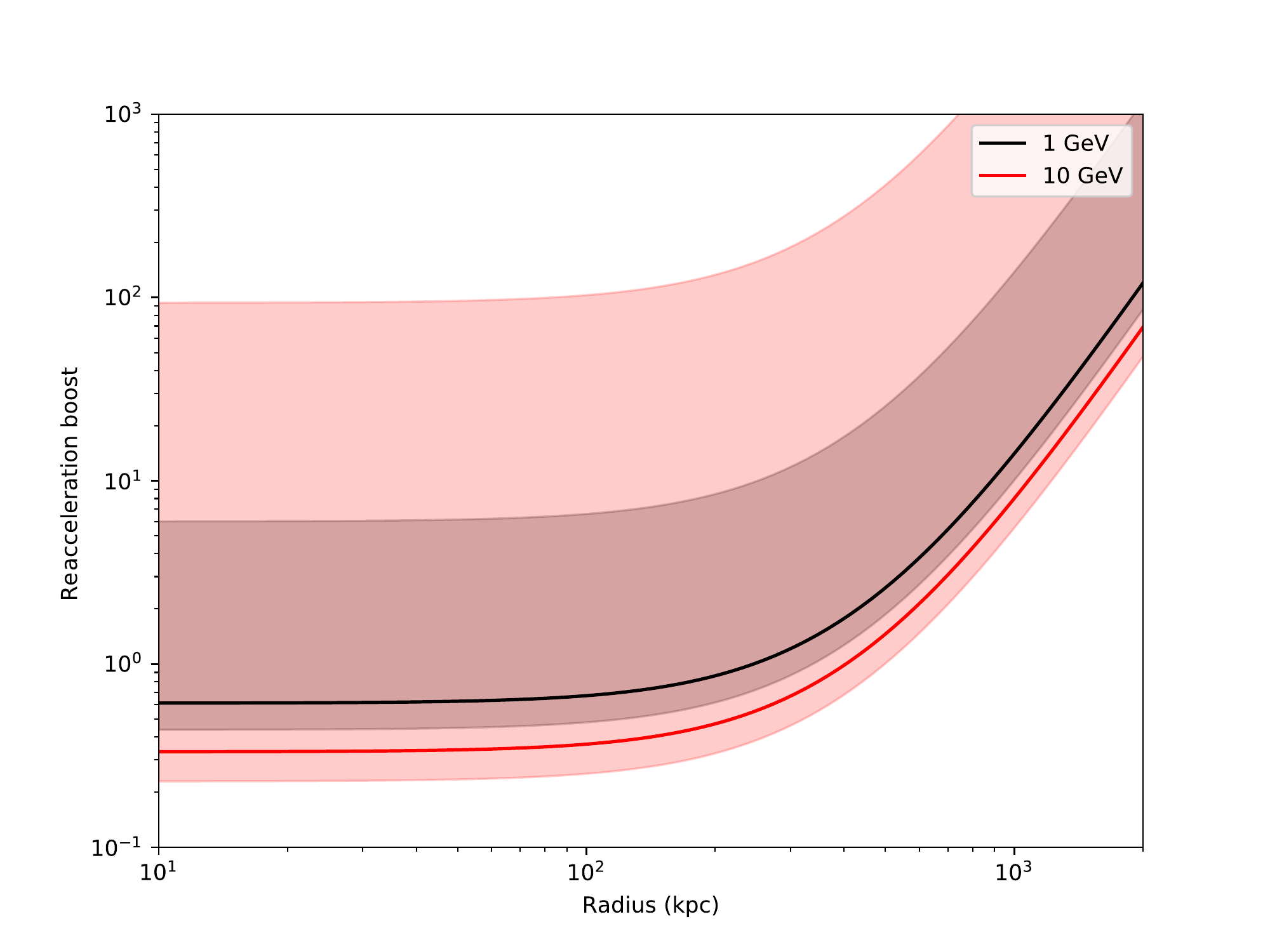}

	\caption{\small Same as Figure~\ref{fig:radio_implication_observable} (rows 1, 2, 3, and 4) and Figure~\ref{fig:radio_implication_observable2} (rows 6 and 7) for the compact model (first column), the flat model (second column), isobaric (third column) and the extended model + point source (fourth column).}
	\label{fig:radio_implication_observable2_app1}
\end{figure*}

This appendix provides the constraints on the CRe populations in the case of the alternative models considered in the case of the spectral modeling of the CRe$_1$ in Figure~\ref{fig:radio_implication_observable2_app2}. Similarly, Figure~\ref{fig:radio_implication_observable2_app1} gives the results for the alternative spatial models for the CRp population. The results are also provided in the case of including 4FGL~J1256.9+2736 in the \textit{Fermi}-LAT sky model (i.e., scenario 3).

\end{document}